%% file: main_expanded.tex
\documentclass[aps,rmp,reprint,longbibliography,unsortedaddress]{revtex4-1} 
\usepackage{bm}
\usepackage{amsmath,amssymb,amsfonts}
\usepackage{dsfont}
\usepackage{tabularx}
\usepackage{graphicx}
\usepackage{multirow}
\usepackage{dcolumn}
\usepackage{color}
\usepackage{mhchem}
\usepackage{tikz}
\usetikzlibrary{patterns,calc,arrows,positioning,decorations.text}
\usepackage{pgfplots}
\pgfplotsset{compat=1.11}
\usepackage{xargs}
\usepackage{varwidth}
\usepackage{xspace}
\usepackage{comment}
\usepackage{tikz-feynman}
\usepackage[normalem]{ulem}
\tikzfeynmanset{compat=1.0.0}

\usepackage{hyperref}
\hypersetup{colorlinks,breaklinks,
            linkcolor=blue,urlcolor=blue,
            anchorcolor=blue,citecolor=blue}

\clearpage{}
\newcommand{\senexp}{\ensuremath{\mathcal{E}}}
\newcommand{\senbkg}{\ensuremath{\mathcal{B}}}

\newcommand{\ctsper} {\ensuremath{\text{events}/(\text{keV}\cdotp\text{kg}\cdotp\text{yr})}}

\newcommand{\molyr} {\ensuremath{\text{mol}\cdotp\text{yr}}}
\newcommand{\evmolyr} {\ensuremath{\text{events}/(\text{mol}\cdotp\text{yr})}}

\newcommand{\nubb}   {\ensuremath{0\nu\beta\beta}}

\newcommand{\nunubb} {\ensuremath{2\nu\beta\beta}}
\newcommand{\nnbb}   {\nunubb}

\newcommand{\mne}    {\ensuremath{m_{\beta}}}
\newcommand{\mbb}    {\ensuremath{m_{\beta\beta}}}

\newcommand{\bb}     {\ensuremath{\beta\beta}}

\newcommand{\A}     {\ensuremath{\alpha}}
\newcommand{\B}     {\ensuremath{\beta}}
\newcommand{\G}     {\ensuremath{\gamma}}

\newcommand{\Qbb}{\ensuremath{Q_{\beta\beta}}}
\newcommand{\Tnubb}   {\ensuremath{T_{1/2}^{\nubb}}}
\newcommand{\Tnnbb}   {\ensuremath{T_{1/2}^{\nnbb}}}
\newcommand{\Thl}     {\ensuremath{T_{1/2}}}

\newcommand{\isot}[2]{\ensuremath{^{\text{#2}}}\text{#1}} 
\newcommand{\Ca}{$^{48}$Ca}
\newcommand{\Ti}{$^{48}$Ti}
\newcommand{\Ge}{$^{76}$Ge}
\newcommand{\Se}{$^{82}$Se}
\newcommand{\Mo}{$^{100}$Mo}
\newcommand{\Te}{$^{130}$Te}
\newcommand{\Xe}{$^{136}$Xe}
\newcommand{\Cd}{$^{116}$Cd}
\newcommand{\Nd}{$^{150}$Nd}

\newcommand{\Zr}{$^{96}$Zr}
\newcommand{\natTeoo}{$^{\text{nat}}$TeO$_2$}
\newcommand{\enrLMO}{Li$_2$$^{\text{enr}}$MoO$_4$}
\newcommand{\enrZS}{Zn$^{\text{enr}}$Se}
\newcommand{\Teoo}{TeO$_2$}
\newcommand{\LMO}{Li$_2$MoO$_4$}
\newcommand{\ZS}{ZnSe}
\newcommand{\U}{$^{238}$U}
\newcommand{\Th}{$^{232}$Th}

\newcommand{\Rn}{$^{222}$Rn}

\newcommand{\an}{\ensuremath{(\alpha,n)}}

\newif\ifcomments

\newcommand{\jd}[1] {\ifcomments{\color{red}[JD: #1]}\else{}\fi}

\newcommand{\smallspace}{\rule{0pt}{10pt}}
\newcolumntype{C}[1]{>{\centering}m{#1}}
\newcolumntype{R}[1]{>{\raggedleft}m{#1}}
\clearpage{}

\begin{document}
\title{Toward the discovery of matter creation with neutrinoless double-beta decay}

\author{Matteo Agostini}
\email{matteo.agostini@ucl.ac.uk}
\affiliation{Department of Physics and Astronomy, University College London, Gower Street, London WC1E 6BT, UK}

\author{Giovanni Benato}
\email{giovanni.benato@lngs.infn.it}
\affiliation{INFN, Laboratori Nazionali del Gran Sasso, 67100 Assergi, L'Aquila, Italy}
\altaffiliation{Now at Gran Sasso Science Institute, 67100 L'Aquila, Italy}

\author{Jason A. Detwiler}
\email{jasondet@uw.edu}
\affiliation{Center for Experimental Nuclear Physics and Astrophysics and Department of Physics, University of Washington, Seattle, WA 98115, USA}

\author{Javier Men\'{e}ndez}
\email{menendez@fqa.ub.edu}
\affiliation{Department  of  Quantum  Physics  and  Astrophysics  and  Institute  of Cosmos Sciences, University of Barcelona, 08028 Barcelona, Spain}

\author{Francesco Vissani}
\email{vissani@lngs.infn.it}
\affiliation{INFN, Laboratori Nazionali del Gran Sasso, 67100 Assergi, L'Aquila, Italy}

\date{\today{}}

\begin{abstract} 
The discovery of neutrinoless double-beta decay could soon be within reach.
This hypothetical ultra-rare nuclear decay offers a privileged portal to physics beyond the Standard Model of particle physics. Its observation would constitute the discovery of a matter-creating process, corroborating leading theories of why the universe contains more matter than antimatter, and how forces unify at high energy scales. It would also prove that neutrinos and anti-neutrinos are not two distinct particles, but can transform into each other, with their mass described by a unique mechanism conceived by Majorana.
The recognition that neutrinos are not massless necessitates an explanation and has boosted interest in neutrinoless double-beta decay. 
The field stands now at a turning point. 
A new round of experiments is currently being prepared for the next decade to cover an important region of parameter space. 
In parallel, advances in nuclear theory are laying the groundwork to connect the nuclear decay with the underlying new physics. 
Meanwhile, the particle theory landscape continues to find new motivations for neutrinos to be their own antiparticle.
This review brings together the experimental, nuclear theory, and particle theory aspects connected to neutrinoless double-beta decay, to explore the path toward --- and beyond --- its discovery.
\end{abstract}

\maketitle

~\\
\cleardoublepage
\tableofcontents

\section{Introduction}\label{sec:intro}
What is ``matter''? Ever since the attempts of the ancient philosophers to conceive matter in terms of a few elements, and the even more radical attempts of the early atomists, humankind has been trying to determine what the ultimate building blocks of nature are and whether they are physically indivisible.  Lavoisier's idea that ``nothing is lost, nothing is created, everything is transformed'' is deeply rooted in our modern way of thinking and has taken a very particular form in the context of the Standard Model of particle physics. Nowadays, we assume that energy can transform into balanced quantities of matter and antimatter, and vice versa, that matter and antimatter can annihilate to produce energy, according to immutable rules. Indeed, in all physical processes observed so far, the creation or destruction of matter particles is always compensated by the destruction or creation of antimatter particles. More precisely, the differences between the number of baryons and antibaryons, and leptons and antileptons, are immutable quantities, i.e., quantum numbers of our canonical field theory.

We now believe that our universe originated in a Big Bang, and that at the beginning of time it was extremely hot, with energy converting into matter-antimatter and vice versa.  Yet, the universe in which we live today contains almost exclusively atoms and not anti-atoms.  This observation creates a strong theoretical appeal for hypothetical ``matter-creating'' or ``antimatter-destroying'' processes, i.e., phenomena that can break the  matter-antimatter balance, and dynamically explain the asymmetry of our universe.  At present, the most promising phenomena of this type for observation in the laboratory are the destruction of a proton --- which could decay by changing the number of baryons while respecting energy conservation --- and the creation of electrons in nuclear decays --- which would change the number of leptons.

The quest to observe the creation of electrons
is being pursued vigorously
in the form of searches for a nuclear decay where the atomic number $Z$ increases by two units
while the nucleon number $A$ remains constant:
$(A, Z) \to (A, Z + 2) + 2e$.
This is commonly known as ``neutrinoless \bb\ decay'' (\nubb\ decay).
Here, the creation of electrons can be enabled by the
``transmutation'' of neutrinos into antineutrinos, which is possible if the neutrino's mass is described by a unique mechanism conceived by Majorana.
Thus the matter-antimatter imbalance and neutrino masses
could have a common origin.

A symmetry between neutrinos and antineutrinos was postulated by Majorana and
further discussed by Racah in 1937.  This led Furry to propose the existence of \nubb\ decay in 1939,
building on Goeppert-Mayer's ideas on two-neutrino double-beta (\nnbb) decay transitions.
Pioneering searches for \nubb\ decay started in the 40s
using time-coincidence counting techniques or visual detection of tracks in
cloud chambers and photographic emulsions. 
Since then, experiments have continued steadily, 
leading to increasingly stronger constraints which at present reach half-lives 
exceeding $10^{26}$\,years.
This means that a nucleus will take on average more than a million billion
times the age of the universe before undergoing \nubb\ decay.
To surpass this sensitivity, experiments must monitor thousands of
moles of atoms for years,
and have the capability to detect the \nubb\ decay of a single one of them.
The rarity of the sought-after signal sets extremely strict requirements
for eliminating other processes that could mimic the decay.

We face a pivotal time for \nubb-decay searches.
The discovery of neutrino mass at the turn of the century brought to the
foreground the question of whether that mass could be of the peculiar
type proposed by Majorana.
This invigorated the effort in \nubb-decay experiments around the world,
covering a variety of \bb-decay nuclei and detection techniques.
These efforts have set the stage for the selection of the most promising methods
for further investment.
The community is currently proposing next-generation experiments as
part of a global enterprise, with the goal for the next decade of extending the
half-life sensitivity in multiple nuclei by two orders
of magnitude beyond the current limits. This could lead to an observation of the transition.

Meanwhile, the theoretical landscape continues to evolve, and has also been deeply 
affected by the neutrino mass discovery.
Most leading theoretical models suggest that neutrinos have a Majorana mass
responsible for lepton number violation, and hence predict \nubb\ decay.  
In fact, multiple lepton-number-violating mechanisms that lead to \nubb\ decay have been
identified, so that there is no definitive prediction of its half-life.
Nevertheless, running experiments are progressively probing the parameter space
available to theoretical scenarios. 
In particular, if the decay is mediated by the exchange of light neutrinos, all anticipated orderings of the neutrino masses are being tested.

A key role in \nubb-decay searches is also played by nuclear theory, which links
the experimentally measurable \nubb-decay half-life with the underlying particle physics
through the modeling of the nuclear behavior.
Sophisticated many-body calculations are required to evaluate the impact
of the structure of the initial and final nuclei on the decay rate.
In addition, the nuclear operators driving the decay need to be
consistent with the treatment of the initial and final nuclei.
The nuclear theory community is placing significant analytical and computational efforts
with the ultimate goal of converting experimental
measurements into constraints on the underlying particle physics mechanisms.
In the opposite direction, only through nuclear theory can we
predict decay half-life values based on selected theoretical scenarios.

In recent years, several review articles have discussed \nubb\ decay,
witnessing the vivid interest of the scientific community in this topic.
Each work emphasizes one or more relevant aspects, such as the experimental part
\cite{Avignone:2007fu,Elliott:2012sp,Giuliani:2012zu,Gomez-Cadenas:2011oep,Schwingenheuer:2012zs,Cremonesi:2013vla},
the nuclear physics \cite{Vergados:2012xy,Vogel:2012ja,Engel17,Ejiri19,Yao21b},
the connection with neutrino masses \cite{Petcov:2013poa,Bilenky:2014uka,DellOro:2016tmg},
other particle physics mechanisms \cite{Rodejohann:2011mu,Rodejohann:2012xd,Deppisch:2012nb,Pas:2015eia,deGouvea13},
or a combination of the above \cite{Dolinski:2019nrj}.
\textcite{Elliott15} discusses Majorana fermions in a broader context.
In the present work, we mostly focus on the first three aspects,
motivated by the intention to follow the theoretical ideas
that describe the most plausible expectations for experiments.
We bring together theory and experiment to give a
comprehensive overview of the field, and explore the path towards a convincing
future discovery and elucidation of the mechanism mediating the decay.

We start our journey in Sec.~\ref{sec:his} with an overview of the history and role of \nubb\ decay. In Sec.~\ref{sec:par}, we revisit the theoretical
motivations to search for this matter-creating process, which 
has a special role in testing the 
foundations of nature that modern theory formulates in terms of 
symmetry principles. 
The reference quantum field theory of particles physics --- i.e., the Standard Model --- 
predicts four global symmetries, with corresponding conserved quantities given by the difference between the
number of baryons and leptons ($B-L$) and the number of leptons of each
flavor ($L_e-L_{\mu}$, $L_{\mu}-L_{\tau}$, $L_{e}-L_{\tau}$).
The observation of neutrino flavor oscillation violates the last three,
forcing us to extend the theory to account for these new phenomena. 
The only residual global symmetry is that related to $B-L$ conservation, as discussed in Sec.~\ref{sec:par:symmetries}.
Testing this symmetry is thus of paramount importance, and \nubb\ decay is its most sensitive direct probe.
Further interest in \nubb\ decay comes from the fact that the transition is plausibly due
to new physics --- beyond the Standard Model --- at an ultra-high energy scale
beyond the reach of current accelerators.
In Sec.~\ref{sec:par:LMJ} and \ref{sec:par:LNV}, we review the mechanisms that give rise to \nubb\
decay, how their contributions can be cast in terms of effective field
theory operators, and what we can learn about them.
The lowest dimension operator --- i.e., the dimension 5 Weinberg operator --- describes 
Majorana masses of the light neutrinos and is one of the better-motivated
mechanisms for \nubb\ decay.
If this is the dominant contribution to the
transition, the half-life of the decay is connected to the neutrino properties
and the origin of neutrino masses (Sec.~\ref{sec:par:mbb}).
This creates an exciting interplay between \nubb-decay searches, neutrino oscillation
experiments, neutrino mass measurements, and cosmology. 
It also implies that the search for \nubb\ decay is a well-defined
scientific target that can be explored in the next years. Finally, in Sec.~\ref{sec:par:cosmo}, we explore the connection between \nubb\ decay and the excess of baryons over antibaryons in the universe.

Section~\ref{sec:nt} reviews recent advances in nuclear theory.
First, Sec.~\ref{sec:eft} introduces 
an effective field theory framework based on the symmetries of the fundamental
theory governing nuclei, i.e., quantum chromodynamics. 
Contributions from different \nubb-decay mechanisms are
organized in terms of effective operators through a master formula
that provides a way to estimate the energy scales constrained by \nubb-decay searches.
Section~\ref{sec:nme_expressions} describes how 
each \nubb-decay mechanism involves at least one nuclear matrix element (NME), as
the decay occurs in a complex many-body nuclear system. We highlight the impact of the recently proposed ``short-range operator'' --- unfortunately with uncertain
coupling --- that could affect significantly the rate of the decay.
In Sec.~\ref{sec:NMEs},
we discuss progress on NME calculations obtained with several
many-body approaches, including recent first-principles studies.
In addition, we discuss NME uncertainties, and
place special importance on recent advances in the understanding of ``$g_A$
quenching'' (Sec.~\ref{sec:quenching}), one of the main sources of theoretical uncertainty.
In single-$\beta$ decay the decades-old puzzle seems mostly solved
thanks to previously-neglected many-body correlations and two-nucleon currents.
However, an extension to higher momentum transfer is needed to estimate the impact on \nubb\ decay.
Finally, Sec.~\ref{sec:NME_tests} presents related nuclear properties and reactions, the tests they place on nuclear theory calculations, and the insights they may provide on \nubb\ decay.

Section~\ref{sec:exp} reviews the experimental aspects of \nubb-decay
searches. 
This decay can be observed in a variety of nuclei, each of them
characterized by specific properties such as Q-value and natural abundance, 
as discussed in Sec.~\ref{sec:exp:iso}.  
Since each isotope enables different detection techniques, the field is very diverse.
We review the main detection principles in
Sec.~\ref{sec:exp:detection}.
Current sensitivities can only be improved with an increase of the active
isotope mass and a concurrent background reduction to unprecedented levels.  
Section~\ref{sec:exp:background}
describes the background sources faced by the various experiments, 
and lists possible new backgrounds arising in future highly sensitive searches.
The available techniques to discriminate a possible \nubb\ decay from background are covered in Sec.~\ref{sec:sec:activebkgsuppression}.
We discuss in Sec.~\ref{sec:exp:stat} the statistical techniques used to extract
the sought-after signal and how two effective parameters ---
the effective background and effective exposure --- can essentially describe the sensitivity of an
experiment.

Finally, in Sec.~\ref{sec:prj}, we present a consistent comparison of recent and future experiments, including projects at the research and development phase.  We describe each experiment's distinctive features, planned developments, and strategies to reach the desired goal sensitivity.

Several questions will be of key importance for \nubb-decay searches
in the upcoming decade.
Are we ready for a discovery?  When can we expect it, and what
will we be able to learn from an observation? 
How will advances in other physics areas influence the \nubb-decay community? 
In Sec.~\ref{sec:dis}, we bring together our expectations for
particle theory, nuclear theory, and experiments in order to
address these questions, and to explore the possible path towards --- and
beyond --- a future discovery of \nubb\ decay.

We hope for this review to become a useful reference for both \nubb-decay experts and nonexperts. With this challenging goal in mind, we alternated introductory and technical sections. We recommend the nonexpert reader to focus on Secs.~\ref{sec:par:symmetries}, \ref{sec:par:LMJ}, \ref{sec:par:mbb} and \ref{sec:par:cosmo} for an overview of the particle theory context, on Secs.~\ref{sec:eft}, \ref{sec:status} and \ref{sec:quenching} for insights on nuclear theory aspects, and on Secs.~\ref{sec:exp} and \ref{sec:prj:land} for an introduction to the experimental techniques and experiments. Experts might also be interested in these sections, as we discuss most topics from a modern point of view, which differs in many aspects from past review works. We also recommend to both experts and nonexperts Sec.~\ref{sec:his}, which gives a historical context for the present-day effort, and Sec.~\ref{sec:dis}, which aims to connect all the dots, bridging theory and experiment, particle and nuclear physics, as well as cosmology and other scientific areas, pointing to a pathway forward toward the discovery of \nubb\ decay and beyond.
 \section{Historical landscape}
\label{sec:his}

In this section, we summarize the role of 
\nubb\ decay in the historical development of particle physics, focusing on its connection with the crucial milestones of neutrino physics, such as: the neutrino postulation (1930-1933); Majorana's hypothesis for the nature of the neutrino (1937);
the role of \nubb\ decay for the neutrino mass (1957-1958); 
neutrinos in gauge theories (1961-present);
and empirical information on the neutrino mass (1967-present).
We also cover the connection between \nubb\ decay and long-standing questions
regarding the basic ingredients of matter and fundamental Standard Model symmetries.
More details on the history of \nubb\ decay are discussed in \textcite{Barabash:2011mf}, \textcite{Tretyak:2011pg}, \textcite{DeBianchi:2018irt}, and \textcite{Vissani:2021gdw}.

The terminology $\alpha$, $\beta$ and $\gamma$ rays introduced by Rutherford at
the turn of the 20th century marked the recognition of new phenomena beyond atomic physics. 
The Bohr-Rutherford model of the atom \cite{Bohr:1913zba} was a milestone for the field, 
but could not and did not claim to explain these new phenomena.
Soon afterwards, \textcite{Harkins-1915,Harkins-1915-2}
inferred a model for the nuclei
describing them as composed of $^4$He, $^3$H and $^1$H nuclei, and
\textcite{RUTHERFORD-1920-nature,Rutherford-1920} discovered through $(\alpha,p)$ reactions that the hydrogen nucleus
was a fundamental component of other nuclei,
and named it the \emph{proton} after Prout's
\emph{protyle} \cite{prout1816correction}.
According to these models, a nucleus with atomic numer $Z$ and mass number $A$
would have been made of $A$ protons and $(A-Z)$ \emph{nuclear} or \emph{inner} electrons,
yielding a nuclear charge $Ze$.
This paradigm could explain the neutrality of atoms, the existence of isotopes
and also radioactivity,
but was still fundamentally non-relativistic,
assuming that particles ``are forever'', i.e., cannot be created or destroyed. 
Moreover, it could not predict the nuclear spin for some nuclei --- for instance $^{14}$N ---
and predicted a monochromatic \B\ radiation spectrum \cite{Ellis:1927opv,Meitner-1930}.

To overcome these problems, \textcite{Pauli:1930pc} proposed to add a new very light and neutral
particle to the nucleus, which was assumed to carry spin and energy.
Thus the neutrino was introduced, albeit 
in a non-relativistic model, similar to the earlier ones.

The discovery of the neutron in 1932--33 \cite{Chadwick:1932ma,Chadwick-1933} was
an important step forward in the formulation of the modern model of the nucleus \cite{Heisenberg-1932-I,Heisenberg-1932-II,Heisenberg-1933-III,Majorana-1933}.
Concurrently, quantum mechanics reached its full maturity, in particular thanks to the 
relativistic quantum theory of the electron \cite{Dirac-1928}.

All these phenomenological and theoretical aspects were merged in Fermi's theory of 
\B\ decay \cite{Fermi-1934}, which introduced the possibility of creation and destruction
of matter particles.
The success of Fermi's theory in describing the observed $\beta$-decay rates and spectra
convinced the scientific community of the existence of the neutrino
and triggered its experimental search.

Shortly thereafter, \textcite{Wick-1934} exploited Fermi's theory to explain $\beta^+$ decay
and electron capture, and
\textcite{PhysRev.61.97} proposed to measure the electron-capture nuclear recoil to
indirectly detect the neutrino.
Between the late thirties and the early fifties, several measurements demonstrated that \B\ decay and electron capture
are subject not only to missing energy, but also to an apparent momentum non-conservation,
thus pointing to the existence of the neutrino \cite{Leipunski-1936,PhysRev.53.789,PhysRev.56.232,PhysRev.61.692,PhysRev.86.976}.
The final confirmation arrived in 1956, with the detection of neutrinos in ``appearance mode''
through inverse $\beta^+$ decay ($\bar{\nu}+p \rightarrow n+e^+$) \cite{Reines:1953pu,Cowan:1992xc},
another process predicted by Fermi's theory.

Other milestones were achieved in those years. 
\textcite{Lee:1956qn} questioned the conservation of parity in weak interactions
and \textcite{Wu:1957my} observed its violation in \B\ decays.
Soon after,
\textcite{Landau:1957tp}, \textcite{Lee-1957}, and \textcite{Salam:1957st} independently came to the conclusion
that, {\em if the neutrino produced by weak interactions was massless,} 
it would have a fixed and opposite helicity compared to the antineutrino,
and parity violation in weak interactions would be maximal.
Experimental evidence in favor of the neutrino's fixed helicity \cite{Goldhaber:1958nb}
and the refinements of Fermi's theory in terms of a $(V-A)$ interaction \cite{Sudarshan:1958vf,Feynman:1958ty}
represented breakthroughs in our understanding of weak interactions.
Unfortunately, it implied that the expected rates of \nubb\ decay were at best much shorter
than originally predicted \cite{Furry:1939qr},
and strengthened the idea that neutrinos were 
massless up to the point that it became regarded as an established
fact. However, this paradigm did not block the discussion entirely;
in fact, the first discussion of \nubb\ decay based on the neutrino mass
hypothesis appears in 1960 \cite{Greuling-1960}.
The history is recounted in \textcite{Vissani:2021gdw}.

In the same decades, the understanding of $\beta$ decay and weak interaction
led to further considerations on the possibility of \emph{double}-$\beta$ decay
and its relevance in connection to the neutrino nature.
In 1935 \textcite{Goeppert-Mayer:1935uil} highlighted the possibility for an isotope to
``change into a more stable one by simultaneous emission of two electrons'',
with a process that would ``appear as the simultaneous occurrence of two transitions,
each of which does not fulfill the law of conservation of energy separately''. She
also used Fermi's theory of $\beta$ decay to predict that such a transition, namely \nnbb\ decay,
would have half-life values exceeding $10^{17}$\,yr.

Two fundamental milestones followed.
\textcite{Majorana:1937vz} introduced an alternative to Dirac's theory where neutral particles can be their own antiparticles,
and explicitly mentioned its possible application to neutrinos,
saying that ``such theory can obviously be modified so that the $\beta$ emission,
both positive and negative, is always accompanied by the emission of a neutrino''.
Shortly thereafter, \textcite{Racah:1937qq} showed that postulating a symmetry between particles and antiparticles in addition to relativistic invariance
leads to a new version of Fermi's theory of \B\ decay,
and demonstrated that the assumption that neutrinos and antineutrino are the same particle
leads directly to Majorana's formalism.
Racah also pointed out that Majorana's theory could not apply to neutrons because
of their non-zero magnetic moment
and because it would imply that a free neutron could undergo both $\beta^+$ and $\beta^-$ decay,
contradicting experiment.
Racah also highlighted the possibility of neutrinos (antineutrinos) inducing
inverse $\beta^+$ ($\beta^-$) decay if they were Majorana particles.

\textcite{Furry:1938zz} pointed out that establishing
which formalism applied to the neutrino, Dirac's or Majorana's, would be 
more difficult than proving the neutrino's existence. 
He also combined Majorana's theory with the \nnbb\ decay proposed by Goeppert-Mayer, and
conceived \nubb\ decay mediated by the emission and re-absorption of virtual Majorana neutrinos \cite{Furry:1939qr}.
The process does not require the presence of Majorana masses, but simply
Majorana neutrinos, which obey Fermi-Racah interactions.
Should the interaction be of scalar type, in the theoretical context at the time
it could have yielded half-life values as short as 10$^{15}$ years.
Furry noted that such a rapid rate would affect the abundance of long-lived isotopes,
opening the possibility of geochemical searches for \nubb\ decay in addition to
direct searches.

Furry's hypothesis motivated the first experimental searches for \nubb\ decay with
rates too rapid to be accommodated by Goeppert-Mayer's proposed mechanism. The
first limit $\Thl\ > 3\cdot10^{15}$\,yr was made with $^{124}$Sn in Geiger
counters \cite{Fireman:1948}. Follow-on direct experiments \cite{Fireman:1949qs,
Lawson:1951, Kalkstein:1952zz, PearceDarby:1952, Fremlin:1952, Fireman:1952qt,
McCarthy:1953, McCarthy:1955} incorporated proportional counters,
scintillators, Wilson chambers, and nuclear emulsions using several isotopes,
and included some positive claims \cite{Fireman:1949qs, Fremlin:1952,
McCarthy:1953, McCarthy:1955} that were disproved in more sensitive
experiments --- a theme that has repeated itself throughout the history of
double-beta decay experiments, see \textcite{Tretyak:2011pg}. 
Meanwhile geochemical searches \cite{Inghram:1949qu, Inghram:1950qv,
Levine:1950qw}, which are sensitive only to the combination of \nubb\ and \nnbb\ decay
and not to each of them separately, yielded very strong limits, as well as the first
observation of \bb\ decay of $^{130}$Te with a half-life of $1.4 \cdot
10^{21}$\,yr \cite{Inghram:1950qv}, consistent with the rate of Goeppert-Mayer's \nnbb\ decay.

In the same period, \textcite{Mayer49} also established the foundations of the nuclear shell model --- an independent particle model at that time ---  which was independently also proposed by Jensen and others \cite{Jensen49}. Together with the interplay between single-particle and collective nuclear motion introduced by \textcite{BohrMottelson53}, these works set up the cornerstones for the theoretical understanding of nuclear structure, which eventually --- after three decades of theory and computing power advances ---
led to the first modern calculations of \nubb-decay nuclear matrix elements.

\begin{figure}[]
  \includegraphics[width=\columnwidth]{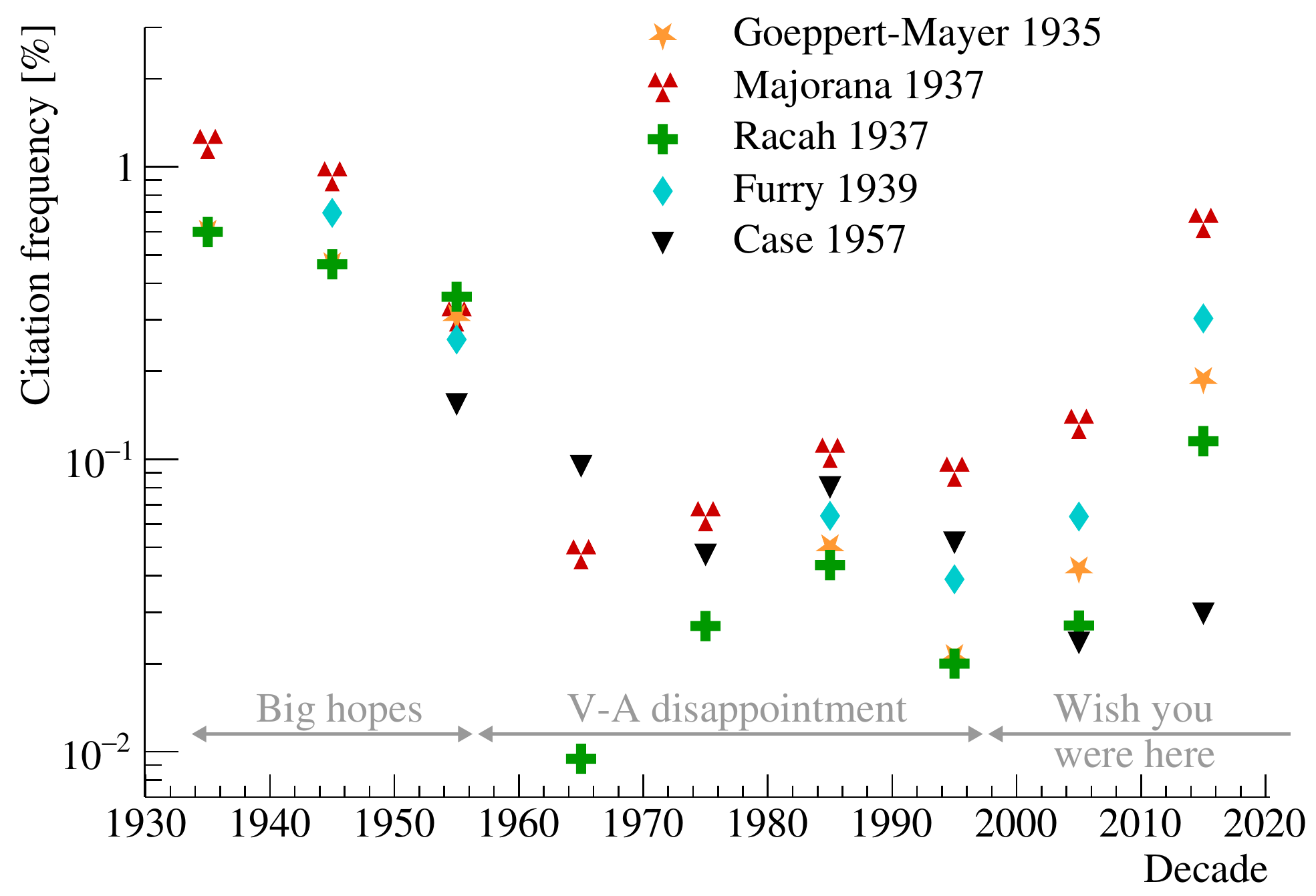}
  \caption{Citation frequency of some seminal papers on \nubb\ decay over time, until 2020.
    The citation frequency is computed as the number of citations per decade divided
    by the total number of papers with $\geq10$ citations published in the same decade.
    Data from \href{https://inspirehep.net}{Inspire}. See also \textcite{Vissani:2021gdw} for further discussion.
  }\label{fig:citations}
\end{figure}

Following the lack of observation of rapid \nubb\ decay, a loss of interest in
the process started when \textcite{Davis:1955bi} did not observe the reactions predicted by Racah's
theory --- e.g., $^{37}$Cl$(\bar{\nu},e^-)^{37}$Ar --- and
the $(V-A)$ theory of weak interactions showed that the \nubb-decay rate did
not depend on just the nature of the neutrino, but also on its mass, as was
elegantly elucidated by \textcite{Case:1957zza}. 
For vanishing Majorana mass, the effect would disappear and the transition would become undetectable, a point made particularly clear by \textcite{radicati-1957}.
In addition, the influential paper by \textcite{Primakoff:1959chj} argued in favor of a Dirac neutrino.
As a result, the enthusiasm for \nubb\ decay declined further,
as testified by the reduction of citations over time shown in Fig.~\ref{fig:citations} of certain fundamental papers on \nubb\ decay
\cite{GoeppertMayer:1935qp,Majorana:1937vz,Racah:1937qq,Furry:1939qr,Case:1957zza}.

The neutrino mass hypothesis was revived by 
ideas on flavor transformations of massive neutrinos 
(1957-1967) \cite{Pontecorvo:1957cp,Pontecorvo:1957qd,Maki:1962mu,Pontecorvo:1967fh},
supported by the first observations of solar neutrinos \cite{Cleveland:1998nv},
and eventually experimentally proven by the discovery of neutrino oscillation \cite{Kajita:2016cak,McDonald:2016ixn}.
Additional interest arrived in the seventies, the ``age of gauge theories'',
with the conception of the  ``seesaw mechanism'' \cite{Minkowski:1977sc,GellMann:1980vs,yanagida:1979as,Mohapatra:1979ia},
in which a heavy Majorana neutrino generates a tiny mass for the light neutrino emitted in \B\ decay.
Furthermore, \textcite{Weinberg:1979sa}, \textcite{Wilczek:1979hc} showed the usefulness of effective operator
analysis to extend the Standard Model of electroweak interactions.
In this context, the rates of new phenomena, e.g., \nubb\ decay or proton decay,
are suppressed by a factor inversely proportional
to the scale of ``Grand Unification''.
If new physics exists at an ultra-high scale,
the leading mechanism for \nubb\ decay would be light neutrino exchange.
The renewed interest in \nubb\ decay,
boosted by the discovery of neutrino oscillations,
was accompanied by an increase in the citation rate of the seminal works, as shown in
Fig.~\ref{fig:citations}.

The community has nowadays a common view on \nubb\ decay, which is a sort of minimal or orthodox vision focused on the supposition that the Standard Model neutrino is a Majorana particle.
There are, however, alternative ideas.
For instance, \textcite{Touschek-1948} showed that
the observation of \nubb\ decay does not directly imply the Majorana nature of the neutrino, unless
the nature of weak interactions is considered to be known.
After the introduction of $(V-A)$ theory, \textcite{Feinberg-1959} pointed out
the possibility of contributions to \nubb\ decay unrelated to neutrino mass.
The understanding of neutrino oscillation, yielding observable phenomena even with very small neutrino masses,
led \textcite{Pontecorvo-1968} to reiterate the point that \nubb\ decay could proceed
through channels other than the Majorana neutrino mass mechanism.
Even today, the possibility of new physics at accelerator or rare process scales,
perhaps involving lepton number violation, allows one to imagine
a \nubb-decay rate significantly greater than that due to Majorana masses. 

The late 1960's to early 1980's also saw a contemporaneous blossoming of
experimental techniques in \nubb\ decay, thanks to inventions such as the Ge(Li)
detector \cite{Freck:1962} and the streamer chamber \cite{Chikovani:1963,
Dolgoshein:1964}. These led
to a leap in half-life sensitivities for direct \nubb-decay searches, with efforts by
Fiorini and Wu yielding limits on the order of
10$^{19-21}$\,yr \cite{Fiorini:1967in, Fiorini:1975, Bardin:1967inh,
Bardin:1970vvi, Cleveland:1975zz}.  This level was also reached with
scintillating crystals \cite{derMateosian:1966qz}.  During this period, the
invention of the high-purity semiconductor Ge (HPGe) detector \cite{Baertsch:1970} and
time-projection chambers (TPCs)~\cite{Nygren:1974nfi} led to new possibilities for the experimental
investigation of \nubb\ decay.

By the mid-1980s the combination of theoretical motivation and experimental
capabilities brought \nubb-decay physics into something of a ``golden era''.
\textcite{Haxton:1984ggj}, Doi, Kotani, and Takasugi \cite{Doi:1985dx} worked out the full
theoretical details of the decay, building on
earlier work by \textcite{Primakoff:1959chj, Primakoff:1969mvu}, and subsequently
refined by \textcite{Tomoda:1990rs}. Nuclear
matrix element calculations also proceeded in earnest. Studies using the quasiparticle random-phase approximation method showed that they could reproduce extremely long \nnbb-decay half-lives once proton-neutron pairing is properly taken into account \cite{Vogel86}. The same physics was found to be relevant for \nubb\ decay \cite{Engel88}. Then, in 1987,
Moe's group reported the first direct observation of \nnbb\ decay in $^{82}$Se 
using a TPC \cite{Elliott:1987kp}. The process was soon after reported in $^{76}$Ge by the ITEP/YePi
experiment using HPGe detectors \cite{Vasenko:1989jv}.   
\textcite{Ejiri:1991dq} observed the decay in $^{100}$Mo using a tracking detector
consisting of a planar source
sandwiched between drift chambers and scintillator
detectors. \nnbb\ decay was also observed in $^{116}$Cd in
multiple tracking and scintillating crystal experiments \cite{Ejiri:1995kd,
NEMO:1995asg, Danevich:1995np}.  TPCs and
tracking detectors made additional observations in numerous
isotopes \cite{Elliott:1991vb, Elliott:1992cf, NEMO:1994sst, NEMO:1996gxj,
Balysh96, DeSilva:1997cp, NEMO:1997rel, Arnold:1999vg}, and an assay of a
sample of enriched Mo powder using HPGe
detectors made the first observation of \nnbb\ decay to an excited
state of the final nucleus, in $^{100}$Mo \cite{Barabash:1995fn}.
The measurement of the half-life of $^{48}$Ca \cite{Balysh96}, the lightest
\nnbb-decay emitter and the one with least complex nuclear structure, was found to be in good agreement with the nuclear shell model prediction \cite{Caurier90,Poves95}, giving confidence to nuclear matrix element calculations.

These experiments achieved exquisite sensitivity also to the \nubb-decay mode,
culminating in half-life limits at the level of 10$^{25}$ years by the
Heidelberg-Moscow and IGEX experiments in
$^{76}$Ge \cite{Klapdor-Kleingrothaus:2000eir, Gonzalez:2000ek}. A subset of the
Heidelberg-Moscow experiment independently published a claimed
observation with half-life on the order of 10$^{25}$\,yr
initially with 3.1$\sigma$ significance \cite{Klapdor-Kleingrothaus:2001oba},
increasing to 4.2$\sigma$ and then $>$6$\sigma$ significance in subsequent
reanalyses \cite{Klapdor-Kleingrothaus:2004yzi, Klapdor-Kleingrothaus:2006zcr}.
This claim was strongly questioned by \textcite{Feruglio02}, \textcite{Aalseth:2002dt}, \textcite{Schwingenheuer:2012zs}, 
and ultimately ruled out by more sensitive experiments, with the first
definitive exclusion at $>$99\% CL coming from the GERDA
experiment \cite{GERDA:2013vls}.

GERDA~\cite{GERDA:2020xhi} along with KamLAND-Zen~\cite{KamLAND-Zen:2022tow} and other experiments from the modern era~\cite{CUORE:2021mvw, Arnquist:2022zrp, EXO-200:2019rkq} (discussed in detail in Sec.~\ref{sec:prj}) have now explored half-lives in the range 10$^{25}$\,yr to a few times 10$^{26}$\,yr. 
At present, major investments are being made in the USA~\cite{Aprahamian:2015qub}, Europe~\cite{Giuliani:2019uno}, and elsewhere (see Sec.~\ref{sec:prj}) to mount experiments capable of reaching 10$^{28}$\,yr and beyond.
A broad class of models predicts high discovery potential for this next-generation of searches.
If nature so chooses, the most exciting chapter in the history of neutrinoless double-beta decay could be about to unfold.

 \section{Particle physics theory and motivations}
\label{sec:par}

Neutrinoless double-beta decay is of fundamental importance for particle physics, and over time became central also to several other fields, including nuclear physics and cosmology.
In this section, we highlight the key aspects of this connection from a modern perspective.  

We first discuss in Sec.~\ref{sec:par:symmetries} the role of global symmetries in particle physics and their associated conserved quantities, and in particular lepton number $L$ and the difference between baryon and lepton number $B-L$, which are both tested by \nubb-decay experiments.
In Sec. \ref{sec:par:LMJ}, we consider the role and meaning of the neutrino's Majorana mass, and of other 
 effective operators which  parameterise possible violations of the global symmetries.
Sec.~\ref{sec:par:LNV} focuses on specific theoretical models that predict lepton number violation phenomena. 
Then, in Sec.~\ref{sec:par:mbb}, we discuss observational neutrino physics, introducing  
the parameter describing the contribution of known neutrinos to \nubb\ decay: 
the effective Majorana neutrino mass \mbb.
Finally, the link between the excess of baryons in the observable universe and the violation of the global symmetries of the Standard Model (SM) is examined in Sec.~\ref{sec:par:cosmo}.

Secs.~\ref{sec:par:symmetries}, \ref{sec:par:LMJ}, \ref{sec:par:mbb} and 
\ref{sec:par:cosmo} are all introductory and contain basic material needed to develop an overview of the field. These parts are intended for nonexpert readers. Section~\ref{sec:par:LNV} covers a wide range of theoretical models connected to \nubb\ decay and, because of its technical nature, it is intended for a more expert audience.

\subsection{Global symmetries}
\label{sec:par:symmetries}
In this section we first examine the role played by global symmetries --- those associated with the conservation of baryon and lepton number --- for the understanding of particle physics (Sec.~\ref{sec:par:symmetries:bl}). Then, we review their meaning in the Standard Model, emphasising the exact (non-anomalous)  symmetries, and in particular the combination $B-L$ (Sec.~\ref{sec:par:symmetries:smbl}). Finally,  we discuss \nubb\ decay in relation to these symmetries (Sec.~\ref{sec:par:symmetries:hwnn}), arguing that it qualifies as a  process in which a net amount of matter particles is created.

\subsubsection{Baryon and lepton number conservation}
\label{sec:par:symmetries:bl}
Nuclear theory was directly based on the idea that the total number of nucleons 
remains the same in any transformation.
This was soon generalized into a ``conservation law for the number of heavy particles''
(\textit{baryon conservation}) by \textcite{Wigner:1949},
who noted that the proton could decay into $p\to e^+ +\pi^0$ unless some law forbids it.
For light matter particles,
namely electrons and neutrinos (leptons), the situation was less clear,
especially in view of the elusive nature of neutrinos \cite{zeldovich,Marx-1953}.
The four-fermion theory of the weak interaction is formulated in a manner that
allows the assignment of a conserved number to the sum of charged and neutral leptons,
where antimatter particles are assigned a negative sign. 
However, after Majorana proposed his theory of massive neutrinos, it became clear that it 
was not even possible to tell {\em a priori} whether a neutrino and an antineutrino
are two distinct particles, or two states of the same particle differing only by
helicity.  Tests of 
the hypothetical decay $(A,Z)\to (A,Z+2) + 2 e$, carried out since the 1940's,
have not yet revealed any hint that the number of leptons could vary.  
Early direct searches for neutrino masses --- such as those conducted by \textcite{Hanna:1949zab} --- and studies of their helicity
suggested that neutrinos are practically massless \cite{Salam:1957st,Landau:1957tp,Lee-1957},
and contributed to a reduced interest in Majorana's proposal.
Moreover, subsequent investigations 
showed that the beam of muon neutrinos from $\pi^+$ decay produces leptons and not antileptons. 
In short, it was hypothesized that also the \textit{number of leptons does not change} in any interaction.
A beautiful summary of the situation can be found in \textcite{Feinberg:1301}.

The discussion deepened with the emergence of the various families of particles. For instance, the question
of why $\mu\to e+\gamma$ is forbidden became as important as that of whether proton decay exists, and motivated the introduction of  separate 
\textit{muon and electron
number conservation laws}. At this point, however, an apparent difference 
between baryons and leptons emerged: the conservation of the hadronic families was violated by weak interactions in transformations between neutrons and protons, while 
that of leptonic families was not.

Nonetheless, the perception of a correspondence between hadrons and leptons remained.
The strengths of their weak interactions were found to be the same
\cite{Pontecorvo:1947vp,Puppi:1948qy}, and mixing among leptons and among quarks
was introduced in the early sixties on theoretical bases \cite{Katayama:1962mx,Maki:1962mu,Cabibbo:1963yz}.
Inspired by the work of \textcite{GellMann:1955jx},
\textcite{Pontecorvo:1957cp} introduced the idea of neutrino transmutation,
noting its connection to neutron-antineutron and 
hydrogen-antihydrogen transmutations, i.e., violations of baryon number. Finally, the seminal work of \textcite{Sakharov:1967dj} on baryogenesis suggested a specific $B-L$ conservation law, and discussed explicitly the possibility of proton decay associated with the Planck mass scale 
$M_{\text{P}}=\sqrt{\hbar c/G_{\text{N}}}$, defined in terms of the speed of light and Planck and Newton's constants. The decay rate is thus strongly suppressed.
\jd{Exactly what is being suppressed by $M_P$?}

\subsubsection{The Standard Model and $B-L$}
\label{sec:par:symmetries:smbl}

Let us come to the age of the Standard Model of particle physics
and its  SU(3)$_{\text{c}}\times$SU(2)$_{\text{L}}\times$U(1)$_{\text{Y}}$ gauge group. 
The renormalizable quantum field theory follows from the conventional choice of 15 quarks $(u,d)$ and leptons $(e,\nu)$ per family,
\begin{equation*}
  \begin{array}{cccccccc}
    u_{\text{r,L}} & u_{\text{g,L}} &u_{\text{b,L}} &\nu_{\text{L}} & u_{\text{r,R}} & u_{\text{g,R}} &u_{\text{b,R}} & \\
    d_{\text{r,L}} & d_{\text{g,L}} &d_{\text{b,L}} &e_{\text{L}} & d_{\text{r,R}} & d_{\text{g,R}} &d_{\text{b,R}} &e_{\text{R}} , \\
  \end{array}
\end{equation*}
with an important feature: baryon number $B$,
the three lepton numbers $L_{\text{e}}, L_{\text{$\mu$}}, L_{\text{$\tau$}}$,  and the total lepton number 
\begin{equation} \label{eccerelle}
  L = L_{\text{e}} + L_{\text{$\mu$}} + L_{\text{$\tau$}}\,,
\end{equation}
are accidentally conserved, i.e., their associated symmetries emerge accidentally without being {\em a priori} required. 
This is in good agreement with experiments. 

Not all of these global symmetries are  expected to be exactly obeyed.
They are all symmetries of the classical Lagrangian density,
but some of them are not symmetries of the full quantum theory, and can hence be violated by quantum fluctuations.
In jargon, these are called anomalous symmetries \cite{Steinberger:1949wx,Adler:1969gk,Bardeen:1969md}.
Indeed, the divergence of  the leptonic and the baryonic currents are not zero  \cite{tHooft:1976rip},
but rather $\partial^\mu J_\mu^{\text{(B)}} =\partial^\mu J_\mu^{\text{(L)}}= 3 g^2/(32 \pi^2) \text{Tr}[ F_{\mu\nu} \widetilde{F}^{\mu\nu} ]$
where $g$ is the  SU(2)$_{\text{L}}$ gauge coupling and $F_{\mu\nu}$ the field strengths, so these currents are not conserved.
The exact (non-anomalous) SM global symmetries are
\begin{equation}
  B-L\ , \quad L_{\text{e}}-L_{\text{$\mu$}}\ , \quad L_{\text{$\mu$}}-L_{\text{$\tau$}}\,,
\end{equation}
along with their linear combinations, e.g., $L_{\text{e}}-L_{\text{$\tau$}}$.
In fact, the SM predicts the existence of  non-perturbative transitions that
violate other combinations, e.g., $B+L$,
as is well-known in ``baryogenesis'' and ``leptogenesis'' theories
that attempt to explain the  cosmic excess of baryons.
Suffice it here to remark the existence of an effective operator formed by the left doublets
$q_{\text{L}}  =(u_\text{L}, d_{\text{L}})^t$ and  $\ell_\text{L}  =(\nu_{\text{L}}, e_{\text{L}})^t$,
that  respects all the anomaly-free symmetries and violates the other ones\footnote{This is $\prod_{i=1}^3  \mathcal{O}^{(i)}$ 
formed by $ \mathcal{O}^{(i)} = \epsilon_{\text{AB}} \epsilon_{\text{CD}}
\epsilon_{\text{abc}} q_{\text{aA}}^{(i)} q_{\text{bB}}^{(i)}
q_{\text{cC}}^{(i)} \ell_{\text{D}}^{(i)}$,
where $i$ is a family index, (a, b, c) are color indices, 
(A, B, C, D) are SU(2)$_{\text{L}}$ indices, and the Levi-Civita tensors $\epsilon$ provide a gauge invariant result.}.

It should be noted that the observation of neutrinos other than those initially produced in ``neutrino appearance'' experiments,
even before invoking an interpretation in terms of massive neutrino oscillation, demonstrated the violation of the anomaly-free symmetries 
$L_{\text{e}}-L_{\text{$\mu$}}$ and $L_{\text{$\mu$}}-L_{\text{$\tau$}}$ \cite{DellOro:2017pgd,DellOro:2018jze}.
For example, SNO observed the appearance of muon and tau neutrinos in the solar
electron neutrino flux \cite{SNO:2001kpb},
and various experiments have seen the appearance of new neutrinos from muon neutrino beams:
electron neutrinos in T2K \cite{T2K:2013ppw} 
and tau neutrinos in the case of OPERA \cite{Agafonova:2018auq}.
A straightforward implication is that {\em the only residual symmetry of the Standard Model is $B - L$.}
If this symmetry is respected, we can distinguish perfectly matter particles from antimatter particles, as described in the Standard Model. However, if $B-L$ is violated,  we should expect transitions between matter and antimatter particles, for example, the transformations between neutrinos and antineutrinos discussed in Sec.~\ref{sec:par:LMJ:mjnu}. 
Thus experimentally investigating $B-L$ is of paramount importance, and 
the process $(A,Z)\to (A,Z+2) + 2 e$ provides a direct test of it.
Note, incidentally, that the observation of the otherwise extremely interesting decay of the proton
via $p\to e^++ \pi^0$ or any other mode induced by dimension-6 operators would not.

\subsubsection{What is a proper name for $(A,Z)\to (A,Z+2) + 2 e$?}
\label{sec:par:symmetries:hwnn}

So far, in this section, we have  avoided referring to the process
$(A,Z)\to (A,Z+2) + 2 e$, as ``neutrinoless double-beta decay''.
We did it intentionally, with the aim of examining first the meaning and the importance of the process at hand. 
Not only is it possible to characterize this decay quite directly as a
``creation of two electrons'', using a terminology accessible even to laypersons,
it is also possible to call it the ``creation of leptons without antileptons'', using jargonic
parlance highlighting specifically the violation of $L$.
Most importantly, considering the SM structure, this term should be associated
with the violation of $B-L$, the only residual global symmetry allowing
the distinction of matter from antimatter particles.
This process can thus be described as the ``creation of matter without antimatter'', or more precisely the creation of particles of matter,
in this case electrons. This is different from usual weak decays, such as normal $\beta$ decays, which produce electrons (matter particles) accompanied by the same number of antineutrinos (antimatter particles), and thus do not change $L$.

The traditional name for the process, ``neutrinoless double-beta decay'', is formally correct but rather obscure as
it defines the process in terms of particles that are {\em not} produced --- something akin to calling a hippopotamus
a ``trunkless elephant''.
Moreover, it uses ``beta-rays'' for electrons, a term that dates back to Rutherford's time
when it was surmised that electrons live in the atomic nucleus.
The standard terminology was introduced to contrast this process with the ``ordinary'' $\beta\beta$ decay of G\"oppert-Mayer, and 
reminds us the theoretical belief that the transition is dominantly triggered by the exchange of virtual Majorana neutrinos, which are valuable points.
However, we think that these are not good reasons to understate the importance
of this process for the current understanding of matter and its interactions
\cite{DellOro:2017pgd,DellOro:2018jze}.

\subsection{Majorana neutrinos and other sources of lepton number violation}
\label{sec:par:LMJ}
In this section we present the main mechanisms that can lead to lepton number violating effects and  
\nubb\ decay. We first introduce the simplest case, in
which ordinary neutrinos are endowed with a Majorana mass and the fermionic spectrum of
the Standard Model is not modified. As we argue in Sec.~\ref{sec:par:LMJ:mjnu}, this assumption means that neutrinos, unlike
all other fermions, are at the same time particles of matter and antimatter.
In Sec.~\ref{sec:par:LMJ:operators}, we then take full advantage of the structure of the
Standard Model, and discuss the numerous effective operators that parameterize all
possible lepton-number-violating effects. Finally, in Sec.~\ref{sec:par:LMJ:numass}, we examine
the simplest renormalizable extension of the Standard Model leading to 
Majorana neutrino masses, namely, the inclusion of right-handed neutrinos.

\subsubsection{Majorana neutrinos: a bridge between matter and antimatter}
\label{sec:par:LMJ:mjnu}
Majorana's neutrinos are both particles and antiparticles. This often-heard statement is far from being trivial. To clarify its meaning, it is useful to remember that neutrinos are particles with spin 1/2, i.e.~fermions. Fermions constitute matter (and antimatter), whereas bosons constitute forces.
In the context of the Standard Model of Glashow, Weinberg and Salam, neutrinos along with all other particles are distinct from their antiparticles. Such a difference is evident for charged fermions, but what about for neutral ones?

In fact, Standard-Model neutrinos are neutral. They have hypercharge but this is broken spontaneously, leaving only two ways to distinguish neutrinos from antineutrinos. The first way concerns the helicity of the particle: it is negative for the neutrino and positive for the antineutrino. The second way is based on the charged lepton that accompanies charged lepton interactions: for example, in all observed $\beta^\mp$-decays, the (anti)neutrino is co-produced with a particle of (negative) positive charge.
 
\begin{figure}[tb]
\begin{center}
\includegraphics[width=\columnwidth]{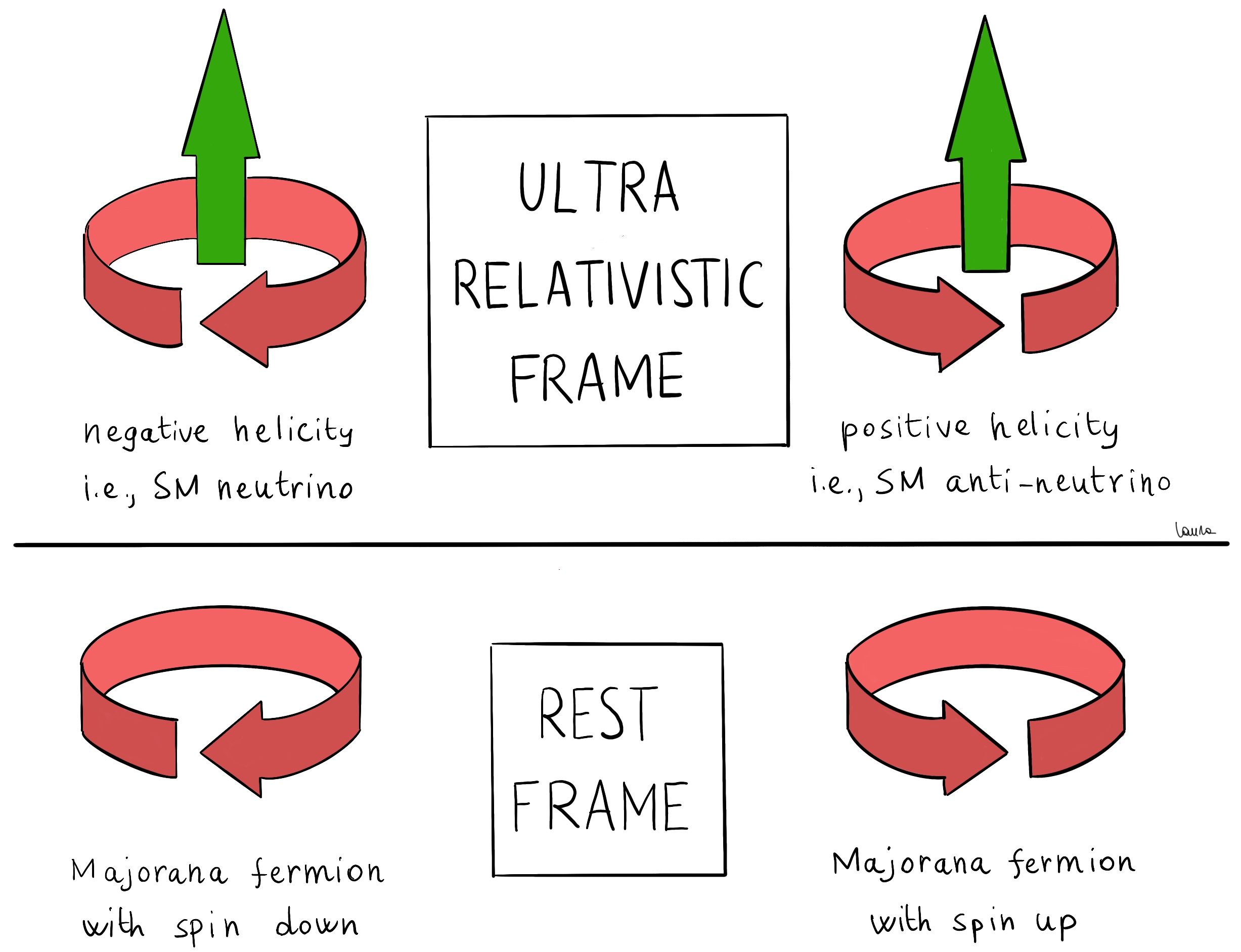}
\caption{Artistic illustration of the relation between the neutrino and antineutrino helicity, which is given by
the projection of the spin (red arrow) onto the momentum (green arrow). The helicity distinguishes neutrinos from antineutrinos in the ultra-relativistic limit (top panel). However, in the rest frame the neutrino and antineutrino are two spin states of the same particle (lower panel). Image courtesy of Laura Manenti.}
\label{spin-elicita}
\end{center}
\end{figure}

The neutrino's helicity is a consequence of the chiral structure of the weak interactions --- formally corresponding to the presence of the $ P_L $ projector in the charged interactions --- but only provided that the neutrino mass is exactly zero. If neutrinos are massive, helicity coincides with chirality only in the ultra-relativistic limit. All experimental observations related to weak interactions have been made, and can be made, only on ultra-relativistic neutrinos. However, as a thought experiment, we can consider observing a neutrino and an antineutrino in their rest frame, whose existence is guaranteed by their tiny masses measured through oscillation experiments. In this frame, the momentum and helicity of the neutrino and antineutrino are both zero and, in the absence of additional quantum numbers, the two particles can differ only by the orientation of their spin. Therefore, symmetry under rotations implies that the two states must be the same particle. 
In conclusion, the structure of the Standard Model, together with the hypothesis that neutrinos have mass, suggests that the neutrino and the antineutrino are the very same particle in the rest frame.
The point is summarised graphically in Fig.~\ref{spin-elicita} and discussed also by \textcite{DellOro:2016tmg}.

A different conclusion can be drawn assuming the existence of some property discriminating the two particles even in the rest frame, for instance lepton number. In this case, two additional neutral particle states must exist in the rest frame,  and they must be ``sterile'', i.e., unable to couple to the Standard Model gauge fields. This possibility is what people refer to when they speak about Dirac neutrinos. 
It should be stressed that Dirac neutrinos require {\em invoking} an ad-hoc property, such as lepton number, as opposed to {\em inferring} such a property from the model structure.  Invoking an ad-hoc property can be perceived as unnatural, in which case one might favor Majorana's neutrinos. 

Majorana neutrinos would be unique among fermions and provide a bridge between matter and antimatter.  
Majorana's hypothesis evidently confronts us with a blatant violation of the $L$ symmetry. Since baryon number is not affected by neutrinos, the $B-L$ symmetry would also be broken.

The previous considerations obviously do not constitute a formal proof that neutrinos are Majorana particles.
There is currently no experimental evidence of $B-L$ violation, except for the indirect cosmological observation that there are more atoms than anti-atoms. However, cosmological observations are unable to test potential lepton asymmetries created by neutrinos, which could compensate the baryon asymmetry. These considerations highlight the crucial importance of experimentally testing the conservation of $B-L$, in particular, through the study of \nubb\ decay.

\subsubsection{Effective operators and energy scale}
\label{sec:par:LMJ:operators}

A general theorem from \textcite{Helset:2019eyc} states that
the variations of lepton number $\Delta L$ and baryon number  $\Delta B$ obey
\begin{equation}
  \frac{\Delta L-\Delta B}{2} = d \text{ mod }2,
\end{equation}
where $d$ is the canonical dimension of the operator causing the transition,  $\mathcal{O}_d$. This operator is  a polynomial of 
SM fields and possibly also right handed neutrinos, i.e., sterile neutrinos under the SM interactions.
As usual, fermionic fields contribute $+3/2$ to $d$ and bosonic fields (or
derivatives) contribute $+1$.
In the case of \nubb\ decay, where baryon number is conserved and $\Delta L=\pm 2$,
the canonical dimension must be  odd,  and the new physics scale $\Lambda$ that parameterizes the operators
$\mathcal{O}_d$  appears as $1/\Lambda^{d-4}$.
After  spontaneous symmetry breaking (SSB),  the electroweak scale
$v=(\sqrt{2} G_{\text{F}})^{-1/2} =246$ GeV,
which is plausibly smaller than  $\Lambda$, is brought into play in the numerator of the operator. Here $G_{\text{F}}$ is the Fermi constant.

A useful introduction to the role of effective operators can be found in the original works of \textcite{Weinberg:1979sa,Weinberg:1980bf}, \textcite{Wilczek:1979hc}, \textcite{Babu:2001ex}, and \textcite{Choi:2002bb}.
The full classification of all operators of dimension 7 
and dimension 9 has been recently completed by \textcite{Lehman:2014jma}, \textcite{Li:2020xlh} and \textcite{Liao:2020jmn}.  
Omitting right handed neutrinos, there is no renormalizable operator that breaks $L$ (or $B$); 
at dimension 5 there is only one operator, the well known Weinberg operator \cite{Weinberg:1979sa,Weinberg:1980bf};
at dimension 7 there are 13 operators that obey $\Delta L=2$; at dimension 9 there are several hundreds of them; we still do not have a systematic study of the number of operators at dimension 11.

The dimension-5 operator leads to a Majorana mass for ordinary neutrinos and can be constructed starting from
the following gauge invariant combination of 
a leptonic doublet $\ell$ and a Higgs doublet $H$:
\begin{align}
  \ell^t_{\text{\tiny L}} \varepsilon H
  & =  \frac{1}{\sqrt{2}} \left(\nu_{\text{\tiny L}}, e_{\text{\tiny L}} \right)
  \begin{pmatrix}
    0   &  1 \\
    -1  & 0  
  \end{pmatrix}
  \begin{pmatrix}
    0  \\
    v+h
  \end{pmatrix}\\
  & = \frac{1}{\sqrt{2}} v\, \nu_{\text{\tiny L }}+\text{interactions},
\end{align}
where $H$ is given in the physical gauge,  
and $\varepsilon=i\sigma_2$
is the invariant matrix of SU(2)$_{\text{\tiny L }}$.
This term behaves just like a spinor field under Lorentz transformations,
so we can use it to form the Minkowski-Weinberg operator,  namely 
the  following Lorentz invariant term of the Lagrangian density 
\begin{equation}\label{d5}
  \delta\mathcal{L} =
  -\frac{1}{2M} ( \ell^t_{\text{\tiny L}a} \varepsilon H)  C^{-1}_{ab}
  (\ell^t_{\text{\tiny L}b} \varepsilon H) + \text{h.c.}
\end{equation}
where $C$ is the charge conjugation matrix and $a,b=1,2,3,4$ are four-spinorial indices. 
After SSB, this yields a bilinear term in $\nu_{\text{\tiny L }}$, i.e., a 
Majorana mass term.  Thus we identify
\begin{equation}
  m =\frac{v^2}{2M}\approx 50\,\text{meV} \times \frac{6\cdot 10^{14}\,\text{GeV}}{M},
  \label{eq:massterm}
\end{equation}
a relation showing that the neutrino 
mass values $m$, which have been discovered by means of neutrino oscillation, correspond to very large masses $M$.
We note that this mass scale  strongly differs from $v = 246$\,GeV,
the electroweak mass scale, and is smaller than the Planck mass: a valuable indication of new physics.

The $d = 7$ operators that after SSB  have a structure $ \mathcal{O}= \bar{e} \nu \bar{u} d v^2/\Lambda^3$ need SM ``dressing'' to specify the \nubb-decay transition; this implies the exchange of virtual neutrinos (the inclusion of a neutrino propagator) but without the need for further lepton number violation.
Moreover, there are dimension-7 operators~\cite{Lehman:2014jma} involving the $W$ boson that
after spontaneous symmetry breaking produce effective operators with structures
$gW \, \bar{e}^2 \,  \bar{u} d  \,  /\Lambda^3$  
and 
$( g W \, \bar{e}  )^2  \, v^2  /\Lambda^3$. Together with the usual SM
 interactions between the $W$ and the quarks, these lead to 
contact operators of the type  $ (\bar{e} \bar{u} d)^2 /(\Lambda^3 v^2)$~\cite{Cirigliano17dim7}.

The dimension $\ge 9$ operators considered above are contact terms and by construction produce $\mathcal{O}\propto ee (u\bar{d})^2$ after SSB; 
they are multiplied by $1/\Lambda^5$ or  $v^2/\Lambda^7$ when the dimension is 9 or $11$, respectively.
Therefore it is quite common to restrict attention to the cases with dimension $\le 9$,
which are expected to provide larger contributions to 
\nubb\ decay (see, e.g., \textcite{Bonnet:2012kh} for dimension-9 operators).

The na\"ive scaling of transition amplitudes for operators of various dimensions are
\begin{align}\label{eq:naive}
  \text{dim 5: }&G_{\text{F}}^2 \frac{v^2}{\Lambda} \frac{1}{p^2}, \nonumber \\
  \text{dim 7: }&G_{\text{F}} \frac{v}{\Lambda^3} \frac{1}{p}, \nonumber \\
  \text{dim 9: }&\frac{1}{\Lambda^5},
\end{align}
where $p\sim200$\,MeV is the virtual momentum of the neutrino, estimated 
as the inverse  of the typical distance between nucleons in nuclei. This suggests a suppression by powers of 
$\epsilon=p / v \Lambda^2 < 10^{-4}$ if $\Lambda\ge 1$\,TeV.
This would indicate that the amplitude decreases with dimension.
These na\"ive expectations are supported by the 
cursory bounds illustrated in \textcite{Choi:2002bb}, 
assuming \nubb-decay half-life values longer than $10^{25}$\,yr.
On the other hand, the above approach neglects the possible presence of small coefficients --- e.g.~Yukawa couplings --- that could in principle 
suppress the lower dimension terms more than the other ones. If for instance we consider  the reasonable value $m_\nu\sim 10$\,meV
suggested by experiments for the Majorana neutrino mass,
rather than estimating the theoretical mass as  $m_\nu\sim v^2/\Lambda $,
we would write the dimension-5 amplitude as $G_{\text{F}}^2 m_\nu/p^2$, which is of the same order as the 
dimension-7 (dimension-9) term if $\Lambda \sim 10^3$\,TeV ($\Lambda \sim 10$\,TeV).
In any case, these estimations are useful for a first orientation at best.
Moreover, considerations of hadronization and nuclear matrix elements can have an impact of orders of magnitude, see the discussion after Eq.~\eqref{eq:mbb_bsm} in Sec.~\ref{sec:nuc}.

\subsubsection{Majorana and right-handed neutrinos}
\label{sec:par:LMJ:numass}

We know that at least two of the three known neutrinos are not massless, and it is usually assumed that no other light neutrinos
mix with them \cite{Dentler:2018sju}.
This simple remark poses a macroscopic theoretical question:
why are the masses of the three ordinary neutrinos so different from
--- so much smaller than --- those of the other SM fermions? 
The answer could be related to the Weinberg operator described in the previous section. This operator was originally  introduced by \textcite{Minkowski:1977sc} in the context of specific models including new ultra-heavy neutrinos that are neutral under the SM interactions.
In this case, the operator is multiplied by a coefficient inversely proportional to the heavy neutrino masses
and directly proportional to the square of Yukawa interactions $Y$ between neutrinos.
In fact, Yukawa interactions guarantee the mixing of ordinary (left-handed)
and new (right-handed) neutrinos, as recalled in Sec.~\ref{sec:par:LNV:rh}.
The general expression of the corresponding Majorana mass, in terms of mass matrices, is
\begin{equation} \label{prngt}
  M_\nu= - M_{\text{D}}\ M_{\text{R}}^{-1}\ M_{\text{D}} ^t\quad
  \text{with}\quad M_{\text{D}} =\frac{1}{\sqrt{2}}Y v,
\end{equation}
where $M_{\text{D}}$ is the Dirac mass matrix, and $M_{\text{R}}$ is that of the heavy
neutrinos.
This mechanism for the generation of ordinary neutrino masses
is called the {\em seesaw mechanism}: in analogy to the children's game 
in which a heavier child lifts a lighter one, the mass of the light neutrino
is inversely proportional to the scale of the heavy neutrino's mass.

The model with ultra-heavy (right-handed) neutrinos illustrates an important and rather general feature: the smallness of the ordinary neutrino masses can be attributed partly or mainly to the occurrence of small (adimensional) coefficients, the Yukawa couplings. In other words, by simply measuring small neutrino masses, it is not possible to deduce that the scale of new physics is large. This is evident for Dirac neutrino masses --- where $M_{\text{R}}=0$ and Eq.~\ref{prngt} does not apply, having $M_\nu= M_{\text{D}}$ --- but it also applies to Majorana neutrino masses.
This kind of difficulty was clear since the beginning.
The very first paper on the topic \cite{Minkowski:1977sc} has the eloquent title  
\emph{$\mu\to e \gamma$ at a rate of one out of $10^{9}$  muon decays?} and 
intentionally assumes 50\,GeV for the heavy neutrino mass,
which shows the awareness of the importance of testing the seesaw hypothesis for ordinary neutrino masses.

If the right handed neutrino masses are not too large, 
a few direct or indirect laboratory tests are possible ---
see \textcite{Alekhin:2015byh} for a fully worked out example. 
It is worth mentioning that a ``hierarchy problem'' occurs 
with new right handed neutrinos heavier than $\sim10^4$\,TeV \cite{Vissani:1997ys},
which could serve as a motivation for supersymmetric models \cite{Barbieri:1987fn}
discussed in Sec.~\ref{sec:par:LNV:su}.
Finally, as mentioned in Sec.~\ref{sec:par:cosmo}, the scenario with ultra-heavy neutrinos can be somewhat subject 
to valuable constraints requiring the validity of specific models for baryogenesis. 

To conclude, 
the only BSM phenomenon observed so far is neutrino oscillation, which requires 
that the masses of at least two ordinary neutrinos are not zero.
This situation resembles that of weak interactions long before the SM, before Fermi's theory.
All we can say is that we have theoretical reasons to suspect that the neutrino masses are due to the dimension-5 operator. 
Despite the simplicity of these statements, 
the essential objectives for real progress are to demonstrate that the neutrino masses have a Majorana character
and that the total number of leptons and $B-L$ are violated.

\subsection{Models for lepton number violation}
\label{sec:par:LNV}
In this section, we review some proposals on how to extend the Standard Model, highlighting their connections to neutrino masses and \nubb\ decay. We start from unified models, based on the gauge principle, just like the Standard Model (Sec.~\ref{sec:par:LNV:gu}). We then discuss the reasons for extending the fermion spectrum and include right-handed neutrinos (Sec.~\ref{sec:par:LNV:rh}). Finally, we consider supersymmetric extensions in Sec.~\ref{sec:par:LNV:su} and close with
a wide range of models compatible with observable signals in the laboratory in Sec.~\ref{sec:par:LNV:fp}.

\subsubsection{Gauge theories and lepton violation at very high energy scales}
\label{sec:par:LNV:gu}

There are various gauge groups that extend the SM and have been regarded with interest for some of their features 
and new predicted phenomena. Among the features are the possibility of
gauge coupling unification (Grand Unification); this can be 
complete or partial, in the sense that it might require the existence of
intermediate scales. 

The new phenomenon predicted by these models and 
which received the greatest emphasis in the 1970s is the occurrence of proton decay, 
but later it was realized that also the existence of non-zero neutrino masses was a generic
consequence of several models \cite{GellMann:1980vs,Mohapatra:1979ia}.
The experimental evidence for non-zero neutrino masses  add motivation for SO(10) \cite{Fritzsch:1974nn}, which can break into SU(5) \cite{Georgi:1974sy} 
or into SU(4)$_{\text{c}}\times$SU(2)$_{\text{L}}\times$SU(2)$_{\text{R}}$ \cite{Pati:1974yy}.
These models are characterized by dimensionless Yukawa couplings $y$, and the scale $\Lambda$ of the new, heavy particles, e.g., heavy right-handed neutrino masses.
In the simplest case, called type I seesaw, Eq.~\ref{eq:massterm} is recovered with scale $1/M$ given by
\begin{equation} \label{pallon}
  \frac{y^2}{\Lambda} \sim \frac{1}{M},
\end{equation}
Other cases besides the type I seesaw are possible and are realised in actual models such as those based on SO(10), as discussed below.
Notice that the same value of $M$ can be obtained with $y$ of order one and $\Lambda\sim M$,
but also with correspondingly smaller $y$ and $\Lambda$.

It is worth mentioning here the fact that proton decay
has still not been found, and that its search continues to be strongly motivated from the theory side. Proton decay, together with neutrino masses, keep drawing attention to  SO(10), a well-defined model for which it is important to keep deriving quantitative predictions and related uncertainties.
Recall that this is a gauge group with only one coupling constant, which includes a right-handed neutrino in each fermion family together with the known leptons and quarks of the Standard Model. Put differently, this is the unification group that overcomes the asymmetry of particle content highlighted just above in Eq.~\ref{eccerelle}, necessarily including --- within its 16-dimensional spinors --- right-handed neutrinos.

\subsubsection{Right-handed neutrinos and the $\nu$SM}
\label{sec:par:LNV:rh}

There are many good reasons to postulate the existence of three right-handed neutrinos. First, they are a plausible mechanism to provide mass to light neutrinos \cite{Minkowski:1977sc,yanagida:1979as}. In addition, as previously discussed, they imply a full symmetry between left and right spinors of the SM \cite{Mohapatra:1979ia}. They also allow the promotion of the $B-L$ symmetry to a non-anomalous gauge symmetry;
indeed they are required in SO(10) and other unification groups \cite{GellMann:1980vs}.
Further, they could explain baryogenesis via leptogenesis, as first argued in \cite{Fukugita:1986hr}, see Sec.~\ref{sec:par:cosmo}.

Right-handed neutrinos can be incorporated in the SM as gauge singlet Weyl fermions $N_i$, with 
Lagrangian terms connecting them to the leptonic weak doublets $\ell_\alpha$:
\begin{equation}
\mathcal{L_{\nu \text{SM}}} = N_i i \partial_\mu \gamma^\mu N_i - Y_{\alpha,i} \ell_\alpha H N_i - \frac{M_i}{2} N_i N_i + h.c.,
\end{equation}
where $H$ is the Higgs weak doublet, the $Y_{\alpha,i}$ are Yukawa couplings, and $M_i$ are Majorana masses for the $N_i$. 
This comprises a minimal, renormalizeable Standard Model extension that accounts
for neutrino masses while remaining consistent with gauge invariance, and is
referred to as the ``$\nu$SM'' \cite{Asaka:2005an,deGouvea:2005er,deGouvea:2007hks}.
The case $M_i = 0$ corresponds to Dirac neutrinos, but
when $M_i \ne 0$, the mass term has the $L$- and $B-L$-violating structure of Eq.~\ref{d5},
and after SSB gives rise to Majorana mass terms for the light neutrinos.

In most models the new neutrinos are heavy and do not have direct implications at low energy scales except for SM neutrino masses. 
In other models right handed neutrinos are lighter, about 1-10 keV, and can
explain dark matter and possibly also the cosmic baryon excess \cite{Asaka:2005an}; interestingly, these models make no new contributions, other than the Majorana masses of light neutrinos, to \nubb\ decay~\cite{Bezrukov:2005mx}.

\subsubsection{Supersymmetry at accelerator energies}
\label{sec:par:LNV:su}

Supersymmetry is a symmetry between fermions and bosons.
The SM extension to a supersymmetric theory is possible but requires the introduction of several new particles,
heavy enough to not have been observed yet. 
The hypothesis that the masses of supersymmetric particles are not too far from
the electroweak scale has been regarded with interest because an approximate
supersymmetry can decouple the high mass scales from the electroweak scale, but 
to date these particles have not been found in direct searches. 

If the gauge principle --- i.e., the principle that 
all terms allowed by the postulated symmetries are present in the Lagrangian density ---
is applied to the supersymmetric SM,  lepton number and/or baryon number
are not automatically conserved.
Usually, 
this situation is felt as a shortcoming of generic 
supersymmetric models to be emended, as it triggers the instability of neutral fermions, which
would otherwise make useful dark matter candidates. 
The usual solution is to postulate a new discrete symmetry,
called R-parity, that amounts to the imposition of lepton and baryon number conservation
and  allows one to recover the dark matter candidate. 
In fact, in the usual parlance, the ``supersymmetric SM'' implicitly assumes R-parity.
At accelerator energies, these types of models have no significant implications for neutrino masses.

\subsubsection{Other new physics near the Standard Model scale}
\label{sec:par:LNV:fp}
To provide a more complete case study, we would like to conclude this overview of models 
by highlighting some of the theoretical scenarios  that are compatible with new contributions to \nubb\ decay
in addition to that due to the masses of light neutrinos.
Without any claim to completeness, and with the aim of illustrating some interesting possibilities, we will focus on R-parity breaking supersymmetry,
on low-scale seesaw,  and on left-right gauge theories
see e.g.~\cite{Deppisch:2015qwa,Alekhin:2015byh,Golling:2016gvc,Agrawal:2021dbo} for a wider discussion.
 A common feature of these models is the appearance of very small couplings, which ensure consistency with available observations and, in particular, allow the smallness of neutrino masses to be explained, replacing the role of the GUT energy scales in the standard theoretical reference frame (seesaw).

\paragraph*{Supersymmetry with broken R-parity.}
Let us begin by returning to consider certain supersymmetric extensions of the Standard Model.
As we have already argued, the supersymmetric extension of the Standard Model does not rule out the existence of violations of lepton number $L$ at the mass scale of the supersymmetry itself. This consideration is evident, noting that the ``superfield'' containing the Higgs doublet $(H^0,H^-)$ has the same quantum numbers as that containing the leptonic doublet $(\nu_e,e)$, and each contains both fermions and bosons.
$L$-violating couplings between these superfields that are sufficiently small can explain the neutrino mass and give rise to new contributions to \nubb\ decay.
They also lead to additional interesting phenomenology for lepton number violation, see for instance 
\cite{Hall:1983id,Ross:1984yg, 
Nilles:1996ij,
Hirsch:2000ef,Hirsch:2000jt,Faessler:2007nz,Bolton:2021hje}. Furthermore, these models include leptoquarks and dileptons, with masses in a region potentially accessible  to direct (accelerator) investigation, and can lead 
to several interesting manifestations.

\paragraph*{TeV scale seesaw.}
The possibility of neutrinos with masses $M_R$ around the TeV scale or even lower 
has been widely discussed, see  \cite{Drewes:2013gca} for a review. 
Electroweak fits are affected and, in some cases, improved by the inclusion of the new heavy neutrino states~\cite{Akhmedov:2013hec}. 
Moreover, these states can have a significant impact on \nubb\ decay and can even constitute the main contribution to the transition rate \cite{Atre:2009rg,Mitra:2011qr,BhupalDev:2013ntw}. In this case, it is a contact contribution whose dimensional fit scales as $G_F^2\ M_{LR}^2/M_R^3$, where $M_{LR}$ denotes the Dirac mass. 
However, the natural neutrino mass contribution from the seesaw, $M_{LR}^2/M_R$, must be suppressed by means of a particular matrix structure; 
this can be achieved without excessive fine-tuning if the right-handed neutrino mass respects an upper limit $\sim 10$ GeV \cite{Mitra:2011qr}.

\paragraph*{Left-right models near the electroweak scale.} In the last decade, 
an elegant, minimal extension of the gauge principle which underlies the ``Standard Model'' has been explored 
in order to realise a predictive theoretical scheme\footnote{However, 
this type of approach is at odds with further steps towards a complete unification of gauge interactions, in 
particular because  neutrinos are treated completely differently from other particles.} 
at a relatively low mass scale \cite{Maiezza:2010ic}, 
in which neutrinos are naturally endowed with mass 
(see in particular \cite{Nemevsek:2012iq,Senjanovic:2018xtu}), 
a situation that could lead to a rich phenomenology.
In fact, the presence of new and relatively large gauge couplings 
would be compatible with the actual production of new particle states at accelerators
(in contrast with the previous class of models, where the production is due to the Yukawa couplings,
that are not expected to be large).
Furthermore, it has been observed that \nubb\ decay would be
a natural manifestation of this type of pattern \cite{Tello:2010am}. 
This research programme has stimulated wide interest and subsequent discussions; we refer to the literature for the progress and insights that have followed 
\cite{Chakrabortty:2012mh,Awasthi:2013ff,Lindner:2016lpp,Li:2020flq}.

\subsubsection{Discussion}
In the earliest theoretical proposals, the physics giving rise to neutrino masses was assumed to be confined to very large energies: this  leads one to expect 
that the (small) Majorana masses of ordinary neutrinos controls the rate of the \nubb-decay transition. Although we believe it is prudent to consider this case the reference one, as illustrated in the previous section, 
we cannot exclude the possibility of significant additional contributions, which could justify even more optimistic expectations.

Moreover, 
we note  that the reference expectation, concerning the leading contribution to \nubb\ decay,
is based on a number of assumptions. In particular, it assumes that the Standard Model is a good approximation of physics at the scales accessible today, and that there are no new light particles that play an important role in lepton number violation.  However, there are indications (albeit indirect and not yet of unambiguous interpretation) of possible experimental anomalies, which depart from the expectations of the Standard Model, and whose interpretation might ultimately require new relatively light particles: e.g., those related to the $g-2$ muon \cite{Muong-2:2021ojo}, the mass of the $W$ boson \cite{CDF:2022hxs} (see also \cite{Cacciapaglia:2022xih}) or flavour physics, 
see e.g.~\cite{DAlise:2022ypp}. 

In addition, there are rather general questions that the Standard Model is unable to address, such as providing a candidate for non-baryonic dark matter, or giving reasons for the origin of the baryonic asymmetry. It cannot be ruled out that also these issues point to the existence of new light particles, which might also play some role for \nubb\ decay.

\subsection{Majorana masses and neutrino phenomenology}
\label{sec:par:mbb}

In this section we analyze the earliest proposed and most straightforward mechanism driving 
a non-zero rate for \nubb\ decay, i.e., Majorana neutrino masses. 
First we recall the experimental evidence for neutrino masses,
provided by neutrino oscillation experiments. 
Then, we introduce the essential formalism and the relevant parameter \mbb, often called
the \emph{effective Majorana neutrino mass}.
Next, the general aspects of the connection between \mbb\ and \nubb\ decay are introduced.
Finally, we discuss the experimental constraints on \mbb, as well as indications (empirical and theoretical) on its value. The quantitative implications for future experiments will be worked out in Sec.~\ref{sec:dis}.

\subsubsection{Neutrino oscillation}
\label{sec:par:mbb:osc}

The definitive evidence of neutrino oscillation implies that neutrinos are massive.
However it does not provide information on either the absolute mass scale \cite{Gribov:1968kq} or the Majorana phases \cite{Bilenky:1980cx}.
In addition, the  observed oscillation phenomena do not probe the Dirac or Majorana nature of neutrino masses,
as the neutrinos and antineutrinos are observed (observable) 
only in the ultra-relativistic regime \cite{Bilenky:1980cx}.
Nevertheless, considering our discussion in Sec.~\ref{sec:par:LMJ} on the importance of testing $B-L$ in addition to the theoretical arguments in favor of Majorana neutrino masses
based on the  SM structure, the recognition that neutrinos have mass
strongly motivates searches for \nubb\ decay.

The parameters of massive neutrinos 
have been quantified by oscillation experiments assuming three-flavor oscillation
\cite{Zyla:2020zbs}.
The squared mass differences are known with 1\%--2\% precision
and the squared sines of the mixing angles relevant for \nubb\ decay are known at the 3\%--4\% level.
One less clear aspect for which progress is expected in the coming years
concerns the arrangement of the neutrino masses, i.e., the neutrino mass
ordering, sometimes also referred to as the neutrino mass hierarchy.
The question concerns the discrimination between the \emph{normal} ordering (NO),
when the three neutrinos have a mass spectrum that resembles the charged fermion spectra,
and the \emph{inverted} ordering (IO), when they do not.
At present, global fits indicate a preference for the NO at the $\sim$3$\sigma$ level
\cite{Esteban:2020cvm,Capozzi:2021fjo}. However, this preference should be taken 
with a heavy grain of salt. Indeed, our current best probes for the mass ordering --- i.e.,  accelerator-based experiments that are directly sensitive to it --- favor the inverted ordering. The overall preference for normal ordering is driven by
the comparison between the neutrino mass squared difference measured in $\nu_\mu$ disappearance at accelerators and $\nu_e$ disappearance at reactors, and is strengthened by the multivariate analysis of Super-Kamiokande atmospheric neutrino data not fully integrated in the global fits.

Finally, various experiments hint at the existence of a new light neutrino with
mass of $O(1\,\text{eV})$ \cite{Dentler:2018sju,Giunti:2019aiy}.  Such a neutrino must be sterile, i.e., non-interacting, 
in view of the measurements done at the Large Electron-Positron (LEP) collider that limit the number of active light neutrinos to three~\cite{Decamp:1989fr,Giunti:2019aiy}.
Updated limits on sterile neutrinos from \nubb\ decay, compared with those from other observational probes, are discussed in \textcite{Bolton:2019pcu}.
However, as repeatedly argued in the literature, see, e.g., \textcite{Dentler:2018sju},
different experiments hint at sterile neutrinos with different parameters, and global fits show tensions among datasets.
Given the absence of strong theoretical arguments favoring such sterile neutrinos,
and the lack of phenomenological support,
we focus here on the scenario with three massive neutrinos.

\subsubsection{Formalism for the \mbb\ parameter}

Due to the absence of electric charge, neutrinos admit a more general type of mass than Dirac's one. 
As described generically in Sections~\ref{sec:par:LMJ:operators} and~\ref{sec:par:LMJ:numass},
a general bilinear term $-\bar{\Psi} M_\nu  \Psi^{\text{C}}/2 + h.c.$
can be added to the SM Lagrangian density, where the charge conjugate 
spinor is $\lambda^{\text{C}} = C \bar{\lambda}^t$, and 
the vector $\Psi$, which includes only left spinors,
can be written as $\Psi^t=(\nu_{\text{Le}} , \nu_{\text{L$\mu$}}, \nu_{\text{L$\tau$}})$
in the SM, or $\Psi^t=(\nu_{\text{Le}} , \nu_{\text{L$\mu$}}
, \nu_{\text{L$\tau$}}, \nu_{\text{Re}}^{\text{C}} ,\nu_{\text{R$\mu$}}^{\text{C}} , \nu_{\text{R$\tau$}}^{\text{C}}  )$
when three right handed neutrinos are assumed.
This Lagrangian density is called a Majorana mass term, and includes Dirac's term as a particular case. 
The mass matrix $M_\nu$ is  complex and symmetric, and can be decomposed as
\begin{equation}
  M_\nu = U\ \text{diag}(m_{\text{1}} ,m_{\text{2}}, \dots, m_{\text{n}})\ U^t,
\end{equation}
where $U^\dagger U = \mathbb{I}_{\text{n$\times$n}}$ and $m_{\text{i}} \ge 0$ are the masses of the neutrinos.
The minimal case includes just the SM neutrinos, with $\text{n} = 3$ and $U$ the 
Pontecorvo–Maki–Nakagawa–Sakata (PMNS) mixing matrix.
It is common practice to define
\begin{equation}
  m_{\beta\beta} =
  \left| \sum_{i=1}^3 | U_{\text{ei}}^2| \ e^{i\varphi_{\text{i}}} \  m_{\text{i}}  \right|,
  \label{eq:mbb_definition}
\end{equation}
where the $\varphi_{\text{i}}$ are called Majorana phases and cannot be probed by oscillation experiments.
\mbb\ is the ee-element of the mass matrix  $| (M_\nu)_{\text{ee} } |$, and thus
is also referred to as ``the effective Majorana mass'' of the electron neutrino.
This Majorana mass term changes the electronic lepton number by two units,
and contributes linearly to the \nubb-decay amplitude.

The free Lagrangian density  for a single neutrino is
\begin{equation}
  \mathcal{L}=i\, \bar{\nu}_{\text{L}}\, \partial_\mu \gamma^\mu\, \nu_{\text{L}} + \frac{m}{2} \nu^t _{\text{L}}C^\dagger  \nu_{\text{L}}
  -\frac{m}{2} \bar{\nu}_{\text{L}} C \bar{\nu}_{\text{L}}^t,
\end{equation}
where $\partial_\mu=\partial/\partial x^\mu$, $\gamma^\mu$ are the $4\times 4$ Dirac matrices and $m$ is a mass parameter that can be chosen to be real and positive by changing the phase of $\nu$.
Adding the total derivative term $-i/2 \partial_\mu( \bar\nu_{\text{L}} \gamma^\mu \nu_{\text{L}} )$
does not change the action, and introducing the Majorana spinor\footnote{Majorana spinors are self-conjugate:
from $\bar\chi= \bar\nu_{\text{L}} - {\nu}_{\text{L}}^t  C^\dagger$,
we find immediately $C  \bar\chi^t=\chi$.
In some sense, the particle and anti-particle nature of a Majorana particle coexist.
}
\begin{equation}
  \chi=\nu_{\text{L}} + C \bar{\nu}_{\text{L}}^t,
\end{equation}
the Lagrangian density reads the same as the usual free case, apart from the factor of 2 because the field is self-conjugated:
\begin{equation}
  \mathcal{L}= \frac{i}{2} \bar\chi\,  \partial_\mu \gamma^\mu\, \chi - \frac{m}{2}  \bar\chi \chi.
\end{equation}
Introducing the left chirality projector $P_{\mbox{\tiny L}}=(1-\gamma_5)/2$, and noting
that $\nu_{\text{L}}=P_{\text{L}} \chi$, 
we find the lepton number violating propagator that describes the exchange of virtual Majorana neutrinos:
\begin{multline}
   P_{\text{L}} \langle 0| T [\chi(x) \bar\chi(y) ] | 0 \rangle P_{\text{L}} =\\ 
   \hspace{21mm} = m\times  \int \mathrm{d}^4q\, \frac{i\,  P_L \, e^{-i 
       q(x-y)}}{q^2-m^2+i 0^+}\\
   = - \langle 0| T [\nu_{\text{L}}(x) \nu_{\text{L}}(y) ] | 0 \rangle C^\dagger,
\end{multline} 
where $|0\rangle$ is the vacuum state and $T$ indicates that the product of the quantised neutrino fields is time-ordered.

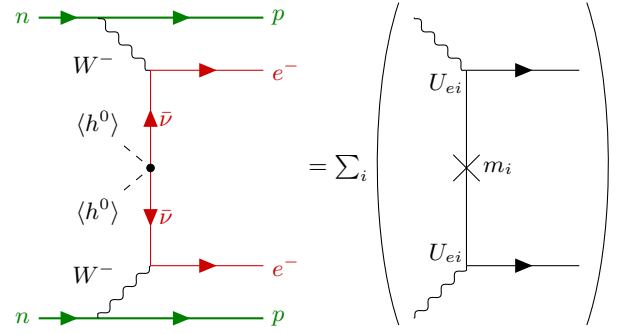
\begin{figure}[tbp]
  \begin{tikzpicture}
  \begin{feynman}
    \vertex[green!50!black](a) {\(n\)};
    \vertex[right=10mm of a] (b);
    \vertex[right=7mm  of b] (c);
    \vertex[right=15mm of c, green!50!black] (d){\(p\)};

    \vertex[below=40mm of a, green!50!black] (e) {\(n\)};
    \vertex[below=40mm of b] (f);
    \vertex[below=40mm of c] (g);
    \vertex[below=40mm of d, green!50!black] (h) {\(p\)};

    \vertex[below=7mm of c] (i);
    \vertex[above=7mm of g] (j);

    \vertex[below=6mm of i, red!80!black] (k) {};
    \vertex[above=6mm of j, red!80!black] (l) {};
    \vertex[right=15mm of i, red!80!black] (m) {\(e^-\)};
    \vertex[right=15mm of j, red!80!black] (n) {\(e^-\)};
    \vertex[below=13mm of i, red!80!black] (p) ;

    \vertex[left=7mm of k] (q){\(\langle h^{0}\rangle\)};
    \vertex[left=7mm of l] (r){\(\langle h^{0}\rangle\)};
    \diagram* {
      (a)-- [fermion, green!50!black, thick] (b)-- [fermion, green!50!black, thick] (d),
      (e)-- [fermion, green!50!black, thick] (f)-- [fermion, green!50!black, thick] (h),
      (b)-- [boson,edge label'=\(W^{-}\)] (i),
      (j)-- [boson,edge label'=\(W^{-}\)] (f),
      (p)-- [fermion, edge label'=\(\bar{\nu}\), red!80!black] (i),
(q)-- [dashed](p),
      (r)-- [dashed](p),
      (p)-- [fermion, edge label=\(\bar{\nu}\), red!80!black] (j),
      (i)-- [fermion, red!80!black] (m),
      (j)-- [fermion, red!80!black] (n),
    };

    \draw[fill=black] (p)circle(0.5mm);
    
    \vertex[right=42mm of b] (A);
    \vertex[right=42mm of f] (B);

    \vertex[right=7mm of A] (C);
    \vertex[below=7mm of C] (D);
    \vertex[right=7mm of B] (E);
    \vertex[above=7mm of E] (F);
    \vertex[right=15mm of D] (G);
    \vertex[right=15mm of F] (H);

    \vertex[below left=-1mm of D] (I){\(U_{ei}\)};
    \vertex[above left=-1mm of F] (J){\(U_{ei}\)};
    \diagram* {
      (A)-- [boson] (D),
      (B)-- [boson] (F),
      (D)-- [fermion] (G),
      (F)-- [fermion] (H),
      (D)-- [insertion={[size=5pt].5}, edge label=\ \(m_i\)] (F)
};

    \vertex[right=25mm of p] (sumvertex);
    \node(sum) at (sumvertex){$=\sum_i$};
    \vertex[right=8mm of sumvertex] (K);
  \vertex[above=22mm of K] (L);
\draw (L)arc (130:230:8mm and 28mm);
\vertex[right=25mm of L] (M);
\draw (M)arc (50:-50:8mm and 28mm);

\end{feynman}
\end{tikzpicture}
   \caption{Feynman diagram of \nubb\ decay with light neutrino exchange.
On the right,  we show the corresponding scheme in terms of neutrino mass eigenstates
    and the PMNS mixing matrix $U$.
    }
  \label{fig:feynman_0nbb}
\end{figure}

Considering the SM electron neutrino, $\nu_{\text{e}} = \sum_i U_{\text{ei}} \nu_{\text{i}}$, 
the only modifications required to describe the propagator
that appears in \nubb\ decay are i)~including the factor $U_{\text{ei}}^2$ and 
ii)~using also $m_{\text{i}}$ for each massive neutrino state. 
Using this propagator to compute the decay rate,
only the absolute value of the parameter matters.
Thus the practical recipe is to replace 
$m\to |\sum_i U_{\text{ei}}^2\ m_{\text{i}} | \equiv m_{\beta\beta}$.
Figure~\ref{fig:feynman_0nbb} shows the Feynman diagram for \nubb\ decay with light neutrino exchange.
 
Note finally that Majorana mass terms violate the SM hypercharge symmetry.
However, this violation can be attributed to the Higgs field vacuum expectation value,
i.e., to SSB of the electroweak group.
  
\subsubsection{Implications for \nubb\ decay}

As previously discussed, several operators can contribute to \nubb\ decay.
Regardless of which are the responsible BSM mechanisms,
the decay rate can be divided into four pieces. 
The first is the phase-space factor $G$ that indicates the feasibility of the decay according to its kinematics.
Its value depends mainly on the energy difference between the initial and final states, or \Qbb.
The second piece is a hadronic matrix element $g$ that encodes the coupling of the weak interaction
to nucleons. In Fermi and Gamow-Teller (GT) transitions this is given by $g_V$ and $g_A$, respectively, while for \nubb\ decay a genuine two-nucleon coupling $g^{\text{NN}}$ needs to be considered as well. 
The third piece is a nuclear matrix element (NME) $M$ that represents the
amplitude for the nuclear transition from the initial to the final state nucleus.
NMEs depend on the nuclear structure of the initial and final nuclei, and also on the nuclear transition operator,
and are covered extensively in Sec.~\ref{sec:nuclear_theory}.
Finally, the decay rate also depends on the responsible BSM mechanism, introducing the scale $\Lambda$ associated with lepton-number violation.
Considering all possible decay channels $i$, the schematic expression for the \nubb-decay rate $\Gamma$ can be written in terms of the half-life $T_{1/2}$ as:
\begin{align}
  \frac{\Gamma_{0\nu}}{\ln{2}} = \frac{1}{T_{1/2}^{0\nu}} = \sum_i G_i\,g_i^4\,|M_i|^2\,f_{i}(\Lambda) + \text{interference terms},
\label{eq:0nbb_rate}
\end{align}
where $f_{i}$ is a dimensionless function encompassing BSM physics. In the case
of light neutrino exchange, $f_{i}$ is conventionally written as the square of
\mbb\ normalized by the square of the electron mass.

The evidence of neutrino masses and the fact that the Weinberg operator has the lowest dimension
suggests that the leading contribution to \nubb\ decay is likely due to Majorana neutrino masses.
From this point of view, the discussion of a full model might be considered premature,
as was the $W$-boson hypothesis right after the discovery of Fermi interactions.
On the other hand, it is not possible to exclude  {\it a priori} the possibility
that the scale of lepton number violation is not far from the one probed with accelerators or rare decays.

In this case a new question arises: how do we avoid an exceedingly large
value of neutrino masses and in particular of \mbb?
A more detailed discussion on this topic is given in \textcite{deGouvea:2007qla} and \textcite{Mitra:2011qr}.
Solving this type of situation is possible if the light neutrino masses
are connected  to small Yukawa couplings, see, e.g., \textcite{Maiezza:2010ic} for a model based on left-right symmetry. 

A very well known consideration is the so-called black box or Schechter-Valle theorem --- even though the term theorem can be disputed and is not used by the authors. 
The original work \cite{Schechter:1981bd} states that 
{\em the observation of \nubb\ decay implies the existence of a Majorana 
  mass term for the neutrino for a ``natural'' gauge theory,}
and further specifies that
{\em one postulates a ``strong-naturality'' in which no global conservation laws are assumed ``a priori''.}
Thus obtaining a quantitative statement  on $m_{\beta\beta}=0$ is possible only within a model.
In a minimal setup, the value of \mbb\ induced by the black box diagram is so small that it lacks any practical interest \cite{Duerr:2011zd}.
Moreover and most simply, it seems  possible to arrange for \mbb$=0$
without contradicting the current knowledge of neutrino masses.
In fact, considering that \mbb\ is the $m_{ee}$ component of $M_\nu$, it is easy to imagine the elements of $M_\nu$ falling into a hierarchy resembling those of the other SM fermions, in which case $m_{ee}$ could be exceedingly small~\cite{DeGouvea2022SM}.
\jd{Francesco check the last sentence. Need to find a (better) reference.}

Considering only the light neutrino exchange contribution to \nubb\ decay,
Eq.~\eqref{eq:0nbb_rate} simplifies to
\begin{align}
  \frac{1}{T_{1/2}^{0\nu}} =& G_{01}\,g_A^4\,\left(M_{\text{light}}^{0\nu}\right)^2\,\frac{m_{\beta\beta}^2}{m_e^2}, \nonumber \\
  M_{\text{light}}^{0\nu}=&M_{\text{long}}^{0\nu}+M_{\text{short}}^{0\nu},
  \label{eq:mbb}
\end{align}
where $G_{01}$ and $M_{\text{light}}^{0\nu}$ are the phase space and NME specific to light neutrino exchange, respectively. Eq.~\eqref{eq:mbb} already reflects a long- and short-range contribution to the NME. For simplicity the dominant coupling of the long-range part, $g_A$, is factored out, but the short-range part is proportional to another two-nucleon coupling $g^{\text{NN}}$.
More details are given in Sec.~\ref{sec:eft}.

\subsubsection{Predictions on \mbb}
\label{sec:par:mbb:predictions}

The definition of \mbb\ given in Eq.~\ref{eq:mbb_definition}
shows how this quantity depends on a total of seven parameters, 
as only $\theta_{12}$ and $\theta_{13}$ enter $U_{ei}$,
and only two Majorana phases are non degenerate. 
Neutrino oscillation experiments are sensitive only to the two mixing angles,
two neutrino mass squared differences, and the mass ordering. 
Thus experimental data can currently bound only four out of seven degrees of freedom, leaving the other three fully unconstrained.
Two of these unconstrained degrees of freedom are naturally associated to the Majorana phases. The third one is related to the three neutrino masses $m_i$, which are constrained by the measurements of only two mass squared differences. 
This freedom raises the question of how to predict the value of 
\mbb, an issue discussed since \textcite{Vissani:1999tu}.

One option is to constrain the remaining parameters
using theoretical considerations of neutrino masses,
but  despite the wide literature  on the subject
we cannot make any definitive statement yet.
Some models have been considered more appealing,
for instance those based on the gauge principle,
or those trying to connect neutrino masses to the masses of other fermions,
or perhaps, to a lesser extent,
those  predicting a more easily explorable parameter space.
The challenge is not the shortage but rather the overabundance of proposals,
and the lack of criteria to identify the correct one, if any.
The history of the theoretical investigation of neutrinos
has produced wrong predictions at almost every turn:
parity was supposed to be respected but is maximally violated in neutrino 
interactions;
$\theta_{12}$ was supposed to be small, but it is about $30^\circ$;
$\theta_{13}$ was thought to be very small until recently, when it was found to
be as large as the Cabibbo angle;
neutrinos were supposed to give a large (or significant) contribution to
the cosmological energy density, 
but apparently they do not;
etc. 
In short, history calls for caution toward a purely theoretical approach 
to making useful predictions on \mbb.

\begin{figure*}[t]
  \centering
  \includegraphics[width=\textwidth]{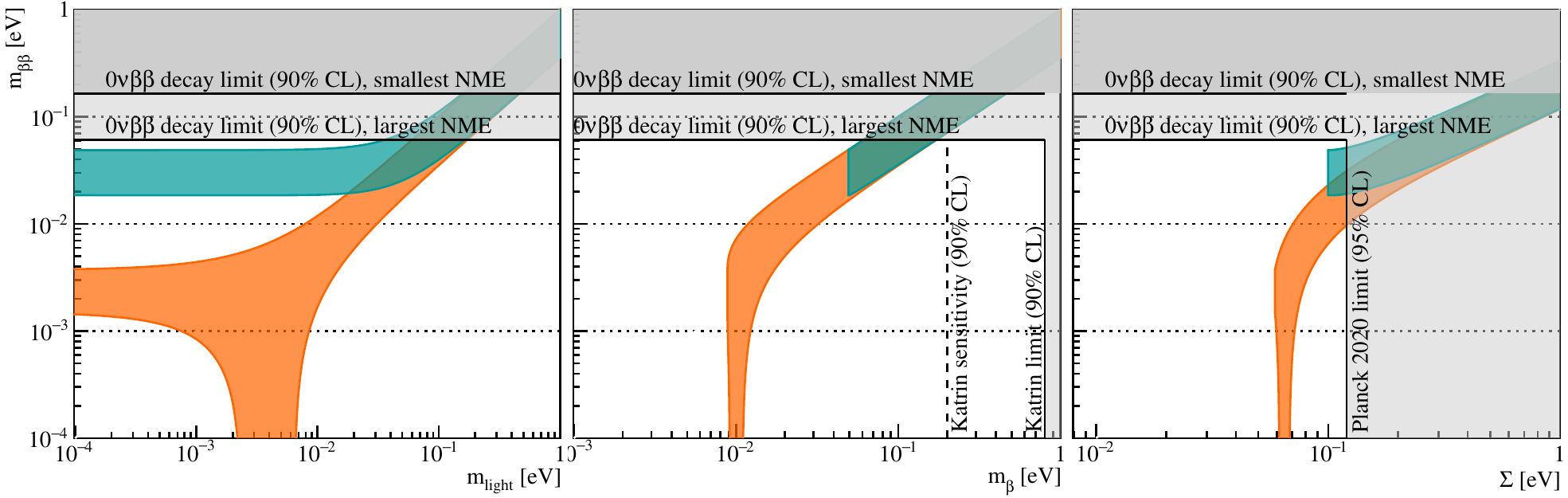}
  \caption{Maximally allowed parameter space for \mbb\  as a function of
  $m_{\text{light}}$, \mne, and $\Sigma$, assuming the
central value of the neutrino oscillation parameters \cite{Zyla:2020zbs}. The
orange and green areas show the parameter space allowed assuming normal and
inverted ordering, respectively.  The shaded areas indicate the regions already
excluded by \nubb-decay experiments \cite{KamLAND-Zen:2016pfg} and cosmological
observations \cite{Planck:2018vyg}; the vertical lines in the middle panel
correspond to the KATRIN limit \cite{KATRIN:2021uub} and
sensitivity \cite{Aker:2019uuj}.}
\label{fig:bikini}
\end{figure*}

In early investigations, predictions for \mbb\ were often obtained by assuming special values for its three unconstrained degrees of freedom. In particular, the Majorana phases were frequently set to be zero, or such as to provide special values of $e^{i \varphi_i}$, e.g., real values.
In recent times, the focus has shifted on the maximally allowed range of \mbb\ values. This is derived by leaving the Majorana phases free to minimize and maximixe \mbb\ for any choice of the last degree of freedom associated to neutrino masses. 
Analytic expressions defining the extreme \mbb\ values are compact \cite{Vissani:1999tu}:
\begin{align}
  m_{\beta\beta}^{\text{max}} &= \sum_{i=1}^3 |  U_{\text{e} i}^2 | m_i,  \nonumber\\ 
  m_{\beta\beta}^{\text{min}} &= \text{max}\!\left\{ 2 |  U_{\text{e} i}^2 | m_i - m_{\beta\beta}^{\text{max}} \ , \ 0\right\}.
  \label{eq:bikini}
\end{align}
The third degree of freedom is often parameterized using the lightest neutrino mass $m_{\text{light}}$ \cite{Vissani:1999tu}. Other conventional options are the observables measured by experiments studying \B\ spectra end-points 
--- i.e., the effective kinematic electron neutrino mass $m_{\beta}=\sqrt{\sum_i
|U_{\text{e} i}^2 | m_i^2}$ ---  or by cosmological surveys --- i.e., the sum of the neutrino masses $\Sigma=\sum_{i} m_i$ \cite{Fogli:2004as}.
Figure~\ref{fig:bikini} shows the maximally allowed range for \mbb\ as a function of these three parameterizations.
The ambiguity in the neutrino mass ordering (NO vs. IO) results in two
distinct regions, which
overlap at high (degenerate) neutrino mass scales, but separate at lower values.
It is within these regions that experiments can test \nubb\ decay via light neutrino exchange.
In view of recent analyses showing some preference for NO, one of the two regions might be favored, but these are still mild indications at the moment as discussed in Sec.~\ref{sec:par:mbb:osc}.

Next-generation \nubb-decay experiments will fully probe the parameter space
allowed for inverted ordered neutrinos, for which the smallest allowed \mbb\
value is 18.4$\pm$1.3\,meV \cite{Agostini:2021kba}. At the same time, these
experiments will also test a significant fraction of the parameter space allowed
for the normal ordering. However, for the normal ordering there is no lower bound on \mbb, which could be extremely small or even null, far out of the reach of conceivable future searches. 
If neutrinos are Majorana particles, data on \mbb\ will indirectly constrain also $m_{\beta}$ and $\Sigma$, and vice versa, creating an exciting interplay among future experiments.

The most stringent constraints on $m_{\beta}$ come from the KATRIN experiment, which 
is designed to kinematically measure the mass of the electron antineutrino with sub-eV precision, by reconstructing the energy distribution of the electrons emitted in tritium $\beta$ decays close to the end-point. 
In the next few years, KATRIN will push the exploration of $m_{\beta}$ values from the current limit of 0.8\,eV \cite{KATRIN:2021uub}
down to 0.2\,eV \cite{Aker:2019uuj}. Any measurement of $m_{\beta}$ in this
range would be incompatible with the existing limits on \nubb-decay unless neutrinos are Dirac particles. In the Majorana neutrino scenario, it would
hint towards non-standard neutrino models (and cosmological models), and/or alternative \nubb-decay mechanisms.

Cosmological data are strongly sensitive to the neutrino radiation density and
the neutrino masses, which affect both Big Bang nucleosynthesis and the large
scale structure of the universe, inducing characteristic signatures in the relative abundance of elements and the cosmic microwave background (CMB) / baryon acoustic oscillation (BAO) power spectra. These effects are covered in several reviews, for instance \textcite{Dolgov:2002wy}, \textcite{Patterson:2015xja}, \textcite{Archidiacono:2016lnv}, and \textcite{Lattanzi:2017ubx}.
Neutrino constraints coming from cosmology are relatively robust, even though they are not as direct as those from laboratory experiments, and need to rely on the Standard Model for cosmology, called $\Lambda$CDM.
The current bound on the sum of the neutrino masses is $\Sigma<120$\,meV \cite{Planck:2018vyg}. 
It stems from the combination of large-scale structure measurements due to
Planck with other measurements at small scales, including lensing and  BAO data.
There exist other sensitive data, such as measurements of the Lyman-alpha forest. Their inclusion 
helps to break some degeneracies, typically yielding stronger constraints on $\Sigma$ \cite{Palanque-Delabrouille:2019iyz,DiValentino:2021hoh}.
The analysis is also relatively robust against standard modifications of $\Lambda$CDM.

\begin{figure}[t]
  \centering
  \includegraphics[width=\columnwidth]{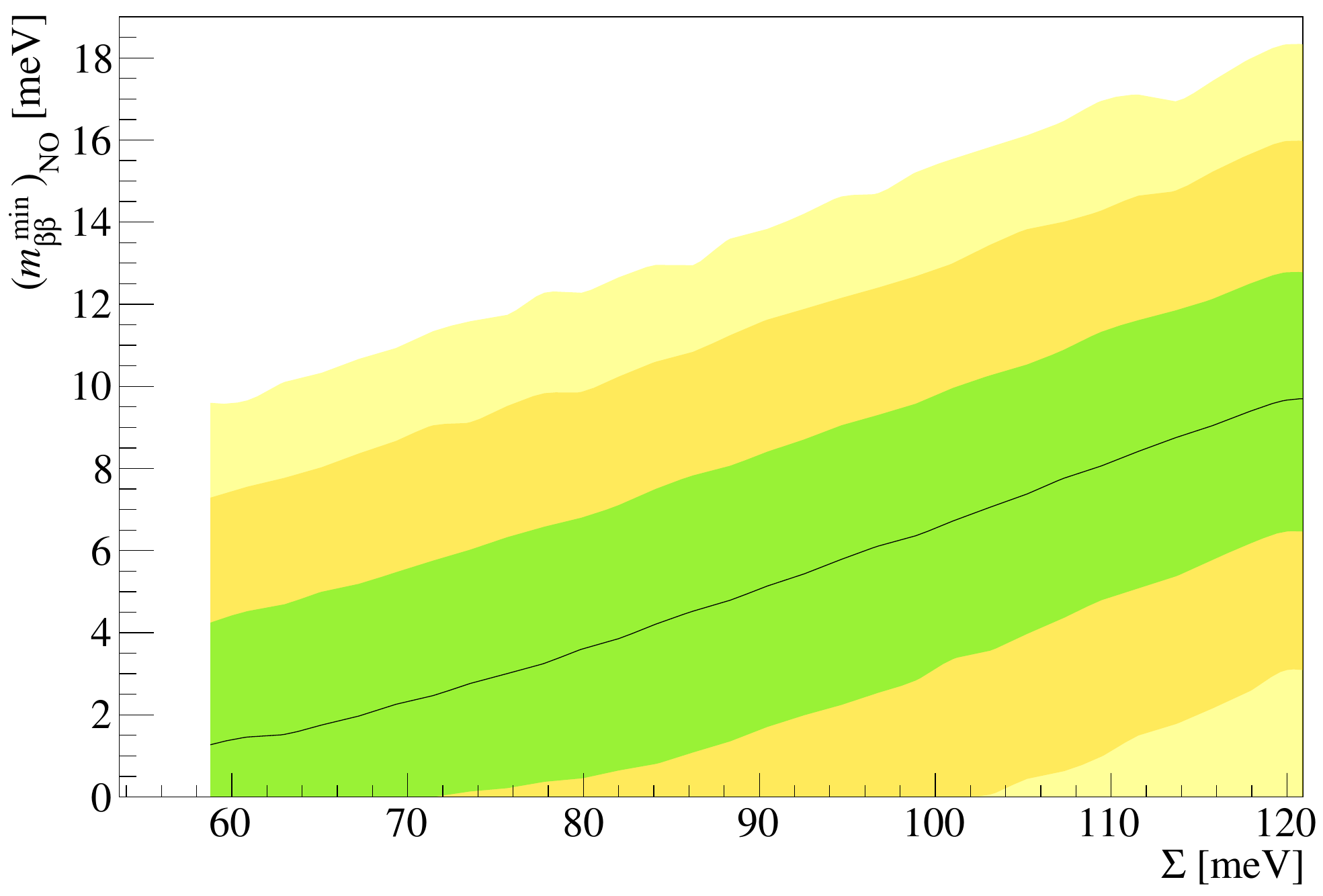}
  \caption{Posterior probability distribution of the lower bound on \mbb\ as a
  function of the true value of $\Sigma$, assuming normal ordering. The
  distribution is constructed by random sampling of the oscillation parameters within their Gaussian uncertainties \cite{Zyla:2020zbs} and assuming that $\Sigma$ will be measured with 20\,meV accuracy. The solid black line shows the median lower bound, while the green, orange and yellow color bands show the 68\%, 95\% and 99\% probability central interval of the distribution. Note that the median limit does not go to zero, not even around 65\,meV when \mbb\ can vanish, as the limit is averaged over an extended $\Sigma$ range accounting for the measurement uncertainty.  The lower bound for the inverted ordering scenario is always larger than that for normal ordering.  }
\label{fig:euclid}
\end{figure}

The next surveys, for instance DESI and EUCLID, aim at measuring $\Sigma$ with an accuracy of 20\,meV \cite{Font-Ribera:2013rwa,Kitching:2014lga}.
Such a measurement will have important implications for \nubb\ decay.
First, the lowest value of $\Sigma$ is bounded by the measurement of the neutrino mass squared differences. This minimum value is $\Sigma>59$\,meV for the normal ordering, and 
$\Sigma>100$\,meV for the inverted one, assuming the central values of the neutrino oscillation parameters \cite{Zyla:2020zbs}. 
This means that the next surveys are guaranteed to resolve a value for $\Sigma$
consistent with these limits if the $\Lambda$CDM paradigm is valid and consistent with
Standard-Model physics.
Further, measurement of $\Sigma$ below 100\,meV would disfavor the inverted ordering hypothesis, as pointed out in \textcite{DellOro:2015kys}.
Moreover, any measurement of $\Sigma$ would naturally set a lower bound on \mbb,
even in the case of the normal ordering. This is already qualitatively visible
in Fig.~\ref{fig:bikini}, but a proper estimation needs to take into account all uncertainties on the oscillation parameters and the anticipated 20\,meV accuracy of the measurement on $\Sigma$.
Figure~\ref{fig:euclid} shows the dependence of the lower bound on \mbb\ on the
true unknown value of $\Sigma$, obtained by propagating all uncertainties via random sampling. 
Should the value of the neutrino mass sum be just below the current limits, $\mbb$ would be bounded to be larger than 10\,meV, a value testable by the coming \nubb-decay experiments assuming favorable NME calculations.

We close this section with an important remark concerning the normal mass ordering parameter space.
Although vanishing \mbb\ values are possible from a mathematical and empirical point of view, the question of whether this is plausible or not is much more subtle.
Figure~\ref{fig:bikini} shows the maximum allowed parameter space on bilogarithmic scales.  This choice under-emphasizes the value of the observational progress and stresses somewhat artificially the role of the lowest values of the masses. In the future, a linear or even bilinear scale might be appropriate; indeed some experiments have begun to plot their results in this way~\cite{KamLAND-Zen:2022tow,Arnquist:2022zrp}.

Recent Bayesian analyses have tried to build a probability distribution for \mbb,
at the price of making assumptions on the (prior) probability distribution for the Majorana phases
and the additional free mass scale parameter, be it $m_{\text{light}}$, \mne\ or $\Sigma.$
If one invokes ``naturalness'' arguments and
parameterize the ignorance on the Majorana phases with a flat prior,  
vanishing \mbb\ values get strongly disfavored as first pointed out by
\textcite{Benato:2015via}, \textcite{Agostini:2017jim}, and \textcite{Caldwell:2017mqu}. One could also try to consider the less favorable value of the Majorana phases and quantify the minimal discovery probabilities \cite{Agostini:2020oiv}.
Finally,  flavor symmetry can also be invoked to constrain at the same time the phases and, e.g, $m_{\text{light}}$, bringing a large part of the parameter space for normal ordering within the reach of the forthcoming experiments \cite{Agostini:2015dna}. These analyses identified several scenarios in which the discovery power for future
experiments is significant, even considering normal-ordered neutrino masses. The more the priors disfavor vanishing values for the lightest neutrino mass and cancelling Majorana phases, the higher the discovery power. The dependence on the prior on the lightest neutrino mass will significantly weaken in the future should the value of $\Sigma$ be measured by cosmological surveys~\cite{Ettengruber:2022mtm}.

Although we have already warned the reader against 
making predictions on \mbb\ using purely theoretical arguments, we want to draw
the attention to the broad class of models examined in a number of articles
\cite{Vissani:1998xg,Vissani:2001im,DellOro:2017pgd,DellOro:2018jze}, which
merely focus on the coarse structure of the neutrino mass matrix without
claiming an understanding of the coefficients of order 1.
This class of mass matrices
correctly anticipated the large mixing angle solution and
the fact that $\theta_{13}$ is of the order of the Cabibbo angle $\theta_{\mbox{\tiny C}}\sim 0.2$,  and they also
 predict the normal ordering scenario currently favored by available data. They were proposed  after the first evidence appeared that the atmospheric neutrino mixing is large, which showed that the neutrino mass matrix  deviates from the hierarchical and quasi-diagonal structure typical of the Yukawa couplings of charged fermions. This consideration leads to the reasonable assumption that the elements of the $\mu-\tau$ block are larger than the others \cite{Vissani:1998xg}.
According to these models, one would expect
\begin{equation}
  \label{atm}
m_{\beta\beta} = \mathcal{O}(1) \times 
\sqrt{ \Delta m^2_{\mbox{\tiny atm}} } \times 
  \theta_{\mbox{\tiny C}}^n \mbox{ with }n=1\mbox{ or }2
\end{equation}
where $\Delta m^2_{\mbox{\tiny atm}}$ is the parameter probed by atmospheric neutrino oscillation, i.e., the mass squared difference $|m^3_3-m^2_1|$ or $|m^3_3-m^2_2|$ depending on the mass ordering.
This leads to $\mbb \approx 10$\,meV  or 2\,meV. Of course, this cannot be considered as a replacement for a complete theory. But it is interesting and gratifying that the explorations that have been conducted on motivated models and in particular those based on SO(10) \cite{Matsuda:2001bg,Bajc:2005zf,Joshipura:2011nn,
Bertolini:2012im,Buccella:2012kc,Dueck:2013gca,Altarelli:2013aqa,Ohlsson:2021lro}
are consistent with these generic expectations. 

Another mass scale of interest for \mbb\ is given by the solar neutrino mass squared difference: 
\begin{equation}
  \mbb\sim\sqrt{\Delta m^2_{\mbox{\tiny sol}}}=8.6\pm0.1\,\mbox{meV},
\end{equation}
with $\Delta m^2_{\mbox{\tiny sol}} = m^2_2-m^2_1$.
This mass scale has been precisely measured by neutrino oscillation measurements, and typical models with NO neutrino masses favor \mbb\ values around this magnitude. Its numerical value is similar to what is obtained by Eq.~\ref{atm} assuming $n=1$, i.e., \mbb$\approx$10\,meV. 

Thus there is an accumulation of theoretical motivation for exploring \mbb\ values around 8--10\,meV.  This scale is interesting also from the experimental point of view: it is almost in the middle of the parameter space remaining after reaching the bottom of the inverted ordering, and can constitute a challenging, and yet conceivable goal for the experimental community.
Future experiments able to explore this parameter space would have exciting discovery opportunities as it does not seem very plausible that \mbb\ is exactly zero.

However, it is clear that we need more precise indications from theory to guide
the experimental program. In particular, it seems more important than ever to bring to full maturity the design of a predictive and motivated model of neutrino and charged fermion masses based on reliable theoretical principles, for example, SO(10).

\subsection{The cosmic baryon excess and models of its origin}
\label{sec:par:cosmo}

Before closing this section on theoretical motivations from particle physics, we return to the question of the cosmic baryon excess and its relationship with \nubb\ decay.

While particles and antiparticles are basically equivalent at the level of fundamental physics, on a cosmic scale, the universe contains only baryons. As discussed in Sec.~\ref{sec:par:cos:cos1}, the Standard Model is unable to account for this observational fact, and this suggests that there may have been some unknown physics at work in the early universe. A large and very interesting class of extensions of the Standard Model succeeds in this task by using the same ingredients that explain the masses of light neutrinos and/or give rise to leptonic number violation phenomena: these are the leptogenesis models, described in Sec.~\ref{sec:par:cos:cos2}. We discuss the connection between these models and \nubb\ decay in Sec.~\ref{sec:par:cos:cos3}, attempting an assessment on the most promising models.

\subsubsection{Observations and theoretical challenges}
\label{sec:par:cos:cos1}

Cosmology has collected evidence that the universe contains only baryons.   
Their amount has been measured in several ways:  
in the present universe, with direct astronomical observations \cite{Tanimura:2017ixt,deGraaff:2017byg};
at recombination time, with  the study of the cosmic microwave background \cite{Aghanim:2018eyx};  and
at much earlier times, with big-bang nucleosynthesis \cite{Pisanti:2020amb}.
These determinations, especially the last two, are rather precise and compatible with each other.
The amount of anti-baryons is insignificant and consistent with secondary production mechanisms.
The lepton asymmetry, stored in the neutrinos produced in the big-bang, is only loosely bounded by observations of  
primordial nucleosynthesis \cite{Mangano:2011ip}. If it is similar in size to the baryonic one, it is practically impossible to measure.

The meaning of the observed baryon excess has been widely discussed in the context of theoretical cosmology. Following \cite{Sakharov:1967dj}, it was discussed which models were able to provide sufficient violations of global symmetries and $CP$, to dynamically generate cosmic baryon asymmetry. 
Recall, the SM predicts non-perturbative processes that violate $B+L$ \cite{tHooft:1976rip,Kuzmin:1985mm,Harvey:1990qw}.
However, when their effect is quantified in the context of cosmological evolution,
they prove insufficient to account for the observed asymmetry \cite{Bochkarev:1987wf,Kajantie:1995kf}.
Thus a dynamical explanation of the origin of the baryon excess is possible  only in a suitable SM extension;
such a theoretical program goes under the name {\em baryogenesis}.
A new source of violation of global symmetries ($B$ and $L$)  from physics beyond the SM is necessary for any successful explanation of the cosmic baryon excess.
The hypothetical observation of lepton number violation in the laboratory would give strong support to this interpretation, even before reaching quantitative predictions.

\subsubsection{Leptogenesis models}
\label{sec:par:cos:cos2}

A specific class of SM extensions, called baryogenesis through leptogenesis or in short 
{\em leptogenesis} models, explain the cosmic baryon density though lepton number violating effects. Most typically, these 
rely on the same ingredients that also explain neutrino masses (see Sec.~\ref{sec:par:LNV:rh}).

The first proposal of \textcite{Fukugita:1986hr} 
is based only on the existence of right-handed neutrinos with very large (GUT scale) Majorana masses.
Their decays out of equilibrium lead to a leptonic asymmetry  $\Delta L$,
due  to interference effects in the decay of the heavy neutrinos  
beyond lowest perturbation order and  due to complex ($CP$-violating) Yukawa couplings.
Subsequently, the non-perturbative SM processes mentioned above that violate $B+L$
convert this leptonic asymmetry into the cosmic baryon excess. The same process also
leaves a comparable asymmetry between neutrinos and antineutrinos, a determination of which is beyond experimental reach.

The issue of model dependence cannot be ignored.
For example, the Grand Unified SO(10) models discussed in Sec.~\ref{sec:par:LNV:gu} contain heavy right-handed neutrinos 
and can thus be considered to be in the class of models required by the original leptogenesis proposal,
but they also contain other sources of lepton number violation, such as SU(2) triplets, which makes it less easy to study the consequences and draw unambiguous conclusions from the theory.
In  fact, the number of variants of leptogenesis models that are formally allowed is very large \cite{Shaposhnikov:2009zzb},
and some of them correspond to very different scenarios. 

Interestingly, it is possible to build models that involve relatively light new particles, potentially within the reach of laboratory experiments. One such model is the mechanism of \cite{Akhmedov:1998qx}, which is compatible with the $\nu$SM \cite{Asaka:2005an},  
but does not change the rate of \nubb\ \cite{Bezrukov:2005mx}. Furthermore, there is a  broad class of low-scale leptogenesis models (mentioned in Sec.~\ref{sec:par:LNV:fp}) that can be verified in the laboratory, especially through the search for \nubb,
as evidenced in a number of papers \cite{Drewes:2016lqo,Hernandez:2016kel,Drewes:2016jae,Drewes:2021nqr,Drewes:2022kap}.

\subsubsection{Provisional assessments}
\label{sec:par:cos:cos3}

The generic scenario described 
above for the origin of cosmic baryons 
is not precise enough to be verifiable, but it can be qualitatively corroborated by laboratory measurements
on, e.g., the Majorana character of neutrino masses
and CP violation in neutrino oscillation.
It has at least been observed that baryogenesis at a high energy scale is hardly compatible with any mechanism causing \nubb\ decay other than the exchange of Majorana neutrinos \cite{Deppisch:2015yqa}.
It is also noteworthy that the long baseline searches for CP violation have recently received strong
support \cite{HyperKFunded, HEPAPSubcommittee:2014bsm}.
Thus although leptogenesis cannot be tested directly by laboratory measurements,
the experimental community is at least poised to deeply probe its key testable predictions.
    
The fact that to date we can only observe the cosmic baryon excess,
but we have very few other possibilities to test our ideas about it, sometimes induces a certain discouragement.
Perhaps, in view of the provisional character of present knowledge, baryogenesis should be regarded not as a theoretical need, but just as a point in favor for SM extensions that can model it.

However, the original models, in which baryon/leptogenesis occurs at very high energy scales, seem much more promising in the perspective of a unified theory. 
For example, in unified theories such as SO(10), one can explain small neutrino masses by the seesaw mechanism, one can incorporate a correspondence between quarks and leptons (which has been a good theoretical guide in the past), and there is no need to invoke strong differences in their Yukawa couplings.
In this spirit, and for the purposes of experimental investigations, 
it seems reasonable to us to consider this hypothesis as the reference one.
 \section{Nuclear physics theory and implications}
\label{sec:nuclear_theory}
\label{sec:nt}
\label{sec:nuc}

Most atomic nuclei are unstable because of the weak interaction.
Nuclei decay by emitting or capturing electrons --- known as \B\ decay or electron capture (EC), respectively --- resulting in a final nuclide more bound than the initial one and with the same number of nucleons.
In \B\ decay a neutron turns into a proton, while the opposite occurs in EC, so that electric charge is conserved.
In addition, either neutrinos (in EC) or antineutrinos (in \B\ decay) are emitted to conserve energy, momentum and lepton number. In a nucleus, $\beta$ decay can also turn a proton into a neutron, but this is disfavored with respect to EC because a positron needs to be produced, reducing the available energy: $Q_{\beta^+}=Q_{\text{EC}}-2m_e$.

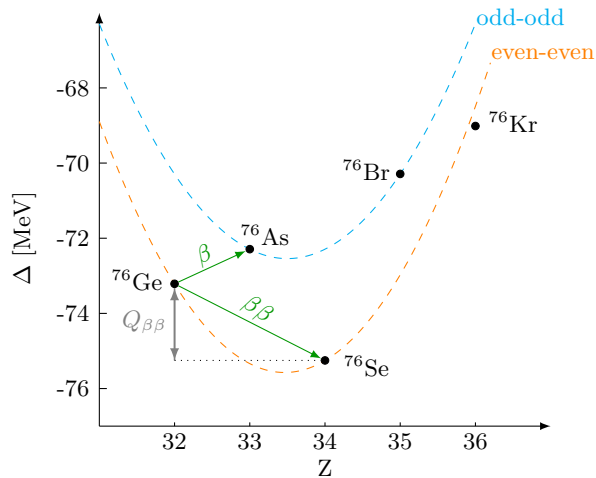
\begin{figure}
  \begin{tikzpicture}[xscale=1,yscale=0.5]
    \definecolor{greenish}{RGB}{00,150,00}
    \definecolor{blueish} {RGB}{00,50,230}
\def\xmin{31}
    \def\xmax{37}
    \def\ymin{-77}
    \def\ymax{-66}
    \draw [-latex] (\xmin,\ymin) -- (\xmax,\ymin);
    \draw [-latex] (\xmin,\ymin) -- (\xmin,\ymax);

\def\tick{0.1}
    \draw (32,\ymin) -- (32,\ymin+\tick);
    \draw (33,\ymin) -- (33,\ymin+\tick);
    \draw (34,\ymin) -- (34,\ymin+\tick);
    \draw (35,\ymin) -- (35,\ymin+\tick);
    \draw (36,\ymin) -- (36,\ymin+\tick);
    \node (Z32) at (32,\ymin-0.4) {$32$};
    \node (Z33) at (33,\ymin-0.4) {$33$};
    \node (Z34) at (34,\ymin-0.4) {$34$};
    \node (Z35) at (35,\ymin-0.4) {$35$};
    \node (Z36) at (36,\ymin-0.4) {$36$};
    \node (Z) at (34,\ymin-1.1) {Z};

    \draw (\xmin,-68) -- (\xmin+\tick,-68);
    \draw (\xmin,-70) -- (\xmin+\tick,-70);
    \draw (\xmin,-72) -- (\xmin+\tick,-72);
    \draw (\xmin,-74) -- (\xmin+\tick,-74);
    \draw (\xmin,-76) -- (\xmin+\tick,-76);
    \node (Y68) at (\xmin-0.35,-68) {-$68$};
    \node (Y70) at (\xmin-0.35,-70) {-$70$};
    \node (Y72) at (\xmin-0.35,-72) {-$72$};
    \node (Y74) at (\xmin-0.35,-74) {-$74$};
    \node (Y76) at (\xmin-0.35,-76) {-$76$};
    \node [rotate=90] (Y) at (\xmin-1,-72) {$\Delta$~[MeV]};

\draw[dashed,color=orange] plot [variable=\t, domain=\xmin:\xmax-0.8, samples=100]
    ( \t, { 1.10073*\t^2 - 73.6674*\t + 1157 } );

    \draw[dashed,color=cyan] plot [variable=\t, domain=\xmin:\xmax-1, samples=100]
    ( \t, { 0.999515*\t^2 - 66.9661*\t + 1049.12 } );

    \node [color=cyan] (ee) at (\xmax-0.4,-66.1){odd-odd};
    \node [color=orange] (ee) at (\xmax-0.1,-67.1){even-even};

\def\xshift{-0.3}
    \node [draw,shape=circle,fill=black,minimum size=1mm, inner sep=0]
    (Ge76) at (32,-73.2128) {};
    \node (Ge76text) at (31.5,-73.1) {$^{76}$Ge};

    \node [draw,shape=circle,fill=black,minimum size=1mm, inner sep=0]
    (As76) at (33,-72.2908) {};
    \node (As76text) at (33.2,-71.9) {$^{76}$As};

    \node [draw,shape=circle,fill=black,minimum size=1mm, inner sep=0]
    (Se76) at (34,-75.2518) {};
    \node (Se76text) at (34.55,-75.4) {$^{76}$Se};

    \node [draw,shape=circle,fill=black,minimum size=1mm, inner sep=0]
    (Br76) at (35,-70.2889) {};
    \node (Br76text) at (34.55,-70.18) {$^{76}$Br};

    \node [draw,shape=circle,fill=black,minimum size=1mm, inner sep=0]
    (Kr76) at (36,-69.0142) {};
    \node (Kr76text) at (36.5,-68.9) {$^{76}$Kr};

\draw [latex-,color=gray,semithick] (Ge76) -- (32,-75.2518);
    \draw [latex-,color=gray,semithick] (32,-75.2518) -- (Ge76);
    \node [color=gray] at (31.6,-74.22) {\Qbb};
    \draw [dotted] (32,-75.2518) -- (Se76);

\draw [latex-,color=greenish] (Se76) -- (Ge76);
    \draw [latex-,color=greenish] (As76) -- (Ge76);
    \node [color=greenish,rotate=-33] (bb) at (33.1,-73.9) {$\beta\beta$};
    \node [color=greenish,rotate=25] (bb) at (32.4,-72.45) {$\beta$};
  \end{tikzpicture}
  \caption{Mass excess $\Delta = (m_A-A)\cdot u$
    for isobars with mass $m_A$ and mass number $A=76$,
    where $u$ is the atomic mass unit.
    Even-even nuclei are distributed on the lower curve,
    odd-odd nuclei on the top one.}\label{fig:parabola}
\end{figure}

When dominant first-order weak processes occur, second-order \bb\ decay or double EC (ECEC) are in practice impossible to observe, due to the small coupling associated with the {\it weak} interaction.
For some selected nuclei, however, \nunubb\ decay and ECEC dominate,
for instance when first-order decays are energetically forbidden while second-order channels are not.
The attractive nuclear pairing interaction brings additional binding to nuclei with even numbers of protons and neutrons, so that some even-even nuclei are more bound than their odd-odd neighbors, 
but less bound than their even-even second-neighbors.
Figure~\ref{fig:parabola} illustrates this by showing the mass excess for isobars with $A=76$ nucleons.
Alternatively, \B\ decays can be very suppressed because of a large
mismatch in total angular momentum between the initial and final nuclei, so that \B- and \nunubb-decay rates are comparable \cite{Alanssari16}.
In these special cases, \bb\ decay or ECEC can be measured.
The nucleus decays into a more bound system with two more protons and two fewer neutrons, or the other way around, emitting or capturing at the same time two electrons and the corresponding (anti)neutrinos.
Such measurements demand extremely sensitive experiments, because of the very long associated half-life values: $T^{2\nu\beta\beta}_{1/2}>10^{18}$\,yr \cite{Barabash20}.

The nuclear transition underlying \nunubb\ decay and ECEC
can be thought of as proceeding via virtual transitions
to excited states of the intermediate odd-odd nucleus, and many body methods can
be used to compute the corresponding nuclear matrix elements (NMEs), albeit with some uncertainty.
The case of \nubb\ decay is fundamentally different in two essential ways. 
First, the mediating mechanism results in significant momentum transfer between the two nucleons involved in the decay.
While \nunubb\ decay and ECEC are restricted to a subset of the intermediate
nuclear states with angular momentum-parity 1$^+$, the high momentum transfer in \nubb\ makes all intermediate states accessible. 
Second, although it is the case for light neutrino exchange, the mediating mechanism is not required to couple to the nucleons via weak interaction vertices, and thus in general the process is not always a second-order weak process.
A more generic framework is thus required to compute \nubb\ decay rates.

In this section, we first summarize in Sec.~\ref{sec:eft} the \nubb-decay rate as given by an effective field theory (EFT) that exploits the separation of scales between particle (BSM), hadron, and nuclear structure scales.
Section~\ref{sec:nme_expressions} presents expressions for the NMEs for \nubb\ decay mediated by the exchange of ``light'' and ``heavy'' particles with respect to the typical momentum transfer $p=|{\bm p}|\sim200$\,MeV, including the recently recognized short-range contribution to light-neutrino-exchange.
Section~\ref{sec:NMEs} discusses current NME calculations, while Sec.~\ref{sec:quenching} is devoted to the so-called ``$g_A$ quenching'' puzzle that could affect NME predictions.
Additional nuclear observables that test calculations and can provide information about the values of the NMEs are outlined in Sec.~\ref{sec:NME_tests}.

The content of Secs.~\ref{sec:eft},~\ref{sec:status}, ~\ref{sec:status2} and~\ref{sec:quenching} is targeted to both nonexperts and experts, while Sec.~\ref{sec:nme_expressions}, the remaining of Sec.~\ref{sec:NMEs} and the final Sec.~\ref{sec:NME_tests} cover somewhat more technical aspects.

\subsection{\nubb-decay rate in effective field theory}
\label{sec:eft}

\begin{figure}[htbp]
  \begin{tikzpicture}
    \begin{feynman}
      \vertex[](a) {\(n\)};
      \vertex[right=20mm of a] (b);
      \vertex[right=20mm of b] (c) {\(p\)};

      \vertex[below=30mm of a] (d) {\(n\)};
      \vertex[below=30mm of b] (e);
      \vertex[below=30mm of c] (f) {\(p\)};

      \vertex[below=15mm of b] (g);
      \vertex[below=10mm of c] (h) {\(e\)};
      \vertex[below=10mm of h] (i) {\(e\)};
      
      \diagram* {
        (a)-- [fermion, thick] (b)-- [fermion, thick] (c),
        (d)-- [fermion, thick] (e)-- [fermion, thick] (f),
        (b)-- [insertion={[size=5pt]1.}, edge label'=\(\nu_{M}\)] (g),
        (g)-- [edge label'=\(\nu_{M}\)] (e),
        (b)-- [fermion] (h),
        (e)-- [fermion] (i),
      };
    \end{feynman}
    \end{tikzpicture}
  \caption{Diagram representing long-range light-neutrino exchange contribution to $0\nu\beta\beta$ decay.}
  \label{fig:light_diagrams}
\end{figure}
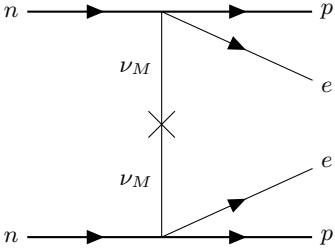

\nubb\ decay is necessarily triggered by BSM physics.
As discussed in Sec.~\ref{sec:par}, the experimentally best motivated and most studied mechanism is the exchange of the known light neutrinos --- if they are Majorana particles --- corresponding to the diagram in Fig.~\ref{fig:light_diagrams}.
This scenario predicts a \nubb-decay rate that only depends on the mass of the lightest neutrino and the neutrino mass ordering, in addition to a NME.
Nonetheless, in general any BSM extension that violates lepton number leads to \nubb\ decay.
Because BSM models are typically defined at higher energy-momentum scales than the electroweak scale ($\sim250$\,GeV), or the relevant scales for hadrons ($\sim1$\,GeV) and nuclei ($\sim m_\pi \sim200$\,MeV), an EFT approach is best suited to organize different \nubb-decay contributions \cite{Prezeau03,Cirigliano17dim7,Cirigliano18master}.
Including information from all these energy scales provides an advantage for assigning the importance of each decay channel, but valuable alternative EFTs usually neglecting chiral ($m_\pi$/GeV) aspects have also been proposed \cite{Pas99,Pas01,Horoi16,Graf18,Deppisch18,Deppisch20}.

\subsubsection{Decay amplitudes}

Above the electroweak scale, lepton number violation and therefore \nubb\ decay are usually considered to be generated by dimension-5 (light-neutrino exchange), dimension-7, or dimension-9 effective operators \cite{Cirigliano17dim7,Cirigliano18master,Graesser17}, see Sec.~\ref{sec:par:LMJ:operators}.
The operators are suppressed by the typical scale $\Lambda$ at which the BSM physics enters: $1/\Lambda$, $1/\Lambda^3$, and $1/\Lambda^5$, for dimension-5, dimension-7, and dimension-9, respectively.
In the standard scenario the scale is set by the light-neutrino masses where $m_{\beta\beta}\propto 1/\Lambda$.

Below the electroweak symmetry breaking SSB scale, heavy fields such as the $W_{L}$, $Z$, and Higgs bosons are integrated out.
This leads to operators with different powers of the Higgs vacuum expectation value $v$, expressed in terms of the Fermi constant as $v=(\sqrt{2}\,G_F)^{-1/2}\approx246$\,GeV.
In terms of Standard Model fields, dimension-3 (light-neutrino exchange), dimension-6, dimension-7, and dimension-9 operators are generated.
The dimension-3 operator is unique, whereas in general multiple operators of a given dimension violate lepton number.
After evolving to the hadronic and nuclear scales, the different contributions to the \nubb-decay amplitude can be organized as follows \cite{Cirigliano18master}
\begin{align}
  T_{1/2}^{-1} = &g_A^4\lbrace G_{01}(|\mathcal{A}_\nu|^2+|\mathcal{A}_R|^2) \nonumber\\
  + & 2G_{04}|\mathcal{A}_{m_e}|^2 +4G_{02}|\mathcal{A}_E|^2 + G_{09}|\mathcal{A}_M|^2 \nonumber \\
  - & 2(G_{01}-G_{04})Re[\mathcal{A}_\nu^*\mathcal{A}_R]
  +2G_{04}Re[\mathcal{A}_{m_e}^*(\mathcal{A}_\nu+\mathcal{A}_R)] \nonumber \\
  - & G_{03}Re[(\mathcal{A}_\nu+\mathcal{A}_R)\mathcal{A}_E^*+2\mathcal{A}_{m_e}\mathcal{A}_E^*]\nonumber\\
  + & G_{06}Re[(\mathcal{A}_\nu-\mathcal{A}_R)\mathcal{A}_M^*]\rbrace,
\label{eq:master}
\end{align}
where the $\mathcal{A}_i$ are transition amplitudes labeled by the lepton structure to which they correspond: $\mathcal{A}_\nu$ corresponds to light-neutrino exchange (besides other operators), $\mathcal{A}_R$ involves lepton right-handed currents, the $\mathcal{A}_{m_e}$ and $\mathcal{A}_{E}$ amplitudes are multiplied by the electron mass and energies, respectively, and $\mathcal{A}_{M}$ by the nucleon mass.
The phase-space factors $G_{0i}$ depend on the electron wavefunctions, and have been calculated accurately \cite{Kotila12,Stefanik15,Neacsu18}.

In general, each amplitude $\mathcal{A}_i$ receives contributions from operators of different dimension (here we refer to the dimension of operators below the electroweak scale).
The amplitude that receives the most contributions is $\mathcal{A}_\nu$. In particular, this is the relevant amplitude for dimension-3, dimension-7, and the majority of dimension-6 operators.
In turn, $\mathcal{A}_{M}$ is dominant for one type of dimension-6 operator and four dimension-9 operators, and $\mathcal{A}_{R}$ gets the dominant contribution from four other dimension-9 operators.
The amplitudes $\mathcal{A}_{E}$ and $\mathcal{A}_{m_e}$ are kinematically suppressed by a factor $m_e/m_N$. Because of this, their importance is relatively minor: $\mathcal{A}_{E}$ is only dominant for one type of dimension-6 operator, and $\mathcal{A}_{m_e}$ is always subleading.

In principle, the angular and energy distributions of the electrons emitted in \nubb-decay can be used to discriminate the leptonic structure responsible for the decay \cite{Ali07,SuperNemo10,Horoi16,Cirigliano17dim7}.
However, most BSM operators have leading contributions that enter into $\mathcal{A}_\nu$, the amplitude related to light-neutrino exchange. Therefore in general it will not always be possible to disentangle the BSM extension responsible for \nubb\ decay by measuring angular and energy distributions.

\subsubsection{The master formula}
\label{sec:master}

The transition amplitudes $\mathcal{A}_i$ include a combination of hadronic and nuclear matrix elements.
They also depend on the Wilson coefficients that couple BSM and Standard Model fields, which depend on the BSM scale $\Lambda$.
In the case of light neutrino exchange, Eq.~\eqref{eq:master} simplifies to Eq.~\eqref{eq:mbb}.
The combination of light neutrino masses \mbb\ sets the scale for lepton number violation.

In a more general scenario, additional contributions emerge, modifying Eq.~\eqref{eq:mbb} as
\begin{align}\label{eq:mbb_bsm}
  T_{1/2}^{-1} = &G_{01}\,g_A^4\,\left(M^{0\nu}_{\text{light}}\right)^{2}
  \frac{m_{\beta\beta}^2}{m_e^2}\nonumber\\
  + &\frac{m_N^2}{m_e^2}\tilde{G}\,\tilde{g}^4\,\tilde{M}^2\left(\frac{v}{\tilde{\Lambda}}\right)^6\nonumber\\
  +&\frac{m_N^4}{m_e^2\,v^2}\tilde{G'}\,\tilde{g'}^4\,\tilde{M'}^2\left(\frac{v}{\tilde{\Lambda}'}\right)^{10}
  +\cdots,
\end{align}
where the second and third terms are typical contributions from dimension-7 and dimension-9 operators, respectively.
For any given BSM extension, several of these contributions are expected.
They can interfere with each other, and also with the light-neutrino exchange channel, as indicated by Eq.~\eqref{eq:master}.
However, interference terms are not expected to dominate \cite{Ahmed17,Ahmed19}.

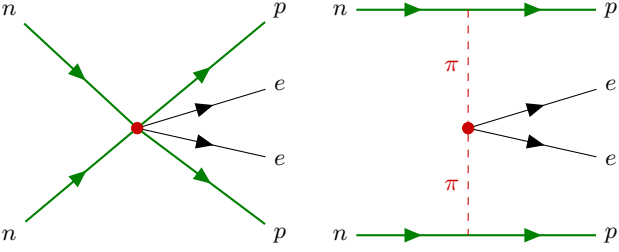
\begin{figure}[htbp]
\begin{tikzpicture}
    \begin{feynman}
      \vertex[               ] (a) {\(n\)};
      \vertex[right=17mm of a] (b);
      \vertex[right=17mm of b] (c) {\(p\)};
      
      \vertex[below=30mm of a] (d) {\(n\)};
      \vertex[below=30mm of b] (e);
      \vertex[below=30mm of c] (f) {\(p\)};

      \vertex[dot, red!80!black, large, below=15mm of b] (g){};

      \vertex[below=10mm of c] (h) {\(e\)};
      \vertex[below=10mm of h] (i) {\(e\)};

      \diagram* {
        (a)-- [fermion, green!50!black, thick] (g)-- [fermion, green!50!black, thick] (c),
        (d)-- [fermion, green!50!black, thick] (g)-- [fermion, green!50!black, thick] (f),
        (g)-- [fermion] (h),
        (g)-- [fermion] (i),
      };

      \vertex[right=44mm of a](A) {\(n\)};
      \vertex[right=17mm of A](B);
      \vertex[right=17mm of B](C) {\(p\)};

      \vertex[below=30mm of A] (D) {\(n\)};
      \vertex[below=30mm of B] (E);
      \vertex[below=30mm of C] (F) {\(p\)};

      \vertex[dot, red!80!black, large, below=15mm of B] (G){};
      \vertex[below=10mm of C] (H) {\(e\)};
      \vertex[below=10mm of H] (I) {\(e\)};
      
      \diagram* {
        (A)-- [fermion, green!50!black, thick] (B)-- [fermion, green!50!black, thick] (C),
        (D)-- [fermion, green!50!black, thick] (E)-- [fermion, green!50!black, thick] (F),
        (B)-- [scalar, red!80!black, edge label'=\(\pi\)] (G),
        (G)-- [scalar, red!80!black, edge label'=\(\pi\)] (E),
        (G)-- [fermion] (H),
        (G)-- [fermion] (I),
      };
    \end{feynman}
  \end{tikzpicture}
  \caption{Diagrams for contact (left) and two-pion exchange (right) contributions to \nubb\ decay.}
  \label{fig:contact_pion_diagrams}
\end{figure}

The factors in front of the dimension-7 and -9 terms are given by EFT \cite{Cirigliano17dim7,Cirigliano18master}.
They capture chiral enhancement factors of the nucleon over the pion mass, $m_N/m_{\pi}$, with respect to the naive analysis in Eq.~\eqref{eq:naive}.
These nuclear effects appear because, when mediated by the exchange of a heavy particle (Fig.~\ref{fig:contact_pion_diagrams}, left), the \nubb-decay amplitude is dominated by the virtual exchange of pions (Fig.~\ref{fig:contact_pion_diagrams}, right).
Each pion exchanged enhances the amplitude by $m_N/m_{\pi}$.

All phase-space factors in Eq.~(\ref{eq:mbb_bsm}), $G_{01}$, $\tilde{G}$, and $\tilde{G}'$, are known, and have typical values $G\sim10^{-14}$\,yr$^{-1}$.
The hadronic matrix elements $g_A$, $\tilde{g}$, and $\tilde{g}'$ can be calculated by lattice QCD or measured.
The present knowledge on $\tilde{g}$, $\tilde{g}'$ values is collected in Ref.~\cite{Cirigliano18master}, and agrees with the EFT expectation that they all are of the same of order.
The NMEs $M^{0\nu}_{\text{light}}$, $\tilde{M}$, and $\tilde{M}'$ can be calculated by nuclear theory, and they are sometimes suppressed or enhanced due to nuclear structure effects, see Sec.~\ref{sec:nme_bsm}.
In addition to the terms explicitly included in Eq.~\eqref{eq:mbb_bsm}, Yukawa couplings can suppress some contributions. These small couplings are the reason that in some models the \nubb\ rate stemming from dimension-9 operators can dominate over light-neutrino-exchange and dimension-7 channels when $\tilde{\Lambda}\sim\tilde{\Lambda}'$.

Therefore a \nubb-decay measurement will constrain, in addition to \mbb, the scales of any given BSM extension, $\tilde{\Lambda}$ and $\tilde{\Lambda}'$.
These new-physics scales are determined by the values of the BSM parameters, typically in terms of small dimensionless Wilson coefficients $C\sim v/\Lambda$, and Yukawa couplings.
For instance, in the left-right symmetric models discussed in Sec.~\ref{sec:par:LNV:fp}, the Wilson coefficients can be related to the heavy mass of the right-handed $W_R$ boson, $C\sim M_{W_L}/M_{W_R}$, or to the small mixing between the right- and left-handed sectors, $C\sim\xi_{LR}$.
Most studies interpret the constraints of \nubb-decay limits on left-right symmetric models in terms of $M_R$ and $\xi_{LR}$ \cite{Stefanik15,Horoi16,Sarkar20,Li21}.

\subsubsection{Experimental constraints on new physics scales}
\label{sec:scales}

Typical constraints by present \nubb-decay experiments, $T^{-1}_{1/2}\gtrsim10^{26}$\,yr, can be estimated from Eq.~\eqref{eq:mbb_bsm}, see also the comparison with the naive expectation in Eq.~\eqref{eq:naive}.
In the light-neutrino exchange mechanism, phase-space factors and typical NMEs lead to $m_{\beta\beta}\lesssim100$\,meV.
Likewise, for dimension-7 and dimension-9 operators, $\tilde{\Lambda}\gtrsim200$\,TeV and $\tilde{\Lambda}'\gtrsim5$\,TeV, respectively \cite{Cirigliano17dim7,Cirigliano18master}.
In contrast, a direct substitution in Eq.~\eqref{eq:mbb_bsm} assuming the EFT expected values for hadronic and nuclear matrix elements anticipates $\tilde{\Lambda}\gtrsim500$\,TeV and $\tilde{\Lambda}'\gtrsim8$\,TeV.
The actual constraints are not as tight because nuclear structure effects suppress $\tilde{M}$ and $\tilde{M}'$ NMEs, as discussed in Sec.~\ref{sec:nme_bsm}.
For dimension-9 operators the impact of the NME cancellation is smaller because $\tilde{\Lambda}'$ enters to a higher power.

Future improvements in \nubb-decay half-life limits of one order of magnitude will tighten the constraints on \mbb\ by about a factor of about 3. BSM scales for dimension-7 operators $\tilde{\Lambda}$ would improve by an additional 50\%, because of their $1/\tilde{\Lambda}^3$ dependence. Constraints for dimension-9 operators would improve $\tilde{\Lambda}'$ by 25\%, since they enter as $(1/\tilde{\Lambda}')^5$.

\subsection{Nuclear matrix elements}
\label{sec:nme_expressions}

In general, each \nubb-decay mechanism needs a particular NME.
However, in practice only few different NMEs are required in the dominant channels for each operator leading to \nubb\ decay.
NMEs encode how the decay occurs within a highly correlated many-body system.
These nuclear structure aspects can enhance or suppress the values of the NMEs.

\subsubsection{Light and heavy neutrino exchange}
\label{sec:op_phenom}
The starting point of most derivations of the \nubb-decay NME for neutrino exchange is the leading weak current for one nucleon \cite{Tomoda:1990rs,Park03}:
\begin{align}
\mathcal{J}^0&=\tau \left[g_V(p^2)\right], \nonumber \\
\bm{J}&= \tau \left[
g_A(p^2) {\bm \sigma}
-g_P(p^2)\frac{{\bm p}\left({\bm p}\cdot{\bm \sigma}\right)}{p^2+m_{\pi}^2}
+ig_M \frac{{\bm \sigma}\times{\bm p}}{2m_N}
\right],
\label{1b_currents}
\end{align}
in terms of the so-called vector (V), axial (A), pseudoscalar (P) and magnetic (M) terms, labeled after the corresponding hadronic couplings $g_V$, $g_A$, $g_P$ and $g_M$.
The vector and axial terms are responsible for Fermi and Gamow-Teller \B\ decays, respectively, while $g_P$ and $g_M$ only contribute to processes with finite momentum transfer (${\bm p}$) such as \nubb\ decay.
The currents also depend on the nucleon isospin $\tau$ and spin ${\bm \sigma}$.

The \nubb-decay rate is then given by the product of two one-body hadronic currents, following second-order perturbation theory in the weak interaction:
\begin{multline}\label{eq:nme_full}
  \sqrt{\Gamma_{0\nu\beta\beta}} = m_{\beta\beta}  \cdot \frac{g_A^2}{R} \cdot    \!\int \!d\bm x \!\int \!d\bm y \;L^{\mu\rho}(\bm x, \bm y) \!\int \!d\bm p\,e^{i\bm p({\bm x -\bm y})}\cdot \\
  \frac{R}{g_A^2} \sum_{n,m,a}
  \langle 0^+_f|
  \frac{\mathcal{J}_n^{\mu\dagger} ({\bm x})\; |J^P_a\rangle  \langle J^P_a|\;\mathcal{J}_m^{\rho\dagger}({\bm y})}
       {\sqrt{m_{\nu}^2+{\bm p}^2}\,[\sqrt{m_{\nu}^2+{\bm p}^2}+E_a^{\text{rel}}]}
       | 0^+_i\rangle,
\end{multline}
where $L^{\mu\rho}$ includes the electrons and $\gamma$ matrices evaluated at positions $\bm x$ and $\bm y$. This term generates the phase space factor, divided by the approximate nuclear radius $R=1.2 A^{1/3}$\,fm introduced to make the NME dimensionless. $m_\nu$ is the mass of the exchanged particle, and the hadronic coupling $g_A^2$ is explicitly factored out to follow the usual convention leading to Eq.~\eqref{eq:mbb}.
The remaining terms in Eq.~(\ref{eq:nme_full}) correspond to the NME, which includes a sum over nucleons $n$, $m$.
The ground states of the initial ($i$) and final ($f$) nuclei have angular momentum and parity $J^P=0^+$, and
the sum is over all states of the intermediate nucleus ($a$) with odd number of protons and neutrons.
$E_a^{\text{rel}}=E_a-(E_i+E_f)/2$, where $E$ denotes the energy of the states.

The momentum transfer in \nubb\ decay is ${p}\sim100-200$\,MeV for the exchange of light neutrinos, and larger for heavy-particle exchange.
Therefore it is common to regard Eq.~(\ref{eq:nme_full}) as practically independent of the intermediate states, because $E_a^{\text{rel}}\sim10$\,MeV$\ll p$, and replace $E_a^{\text{rel}}$ with an average $\langle E \rangle$.
This is called the closure approximation.
Explicit QRPA and shell model calculations estimate that the closure approximation is good to $10\%$~\cite{Muto94,Senkov13,Senkov16}.
Evaluating Eq.~(\ref{eq:nme_full}) for $m_\nu\ll p$ and $m_\nu\gg p$ allows one to define a long-range NME for light-neutrino exchange and a heavy-neutrino exchange NME, respectively: 
\begin{multline}\label{eq:nme}
  M^{0\nu}_\text{long} = \frac{1.2 A^{1/3}\,\text{fm}}{g_A^2}\cdot\\
  \langle 0^+_f|
  \sum_{n.m}
  \tau^-_m\tau^-_n\big[
    H_F^\nu(r)\mathds{1}+H_{GT}^\nu(r){\bm \sigma}_n\cdot{\bm \sigma}_m+H_T^\nu(r)\,S_{nm}
    \big]
  | 0^+_i\rangle, 
\end{multline}
\begin{multline}\label{eq:nme_N}
  M^{0\nu}_\text{heavy}\equiv\frac{m_\nu^2}{m_\pi^2}M^{0\nu}_{m_\nu}
  =\frac{1.2 A^{1/3}\,\text{fm}}{g_A^2} \frac{1}{m_{\pi}^2}\cdot\\
  \langle 0^+_f|
  \sum_{n.m}
  \tau^-_m\tau^-_n\big[
    H_F^N(r)\mathds{1}+H_{GT}^N(r){\bm \sigma}_n\cdot{\bm \sigma}_m+H_T^N(r)\,S_{nm}
    \big]
| 0^+_i\rangle,
\end{multline}
where $r=|{\bm r}_n-{\bm r}_m|$ is the distance between nucleons.
The three spin structures are denoted as Fermi (F), Gamow-Teller (GT) and tensor (T), with the latter operator defined as
$S_{nm}=3({\hat r}\cdot{\bm \sigma}_n)({\hat r}\cdot{\bm \sigma}_m)-{\bm \sigma}_n\cdot{\bm \sigma}_n$.
Compared to Eq.~\eqref{eq:nme_full}, Eq.~\eqref{eq:nme_N} has been multiplied by a factor $m_\nu^2/m_\pi^2$, which allows a better comparison because then $M^{0\nu}_\text{long}\sim M^{0\nu}_\text{heavy}\sim 1$ \cite{Cirigliano17dim7,Cirigliano18master}.
This definition differs by a factor $m_\pi^2/(m_N\,m_e)$ from the standard one in the literature.

Since in \nubb\ decay the exchanged particles are not emitted, they become part of the transition operator, and thus the NME.
The so-called neutrino potentials for the exchange of light $H^\nu(r)$ and heavy $H^N(r)$ particles are given by   
\begin{align}\label{eq:nu_potentials}
  H_\text{spin}^\nu(r) & =\frac{2}{\pi}\int j_\text{spin}(pr)\frac{h_\text{spin}(p)}{p(p+\langle E \rangle)}p^2 d p,\nonumber\\
  H_\text{spin}^N(r) &= \frac{2}{\pi}\int j_\text{spin}(pr)h_\text{spin}(p)p^2 d p\quad,
\end{align}
where the subindex distinguishes spin structures.
The spherical Bessel function $j_0$ applies to Fermi and GT potentials, while the tensor goes with $j_2$.
The neutrino potentials $h_\text{spin}(p)$ are given by
\begin{align}\label{nu_potentials_compact}
  h_F    & = h^{VV}_{F}, \nonumber \\
  h_{GT} & = h^{AA}_{GT}+h^{AP}_{GT}+h^{PP}_{GT}+h^{MM}_{GT}, \nonumber \\
  h_{T}  & = h^{AP}_{T}+h^{PP}_{T}+h^{MM}_{T},
\end{align}
with
\begin{align}\label{nu_potentials_explicit}
h^{VV}_{F}&=g_V^2\;f^2(p/\Lambda_V),\\
h^{AA}_{GT}&=g_A^2\;f^2(p/\Lambda_A), \nonumber \\
h^{AP}_{GT}&=-h^{AP}_{T}=-g_A^2\,\frac{2}{3}\frac{p^2}{p^2+m_\pi^2}\;f^2(p/\Lambda_A), \nonumber \qquad\\
h^{PP}_{GT}&=-h^{PP}_{T}=g_A^2\,\frac{1}{3}\frac{p^4}{(p^2+m_\pi^2)^2}\;f^2(p/\Lambda_A), \nonumber \\
h^{MM}_{GT}&=2\,h^{MM}_{T}=\frac{g_M^2}{6}\frac{p^2}{m_N^2}\;f^2(p/\Lambda_V), \nonumber 
\end{align}
where the superscripts correspond to the terms in the one-body current in Eq.~\eqref{1b_currents} leading to each neutrino potential.
The magnetic coupling, $g_M=1+\kappa_1=4.71$, depends on the anomalous isovector nucleon magnetic moment $\kappa_1$.
The standard phenomenological derivation includes a momentum-dependent dipole form factor $f(x)=1/(1+x^2)^2$ for all terms, with axial and vector regulators $\Lambda_{A,V}\sim1$\,GeV.

Organizing by spin structure, the NMEs for light- and heavy-neutrino exchange can thus be written as
\begin{multline}\label{eq:nme_lightheavy}
  M^{0\nu}_\text{long} =M^{AA}_{GT}+M^{VV}_{F} +M^{AP}_{GT}+M^{PP}_{GT}+M^{MM}_{GT}\\
  +M^{AP}_{T}+M^{PP}_{T}+M^{MM}_{T}\quad,
\end{multline}
\begin{multline}
  M^{0\nu}_\text{heavy} = M^{AA}_{GT,\text{h}}
  +M^{VV}_{F,\text{h}}+M^{AP}_{GT,\text{h}}+M^{PP}_{GT,\text{h}}\\
  +M^{MM}_{GT,\text{h}}+M^{AP}_{T,\text{h}}+M^{PP}_{T,\text{h}}+M^{MM}_{T,\text{h}}\quad, \nonumber
\end{multline}
where the superscripts have the same meaning as in Eq.~\eqref{nu_potentials_explicit}.
NMEs are also available for $m_\nu\sim p$ \cite{Blennow10,Barea15a,Faessler14}.

\subsubsection{Short-range operator for light neutrino exchange}
\label{sec:op_eft}

A more systematic derivation can be obtained within the EFT for \nubb\ decay \cite{Cirigliano17light,Cirigliano18contact,Cirigliano18master}.
The EFT replicates all terms given in Sec.~\ref{sec:op_phenom}, with small differences only.
In addition, the EFT provides an expansion, or counting, of the different contributions that determines which of them should be considered at a given EFT order.
For instance, in the EFT, including the closure energy $\langle E \rangle$ in Eq.~\eqref{eq:nu_potentials} is a higher-order effect.
Likewise, the momentum dependence of the axial and vector form factors in $h_\text{spin}(p)$, besides quadratic terms, appear also at higher order in the EFT.
However, the numerical impact of the differences introduced by the EFT with respect to the expressions used by most NME calculations is about few percent \cite{Rodin06,Menendez11}.
In addition, the EFT also predicts additional contributions not considered in practical calculations yet.
Preliminary estimations suggest that the additional terms are numerically small corrections to the light-neutrino exchange $M^{0\nu}$ \cite{Cirigliano17light}, with one exception.

A novel, potentially relevant term was introduced by \textcite{Cirigliano18contact}, and described in detail in \textcite{Cirigliano19long}.
The main idea is that the exchange of high-energy light neutrinos, which is naively expected to be a high-order correction, may be in fact a leading-order contribution.
The NME associated with this new contact diagram can be defined as
\begin{multline}\label{eq:contact}
  M^{0\nu}_\text{short} = \frac{1.2 A^{1/3}\,\text{fm}}{g_A^{2}}\\
\langle 0^+_f|
\sum_{n.m}
\tau^-_m\tau^-_n\,\mathds{1}\,\big[
\frac{2}{\pi}\int j_0(qr)h_S\,q^2 d q\big]
| 0^+_i\rangle,
\end{multline}
which follows the structure of Eqs.~\eqref{eq:nme}--\eqref{eq:nu_potentials}.
The neutrino potential $h_S=2g^{\text{NN}}_\nu\,f_S(p/\Lambda_S)$
depends on a two-nucleon coupling expected to scale as $g^{\text{NN}}_\nu\sim1/m_{\pi}^2$, with regulator $f_S$ and scale $\Lambda_S$. The momentum dependence of $f_S$ can be more general than the momentum transfer $p$.
The new matrix element depends on the nuclear structure of the initial and final nuclei, and on the contact coupling $g^{\text{NN}}_\nu$, satisfying $M^{0\nu}_\text{short}/(g^{\text{NN}}_\nu m_\pi^2)\sim M^{0\nu}_{\text{heavy}}\sim M^{0\nu}_\text{long}$.
In fact, $M^{0\nu}_\text{short}$ is related to the short-range NME for heavy-neutrino exchange, sharing the same spin structure as $M^{VV}_{F, \text{h}}$.
The short-range term cannot be derived by the product of two one-nucleon weak currents as in Eq.~\eqref{eq:nme_full}, which explains why $g^{\text{NN}}_\nu$ appears linearly in $h_S$, in contrast to $g_A$ which is always squared.

The contact coupling $g^{\text{NN}}_\nu$ is not known experimentally. Because both the value and sign of $g^{\text{NN}}_\nu$ are unknown, the new short-range term could either enhance or reduce expected \nubb-decay rates, but it could also have a small impact if $g^{\text{NN}}_\nu\ll 1/m_{\pi}^2$.
Lattice QCD calculations of the neutrinoless two-nucleon decay can determine $g^{\text{NN}}_\nu$ and efforts in this direction are ongoing \cite{Davoudi21,Davoudi:2020ngi,Davoudi22}.
Alternatively, $g^{\text{NN}}_\nu$ can be inferred from an approximated calculation of the same process using perturbative QCD methods \cite{Cirigliano20,Cirigliano21} that describe well the related charge-independence-breaking in the electromagnetic sector.
This avenue has been used to obtain $M^{0\nu}_\text{short}$ in \Ca, suggesting a positive contribution that enhances the long-range NME by about $40\%$~\cite{Wirth21}. A similar enhancement around $30\%-50\%$ has been found for transitions in nuclei from \Ca\ to \Xe\ \cite{Jokiniemi21}, assuming $g^{\text{NN}}_{\nu}$ values taken from the charge-independence-breaking term of different nuclear Hamiltonians, an assumption supported by \cite{Cirigliano21} and \cite{Richardson21}.
Given the potential impact of this contribution, a more robust determination of $g^{\text{NN}}_\nu$ should be pursued.

Including the short-range term, the light-neutrino exchange rate in Eq.~\eqref{eq:mbb} is modified as
\begin{align}
T_{1/2}^{-1}&=G_{01}\,g_A^4\big(M^{0\nu}_\text{long}
+M^{0\nu}_\text{short}\big)^2\,
\frac{m_{\beta\beta}^2}{m_e^2}, 
\label{eq:t12_contact}
\end{align}
leading to the light-neutrino exchange NME
\begin{align}
M^{0\nu}_\text{light}&=M^{0\nu}_\text{long}+M^{0\nu}_\text{short}.
\label{eq:nme_contact}
\end{align}

Likewise, a short-range contribution is expected for the exchange of heavy neutrinos discussed in Sec.~\ref{sec:op_phenom}~\cite{Dekens20}. In addition, in this case the contact coupling $g^{\text{NN}}_\nu$ depends on the neutrino mass in a non trivial way. Analyses of BSM scenarios with heavy sterile neutrinos thus need to complement the NME dependence on the neutrino mass with this additional dependence.

\subsubsection{Two-body currents}
\label{sec:op_2b}

Nucleons are composite particles.
Nuclear structure calculations, however, ignore that nucleons are formed by quarks and gluons and thus exhibit possible nucleon excitations.
To compensate for the missing degrees of freedom and other high-energy effects, the one-body current in Eq.~\eqref{1b_currents} needs to be complemented with two-body or meson-exchange currents (2bc). 
In the EFT, 2bc are associated with hadronic couplings, denoted by $c_i$, that also appear in the nucleon-nucleon forces that describe the same physics \cite{Park03,Krebs16,Baroni16}.

The importance of 2bc has been appreciated for decades \cite{Towner87,Brown87}.
However only EFT identifies the leading 2bc diagrams and predicts the value of the couplings.
While 2bc appear at higher EFT order than the terms introduced in Sec.~\ref{sec:op_phenom}, long-range 2bc are enhanced because they encode $\sim300$\,MeV nucleon excitations to the $\Delta$-isobar \cite{vanKolck94,Bernard96}. In fact, an EFT with explicit $\Delta$'s places 2bc at next-to-next to leading order, which is the same order as other contributions in Eq.~\eqref{eq:nme_lightheavy} \cite{Epelbaum07}.
EFT weak 2bc play a limited ($\lesssim 5\%$) but key role to reproduce experimental \B-decays half-lives \cite{Gazit09,Pastore18b} and neutrino scattering cross-sections\cite{Butler01,Nakamura01} in light $A\leq14$ nuclei.
In heavier $20\lesssim A\leq100$ systems, 2bc reduce \B\ decay matrix elements by $\sim10\%-20\%$ \cite{Gysbers19,Ekstrom14}, as discussed in detail in Sec.~\ref{sec:quenching}.

\bb\ decay involves the product of two weak currents, like in Eq.~\eqref{eq:nme_full}, so that 2bc generate three- and four-body transition operators.
Approximating 2bc as effective one-body currents via normal-ordering with respect to a symmetric nuclear matter reference state gives the estimate \cite{Menendez11}
\begin{multline}
\bm{\mathcal{J}}^{1b}+\bm{\mathcal{J}}^{2b}_\text{eff}=
\tau
\bigg[g_A{\bm \sigma}-
{\bm \sigma}\frac{2k_F^3\,g_A}{3\pi^2 F_\pi^2}\cdot\\
\bigg(
\frac{2c_4-c_3}{3}
\left[1-\frac{3m_\pi^2}{k_F^2}+\frac{3m_\pi^3}{k_F^3}\text{atan}\!\left(\frac{k_F}{m_\pi}\!\right)\!\right]
-\frac{c_D}{4g_A\Lambda_\chi}
\bigg)
\bigg],
\label{2bFermi}
\end{multline}
which modifies the GT term in Eq.~\eqref{1b_currents}. The $c_i$ couplings and Fermi momentum $k_F\sim200$\,MeV lead to a reduction of the GT operator $\sim20\%$ \cite{Gysbers19}, suggesting that 2bc contribute to ``$g_A$ quenching'', see Sec.~\ref{sec:quenching}. An improved expression is given by \textcite{Ney:2021anu}, which shows that the impact is reduced on neutron-rich nuclei.
Similar expressions modify the pseudoscalar and magnetic terms in Eq.~\eqref{1b_currents} \cite{Hoferichter20}.

The EFT 2bc in Eq.~\eqref{2bFermi}, when extended to finite momentum transfer, reduce \nubb-decay NMEs by $\sim30\%$ \cite{Menendez11,Engel14}.
This is less than doubling the reduction in \B\-decay matrix elements, because 2bc predict a milder reduction of the GT operator at  $p\sim200$\,MeV.
An improved treatment including three-body operators found only a $\sim10\%$ NME reduction for \Ge\ \cite{Wang18}, but a short-range term similar in nature to the one discussed in Sec.~\ref{sec:op_eft} could not be evaluated because of the unknown coupling.
In sum, 2bc could moderately modify \nubb-decay NMEs, perhaps similarly to or less than GT \B\ decays. Calculations with exact 2bc will provide an answer.

\subsubsection{Other exchange mechanisms}
\label{sec:nme_bsm}

BSM physics is typically mediated by a heavy particle.
Nevertheless, whenever permitted by symmetries of the operator,  the EFT predicts \cite{Prezeau03} that the dominant contribution to the \nubb-decay rate will be through the pion-exchange diagrams shown in Fig.~\ref{fig:contact_pion_diagrams}, enhanced by a factor $(m_N/m_\pi)^2$ as discussed in Sec.~\ref{sec:master}.
On the other hand, for dimension-7 operators, contact and pion-exchange diagrams compete with the short-range coupling to the nucleon magnetic moment, proportional to $g_{M}$ in Eq.~\eqref{1b_currents}.
The latter is enhanced with respect to the naive estimate because of the large coupling $g_M=4.71$. 

In general, many nuclear matrix elements contribute to \nubb\ decay mediated by BSM physics \cite{Cirigliano18master}.
The relevant combinations additional to $M^{0\nu}_\text{long}$ and $M^{0\nu}_\text{short}$ are
\begin{align}\label{eq:nme_bsm}
  M^{PS}&=\frac{1}{2}M^{AP}_{GT}+M^{PP}_{GT}+\frac{1}{2}M^{AP}_{T}+M^{PP}_{T}, \nonumber \\
  M^{M} &=  M^{MM}_{GT}+M^{MM}_{T}, \nonumber \\
  M^{PS}_\text{heavy}&=\frac{1}{2}M^{AP}_{GT,\text{h}}+M^{PP}_{GT,\text{h}}+
  \frac{1}{2}M^{AP}_{T,\text{h}}+M^{PP}_{T,\text{h}}, \nonumber\\
  M^{AP}_\text{heavy}&= M^{AP}_{GT,\text{h}}+M^{AP}_{T,\text{h}},
\end{align}
where the superscripts on the left-hand side NMEs indicate pseudoscalar (PS), magnetic (M) and axial-pseudoscalar (AP) in reference to the one-nucleon terms in Eq.~\eqref{1b_currents}. Tensor contributions are usually much smaller than GT ones, according to NME calculations in \bb\ emitters \cite{Menendez09,Hyvarinen15,Barea15}.
All six NMEs are combinations of the contributions to the light- and heavy-neutrino exchange matrix elements introduced in Eqs.~\eqref{eq:nme_lightheavy}.

The NMEs $M^{PS}$ and $M^{MM}$ are dominant for dimension-7 operators, while $M^{PS}_\text{heavy}$, $M^{AP}_\text{heavy}$, and $M^{0\nu}_\text{short}$ are the most relevant for dimension-9 operators.
The naive EFT counting that neglects nuclear structure effects predicts $M^{PS}\sim M^{PS}_\text{heavy}\sim M^{AP}_\text{heavy}\sim M^{0\nu}_\text{long}$.
However, calculations \cite{Menendez09,Hyvarinen15,Barea15} show that the two terms in $M ^{PS}$ have opposite sign and mostly cancel, in both GT and tensor parts, so that $M^{PS}\sim M^{PS}_\text{heavy}\sim M^{0\nu}_\text{long}/10$.
This nuclear-structure-based suppression is responsible for the reduced sensitivity of \nubb-decay experiments to the physics scale of typical dimension-7 and dimension-9 operators compared to naive EFT expectations, discussed in Sec.~\ref{sec:scales}. 
On the other hand, EFT indicates that $M^{M}\sim (m_\pi^2/m_N^2)\,M^{0\nu}_\text{long}$.
In contrast, the magnetic term is enhanced by the large hadronic coupling $g_M$, leading to $M^{M}\sim M^{0\nu}_\text{long}/10$, so that it competes with $M^{PS}$ as the dominant NME for dimension-7 operators. For a discussion on how these cancellations impact the BSM physics sensitivities of \nubb-decay experiments compared to LHC searches, see \textcite{Graesser22}.

Different NMEs for BSM \nubb-decay mechanisms have also been proposed and calculated in \textcite{Vergados:2012xy,Doi:1985dx,Tomoda:1990rs,Kotila21}.

\subsection{Many-body methods}
\label{sec:NMEs}

In the absence of a \nubb-decay observation, and as long as the light-neutrino masses, their ordering, or the BSM parameters responsible for the decay are not known, NMEs need to be obtained from theoretical nuclear structure calculations.
Here we present updated NME results and describe briefly the nuclear many-body methods used to obtain them.
A more thorough discussion of NMEs and nuclear many-body methods can be found in \textcite{Engel17}.

\subsubsection{Current status for long-range nuclear matrix elements}
\label{sec:status}

\begin{figure}[t]
	\centering
\includegraphics[width=\columnwidth]{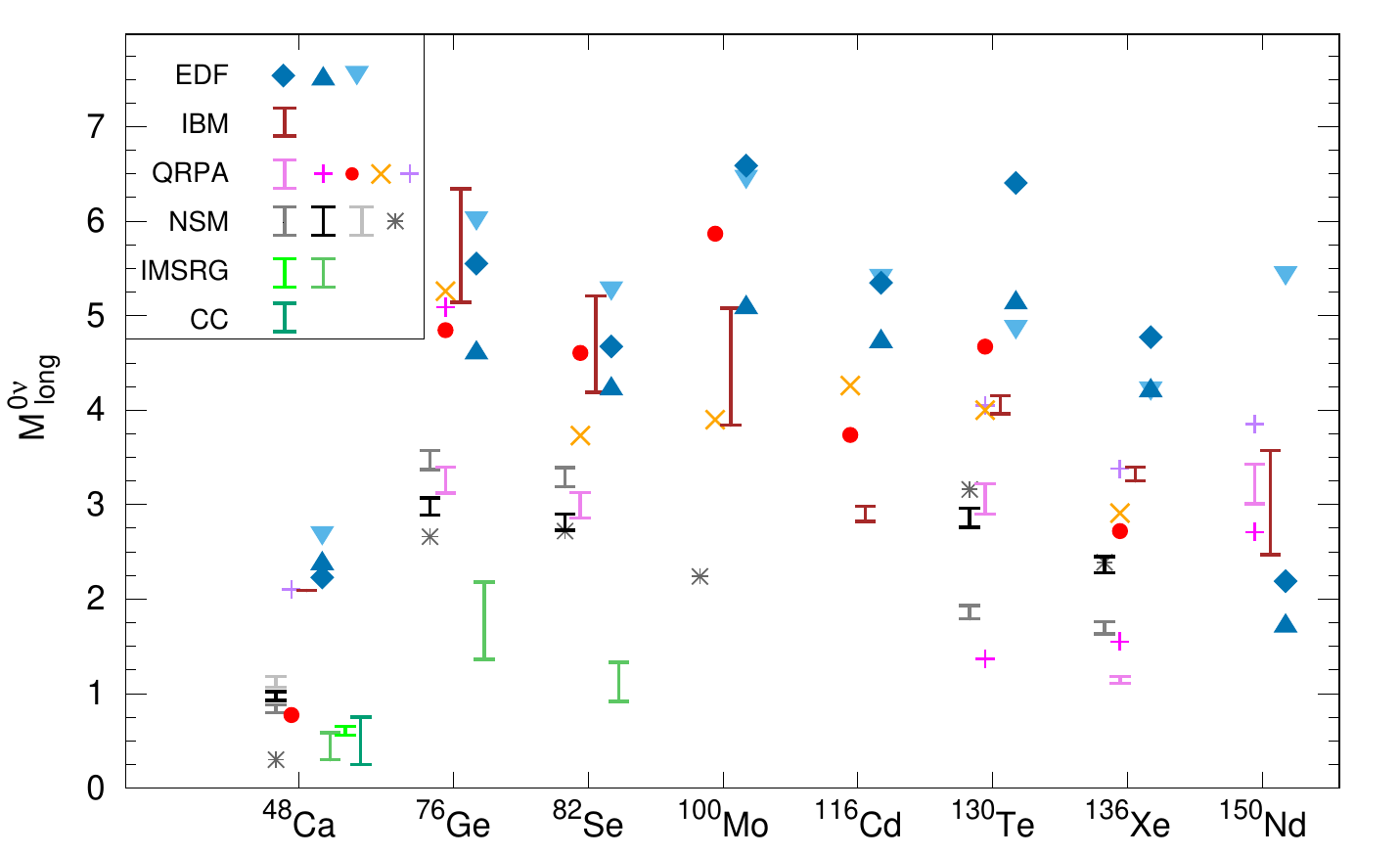}
	\caption{Nuclear matrix elements $M^{0\nu}$ for light-neutrino exchange from different many-body methods. NSM: black \cite{Menendez18}, grey \cite{Horoi16b}, light-grey \cite{Iwata16} bars and grey stars \cite{Coraggio20,Coraggio22}; QRPA: deformed in violet bars \cite{Fang18}), and spherical in magenta \cite{Mustonen13} and purple \cite{Terasaki15,Terasaki19,Terasaki20} crosses, red circles \cite{Simkovic18}, and orange multiplication signs \cite{Hyvarinen15};
	IBM: brown bars \cite{Barea15,Deppisch20};
	EDF theory: nonrelativistic in blue diamonds \cite{Rodriguez10} and blue up-triangles \cite{Vaquero14}), and relativistic in light-blue down-triangles \cite{Song17};
	IMSRG: IM-GCM in the light green \Ca\ bar \cite{Yao20}, and valence space in green bars \cite{Belley21};
	and CC theory: dark green \Ca\ bar \cite{Novario21}.}
	\label{fig:nme_light}
\end{figure}

Comparisons of NMEs obtained with different many-body approaches are common in the \nubb-decay literature \cite{Bahcall04,Vogel12,GomezCadenas12,Engel17,Feruglio02}. Figure~\ref{fig:nme_light} shows updated results for \nubb-decay NMEs of eight \bb\ emitters, covering calculations from the nuclear shell model (NSM), the quasiparticle random-phase approximation (QRPA) method, the interacting boson model (IBM) and energy-density functional (EDF) theory. Also included are recent ab initio \Ca\ NMEs obtained with the in-medium generator coordinate method (IM-GCM), a multi-reference version of the similarity renormalization group (IMSRG), and coupled-cluster (CC) theory, and \Ca\, \Ge\ and \Se\ NMEs from the valence-space (VS) IMSRG method.
Table~\ref{tab:NMEs} collects the NMEs for the five nuclei most relevant for next-generation experiments, and indicates the range of NMEs for each nuclear structure method, obtained by combining the results of different calculations for each approach.

\addtolength{\tabcolsep}{3pt}
\begin{table*}[t]
		\caption{Nuclear matrix elements $M ^{0\nu}$ for light neutrino exchange calculated with the shell model, QRPA, EDF theory and IBM methods, for the \nubb\ decay of nuclei considered for next-generation  experiments.
			The combined NME range for each many-body method with several NME calculations is also shown.
			All NMEs were obtained with the bare value of $g_A$ and do not include the short-range term proportional to $g^{\text{NN}}_\nu$.}
			\label{tab:NMEs}
\begin{tabular}{llccccc}
			\toprule
                        \smallspace
&& \Ge\ & \Se\ & \Mo\ & \Te\ & \Xe\ \\ \colrule
                        \smallspace
                        \multirow{4}{*}{Shell model}
                        &\textcite{Menendez18} & $2.89, 3.07$ & $2.73, 2.90$ & $-$ & $2.76, 2.96$ & $2.28, 2.45$ \\
			&\textcite{Horoi16b}   & $3.37, 3.57$ & $3.19, 3.39$ & $-$ & $1.79, 1.93$ & $1.63 ,1.76$ \\
			&\textcite{Coraggio20,Coraggio22} & $2.66$ & $2.72$ & $2.24$ & $3.16$ & $2.39$ \\
                        & min--max             & $2.66-3.57$ & $2.72-3.39$ & $2.24$ & $1.79-3.16$ & $1.63-2.45$ \\ \colrule
                        \smallspace
                        \multirow{6}{*}{QRPA}
			&\textcite{Mustonen13}        & $5.09$ & $-$    &  $-$   & $1.37$ & $1.55$ \\
			&\textcite{Hyvarinen15}       & $5.26$ & $3.73$ & $3.90$ & $4.00$ & $2.91$ \\
			&\textcite{Simkovic18}        & $4.85$ & $4.61$ & $5.87$ & $4.67$ & $2.72$ \\
			&\textcite{Fang18}            & $3.12, 3.40$ & $2.86, 3.13$ & $-$ & $2.90, 3.22$ & $1.11, 1.18$ \\
			&\textcite{Terasaki20}        & $-$ & $-$ & $-$ & 4.05 & 3.38 \\
			& min--max                    & $3.12-5.26$ & $2.86-4.61$ & $3.90-5.87$ & $1.37-4.67$ & $1.11-3.38$ \\ \colrule
                        \smallspace
                        \multirow{4}{*}{EDF theory}
			&\textcite{Rodriguez10} & $4.60$ & $4.22$ & $5.08$ & $5.13$ & $4.20$ \\
			&\textcite{Vaquero14}   & $5.55$ & $4.67$ & $6.59$ & $6.41$ & $4.77$ \\
			&\textcite{Song17}      & $6.04$ & $5.30$ & $6.48$ & $4.89$ & $4.24$ \\ 
			& min--max              & $4.60-6.04$ & $4.22-5.30$ & $5.08-6.59$ & $4.89-6.41$ & $4.20-4.77$ \\ \colrule
                        \smallspace
                        \multirow{3}{*}{IBM}
			&\textcite{Barea15} \footnote{With the sign change in the tensor part indicated in \textcite{Deppisch20}.}           & $5.14$ & $4.19$ & $3.84$ & $3.96$ & $3.25$ \\
			&\textcite{Deppisch20}         & $6.34$ & $5.21$ & $5.08$ & $4.15$ & $3.40$ \\ & min--max                & $5.14-6.34$ & $4.19-5.21$ & $3.84-5.08$ & $3.96-4.15$ & $3.25-3.40$ \\
			\botrule
		\end{tabular}
	\end{table*}

The variation in $M^{0\nu}$ in Fig.~\ref{fig:nme_light}, about a factor three, highlights the uncertainties introduced by the approximate solutions of the nuclear many-body problem.
With few exceptions among the \bb\ emitters considered, the NMEs follow a similar trend: shell model NMEs tend to be smallest, and EDF theory ones largest, with the IBM and QRPA somewhere in between.
Recent QRPA calculations by \textcite{Fang18} including deformation (violet bars), however, modify this picture as they find smaller NMEs than spherical QRPA calculations, close to the shell model NMEs.
These results follow a tendency of smaller QRPA NMEs hinted by the sophisticated QRPA of \textcite{Mustonen13} --- magenta crosses. 
Nevertheless, the deformed QRPA likely underestimates NMEs because the current calculation misses the effect of configuration mixing that enhances their value~\cite{Rodriguez10}.
Finally, the \Ca\ NMEs from the IM-GCM \cite{Yao20}, VS-IMSRG \cite{Belley21}, and CC \cite{Novario21} theory are consistent with each other and smaller than the shell model ones.
The VS-IMSRG \Ge\ and \Se\ NMEs are also smaller than in other calculations, but currently the ab initio description of these nuclei is of lower quality than for \Ca, see Sec.~\ref{sec:NME_tests}.

Overall, the smaller ab initio NMEs suggest that phenomenological NMEs might be overestimated. This is consistent with the fact that, as discussed in the following sections, the many-body methods predicting larger NMEs, energy-density functional theory and the IBM, do not include explicitly proton-neutron pairing correlations which are known to reduce the value of the NMEs. Further, especially for \Ca\ and \Ge\ ab initio results are not far from shell-model and some of the QRPA ones, the only two-body methods which so far have predicted \nnbb\ or $2\nu$ECEC half-lives before their measurement (see Sec. \ref{sec:2nbb}).
Nonetheless, especially compared to concerns related to a dramatic reduction of NMEs due to ``$g_A$ quenching'' (see Sec.~\ref{sec:quenching}), the overestimation of the more phenomenological NMEs appears relatively moderate, taking into account that the ab initio methods used for \Ca\ reproduce well $\beta$-decay matrix elements without any adjustments.

\subsubsection{Uncertainties and other nuclear matrix elements}
\label{sec:status2}

Beyond these main features, Fig.~\ref{fig:nme_light} highlights that more calculations are available for some \nubb\ decays than others. On the one hand, \Ca\ has been studied by all many-body methods, including three ab-initio ones. This is because \Ca\ is doubly-magic, and therefore can be described with relatively simple nuclear correlations. Indeed, most of the latest calculations roughly converge to rather small NME values. On the other hand, neither ab initio nor shell model NMEs are available for \Cd, or \Nd, and the only \Mo\ shell-model NMEs are very recent \cite{Coraggio22}. The difficulty is that these nuclei have a very complex nuclear structure with several neutrons and protons away from closed shells. In fact, for \Nd\ EDF results, which typically agree with each other, disagree by a factor of three, indicating the challenge of the calculations. The remaining decays lie in between, even though for instance the $A=76$ nuclear structure might include subtleties due to deformation, see Sec.~\ref{sec:NME_tests}.

Unfortunately, the phenomenological character of most NME calculations prevents a reliable estimation of theoretical uncertainties.
For instance, the impact of enlarging the configuration space in the shell model, or the effect of including explicit proton-neutron pairing correlations in EDF theory, are hard to quantify.
Part of the theoretical uncertainties, however, are easier to evaluate.
For instance, the difference in the shell model results in Fig.~\ref{fig:nme_light} (black and grey bars and stars), or the EDF theory calculations (diamonds and up and down triangles) give an estimate of the uncertainty of each approach when the parameters of the model, typically the nuclear Hamiltonian, are varied.
Likewise, the difference between spherical QRPA NMEs (red circles, magenta and purple crosses, orange multiplication symbols) and the IBM uncertainty (brown error bar) estimate this kind of theoretical uncertainty.
On the other hand, smaller uncertainties shown as error bars in Fig.~\ref{fig:nme_light} explore a very small part of this uncertainty, because only a very limited subset of the parameters of the model --- typically those associated with short-range correlations (SRC), see below --- is varied.
Symbols without error bars in Fig.~\ref{fig:nme_light} indicate that no parameter variation was explored.
These kind of uncertainties have been recently evaluated more systematically in the shell model for energy levels \cite{Yoshida18} and electroweak matrix elements in light nuclei \cite{Fox19}, and in heavier systems with EDF theory \cite{Neufcourt19}. First efforts based on systematic calculations to assign these kind of theoretical uncertainties to \nubb-decay NMEs are available for \Ca\ in the shell model \cite{Horoi22} and for all \nubb-decay candidates in the shell model and QRPA \cite{Jokiniemi22}. The latter give uncertainties for these methods comparable to the min-max ones in Table~\ref{tab:NMEs}.

Ab initio calculations in principle allow for a quantification of the theoretical uncertainties \cite{ProjectReport22}.
The error bars in the ab initio results in Fig.~\ref{fig:nme_light} are dominated by the uncertainty from the nuclear Hamiltonians used, except for CC theory, where the dominant error stems from the many-body method, which had to be extended to deal with \nubb\ decay, see Sec.~\ref{sec:abinitio}. Nonetheless, even the ab initio NME uncertainties in Fig.~\ref{fig:nme_light} are underestimated, because a relevant ingredient, two-body currents at finite momentum transfers, is not yet included in the calculations.

\begin{figure}[t]
	\centering
	\includegraphics[width=\columnwidth]{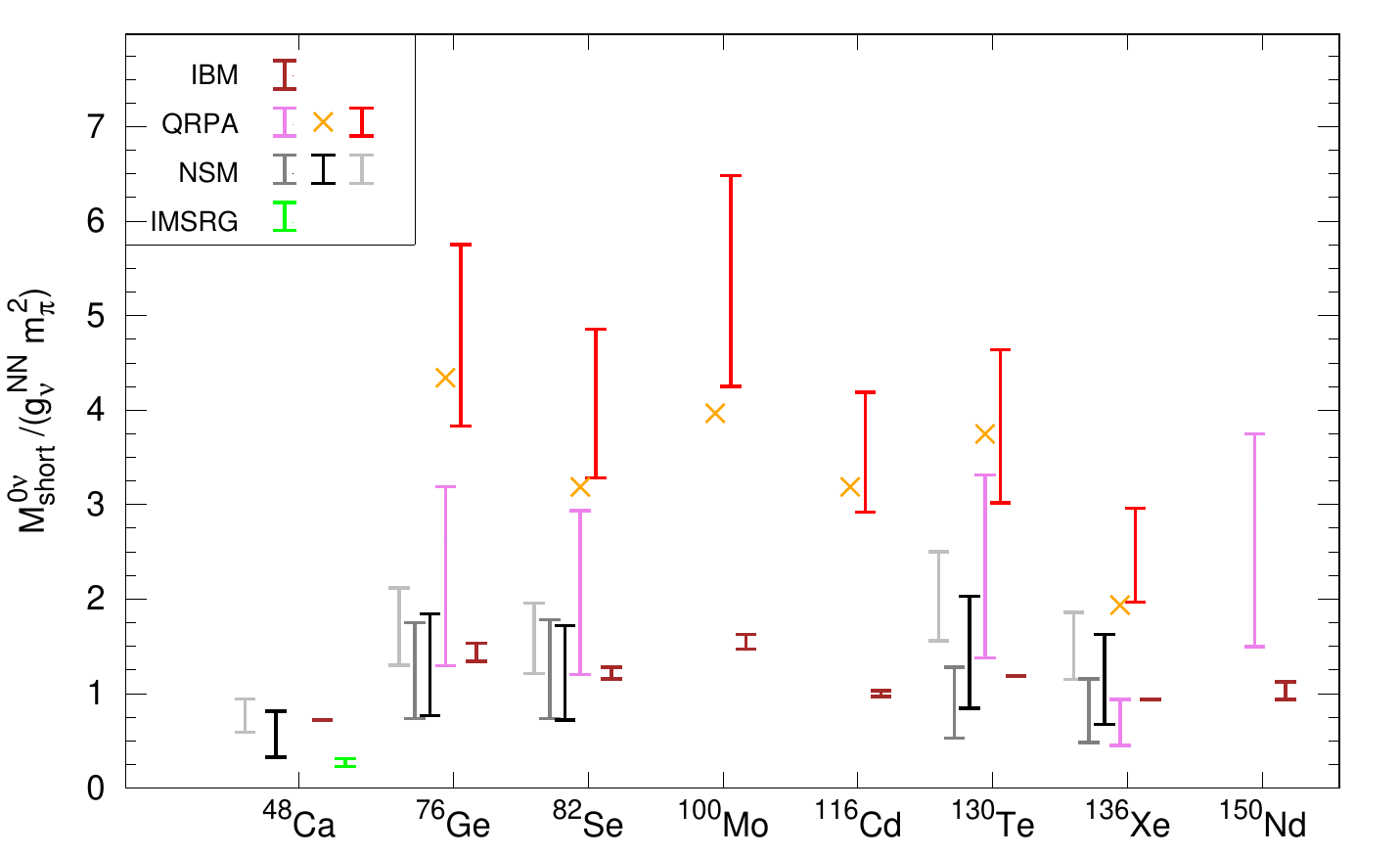}
	\caption{Short-range light-neutrino exchange nuclear matrix elements $M^{0\nu}_\text{short}$ without the coupling $g^{\text{NN}}_\nu$. Results from the NSM: black \cite{Menendez18}, grey \cite{Neacsu15,Senkov14,Senkov16}, and light grey \cite{Jokiniemi21} bars; the QRPA: deformed in violet bars \cite{Fang18} and spherical in orange mulitplication signs \cite{Hyvarinen15} and red bars \cite{Jokiniemi21}); the IBM: brown bars \cite{Barea15,Deppisch20}; and the IM-GCM: light green bars \cite{Wirth21}.}
	\label{fig:nme_contact}
\end{figure}

\begin{figure}[t]
	\centering
	\includegraphics[width=\columnwidth]{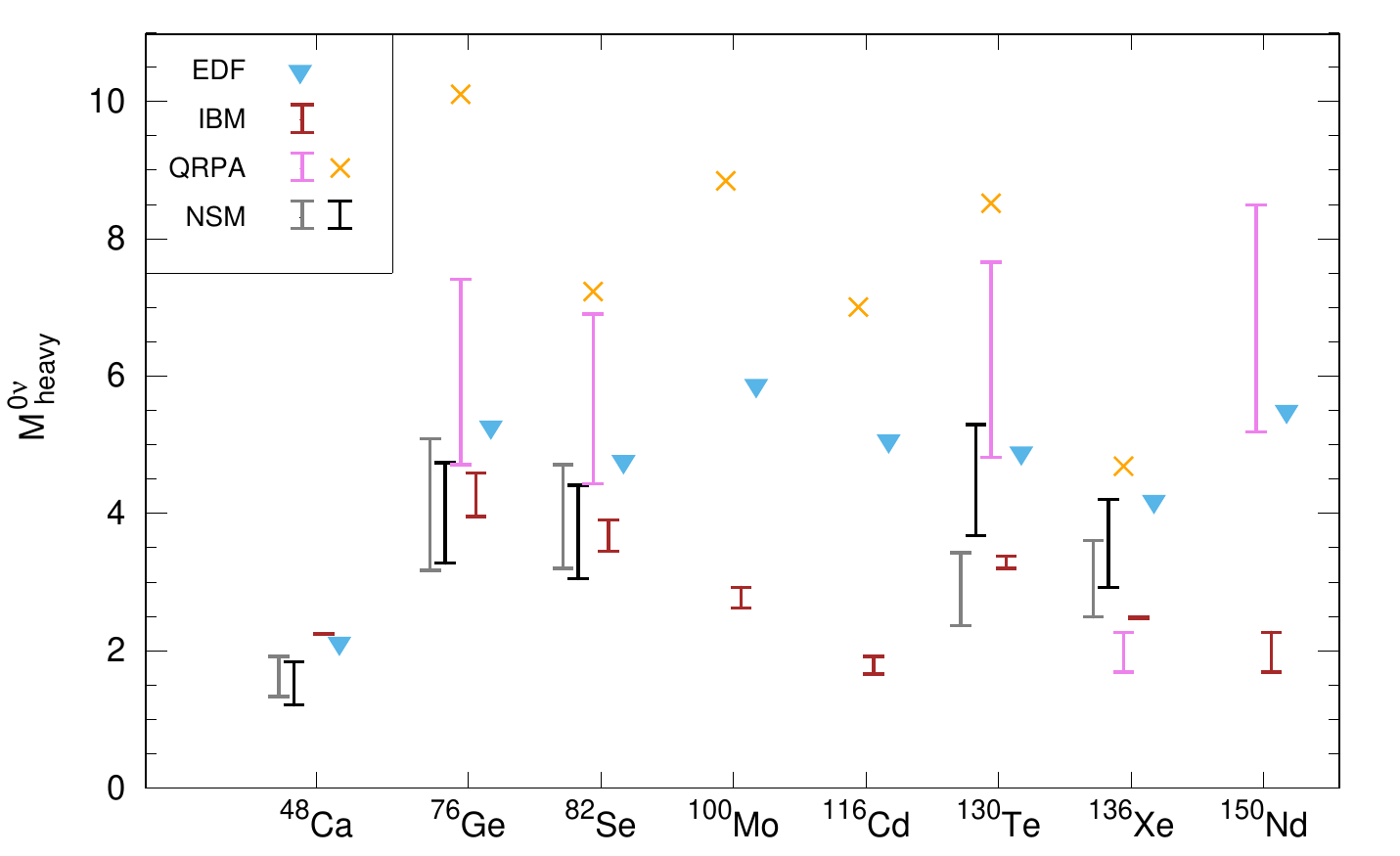}
	\caption{Nuclear matrix elements $M^{0\nu}_\text{heavy}$ for the heavy-neutrino exchange \nubb\ decay. Results from the NSM: black \cite{Menendez18} and grey \cite{Horoi16b} bars; the QRPA: deformed in violet bars \cite{Fang18} and spherical in orange multiplication signs \cite{Hyvarinen15};
		the IBM: brown bars \cite{Barea15,Deppisch20};
		and relativistic EDF theory: light-blue down-triangles \cite{Song17}.
		Note that $M^{0\nu}_\text{heavy}$ includes an additional factor $(m_N\,m_e)^2/m_\pi^2$ with respect to the standard definition.}
	\label{fig:nme_heavy}
\end{figure}

An additional uncertainty not immediately apparent in Fig.~\ref{fig:nme_light} concerns the possible reduction of the NMEs, usually known as ``$g_A$ quenching''.
This effect was proposed to compensate the finding that calculated GT \B\ matrix elements tend to overpredict measured values by a roughly uniform factor.
This introduces a potentially large uncertainty, because a naive direct quenching of the axial coupling constant $g_A^\text{eff}=0.7 g_A$, as has been suggested often in the literature, would reduce the \nubb-decay NMEs by $(0.7)^2\sim 1/2$, and decay rates by $(0.7)^4\sim1/4$.
The ``$g_A$ quenching'' highlights deficiencies in the nuclear theory calculations, but it is not clear how to scale them from \B\ to \nubb\ decays.
For this reason, Fig.~\ref{fig:nme_light} assumes the unquenched $g_A=1.27$.
Recent ab initio calculations that reproduce \B\ decays without any ``$g_A$ quenching'' pave the way to solve this puzzle \cite{Gysbers19}.
We address this issue in detail in Sec.~\ref{sec:quenching}.

In addition to the nuclear structure of the initial and final nuclei, the range of the \nubb-decay operator has a strong impact on the NMEs.
Figures~\ref{fig:nme_contact} and \ref{fig:nme_heavy} compare $M^{0\nu}_\text{short}/(g^{\text{NN}}_\nu m_\pi^2)$ and $M^{0\nu}_\text{heavy}$, corresponding to the short-range light-neutrino exchange term (without coupling) and the exchange of heavy neutrinos, discussed in Secs.~\ref{sec:op_eft} and~\ref{sec:op_phenom}, respectively.
Except for the QRPA, short-range and heavy-neutrino NMEs are close.
This suggests that differences in $M^{0\nu}_\text{long}$ are due to how longer-range nuclear correlations are treated differently in the various many-body methods \cite{Menendez18}.

As for the contact term, combining the short-range NMEs in Fig.~\ref{fig:nme_contact} with $g^{\text{NN}}_\nu$ values from charge-independent-breaking Hamiltonians leads to sizable contributions with respect to $M^{0\nu}_\text{long}$ \cite{Jokiniemi21}, both for the shell model (light grey bars, $\sim30\%$ impact) and for the QRPA (red bars, $\sim50\%$ effect). These NMEs are consistent with other shell model and QRPA estimations in Fig.~\ref{fig:nme_contact}; the main difference is that the latter use a dipole $f_S$ instead of a gaussian. The value of $g^{\text{NN}}_\nu$ is found to be positive in \Ca\ and other lighter nuclei in \cite{Wirth21}. Therefore, Fig.~\ref{fig:nme_contact} suggests that the difference between NMEs in Fig.~\ref{fig:nme_light} will persist, with QRPA continuing to prefer larger $M^{0\nu}_\text{light}$ values.

The large error bars in Figs.~\ref{fig:nme_contact} and \ref{fig:nme_heavy} are due to SRCs, typically ignored because doing so simplifies computations and does not affect much most nuclear structure properties.
However, for \nubb-decay NME SRCs are extracted from calculations which include SRCs explicitly \cite{Kortelainen07,Simkovic09,Cruz-Torres18} typically via prescriptions used in other many-body calculations.
The error bars in Fig.~\ref{fig:nme_light}, \ref{fig:nme_contact}, and \ref{fig:nme_heavy} indicate a higher sensitivity to SRCs in $M^{0\nu}_\text{heavy}$ and $M^{0\nu}_\text{short}$ than in $M^{0\nu}_\text{long}$, where the impact is relatively small as also indicated by \textcite{Engel:2009gb}.
Nonetheless, very recently, the SRCs captured by an ab initio method have been combined with the shell model using an effective theory for SRCs validated in comparisons to SRC measurements~\cite{Cruz-Torres21}. 
The results suggest a larger $\sim30\%$ reduction in $M^{0\nu}_\text{long}$ due to SRCs~\cite{Weiss21}, which is similar to the effect found by \textcite{Benhar:2014cka}.

Finally, $M^{PS}_\text{heavy}$ and $M^{AP}_\text{heavy}$ matrix elements defined in Sec.~\ref{sec:nme_bsm} calculated with the shell model \cite{Horoi16b,Menendez18} and the QRPA \cite{Hyvarinen15} show agreement similar to that in Fig.~\ref{fig:nme_heavy}.
Likewise, shell model and QRPA $M^M$ and $M^{PS}$ matrix elements compare similarly to $M^{0\nu}_\text{long}$ in Fig.~\ref{fig:nme_light}.
Therefore, the nuclear matrix elements needed for light neutrino exchange and any other mechanism appear to have similar uncertainties.

\subsubsection{The nuclear shell model}
\label{sec:shell model}

The nuclear shell model is the primary method used to describe nuclear structure \cite{Caurier05,Otsuka19,Brown01,Poves17}.
Modern shell-model calculations are based on mixing nuclear configurations within a given space.
Usually the configuration space comprises one major harmonic oscillator shell for protons and neutrons, but due to advances in computing power two-shell calculations are increasingly more common.
Within the configuration space, the shell model includes the most general nuclear correlations.
This is sufficient to describe well the spectroscopy of nuclei from oxygen to lead.

Most calculations of \nubb-decay NMEs are currently limited to one shell \cite{Menendez09,Senkov13,Senkov14,Neacsu15,Senkov16,Menendez18,Coraggio20,Coraggio22}.
So far the only two-shell calculation is for \Ca\ \cite{Iwata16},
which results in a moderate $\sim20\%$ NME enhancement over the one-shell NME --- light-grey bars in Fig.~\ref{fig:nme_light}. 
It also reveals a subtle competition: pairing-like excitations enhance NMEs~\cite{Caurier08a}, while particle-hole-like ones reduce NME values \cite{Horoi13}.
The overall effect of larger configuration spaces is thus expected to be limited.
Two-shell calculations in heavy nuclei demand approximate solutions, for instance using the GCM with collective degrees of freedom --- deformation, isoscalar and isovector pairing --- as coordinates \cite{Jiao18,Jiao19,Hinohara14,Menendez16,Vaquero14}.
For \Ge\ a GCM two-shell calculation \cite{Jiao17} finds a slight NME reduction. Likewise, studies that explore the impact of larger configuration spaces with perturbation theory --- grey stars in Fig.~\ref{fig:nme_light} --- also suggest a $~20\%-30\%$ change on NMEs at most \cite{Holt13,Coraggio20}.
The only exception is $^{48}$Ca, which as a doubly-magic nucleus probably needs additional refinement in this framework.

The Monte Carlo shell model is a novel approach that aims to capture the most relevant correlations when handling multi-shell configuration spaces \cite{Shimizu17,Otsuka16}.
A relatively small number of angular-momentum-projected deformed basis states is sufficient to explore the most relevant configurations, tackling spaces with $\gg10^{20}$ Slater determinants \cite{Marsh18,Ichikawa19} --- the standard shell model is limited to $\sim10^{11}$ explicit configurations. A related strategy based on the superposition of quasiparticle states is more suited to \nubb-decay NMEs, and may enable calculations for e.g.~\Nd\ \cite{Shimizu21}.

The success of the shell model is based on effective nuclear Hamiltonians adapted to each configuration space \cite{Caurier05}.
High quality Hamiltonians are important for \nubb-decay studies, because schematic interactions can lead to NMEs outside the shell-model range discussed in Sec.~\ref{sec:status} \cite{Yoshinaga18,Higashiyama20}.
Nonetheless, even effective Hamiltonians derived from nucleon-nucleon potentials demand phenomenological adjustments, mainly in the part that describes single-particle degrees of freedom, i.e.~the monopole component. 
Due to this, shell model NMEs have a phenomenological component.
This limitation is lifted by effective Hamiltonians built by ab initio methods.
They are derived without phenomenological adjustments from chiral EFT nucleon-nucleon and three-nucleon interactions \cite{Bogner14,Stroberg16a,Jansen14,Dikmen15} connected to the underlying theory of the nuclear force, QCD.
Ab initio methods are described in Sec.~\ref{sec:abinitio}.

\subsubsection{The QRPA and its variants}
\label{sec:QRPA}

The QRPA was the first many-body method to reliably address \bb\ decay \cite{Vogel86,Engel88}.
Contrary to the nuclear shell model, the QRPA uses large configuration spaces encompassing several harmonic oscillator shells.
On the other hand, the nuclear correlations included in the QRPA are more limited than the ones the shell model captures.
The QRPA relies on small amplitude nuclear correlations, and has been reviewed for instance in \textcite{Suhonen98}, \textcite{Avignone08}, and \textcite{Engel17}.

One aspect particularly relevant for QRPA \nubb-decay studies is the strength of the proton-neutron pairing interaction.
Several prescriptions have been proposed to fix its value,
for instance, using \B\ decay data involving the intermediate, initial or final \bb-decay nuclei \cite{Engel88},
or using \nnbb\ decay \cite{Rodin03} --- the latter strategy is used in the orange multiplication signs in Fig.~\ref{fig:nme_light}.
These approaches share the disadvantage that the proton-neutron pairing interaction is difficult to disentangle from a possible ``$g_A$ quenching'' needed by the QRPA, see Sec.~\ref{sec:quenching}.
Recently, two alternatives have been proposed.
The first imposes SU(4) symmetry and therefore a vanishing double GT matrix element \cite{Simkovic18} --- red circles in Fig.~\ref{fig:nme_light}.
The second demands the equivalence, in the closure approximation explained above Eq.~\eqref{eq:nme}, of the NMEs through intermediate $(N-1,Z+1)$ and $(N-2,Z)$ nuclei, with respect to the $(N,Z)$ initial one \cite{Terasaki15} --- purple crosses in Fig.~\ref{fig:nme_light}.
These choices lead to mildly different NMEs.
On the other hand, the QRPA fixes the isovector part of the proton-neutron interaction by demanding that \nnbb-decay Fermi matrix elements vanish \cite{Simkovic13,Hyvarinen15}.
This condition effectively restores isospin symmetry, which is a robust symmetry in nuclei.

Most QRPA calculations assume spherical initial and final nuclei.
This simplification may not be justified in some cases, leading to overestimated \nubb-decay NMEs, as suggested by EDF theory, shell model and IMSRG studies \cite{Menendez08-2,Menendez10-2,Rodriguez10,Yao20}.
Recently, \textcite{Fang18} calculated QRPA NMEs including deformation --- violet bars in Fig.~\ref{fig:nme_light}.
The deformed QRPA NMEs are much smaller than in most spherical QRPA calculations, in fact they are comparable to shell model NMEs.
The main reason is the suppression due to the small overlap between the initial and final nuclei, which is reduced for states with different deformation.
This overlap, usually neglected in QRPA calculations, has been shown to lead to very small NMEs \cite{Mustonen13} --- magenta crosses in Fig.~\ref{fig:nme_light}.
However, \textcite{Fang18,Mustonen13} probably underestimate NMEs because they assume only one deformation for each nuclear state.
A more realistic description should consider the mixing between different configurations, for instance via the GCM \cite{Rodriguez10,Hinohara14}, which enhances NME values.

\subsubsection{Energy-density functional theory}
\label{sec:EDF}

The largest NMEs in Figs.~\ref{fig:nme_light} and \ref{fig:nme_heavy} derive from EDF theory.
This approach is used extensively, and describes very well the ground state properties and spectroscopy of medium-mass and heavy nuclei \cite{Bender03}.
Based on a mean-field description, EDF theory calculations incorporate additional correlations beyond the mean field via restoration of symmetries --- notably particle number and angular momentum --- and configuration mixing in terms of the GCM \cite{Robledo19,Egido16}.
The variational solution of the Schr\"odinger equation is obtained self-consistently
in configuration spaces of about a dozen harmonic oscillator shells.
Unlike other many-body methods, EDF theory can calculate any nucleus with a common nuclear functional (or interaction).

EDF \nubb-decay NMEs are computed in the closure approximation.
However, the same level of sophistication in odd-odd nuclei can only be achieved at a much larger computational cost, and is only feasible in lighter nuclei \cite{Bally14}.
This prevents tests of \B\ and \nnbb-decay EDF matrix elements.
Two EDF versions have been applied to \nubb\ decay: using nonrelativistic \cite{Rodriguez10,Vaquero14} and relativistic \cite{Song17,Yao15} functionals, both including the GCM \cite{Yao21b}.
The two sets of NMEs are quite similar except in $^{150}$Nd, see Fig.~\ref{fig:nme_light}.
The significantly larger NMEs of EDF with respect to the nuclear shell model can be traced back to nuclear correlations: a comparison of NMEs for calcium isotopes calculated with uncorrelated nuclear states found NME agreement as good as $\sim30\%$ \cite{Menendez14} instead of the factor of $\sim$3 difference in Fig.~\ref{fig:nme_light}.
Unfortunately, actual \bb\ emitters are strongly correlated nuclei.

Possible explanations for the large EDF theory NMEs are high-seniority components of the nuclear states beyond the reach of EDF theory, and proton-neutron pairing correlations \cite{Menendez16} not explicitly taken into account by current calculations.
Shell model and GCM studies suggest that both effects reduce NME values \cite{Hinohara14}.
The precise impact, however, needs to be checked in actual EDF theory calculations.
An extension to handle nuclear Hamiltonians instead of functionals, so that proton-neutron pairing can be accommodated explicitly, has been proposed recently \cite{Bally20}.

\subsubsection{The interacting boson model}
\label{sec:IBM}

The IBM \cite{Arima76,Arima78} exploits symmetry arguments to model nuclei as a collection of bosons, called $s$-, $p$-, $d$-bosons... according to their angular momentum.
Bosonic operators are then mapped to nucleon degrees of freedom \cite{Otsuka78}, typically using the shell model as a reference.
However, effective operators adapted to the collective subspace where the IBM degrees of freedom operate have also been proposed \cite{VanIsacker17}.

IBM calculations of \nubb\ decay use the closure approximation.
Typical IBM configuration spaces encompass one harmonic oscillator shell for neutrons and protons --- like the shell model.
On the other hand, like EDF theory, calculated IBM NMEs for \bb\ emitters \cite{Barea09,Barea15}
do not explicitly include proton-neutron pairing correlations, which could lead to an overestimation of the NMEs, as discussed in Sec.~\ref{sec:EDF}.
Recently, $p$-bosons that capture explicitly proton-neutron pairing correlations have been introduced in NME calculations for isotopes around \Ca\ \cite{VanIsacker17}.
For light-neutrino exchange, IBM NMEs take intermediate values with respect to other NME calculations, see Fig.~\ref{fig:nme_light}, while IBM NMEs are similar to most other NMEs when \nubb\ decay is mediated by the exchange of a heavy particle, see Figs.~\ref{fig:nme_contact} and \ref{fig:nme_heavy}.

\subsubsection{Ab initio methods}
\label{sec:abinitio}

Ab initio or first principles nuclear structure calculations solve the many-body problem by treating explicitly all nucleons in the nucleus, interacting though realistic nuclear forces.
Ab initio methods handle nucleon-nucleon and three-nucleon forces and, likewise, they can accommodate one-body operators as well as 2bc.
They yield in general excellent agreement for the nuclear properties of light and medium-mass nuclei \cite{Carlson15,Barrett13,Navratil16,Hebeler15,Hergert16,Hagen14,Lee18}.
Here we briefly review the most common ab initio approaches applied to \B\ and \bb\ decays.

Quantum Monte Carlo (QMC) techniques are one of the most accurate ab initio methods in very light $A\lesssim12$ nuclei \cite{Carlson15}, with promising extensions proposed for medium-mass systems \cite{Lonardoni18}.
The QMC approach is based on the time evolution of a trial nuclear state, according to the nuclear Hamiltonian, towards the lowest-energy configuration.
With sufficiently long evolution, the exact properties of the ground state can be obtained.
QMC \B\ decay calculations are discussed in Sec. \ref{sec:beta}.
More interestingly, \textcite{Pastore18} and \textcite{Weiss21} have studied \nubb-decay NMEs in $A\leq12$ nuclei.
While these isotopes are not of experimental interest, QMC NMEs provide benchmarks for other approaches that can also cover heavier nuclei.
Compared to shell model NMEs for $^{10,12}$Be, QMC ones are $\sim20\%$ smaller \cite{Wang19}, but the ratio between QMC $M^{0\nu}_\text{short}$ and $M^{0\nu}_\text{long}$ NMEs is consistent with Figs.~\ref{fig:nme_light} and \ref{fig:nme_contact} \cite{Cirigliano19long}.
QMC nuclear states include reliable SRCs, which can be combined with the shell model via the generalized contact formalism \cite{Weiss21}. This results in NMEs for heavy $\beta\beta$ emitters reduced by about $30\%$ with respect to the shell model ones in Fig.~\ref{fig:nme_light}.

The no-core shell model (NCSM) is the ab initio extension of the nuclear shell model to very large configuration spaces \cite{Barrett13,Navratil16}.
Unlike the nuclear shell model, the lowest-energy nucleons are treated explicitly, which implies the absence of a core.
On the other hand, high-energy orbitals are added to the configuration space until reaching convergence.
Because of the combinatorial scaling of the shell model framework, the NCSM is limited to very light nuclei $A\lesssim22$, and reaching these nuclei actually requires strategies to select the most relevant configurations \cite{Abe12,Roth09}.
Section~\ref{sec:beta} presents NCSM \B\ decay results in light systems, and \textcite{Basili19,Yao21} gives benchmark NCSM \nubb-decay NMEs from $^{6}$He to $^{22}$O that include full nuclear correlations. 

The IMSRG introduced in Sec.~\ref{sec:status} is based on unitary transformations that simplify the solution of the many-body problem \cite{Hergert16}.
Transition operators, including \bb-decay ones, are transformed consistently.
The IMSRG relies on adding correlations on top of a reference state, which needs to be a reasonable approximation, sufficiently close to the exact solution.
The advantage of the IMSRG over NCSM or QMC is the polynomial, rather than exponential, scaling with the number of nucleons, 
making extensions to \bb\ decay feasible.
Two versions of the IMSRG have been applied to \bb\ decay: the IM-GCM described here and the VS-IMSRG discussed in the last paragraph of this section.
Since the initial and final \bb\ nuclei typically involve substantial nuclear correlations, the IM-GCM uses a combination of various reference states \cite{Yao18} and then exploits the GCM to explore additional nuclear correlations such as deformation and proton-neutron pairing.
The IM-GCM NMEs agree well with NCSM benchmarks in light systems \cite{Basili19,Yao21}.
\textcite{Yao20} have obtained a \Ca\ $M^{0\nu}_\text{long}$ smaller than other calculations --- see Fig.~\ref{fig:nme_light} --- complemented by a $M^{0\nu}_\text{short}$ NME which enhances $M^{0\nu}_\text{light}$ by about 40\%~\cite{Wirth21}. Strategies to study heavier $\beta\beta$ emitters are in progress \cite{Romero21}.

The CC method is also based on adding nuclear correlations to a reference state \cite{Hagen14}.
Such correlations can be singles, doubles, triples, etc., according to the number of creation-annihilation operators allowed.
Similar to the IMSRG, CC calculations scale polynomially.
At present, however, CC studies are mostly limited to spherical nuclei in the vicinity of magic or semimagic isotopes --- those nuclei for which nuclear correlations are especially small \cite{Hagen17,Morris18}.
CC \B\ decays in heavy nuclei are discussed in Sec.~\ref{sec:beta}.
Very recently, \textcite{Novario21} have calculated the \Ca\ \nubb-decay NME, see Fig.~\ref{fig:nme_light}. An extension of
the CC framework breaking rotational invariance was necessary to take into account the deformation of \Ti\ \cite{Novario21}. The NME is consistent with the IM-GCM one, but with larger uncertainty. More recent CC nuclear structure calculations restore rotational symmetry through angular momentum projection~\cite{Hagen:2022tqp}.

The NCSM \cite{Dikmen15}, CC \cite{Jansen14} and IMSRG \cite{Bogner14,Stroberg16a} can be formulated to yield an effective Hamiltonian in a shell model space.
At the same time, they solve the energy of the shell model core.
Therefore the ab initio calculation can be separated in two steps: first, the energy of the shell model core and an effective shell model interaction are obtained.
Second, the shell model techniques described in Sec.~\ref{sec:shell model} are used to calculate observables such as nuclear energies or NMEs.
In particular, the valence-space version of the IMSRG method (VS-IMSRG) has been used extensively, with good agreement on nuclear properties up to tin \cite{Stroberg19,Taniuchi19}.
The VS-IMSRG \Ca\ \nubb-decay NME is in good agreement with the IM-GCM and CC ones \cite{Belley21}, see Fig.~\ref{fig:nme_light}. Furthermore, first VS-IMSRG NMEs have been obtained for the heavier \Ge\ and \Se\, see Sec.~\ref{sec:NME_tests}.

\subsection{``$g_A$ quenching''
\label{sec:quenching}}

The so-called ``$g_A$ quenching'' is a potential source of uncertainty in \nubb-decay NMEs.
Most calculations of GT \B\ decay matrix elements overpredict experiment, indicating the need of a correction, sometimes attempted by quenching the value of the axial coupling $g_A$.
Very recently \B\ decay has been studied with the ab initio methods introduced in Sec.~\ref{sec:abinitio}.
These calculations suggest that the overprediction of matrix elements is more likely related to the GT \B\ decay operator than to $g_A$.
Ab initio \nubb-decay studies including 2bc are needed to assess whether the NMEs discussed in Sec.~\ref{sec:NMEs} require a compensation similar to GT \B\ decay ones, less compensation, or none at all.

\subsubsection{\B\ decay half-life values}
\label{sec:beta}

Theoretical nuclear structure typically does not reproduce well \B-decay half-life values in GT transitions of nuclei with masses similar to those of \bb\ emitters.
Calculated GT decay half-lives tend to underestimate data, which means that theoretical matrix elements are overestimated.
As a pragmatic fix to this deficiency, a quenching factor is usually introduced to reduce the strength of the GT operator, and consequently the calculated GT matrix elements.
Remarkably, in the nuclear shell model a common quenching factor $\bm{\sigma}\tau\to q\bm{\sigma}\tau$ with $q\sim0.7-0.8$ is sufficient to bring agreement with experiments for GT matrix elements across and entire mass range \cite{MartinezPinedo96,Wildenthal83,Chuo93}.
Nevertheless, to the extent that the need of quenching reflects the deficiency of a given nuclear many-body method to describe GT transitions, each nuclear structure model can be expected to need its own quenching factor \cite{Ejiri19}.
In general, more sophisticated approaches require less severe quenching.

An alternative view expresses the phenomenological modification required in GT transitions as a ``quenching" of the axial coupling constant, $g_A$ \cite{Suhonen17}.
The corresponding label ``$g_A$ quenching'' is used widely in the literature.
However, similar phenomenological adjustments have been advocated in nuclear electromagnetic transitions --- in particular, magnetic dipole transitions --- which do not depend on $g_A$ \cite{vonNeumannCosel98}.
Therefore it may be more appropriate to associate the quenching factor to the transition operator instead of the hadronic coupling $g_A$.

The origin of the quenching has been debated extensively, with two primary explanations.
One possibility is missing nuclear correlations, because calculations are performed in limited configuration spaces \cite{Bertsch82,arima87}.
Another possibility is corrections to the transition operator, such as 2bc --- meson-exchange currents --- presented in Sec.~\ref{sec:op_2b}.
They reflect neglected degrees of freedom, such as nucleon excitations to the $\Delta$ isobar \cite{Menendez11}.
Even though both effects were investigated for decades \cite{Towner87,Brown87}, the outcome was not conclusive.

Nuclear theory is finally in a position to address \B\ decays in not only light but also medium-mass and even heavy nuclei with ab initio methods that correct for both of the aforementioned deficiencies.
For $A\lesssim12$ systems that undergo \B\ decay the experimental rate can be confronted with ab initio QMC and NCSM calculations.
The theoretical predictions of GT matrix elements are in excellent agreement --- within a few percent --- with experiment, without the need of any adjustment \cite{Pastore18b,Gysbers19}.
\textcite{Gysbers19} also studied GT transitions of nuclei with mass number  $A\sim30$ and $A\sim50$ with the VS-IMSRG.
In contrast to standard shell model calculations that need sizable quenching, the VS-IMSRG reproduces measured GT transitions to better than 10\%.
\textcite{Gysbers19} also presented a detailed ab initio CC study of the GT decay of the doubly-magic $^{100}$Sn --- the largest GT transition observed in the nuclear chart.
The CC result agrees very well with the measured GT matrix element, without any adjustment.
The VS-IMSRG and CC analyses both conclude that nuclear correlations not included in previous calculations and 2bc contribute in a similar amount to the GT matrix element.
Further, the relative importance of 2bc and correlations depends on the nuclear interaction used: two body effects are larger for interactions with a less pronounced short-range character. That is, these two effects are intertwined. For instance, in QMC GT matrix elements obtained with ``hard'' potentials with marked short-range repulsion, the effect of 2bc is very small. In contrast, in ``softer'' potentials with less rich short-range correlations, the impact of 2bc is more relevant.  
In general, there may not be a dominant contribution to quenching, but two entangled ones with relative impact dependent on the nuclear interaction used.

The same considerations apply when comparing calculations to GT transitions extracted from charge-exchange reactions \cite{Frekers18,Ichimura06,Fujita11}. The shell model reproduces data well once the same quenching as in \B\ decay is included \cite{Caurier12,Iwata14}, perhaps because the normalization of GT transitions extracted from experimental cross-sections involves \B-decay half-lives. Using perturbation theory to obtain a GT operator that captures correlations beyond the configuration space also leads to good agreement with experiment \cite{Coraggio18}.

The findings of \textcite{Gysbers19} bring immediate implications.
Since 2bc are partially responsible for GT quenching, the expectation that 2bc are less important in \nubb\ than in GT decay,
as discussed in Sec.~\ref{sec:op_2b}, suggests that assuming a quenching $q^2$ in \nubb\ decay relative to $q$ for \B\ decay is not justified.
The first ab initio calculations have also explored the impact of missing nuclear correlations in \nubb\ decay. In \Ca\, they suggest that the value of the \nubb-decay NME is only moderately reduced \cite{Belley21,Yao20,Novario21}.
Perturbation theory studies also find a milder impact of additional correlations in \nubb\ decay than in GT transitions \cite{Coraggio20}.

\subsubsection{\B\ decay spectra}
\label{sec:e_spectrum}

The energy spectrum of the emitted electron is fixed by kinematics because a single nuclear matrix element dominates GT transitions.
By contrast, several matrix elements contribute to non-unique forbidden \B\ decays, and the electron spectrum is related to their relative impact \cite{Behrens71}.

This idea has been exploited to show that the shape of the \B\ spectrum of for example $^{113}$Cd depends on the relative value of nuclear matrix elements divided in two groups: those proportional to the vector and axial couplings $g_V$ and $g_A$ \cite{Haaranen16,Haaranen17} --- the groups stem from the leading terms in Eq.~\eqref{1b_currents}. Assuming that all axial and all vector matrix elements need to be corrected by the same quenching, a fit to the \B\ spectrum leads to a preferred value of the ratio $g_A/g_V$.
Higher sensitivities appear if competing contributions from different matrix elements partially cancel, a feature identified in other non-unique \B\ decays as well \cite{Kostensalo17,Kostensalo17b,Kumar20}.

A comparison to measurements of the $^{113}$Cd \B\ spectrum suggests a ratio of about $g_A/g_V\sim0.9$, valid for the shell model and other many-body methods \cite{COBRA19}.
The ratio $g_A/g_V\sim0.9$ is roughly consistent with the ``$g_A$ quenching'' observed in \B\ decay but does not reproduce the $^{113}$Cd half-life.
This inconsistency, also found in other \B\ decays \cite{Kumar21}, could be explained if each axial or vector matrix element is affected by a different deficiency, and therefore needs its own quenching factor.
Even though at least some of the matrix elements may require a similar quenching \cite{EXO_spectrum20}, a different behavior has been indeed observed in shell model studies of non-unique \B\ decays \cite{Warburton88,Yoshida18b,Zhi13}.
In summary, \B-decay spectra of non-unique forbidden decays provide complementary tests of the quality of nuclear theory calculations, and are will help to determine whether ``$g_A$ quenching'' can be resolved by simple scaling of the axial coupling.

\subsubsection{\nnbb\ decay and 2$\nu$ECEC}
\label{sec:2nbb}

\bb\ decay and ECEC with the emission of two (anti)neutrinos have been measured in a dozen nuclei \cite{Barabash20,XENON19}.
Since the initial and final nuclei are common to two-neutrino and neutrinoless decays, a good description of \bb\ and ECEC decay is a key test of \nubb-decay predictions.
The calculation of the corresponding nuclear matrix elements is, however, more challenging than for the neutrinoless mode. This is because the emission of neutrinos reduces the momentum transfer below typical nuclear energy differences, and the closure approximation leading to Eq.~\eqref{eq:nme} is not always justified (closure is, nonetheless, typically used by the IBM \cite{Barea15}).
Thus the intermediate nucleus with an odd number of neutrons and protons needs to be calculated explicitly. 

The remarkable nuclear shell model prediction of the \Ca\ decay rate \cite{Caurier90,Poves95} before its measurement \cite{Balysh96} highlighted the power of the use of this many-body method to predict \bb\ decay rates.
These works assumed that the same deficiency present in GT matrix elements in the vicinity of \Ca\ was also present in \bb\ decay, so that the quenching needed for \B\ decay was used in \bb\ decay.
Following the same strategy, the $^{124}$Xe two-neutrino ECEC was predicted \cite{Coello18} in very good agreement with the subsequent, recent observation \cite{XENON19}.
Likewise, shell model \bb-decay matrix elements in other nuclei reproduce measured decay rates when corrected by quenching factors which are in reasonable agreement with those needed for GT transitions \cite{Caurier12,Neacsu15,Senkov16,Kostensalo20}.
Only in \Xe\ the quenching needed in \bb\ decay may be more pronounced \cite{Caurier12}. Matrix elements obtained with perturbation theory on top of the shell model also reproduce the measured half-life well \cite{Coraggio18,Coraggio22}.

Other many-body methods can also access \bb\ decays.
The QRPA often uses \nnbb\ decay to fix the strength of the proton-neutron pairing interaction \cite{Rodin03}, but when alternative schemes are adopted predicted \bb-decay rates are qualitatively good, both when the QRPA is used with phenomenological Hamiltonians \cite{Simkovic18} and with energy-density functionals \cite{Mustonen13,Hinohara22}.
In fact, the QRPA half-life for $^{124}$Xe \cite{Suhonen13,Pirinen15} well-predicted its subsequent measurement, albeit
with a larger uncertainty than the shell model.
The larger error arises from the difficulty of disentangling quenching from the strength of the proton-neutron pairing in the QRPA, see Sec.~\ref{sec:QRPA}.
An effective theory for \bb\ and ECEC decay, based on \B\ and EC data \cite{Coello17}, also predicted well the $^{124}$Xe \nnbb-decay half-life, including a quantified theoretical uncertainty \cite{Coello18}. The same method very recently has given predictions for \nubb\ NMEs~\cite{Brase21} with quantified uncertainties, favoring smaller values than all methods in Table~\ref{tab:NMEs}. Very recently, IBM \nnbb-decay calculations have been performed beyond the closure approximation \cite{Nomura22}.

Ab initio methods can calculate \bb-decay matrix elements as well, but this is more challenging because of the relevance of the intermediate states. For \Ca, CC theory mildly overestimates the experimental matrix element even when including the effect of 2bc, as it does in \B\ decay~\cite{Novario21}. In turn, the VS-IMSRG \Ca\ matrix element is too small even without 2bc \cite{Belley21}.

Similarly to \B\ decay, measured \bb-decay spectra further test nuclear theory. Even if only one nuclear matrix element dominates the decay rate, precisely measured spectra can be sensitive to small deviations caused by subleading matrix elements \cite{Simkovic18b}.
A precision analysis of the \Xe\ summed electron energy spectrum provides limits which confront shell model and QRPA predictions \cite{KamLAND-Zen19}.
The results constrain both the quenching needed to reproduce the half-life --- different for each calculation --- and the ratio of the leading and subleading matrix elements.
The analysis is consistent with most of the theoretical predictions, but excludes part of the QRPA results.

In addition, a precise \bb-decay spectrum measurement can inform the distribution of the leading \bb-decay matrix element as a function of the virtual states in the intermediate odd-odd nucleus through which the decay proceeds~\cite{Simkovic01}, see Eq.~\eqref{eq:nme_full}.
Recent analyses in \Mo\ \cite{NEMO19,lumineu-2nbb} and \Se\ \cite{Se2nbb} suggest that only the lowest $J^P=1^+$ state contributes, the so-called single-state dominance.
Charge-exchange reactions also hint at single-state dominance in the \nnbb\ decay of \Zr\ \cite{Thies12}.
This behavior should be reproduced by all theoretical calculations.

\subsection{Connections to nuclear structure measurements}
\label{sec:NME_tests}

Besides \B\ and \bb\ decays, a good description of the main properties of the nuclear states is a necessary requirement for trustworthy \nubb-decay NME calculations.
On the other hand, processes with similar momentum transfer --- muon capture and neutrino-nucleus scattering --- can give additional insights.
Double Gamow-Teller (DGT) and second-order electromagnetic transitions may offer a unique opportunity due to their relation to \nubb-decay NMEs.

\subsubsection{Spectroscopy and charge exchange}
\label{sec:spectroscopy}

Nuclei involved in \bb\ decay have even numbers of protons and neutrons.
Due to the attractive nuclear pairing interaction, they have $J^P=0^+$ ground states, with vanishing quadrupole and magnetic moments.
Theoretical calculations, therefore, need to be confronted with other ground state properties. 
A valuable source of information comes from orbital occupation probabilities deduced from analyses of low-energy nucleon adding and removing reactions \cite{Freeman12,Entwisle16,Szwec16,Freeman17}.
In fact, various studies have used the experimental results to improve the description of the initial and final \bb-decay nuclei \cite{Menendez09b,Suhonen08,Suhonen10,Kotila16,Deppisch20}. The impact on the NMEs is illustrated by the IBM error bar in Fig.~\ref{fig:nme_light}. This moderate effect proved beneficial to bring QRPA and shell model NME predictions in better agreement to each other.

In addition, the quality of nuclear structure calculations is assessed by comparing excitation energies of low-lying states \cite{ENSDF} and their electromagnetic transitions \cite{XUNDL}.
In particular, the shell model and EDF theory agree with data very well \cite{Rodriguez10,Song14,Neacsu15,Horoi16b,Vietze14,Hoferichter19,Coraggio18}, even though more subtle aspects such as pairing correlations \cite{Sharp19,Roberts13} or the triaxial character of $A=76$ \cite{Toh13,Henderson19,Ayangeakaa19} and $A=130$ \cite{Morrison20, Hicks22} nuclei have also been explored experimentally and challenge all theoretical studies.
Further, the shape of GT strength distributions as a function of energy is also sensitive to the nuclear structure of the nuclei involved, and has been used to test and improve nuclear interactions \cite{Alanssari16}. Thus, the comparison to the experimental GT strength at low energies is another stringent test for calculations \cite{Caurier12,Coraggio18,Iwata14} and for instance shows that $^{100}$Mo is currently more difficult to describe than other $\beta\beta$ nuclei \cite{Coraggio22}. While nuclear structure data for otherwise-stable $\beta\beta$ nuclei has been collected over decades, modern experiments keep illuminating new aspects that test theoretical predictions, for instance recent work on heavy $A=136$ \cite{Rebeiro21, NzobadilaOndze20} and $A=150$ \cite{Basak21} nuclei. It is very important to pursue further studies of this kind, as they may indicate physics missing in the calculations but relevant for \nubb\ decay. For instance, recent measurements on magnetic dipole transitions in $A\sim150$ nuclei have been used to fix IBM parameters, giving significant changes in NMEs to excited states \cite{Beller:2013eha,Kleemann:2021wkd}.

Similar benchmarks are demanded for ab initio calculations. In fact, the CC theory framework had to be extended by breaking rotational invariance to describe \Ca\ decay \cite{Novario21} due to the deformation of \Ti. The IM-GCM and VS-IMSRG calculations of \textcite{Yao20} and \textcite{Belley21} describe \Ca\ and \Ti\ in good agreement with experiment.
In contrast, the VS-IMSRG excitation spectra for the heavier nuclei \Ge, $^{76}$Se, \Se, and $^{82}$Kr are too stretched in energy \cite{Belley21}.

Unfortunately, two methods that describe well the nuclear structure properties of \bb-decay nuclei can differ significantly in their \nubb-decay NME predictions. For example, both relativistic \cite{Song14} and non-relativistic EDF theory \cite{Rodriguez10} describe the nuclear structure of \Nd\ and $^{150}$Sm well, but predict NMEs a factor three apart. In fact, a recent statistical shell-model analysis in the decay of \Ca\ to $^{48}$Ti finds that the nuclear structure properties of these nuclei are in general modestly correlated with the \nubb-decay NME~\cite{Horoi22}.
Nonetheless, nuclear structure is relevant: the energy of the lowest $2^+$ state in $^{48}$Ti, which is a measure of the deformation of that nucleus, is a property with a higher correlation. In the same fashion, a demand for the IM-GCM ab initio calculation was to describe well the low-lying electric quadrupole transition in $^{48}$Ti \cite{Yao18}.
The  consistent \Ca\ NMEs obtained with three ab initio approaches brings hope for more confident \nubb-decay NME results in the future.

\subsubsection{Muon capture and neutrino scattering}
\label{sec:mcns}
Nuclear structure or \B\ decay measurements do not probe, however, momentum transfers $p\sim100$\,MeV similar to \nubb\ decay.
Two other processes offer the opportunity to do so.
The first is muon capture, mostly explored with the QRPA \cite{Zinner06,Jokiniemi19,Jokiniemi19b}.
An ideal comparison would involve capture branching ratios to low-energy excited states, which can also be computed with the shell model and VS-IMSRG \cite{Jokiniemi21b}.
The second process is inelastic neutrino-nucleus scattering.
In the very few nuclei, such as $^{12}$C, for which data are available~\cite{Formaggio:2012cpf}, different shell model studies disagree on whether matrix elements at finite momentum transfer are overpredicted, like in \B\ decay, or not \cite{Hayes00,Volpe00,Hayes03,Suzuki06}.
Given the relevance of large momentum transfer observables to test calculations of \nubb-decay NMEs, it would be important to get more data on both muon capture and neutrino-nucleus scattering.

\subsubsection{Two-nucleon processes: $\beta\beta$ decay, pair transfers, double Gamow-Teller, and $\gamma\gamma$ transitions}
\label{sec:DGT}

\nubb\ decay is also special from the nuclear structure point of view.
None of the observables discussed in Sec. \ref{sec:spectroscopy} or \ref{sec:mcns} have been found to be well correlated to \nubb\ decay. Nuclear processes involving two nucleons are more promising.

Until very recently, no clear correlation had been observed between \nnbb- and \nubb-decay NMEs, other than an analytical relation between the corresponding transition densities \cite{Simkovic11}. However, \textcite{Horoi22} find that the \Ca\ \nnbb-decay NME is the quantity best correlated with this nucleus' \nubb-decay NMEs from a set of 24 nuclear structure properties of \Ca\ and $^{48}$Ti. This result is supported by subsequent work by \textcite{Jokiniemi22}, which finds a good linear correlation between the two $\beta\beta$-decay NMEs across the nuclear chart for shell model and QRPA calculations. Using the correlation, \nubb-decay data can be used to predict \nubb-decay NMEs.

Two-nucleon transfer amplitudes have also been related to \nubb\ decay \cite{Brown14}. A recent experiment involving a two-neutron transfer from $^{138}$Ba to $^{136}$Ba found a larger contribution of pairs of neutrons coupled to angular momentum $J=0$ than predicted by the shell model \cite{Rebeiro20}. The size of the missing contributions is about $50\%$. This result suggests that the $J=0$ contribution to \nubb-decay NMEs could also be underestimated. This experimental finding is consistent with theoretical work finding a $\sim25\%$ enhancement when increasing the shell model configuration space \cite{Iwata16}, but which also predicts more contributions from $J>0$ neutron pairs which suppress the NME. The latter cancellation is still to be confirmed by experiments.

Double charge-exchange reactions can also provide insights on NMEs, in a similar connection to the one between \B\ decay and (single) charge-exchange reactions.
This is in spite of the fact that charge-exchange experiments probe the strong instead of the weak interaction. An experimental program pursues this approach \cite{Cappuzzello18}, which demands developments in reaction theory \cite{Lenske19,Bellone19}.

Connections between DGT transitions and \bb\ decay have been indicated for decades \cite{Vogel88,Auerbach89}. DGT transitions can be explored with double charge-exchange reactions \cite{Takaki15,Takahisa17,Uesaka15}.
Most works, however, focus on sum rules or the DGT giant resonance \cite{Sagawa16,Auerbach18,Roca-Maza19}. \textcite{Shimizu18} studied DGT transitions to the ground-state of the final nucleus, i.e., between the initial and final \bb-decay nuclei.
Remarkably, a comparison of shell model DGT and \nubb-decay NMEs shows a very good linear correlation, valid from calcium to xenon~\cite{Brase21}.
The same correlation is fulfilled for EDF theory \cite{Rodriguez13}, even though for any \bb\ emitter EDF NMEs are much larger than shell model ones, see Fig.~\ref{fig:nme_light}.
Further, the IBM also finds a linear correlation \cite{Santopinto18,Barea15}. 
The QRPA in general does not observe a correlation \cite{Simkovic18}, but it does so when exploring different values of the proton-neutron pairing \cite{Jokiniemi22}.
The origin of the linear correlation could be explained by
the relatively short-range character of both DGT and \nubb-decay NMEs \cite{Anderson10,Bogner12} in the shell model --- neutrons more than $\sim3$~fm apart almost do not contribute to these processes --- in contrast to the QRPA where DGT transitions receive contributions from nucleons separated by long distances.
Further work is needed to establish the robustness of the correlation between DGT and \nubb\ decay, and to connect experimental cross-sections with DGT matrix elements. A very recent {\it ab initio} study also finds a linear correlation between DGT and \nubb-decay NMEs, albeit somewhat weaker than in the shell model~\cite{Yao22}. The likely reason are the additional nuclear correlations included in the {\it ab initio} calculations.

Second order electromagnetic transitions have been measured recently in competition
with the much faster single $\gamma$ decays \cite{Walz15,Soderstrom20}.
Electromagnetic decays connect states in the same nucleus,
so that a relation with \nubb\ decay can only be expected in the final \bb-decay system,
when the initial state is the double isobaric analogue --- the state
with the same nuclear structure but rotated in isospin space --- of the initial \bb-decay state.
A recent study finds a linear correlation between $\gamma\gamma$ magnetic dipole and \nubb-decay NMEs
in the shell model framework \cite{Romeo21}, opening the door to exploring \nubb\ decay with nuclear spectroscopy.

 \section{Experimental aspects and methods}
\label{sec:exp}

Neutrinoless double-beta decay can be observed in a variety of isotopes, each of them
characterized by specific features, such as the $Q$-value, the natural abundance, or material properties.
Because of this, each isotope enables different detection techniques, with their own strengths and technical challenges.
This makes the experimental field extremely diverse and always in evolution. 

This section presents the modern experimental methods used to search for \nubb\ decay.
We summarize the \nubb-decaying candidate isotopes in Sec.~\ref{sec:exp:iso} and their
related detection concepts and event reconstruction techniques in Sec.~\ref{sec:exp:detection}.
Section~\ref{sec:exp:background} describes the background interfering processes which can mimic \nubb-decay events in recent and future experiments, while the techniques to discriminate them are reviewed in Sec.~\ref{sec:sec:activebkgsuppression}.
Finally, the statistical techniques used to extract the sought-after signal are
covered in Sec.~\ref{sec:exp:stat}, where we find that the sensitivity of these
experiments is driven not only by the amount of deployed isotope, but critically
also by the background rate, with a distinct advantage for those experiments
that are at or near the ``background-free'' regime.

While all of these sections are written to be accessible for both expert and nonexpert readers, Sec.~\ref{sec:exp:iso}, Sec.~\ref{sec:exp:detection}, and Sec.~\ref{sec:sec:activebkgsuppression} are more general in nature, while Sec.~\ref{sec:exp:background} and Sec.~\ref{sec:exp:stat} are more technical.
The detailed aspects of specific experiments that might be of higher interest for experts in the field are the subject of Sec.~\ref{sec:prj}, which makes extensive use of Sec.~\ref{sec:exp:stat} to present each project on an equal footing.

\subsection{Isotopes}
\label{sec:exp:iso}

\nubb\ decay is observable in isotopes for which the single $\beta$ decay is energetically forbidden and the only allowed decay channel is \bb\ decay.
Nature provides us 35 such isotopes that can undergo $\beta^-\beta^-$,
and 34 that can undergo $\beta^+\beta^+$, $\varepsilon\beta^+$, or
$\varepsilon\varepsilon$ \cite{Tretyak:2002dx}\footnote{For a review of
$\beta^+\beta^+$, $\varepsilon\beta^+$, and $\varepsilon\varepsilon$ processes,
see \cite{Maalampi:2013mba}.}.
The candidate isotopes for experimental searches are those
readily available at the level of thousands of moles (i.e., hundreds of kg) or more,
with a high Q-value and thus a large decay rate,
and compatible with existing detection technologies.
A number of the key isotopes meeting these criteria is listed in Tab.~\ref{tab:isotopes}.
\begin{table*}[tb]
  \setlength{\tabcolsep}{4pt} \caption{Target isotopes currently being pursued by leading \nubb-decay experiments.
    The reported \nnbb-decay half-life values are the most precise available in literature.
    The \nubb-decay half-life values are the most stringent 90\% C.L. limits. 
  }
  \label{tab:isotopes}
  \begin{tabular}{llD{.}{.}{4.8}D{.}{.}{2.6}lll}
    \toprule
    \smallspace
    Isotope & Daughter
    & \multicolumn{1}{l}{\Qbb\footnote{Values from \cite{48CaQ,76GeQ,82SeQ,96ZrQ,100MoQ,110PdQ,116CdQ,130TeQ,136XeQ,150NdQ}.}}
    & \multicolumn{1}{l}{$f_{\text{nat}}$\footnote{Values from \cite{natabund}.}}
    & $f_{\text{enr}}$\footnote{Values from \cite{Ca_enr2,LEGEND:2021bnm,Se_enr,Zr_enr1,Mo_enr,Pd_enr,Cd_enr1,Te_enr,Xe_enr,Nd_enr}.
      Enrichment is performed via gas centrifuge for all isotopes except for \Ca, for which the unpublished report in \cite{Ca_enr2} used electrophoresis \cite{Ca_enr1}.
      For \Zr, 86\% is commercially available \cite{Zr_enr1}, however a 91\% enrichment was achieved at smaller scale \cite{Zr_enr2}.
      For \Cd, 82\% is the highest value used in a \nubb-decay experiment\cite{Cd_enr1},
      however enrichment up to 99.5\% is possible\cite{Cd_enr2}.
      For \Nd, 91\% is the highest value used in a \nubb-decay experiment\cite{Barabash:2018yjq},
      however enrichment up to 98\% is possible\cite{Nd_enr}.}
    & \Tnnbb\footnote{Values from \cite{Ca2nbb,Gerda2nbb,Se2nbb,Argyriades:2009ph,lumineu-2nbb,Barabash:2018yjq,Cuore0-2nbb,Exo2nbb,Arnold:2016qyg}.}
    & \Tnubb\footnote{$90\%$ C.L. limits from \cite{GERDA:2020xhi,cupid0-final,Argyriades:2009ph,Barabash:2018yjq,Adams:2021xiz,CUORE:2021mvw,KamLAND-Zen:2016pfg,Arnold:2016qyg,Umehara:2008ru,Armengaud:2020luj}. } \\
    & & \multicolumn{1}{l}{[keV]} & \multicolumn{1}{l}{$[\%]$}                         & $[\%]$           & [yr]   & [yr]   \\
    \colrule
    \smallspace
    $^{\phantom{1}48}$Ca & $^{\phantom{1}48}$Ti  & 4\,267.98(32)  & \phantom{3}0.187(21) & 16   & $\left(6.4^{+0.7}_{-0.6}(\text{stat})^{+1.2}_{-0.9}(\text{syst})\right)\cdot10^{19}$     & $>5.8\cdot10^{22}$ \\
    $^{\phantom{1}76}$Ge & $^{\phantom{1}76}$Se  & 2\,039.061(7)  & \phantom{3}7.75(12)  & 92   & $\left(1.926\pm94 \right)\cdot10^{21}$                                                               & $>1.8\cdot10^{26}$ \\
    $^{\phantom{1}82}$Se & $^{\phantom{1}82}$Kr  & 2\,997.9(3)    & \phantom{3}8.82(15)  & 96.3 & $\left(8.60\pm0.03(\text{stat})^{+0.19}_{-0.13}(\text{syst})\right)\cdot10^{19}$         & $>3.5\cdot10^{24}$ \\
    $^{\phantom{1}96}$Zr & $^{\phantom{1}96}$Mo  & 3\,356.097(86) & \phantom{3}2.80(2)   & 86   & $\left(2.35\pm0.14(\text{stat})\pm0.16(\text{syst})\right)\cdot10^{19}$                             & $>9.2\cdot10^{21}$ \\
    $^{100}$Mo           & $^{100}$Ru            & 3\,034.40(17)  & \phantom{3}9.744(65) & 99.5 & $\left(7.12^{+0.18}_{-0.14}(\text{stat})\pm0.10(\text{syst})\right)\cdot10^{18}$         & $>1.5\cdot10^{24}$ \\
$^{116}$Cd           & $^{116}$Sn            & 2\,813.50(13)  & \phantom{3}7.512(54) & 82   & $\left(2.63^{+0.11}_{-0.12} \right)\cdot10^{19}$                                                       & $>2.2\cdot10^{23}$ \\
    $^{130}$Te           & $^{130}$Xe            & 2\,527.518(13) & 34.08(62)            & 92   & $\left(7.71^{+0.08}_{-0.06}(\text{stat})^{+0.12}_{0.15}(\text{syst})\right)\cdot10^{20}$ & $>2.2\cdot10^{25}$ \\
    $^{136}$Xe           & $^{136}$Ba            & 2\,457.83(37)  & \phantom{3}8.857(72) & 90   & $\left(2.165\pm0.016(\text{stat}\right)\pm0.059(\text{syst}))\cdot10^{21}$               & $>1.1\cdot10^{26}$ \\
    $^{150}$Nd           & $^{150}$Sm            & 3\,371.38(20)  & \phantom{3}5.638(28) & 91   & $\left(9.34\pm0.22(\text{stat})^{+0.62}_{-0.60}(\text{syst})\right)\cdot10^{18}$         & $>2.0\cdot10^{22}$ \\
    \botrule
  \end{tabular}
\end{table*}

Acquiring isotope is feasible if the market can supply it in large amounts
at an affordable cost on the timescale of years or less.
Isotopic enrichment drives the total cost for the material but
allows a minimization of the \bb-inactive material,
which is mandatory for most detector technologies.
Isotopes with a high natural abundance and with low-abundant neighboring isotopes
are easier, and thus cheaper, to enrich.
The cost also depends on the viable enrichment technologies
(gas ultracentrifuge is a cost-effective, high-throughput technique used for
nearly all \bb\ isotopes),
on the chemical processes involved, on the level of enrichment,
and on the required purity of the final material.
Finally, isotopes of elements used in commercial applications
are typically cheaper due to their mature supply chains.
On the other hand, when an experiment requires a quantity of material
that is of the order of the yearly global supply, competing commercial demands
lead to higher costs, and if
significant quantities of depleted material enter the commercial supply chains,
independent supply chains must anyway be pursued.

The \nubb-decay rate scales as $Q_{\beta\beta}^5$ for light neutrino exchange,
and $Q_{\beta\beta}^7$ for other exchange mechanisms \cite{Haxton1985}.
Higher $Q$-values thus lead to a more rapid decay, yielding higher sensitivity.
Moreover, higher $Q$-values (greater than $\sim$2\,MeV) are advantageous because fewer processes
can mimic the \nubb-decay signal.

The candidate isotope must be suitable for use with a detection technology
capable of identifying a single \nubb-decay signal in thousands of moles of material.
Thus the detector must be able to distinguish the signal from mimicking processes.
Consolidated detector technologies have been available for decades
for some isotopes, or have lately become available for others.
Recent promising developments might allow exploiting further isotopes in the future.
Finally, some isotopes lend themselves to advantageous detection techniques.
For example, some sources can be made directly into detectors, such as $^{76}$Ge and $^{136}$Xe,
minimizing the amount of inactive, background-generating material near or within the detector.

\subsection{Signal detection}
\label{sec:exp:detection}

\nubb\ decay is a nuclear decay, and thus is a random process obeying Poisson statistics.
Given that \nubb-decay half-life values are much longer than the age of the universe,
the expected signal rate is homogeneous in time for the entire duration of an experiment.
\nubb\ decay is a three body process with the final state composed of the nuclear recoil
plus the two emitted electrons. Since the electron mass is orders of magnitudes
smaller than that of the daughter nucleus,
the nuclear recoil energy is negligible ($<$0.1\,keV), and the sum of the electron energies
is practically equivalent to the available energy, i.e., to \Qbb.
The daughter nucleus can be produced either in its ground state or in some excited state,
and then relax down to its ground state emitting $\gamma$ rays.

In principle, the measurable quantities in \nubb\ decay are the kinetic energies and
momenta of the emitted electrons, as well as the position and time of the decay.
Additionally, any $\gamma$ ray emitted in \nubb\ decay to excited states can be measured,
and the daughter nucleus can be tagged via atomic or molecular means as well.

\begin{figure}

  \begin{tikzpicture}[xscale=3,yscale=1.5]

    \def\qvalue{2.039}

\fill[fill=gray!50, variable=\t, domain=0:0.9*\qvalue, samples=100]
    (0,0)
    -- plot (\t, { ( \t^4 +10*\t^3  + 40*\t^2 + 60*\t + 30 ) * \t * ( \qvalue - \t )^5/200 })
    -- (\qvalue,0)
    -- cycle;

    \draw plot [variable=\t, domain=0:\qvalue, samples=100]
    (\t, { ( \t^4 +10*\t^3  + 40*\t^2 + 60*\t + 30 ) * \t * ( \qvalue - \t )^5/200 });

\def\sigma{0.015}
    \draw[fill=gray!50, variable=\t, domain=0.95*\qvalue:1.05*\qvalue, samples=100]
    (0.95*\qvalue,0)
    --plot ( \t, {0.5*exp( -(\t-\qvalue)^2/2./(\sigma)/(\sigma) )} )
    (1.05*\qvalue,0)
    --cycle;

\draw [-latex] (0,0) -- +(\qvalue*1.1,0);
    \draw [-latex] (0,0) -- +(0,1.8);

\node (A) at (\qvalue,-0.2) {\Qbb};
    \node (B) at (\qvalue/2.,-0.2) {Energy};
    \node[rotate=90] (C) at (-0.1,0.9) {Events};
    \node (D) at (0.55,1.67) {\nnbb};
    \node (D) at (\qvalue,0.62) {\nubb};

  \end{tikzpicture}
  \caption{Theoretical spectra of \nnbb\ and \nubb\ decays
    with $1.5\%$ energy resolution (FWHM).
    The relative normalization  is for illustrative purpose only.}\label{fig:signature}
\end{figure}
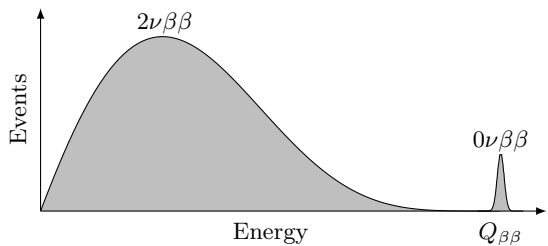

For all isotopes, \nubb\ decay competes with its \nnbb-decay mode,
a five body decay with two electrons and two anti-neutrinos emitted.
The anti-neutrinos escape undetected, hence the sum energy of the two electrons is $\leq$\Qbb.
The electron momenta in both modes vary statistically, and
the daughter nucleus and any $\gamma$ ray emitted by the daughter de-excitation
are common between the \nubb- and \nnbb-decay modes.
Thus measurement of the sum electron energy is a necessary condition for 
discovery: the \nubb\ decay will feature a peak at \Qbb,
the \nnbb-decay mode a continuum from zero to \Qbb\ (Fig.~\ref{fig:signature}).
In a high-resolution experiment free of other background sources, an energy
measurement is also a sufficient condition for discovery.

The measurement of energy is optimal if the candidate isotope is part of the detector itself.
This condition simultaneously maximizes the detection efficiency (by optimizing
containment) while minimizing any energy loss, providing a clear signature for the signal as a \nubb-decay
peak over the background, with shape governed by the energy resolution function
of the detector. The resolution function is characterized by its 
full width at half-maximum (FWHM), which is given by $2 \sqrt{2 \ln 2} \, \sigma$ for a
Gaussian resolution function of width $\sigma$, but can also be used to
characterize and compare less ideal detector responses.
A \nubb-decay event reconstructs at \Qbb\ for those nuclei within the
active volume of the detector with a fully calibrated non-zero energy response,
and for those events whose ejecta are fully contained within the active volume.

In many detectors, the measurement of energy is accompanied by identification of
the time and sometimes also the position of the energy deposition within the
detector. These observables further improve the \nubb-decay signal identification by discriminating 
correlated or time-varying backgrounds as well as background contributions with
spatial distributions distinct from that of the parent isotope.
For large monlithic detectors with strong self-shielding, the discrimination of
external backgrounds can be captured with a fiducial volume cut that removes
high-background regions near the detector edges that do not contribute to the sensitivity.

Particle tracking allows to independently measure the single electron momenta and directions
and consequently their angular correlation.
Precise tracking of electrons with MeV-scale energies,
including the measurement of the decay location,
is only achievable in low-pressure gaseous detectors\footnote{In this context,
we define pressures $\sim1$\,bar as low,  and in the $10-20$\,bar range as high.}
or highly pixelated solid detectors at present.
For the former, the quest to maximize the isotope mass motivates the use of composite detectors
with solid sources sandwiched between gaseous tracking detectors. Pixelated
detectors on the other hand require small surface to volume ratios.  In either case, the passage of
the decay electrons through passive material near the detection medium induces an unavoidable energy loss
and distorts the expected \Qbb\ peak in the sum energy spectrum.
In monolithic solid or liquid detectors the electrons emitted in \nubb\ decay
scatter multiple times within a few mm$^3$ before being absorbed,
making precise tracking of the decay electrons and identification of the decay
vertex impractical.
In high-pressure gas detectors a \nubb-decay event will feature
two electron tracks of several cm length originating from the same unknown location.
The single electron momenta and angular correlation cannot be measured
unambiguously, but the single electron energies can be estimated.

The presence of the final state nucleus at the event vertex is a nearly
unique feature of \bb\ decays. The first experimental discovery of
$\nnbb$ decay was made using geochemical methods in which 
trace levels of \bb\  decay daughters were detected in materials
containing the parent isotope \cite{Inghram:1950qv}.
The tagging of de-excitation gammas in the final state can provide such
identification in real-time but requires the phase-space-suppressed decay to an
excited state of the daughter nucleus. Nevertheless, such excited state decays have been
observed in a number of \bb\  nuclei \cite{Belli:2020dxt},
and for some nuclei, \nnbb\ decay has been probed unambiguously
only via excited state decays, e.g., $^{110}$Pd and $^{102}$Pd \cite{Lehnert:2016vfq}.
Modern efforts to perform real-time tagging of the daughter nucleus in its
ground state are based on its atomic features, as first proposed by \textcite{Moe1991},
and are advantageous if the background reduction outweighs the \nubb-decay signal loss
due to the tagging inefficiency. If methods can be developed to perform such
tagging with high efficiency, with sufficient resolution such a search would be
effectively background-free.

\subsubsection{Detector concepts}

\begin{figure}[tb]
  \centering
  \includegraphics[width=\columnwidth]{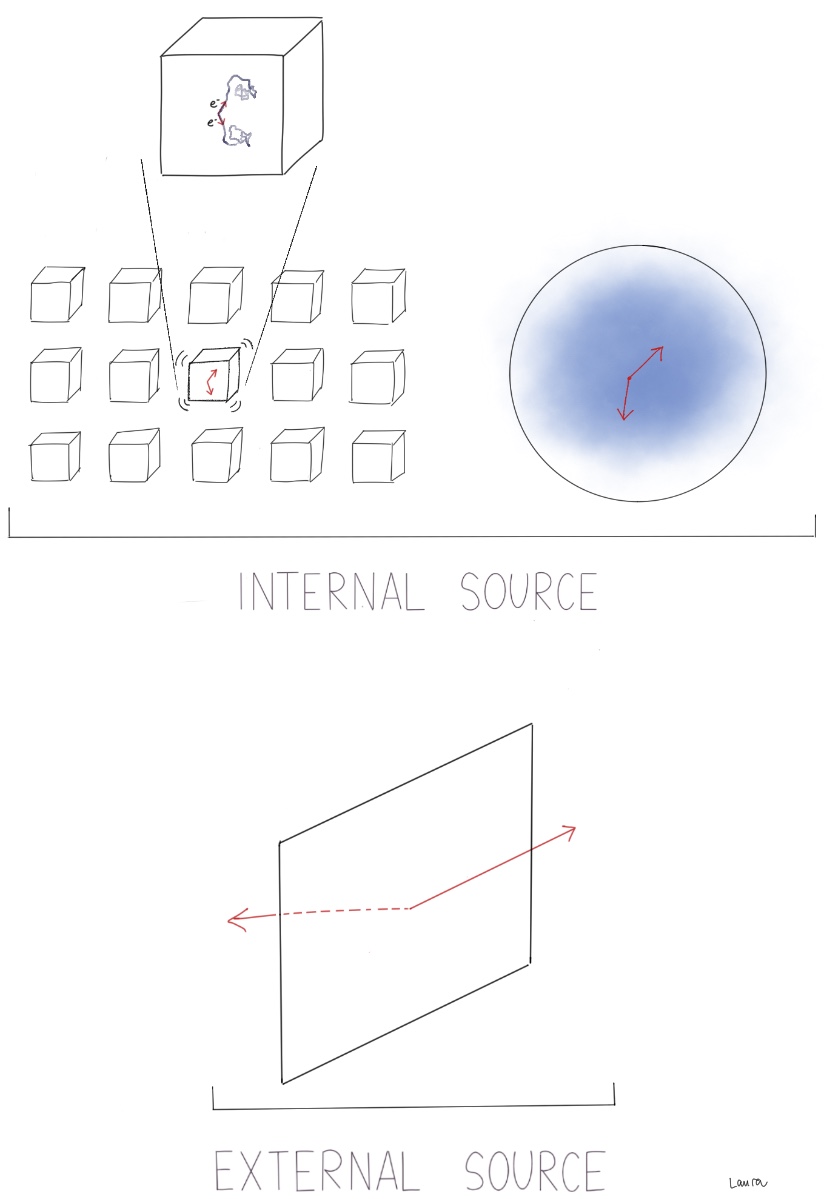}
  \caption{Artistic rendition of the three detector concepts used to search for \nubb\ decay: internal-source experiments using solid state detectors (left) or monolithic liquid or gas detectors (center), and composite experiments (right) for which the source is external to the detection apparatus. Image courtesy of Laura Manenti.}
  \label{fig:det-concept}
\end{figure}

Fig.~\ref{fig:det-concept} shows the three detector concepts used to search for \nubb-decay: solid state detectors with an embedded source, monolithic
liquid or gas detectors with an embedded or dissolved source,
and composite detectors with external sources.

Solid state detectors consist of crystals grown from material containing the \bb\ isotope.
The crystal mass typically ranges from a few hundred grams to few kilograms, depending on the material.
The crystal volumes are up to hundreds of cubic centimetres: they can fully contain electrons of a few MeV emitted at their center,
but can miss a fraction of the energy for those emitted near the borders.
Typical containment efficiencies for solid detectors are in the 70--95\% range, depending on the material and detector dimensions.
The energy released by the two electrons cannot be distinguished,
thus crystal-based experiments mainly perform calorimetric measurements.
The primary readout channels are ionization and phonons,
yielding energy resolutions up to the per-mill level. Scintillating
detectors are also pursued.
A main feature of these experiments is granularity,
allowing a staged approach where the total detector mass can be increased in steps
using the same infrastructure.
On the other hand, the production and operation of a large number of detectors 
can be challenging.

Monolithic liquid and gas based experiments are single detector systems
where the \bb\ isotope either coincides with or is dissolved in the active material.
Typical linear dimensions range from 1 to 10 meters.
Liquid detectors of this size are larger than both the range of electrons
and the attenuation length of $\gamma$ rays with few MeV of energy.
This guarantees a containment efficiency close to 100\%,
and yields an increasing sensitivity to a \nubb-decay signal towards the detector core,
where the presence of background events is suppressed (see Sec.~\ref{sec:exp:background}).
Gas detectors can have linear dimensions up to a few meters, yielding a 
 $\gtrsim75\%$ containment efficiency.
The possible readout channels are scintillation light and ionization (see
Sec.~\ref{sec:exp:detection:eventreconstruction}), so
the active material is surrounded (fully or partly) by light or charge detectors.
Liquid and gas detectors are primarily used for calorimetry, but with sufficient
spacial resolution they can provide some event topology and electron tracking
reconstruction capability, particularly in gas detectors.  Given that the \bb\ isotope is homogeneously
distributed in the active material, in these detectors it is not possible to
unambiguously identify the starting point of the electron tracks. 
Thus measurements of single electron energies and emission angle distributions
can be estimated only with significant uncertainties.
Due to self-shielding, in monolithic experiments the background from external
$\gamma$ rays decreases exponentially as the linear dimension increases. 
Meanwhile, backgrounds from the readout scales with the instrument area, and isotope mass
scales with the volume. These qualities 
make them among the most easily scalable technology in terms of signal-to-background
ratio.
If the \bb-decaying isotope is dissolved in the active material, a staged approach is possible 
by increasing the isotope concentration in phases. On the other hand, if the source coincides
with the active material, an increase in mass will require the deployment,
and thus the construction, of a new, larger infrastructure.

In composite experiments, the \bb-decaying isotope is embedded in a sub-millimetre thin foil
to allow the electrons to escape with minimal energy losses.
The source is surrounded by low-pressure gas detectors that measure the single
electron momenta. 
The full reconstruction of the decay kinematics allows efficient discrimination
of \nubb-decay events from other processes.
Composite experiments also present several challenges.
The energy reconstruction is biased by the energy losses, and
the composite detector system yields a low detection efficiency.
Both the isotope mass and number of readout detectors are proportional to the foil area,
thus mass scaling is less advantageous than for other technologies.
On the other hand, composite systems are not bound to the measurement of a single isotope,
and offer uniquely precise measurement of the decay vertex and angular
correlation, providing the possibility to distinguish between different \nubb-decay mechanisms
through the measurement of the electron angular correlation.

\subsubsection{Event reconstruction}
\label{sec:exp:detection:eventreconstruction}

\begin{figure}[tb]
  \centering
  \includegraphics[width=\columnwidth]{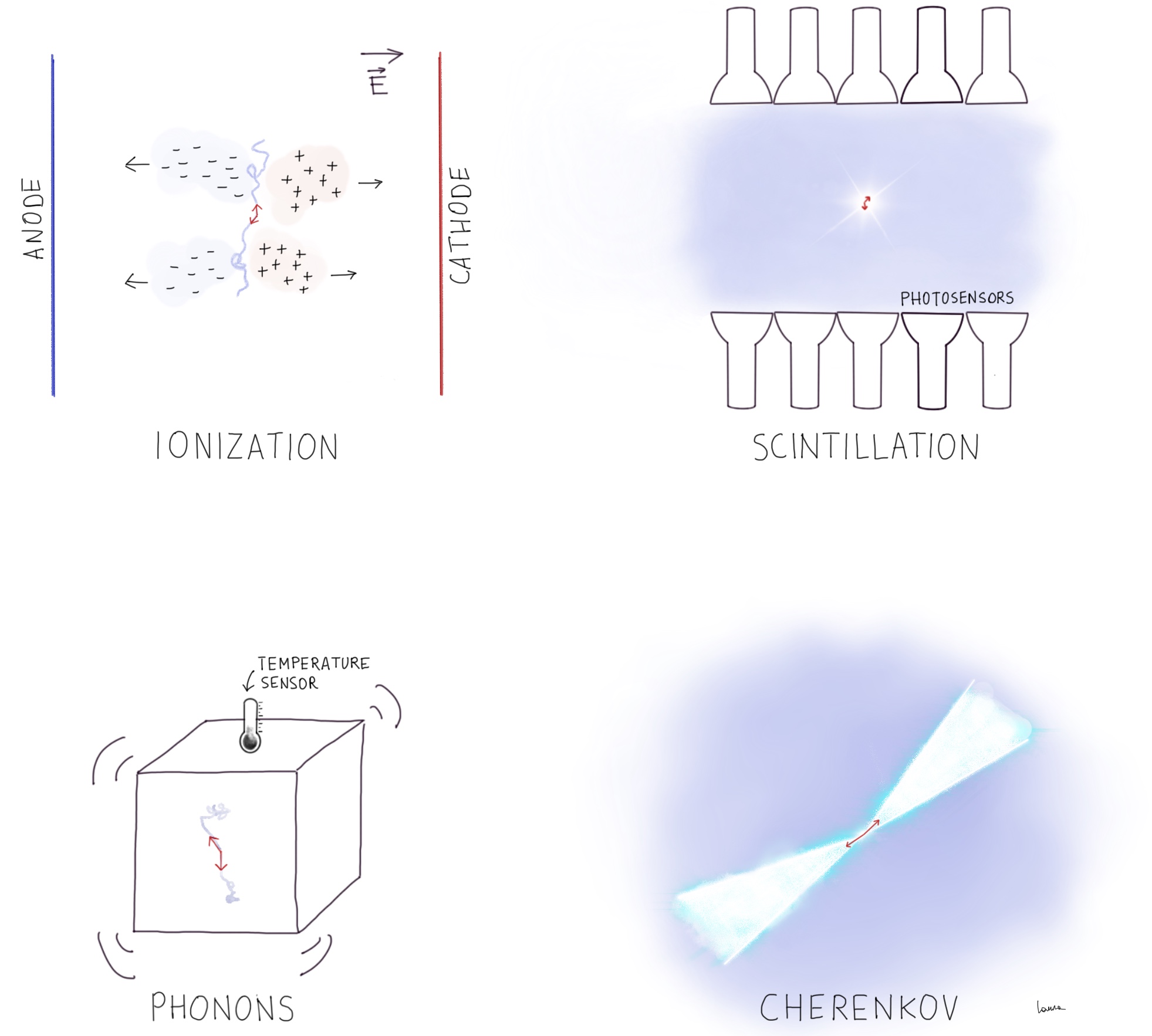}
  \caption{Artistic rendition of the four channels exploited by the experiments to detect the electrons emitted in a \nubb\ decay. In left-to-right reading order: ionization, scintillation, phonons, and Cherenkov radiation.
Image courtesy of Laura Manenti.}
  \label{fig:det-channels}
\end{figure}

The event reconstruction in \nubb-decay experiments can exploit four primary
detection channels: ionization, phonons, scintillation light, and Cherenkov
light. These channels are summarized in Fig.~\ref{fig:det-channels} and discussed in this section. We also address briefly
methods being pursued for real-time daughter nucleus tagging.

Energetic charged particles traversing a material lose energy due to ionization
processes in which charge carriers (e.g., ions, electrons, holes) are produced.
The charge carriers can be collected via an electric field,
and read out as an electrical signal.
The number of produced charge carriers is inversely proportional to the ionization energy
for gas and liquids, or to the mean energy necessary for the creation of an electron-hole pair
in semiconductor crystals. 
The best achievable energy resolution 
is determined by the variance in the number of charge carriers, which exhibit
sub-Poisson fluctuations characterized by the Fano factor ($F$) \cite{Fano:1947zz}. The optimal
resolution for measuring energy deposition $E$ is thus
\begin{equation}
  \text{FWHM} = 2.355 \sqrt{F \, w \, E},
\end{equation}
where $w$ is the mean energy required to produce a charge carrier, and we have used 
the Gaussian approximation with $2 \sqrt{2 \ln 2} \approx 2.355$.  
Typical Fano factors for ionization detectors are $\sim$0.1-0.2.  The value of $w$ 
ranges from a few eV in semiconductor detectors to tens of eV in noble elements.
In practice, the energy resolution is further limited by the charge collection efficiency,
which strongly depends on the detector technology. For instance, energy resolution in a xenon time projection chamber (TPC) can be optimal in the gas phase, but is
degraded in the liquid phase due to charge recombination \cite{BOLOTNIKOV1997360}.
An important aspect of the ionization channel is that the charge collection
is typically slower than the electronic readout.
Hence, the charge arrival time allows reconstruction of the spatial distribution
of the ionization, and thus provides a handle in the identification of different
event topologies.

Energy released in a crystal results also in the production of
phonons, collective excitations of the crystal lattice.
Phonons can be detected by sensors capable of collecting and transforming them
into an electrical signal proportional to the deposited energy.
Since phonons do not leave the crystal, they eventually thermalize, and can thus
be detected from the difference in temperature they induce in crystals cooled to
cryogenic temperatures ($\sim$10\,mK), for example using neutron transmutation doped germanium (NTD) sensors \cite{ntd1,ntd2},
superconducting transition-edge sensors (TES) \cite{tes1,tes2},
Metallic Magnetic Calorimeters (MMC) \cite{mmc},
or Kinetic Inductance Detectors (KID) \cite{kid1,kid2,kid3}.
NTDs have a volume of $O(10)$\,mm$^3$ and resistances in the 1-100\,M$\Omega$ range,
provide signals of few seconds length and feature a large dynamic range,
which makes them suitable for measuring energies up to $\sim10$\,MeV.
TESs, MMCs and KIDs have lower noise and thresholds than NTDs but a narrower dynamic range,
thus they are typically employed
for detecting smaller signals where a low threshold is crucial.
To a rough approximation, the energy resolution for phonon detection from the deposition of energy
$E$ in a crystal at temperature $T$ is:
\begin{equation}
  \text{FWHM} = 2.355 \sqrt{ \varepsilon_a [F\, E +  C(T) \cdot T]  + \sigma_n^2}\quad,
\end{equation}
where $\varepsilon_a=k_BT$ is the average phonon activation energy, the
second term involving the heat capacity $C(T) \propto T^3$
accounts for fluctuations from phonon exchange with the thermal bath, and
$\sigma_n$ is the contribution from noise. The massive devices required for \nubb-decay searches typically have long thermalization time scales
that make the readout sensitive to noise in the vibrational frequency range,
so in practice the contribution from $\sigma_n$ has dominated.
In general, crystals employed in \nubb-decay searches feature an energy resolution
which can be as good as 5\,keV. 
As the name suggests, cryogenic calorimeters excel at measuring energy.
Nevertheless, for some crystals different particles induce slightly different
signal shapes, thus allowing --- to some extent --- particle identification techniques.

Following the incidence of ionizing radiation, certain organic materials, inorganic
crystals, and noble elements de-excite by scintillation light emission.
The light yield depends on the material, and exhibits non-linearities due to
effects such as scintillation quenching (ionization density dependence), which must
be characterized and calibrated in-situ.
Typical light yield values for organic materials and noble elements are
$\sim$10 photons per keV of deposited energy, but can be as high as $\sim$70\,photons/keV.
The emission spectrum is continuous, material-dependent,
and goes from the ultraviolet to the visible range.
The light is detected via the photo-electric effect using optical sensors, such
as photo-multiplier tubes (PMT),
silicon photo-multipliers (SiPM), or avalanche photo-diodes (APD).
Each light sensor is characterized by a quantum efficiency,
defined as the detection probability for an incoming photon.
The quantum efficiency is also a function of photon frequency, and typically has
a maximum of 30--40\% for PMTs, but approaches $\sim$100\% for the other technologies.
If the scintillation spectrum does not match well with the quantum efficiency profile,
a wavelength shifter is placed between the main scintillator and the detector.
Wavelength shifters are scintillator materials that absorb higher frequency (e.g., ultraviolet) photons
and emit lower frequency ones.
In the end, the detected number of photons thus depends on the scintillation spectrum,
the quantum efficiency profile, the wavelength shifter transmission spectrum (if present),
and the probability for a photon to travel from the scintillator to the detector,
during which a photon can be reflected, refracted, scattered, or absorbed.  
In many liquid organic scintillators the emission and absorption spectra
overlap, and so a photon can also be re-emitted multiple times before being detected.
The light yield can be tuned by adding as a solute a second scintillator
that shifts the photons to higher wavelength, where the primary scintillator is transparent.
The energy resolution is given by
\begin{equation}
  \text{FWHM} = 2.355 \sqrt{E \,Y \, \langle P_t \rangle\,  f_\Omega
  \epsilon_q},
\end{equation}
where $Y$ is the light yield, $\langle P_t \rangle$ is the average transmission
probability, $f_\Omega$ is the fractional solid angle instrumented with
photosensitive surfaces, and $\epsilon_q$ is the quantum efficiency of the light
detector. The product of these four factors yields the number of
photoelectrons collected per unit energy deposition, and has 
typical values on the order of one photon per keV or less.
The relatively small number of detected quanta, 
combined with a Fano factor of $\sim$1 due to the small fraction of
$E$ that ends up as detected scintillation light,
results in a FWHM that is an order of magnitude
larger than the one obtained in the ionization channel.
A crucial aspect of scintillators is the time profile of their light emission.
The de-excitation typically follows a double-exponential profile
with decay times differing by over an order of magnitude.
The fast component provides a precise measurement of the event time.
In large scintillator experiments, the measurement of the fast component for the same event
at opposite sides of the detector also provides the location of the energy deposit,
via time-of-flight measurement.
Moreover, in many scintillators the ratio between the amount of light in the fast
and the slow component depends on the interacting particle,
thus allowing particle identification. 

Cherenkov radiation is emitted when a charged particle travels in a medium at a speed
higher than the phase velocity of light in the same medium.
The Cherenkov spectrum is a continuum that is more intense at short wavelengths (ultraviolet)
but ranges up into the red.
A 1\,MeV electron
produces hundreds of photons, depending on the refractive index of the medium \cite{Aberle:2013jba}.
Thus Cherenkov radiation effectively cannot be used for calorimetry in
\nubb-decay experiments, but can provide some information on the identity and
the initial directions of the emitted electrons.
Its mere presence identifies the particles as electrons as opposed to alphas or nuclear recoils.
Cherenkov light is also emitted on a cone pointed along the particle direction.
The electrons do not follow a straight trajectory in a solid or liquid,
but a large fraction of the Cherenkov photons are produced at the beginning of the track,
when the electron direction is still aligned with its emission direction.
The Cherenkov cone is hence smeared by the electron scattering,
but can be used to some extent for event topology reconstruction.

Multiple channel readouts are beneficial to improve the reconstruction of event topology
or to discriminate electrons from other ionizing particles.
For example, the ionization or phonon channels can be used for calorimetry,
while scintillation can be exploited for distinguishing between $\beta/\gamma$ and $\alpha$ particles,
and to provide a more accurate event timing, improving the spatial reconstruction
performed with the ionization signal.
In liquid noble TPCs, the collection of scintillation light along with
ionization can also improve the energy reconstruction \cite{EXO-200:2019bbx}, as fluctuations in
charge recombination that quench ionization also result in increased scintillation.
The simultaneous readout of scintillation and Cherenkov light is possible even if more complicated,
as their emission spectra and time profiles partly overlap.
Cherenkov and scintillation light can be distinguished by timing \cite{Caravaca:2016fjg,Gruszko:2018gzr,Caravaca:2020lfs,Land:2020oiz}
provided that the light detector has a sub-ns response time.
The scintillation light emission can be slowed down and/or wavelength-shifted \cite{Graham:2019zqb},
or suppressed with optical filters \cite{Kaptanoglu:2018sus}, albeit at the cost
of reduced light yield,
leading to sub-optimal energy resolution.
This technique was recently demonstrated in neutrino detectors~\cite{BOREXINO:2021efb} and is proposed for use in future\nubb\ experiments~\cite{Askins:2019oqj}. 

Finally, we briefly mention attempts to reconstruct the identity of the final-state nucleus after the decay.
Real-time tagging of the daughter nucleus is being pursued in liquid and gas Xe TPC
experiments, in which the final state nucleus is the alkaline earth metal Ba.
Tagging of single atoms of Ba can be achieved using, e.g., fluorescence
imaging \cite{McDonald:2017izm, nEXO:2018nxx, Rivilla:2020cvm}. 
The Ba ion following a decay can either be probed in-situ or transported to an
imaging stage via drifting in static or dynamic electric fields, or by the
physical motion of a collection probe \cite{Twelker:2014zsa, Brunner:2014sfa}. 
Efforts are underway to realize these techniques.

\subsection{Mimicking processes}
\label{sec:exp:background}

\nubb-decay events can be mimicked by a plethora of other physics processes,
which can be induced by cosmic rays, elements in the actinide decay chains,
anthropogenic radioactive isotopes, neutrinos, and \nnbb\ decay. While few of these 
create a peak at or near \Qbb, continuum backgrounds also pose a
problem since more signal counts are then required to observe a peak exceeding
the level of fluctuations.  Hence these background sources must also be either
eliminated or mitigated and minimized.

\subsubsection{Cosmic-ray induced processes}
\label{sec:exp:background:cosmic}

\begin{figure}[tbp]
  \centering
  \includegraphics[width=\columnwidth]{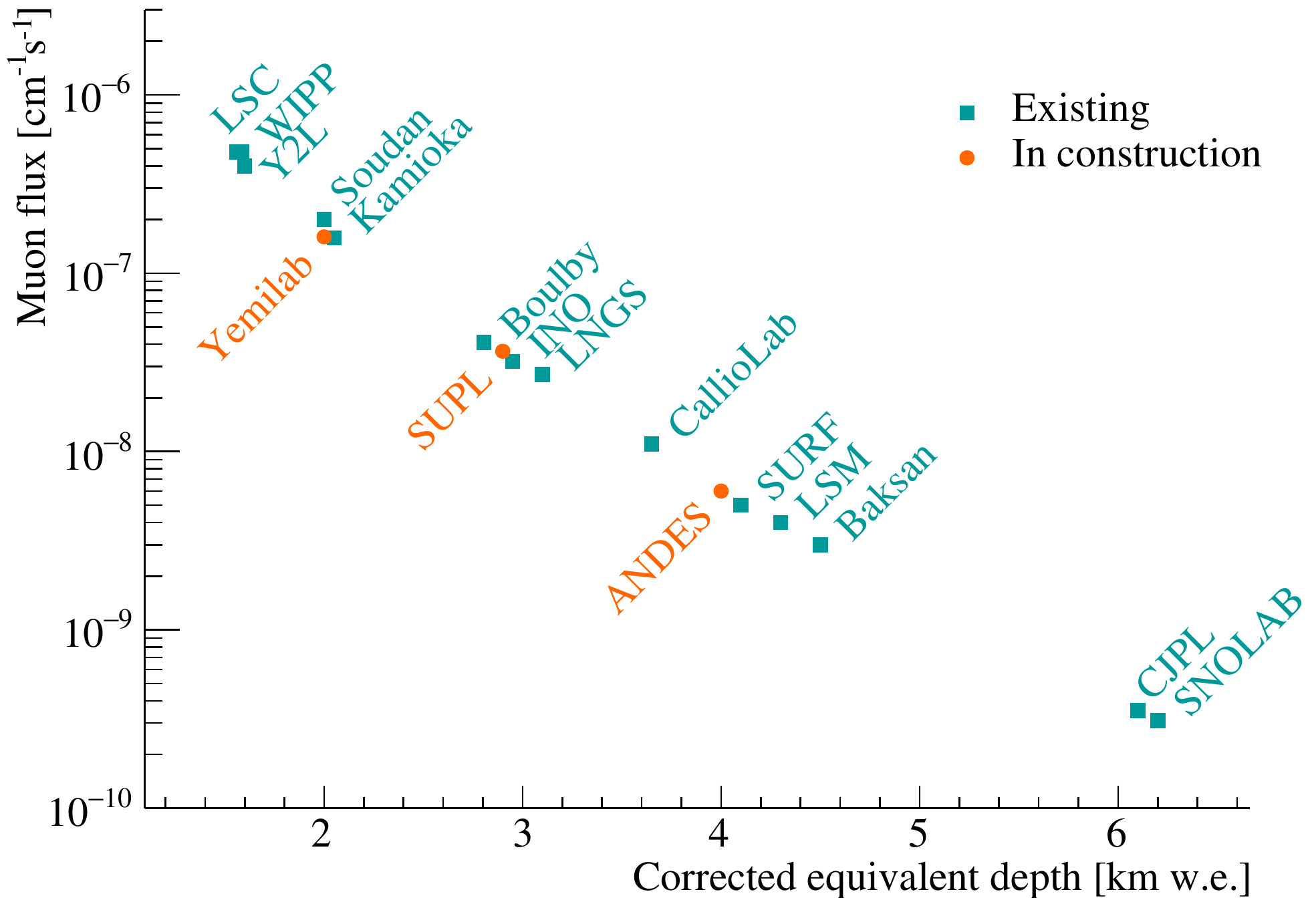}
  \caption{Muon flux as a function of kilometers of water-equivalent depth (km w.e.) for a selection of deep underground laboratories worldwide.
    The actual depth is corrected for the overburden shape, if it is not flat.
    Thus laboratories located under a mountain have a slightly lower equivalent depth 
  than the actual one. Figure adapted from \textcite{Ianni:2020rse}.
  }
  \label{fig:labs}
\end{figure}

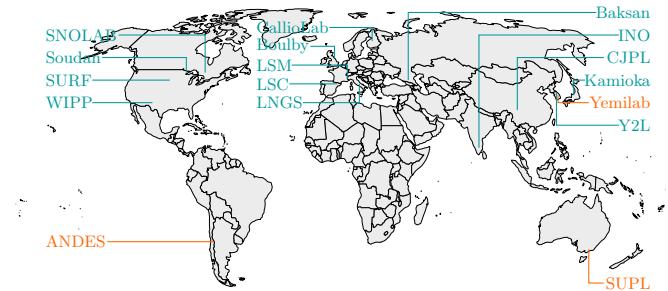
\begin{figure}[htbp]
  \centering
  \begin{tikzpicture}
    \definecolor{darkcyan}{RGB}{00,153,153}
    \definecolor{myorange}{RGB}{255,102,0}
    \node at(0,0){\includegraphics[width=\columnwidth]{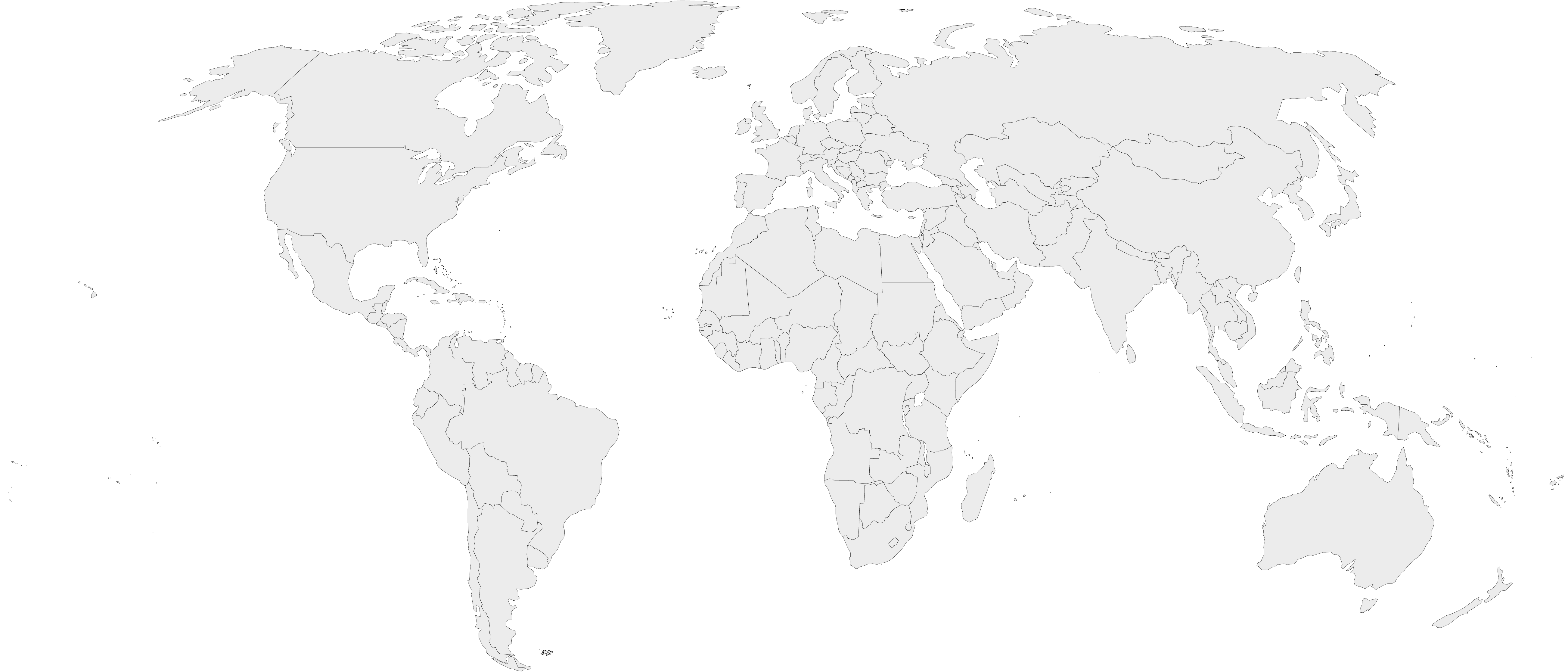}};
    \def\scale{0.7}
\node[anchor=west, color=darkcyan, scale=\scale] at(-4,1.5){\href{https://www.snolab.ca/}{\color{darkcyan}SNOLAB}};
    \draw[darkcyan] (-2.95,1.5)--(-1.8,1.5)--(-1.8,1.0);
    \node[anchor=west, color=darkcyan, scale=\scale] at(-4,1.2){\href{https://www.soudan.umn.edu/}{\color{darkcyan}Soudan}};
    \draw[darkcyan] (-3.17,1.2)--(-2.05,1.2)--(-2.05,1.);
    \node[anchor=west, color=darkcyan, scale=\scale] at(-4,0.9){\href{https://sanfordlab.org/}{\color{darkcyan}SURF}};
    \draw[darkcyan] (-3.3,0.9)--(-2.27,0.9);
    \node[anchor=west, color=darkcyan, scale=\scale] at(-4,0.6){\href{https://wipp.energy.gov/science-at-wipp.asp}{\color{darkcyan}WIPP}};
    \draw[darkcyan] (-3.3,0.6)--(-2.5,0.6);
\node[anchor=west, color=myorange, scale=\scale] at(-4,-1.25){\href{https://andeslab.org}{\color{myorange}ANDES}};
    \draw[myorange] (-3.1,-1.25)--(-1.68,-1.25);
\node[anchor=east, color=myorange, scale=\scale] at(4.2,-1.8){\href{https://www.supl.org.au/}{\color{myorange}SUPL}};
    \draw[myorange] (3.5,-1.8)--(3.3,-1.8)--(3.3,-1.35);
\node[anchor=west, color=darkcyan, scale=\scale] at(-1.2,1.6){\href{https://calliolab.com/}{\color{darkcyan}CallioLab}};
    \draw[darkcyan] (-0.15,1.6)--(0.42,1.6)--(0.42,1.42);
    \node[anchor=west, color=darkcyan, scale=\scale] at(-1.2,1.35){\href{https://www.boulby.stfc.ac.uk/}{\color{darkcyan}Boulby}};
    \draw[darkcyan] (-0.4,1.35)--(-0.08,1.35)--(-0.08,1.19);
    \node[anchor=west, color=darkcyan, scale=\scale] at(-1.2,1.1){\href{http://www.lsm.in2p3.fr/}{\color{darkcyan}LSM}};
    \draw[darkcyan] (-0.65,1.1)--(0.09,1.1)--(0.09,0.92);
    \node[anchor=west, color=darkcyan, scale=\scale] at(-1.2,0.85){\href{https://lsc-canfranc.es/}{\color{darkcyan}LSC}};
    \draw[darkcyan] (-0.68,0.85)--(-0.05,0.85);
    \node[anchor=west, color=darkcyan, scale=\scale] at(-1.2,0.6){\href{https://www.lngs.infn.it/en}{\color{darkcyan}LNGS}};
    \draw[darkcyan] (-0.5,0.6)--(0.25,0.6)--(0.25,0.85);
\node[anchor=east, color=darkcyan, scale=\scale] at(4.2,1.8){\href{https://www.inr.ru/eng/ebno.html}{\color{darkcyan}Baksan}};
    \draw[darkcyan] (3.4,1.8)--(0.9,1.8)--(0.9,0.9);
    \node[anchor=east, color=darkcyan, scale=\scale] at(4.2,1.5){\href{https://www.ino.tifr.res.in/ino/}{\color{darkcyan}INO}};
    \draw[darkcyan] (3.7,1.5)--(1.84,1.5)--(1.84,0.);
    \node[anchor=east, color=darkcyan, scale=\scale] at(4.2,1.2){\href{https://cjpl.tsinghua.edu.cn/column/home}{\color{darkcyan}CJPL}};
    \draw[darkcyan] (3.5,1.2)--(2.35,1.2)--(2.35,0.5);
    \node[anchor=east, color=darkcyan, scale=\scale] at(4.2,0.9){\href{https://www-sk.icrr.u-tokyo.ac.jp/en/about/}{\color{darkcyan}Kamioka}};
    \draw[darkcyan] (3.25,0.9)--(3.1,0.9)--(3.1,0.7);
    \node[anchor=east, color=myorange, scale=\scale] at(4.2,0.6){\href{https://cupweb.ibs.re.kr/en/facilities-and-equipment/underground-labs/yemilab/}{\color{myorange}Yemilab}};
    \draw[myorange] (3.3,0.6)--(2.89,0.6)--(2.89,0.7);
    \node[anchor=east, color=darkcyan, scale=\scale] at(4.2,0.3){\href{https://cupweb.ibs.re.kr/en/facilities-and-equipment/underground-labs/y2l/}{\color{darkcyan}Y2L}};
    \draw[darkcyan] (3.7,0.3)--(2.87,0.3)--(2.87,0.74);
  \end{tikzpicture}
  \caption{Worldwide map of deep underground laboratories.
    The existing laboratories are marked in teal,
    while the laboratories that are planner or under construction are in orange.
    The labels are linked to the laboratory webpages.}\label{fig:labmap}
\end{figure}

\nubb-decay experiments are conducted in deep underground laboratories
where they are shielded from the otherwise overwhelming background
due to cosmic rays generated in the Earth's atmosphere. 
While most of the generated particles are absorbed by a small amount of material,
muons can penetrate kilometers of rock and create background either directly, by interacting within the detector,
or indirectly, by producing protons and neutrons or showers of particles in the material surrounding the experimental setup. 
The muon flux decreases by roughly an order-of magnitude for every $\sim$1.5\,km of water or $\sim$0.5\,km or rock.
The muon flux attenuation for a selection of deep underground laboratories around the world is shown in Fig.~\ref{fig:labs}.
The corresponding laboratory location is shown in Fig.~\ref{fig:labmap}.
We refer the reader to \textcite{Ianni:2020rse} for a recent review of operating and planned undeground laboratories.

The muons reaching a deep underground laboratory have energies up to several TeV
and an angular distribution that depends on depth, density and profile 
of the rock surrounding the laboratory \cite{Ambrosio:1995cx}.
While large monolithic experiments can directly reconstruct muons crossing the detector active volume, TPCs and granular experiments are typically immersed in water tanks equipped with PMTs to detect the muon-induced Cherenkov light, or surrounded by plastic scintillator panels.
Without these precautions, cosmic rays would be a major background for most of the experiments \cite{Freund:2016fhz}.

Cosmic rays can also induce spallation as they traverse material.
The nucleons emitted by spallation\footnote{With some loose terminology, by spallation
we mean spallation and evaporation, as well as any associated/subsequent hadronic
showering.}
have energies up to the GeV scale
and can cause a variety of secondary nuclear processes, including further
spallation and fission. 
The relevance of these processes is threefold. First,
they can activate unstable ``cosmogenic'' nuclei in the experiment materials
prior to their deployment underground~\cite{Avignone:1986gd,Brodzinski:1990zz}.
Cosmogenic nuclei are worrisome when their decay can mimic \nubb-decay events,
e.g., if they undergo $\beta$ decay with a high end-point and have a half-life comparable
to run-time of the experiment. Thus it is common practice to minimize the above-ground exposure of all materials
that constitute or are near the detector,
and store them underground before the construction of the experiment to reduce the contamination due to short-lived isotopes \cite{Abgrall:2015jfa}.
In some cases, selected materials are directly fabricated underground \cite{Aalseth-2005,Hoppe:2014nva,Bandac:2017eks}.
Secondly, spallation from residual underground muons can induce the same activation in-situ.
Its occurrence is obviously much more rare than on the surface,
but it can be relevant for liquid scintillator experiments, where the amount of active material
is much larger than that of the isotope only.
If the isotopes activated in situ have a half-life of up to some minutes,
the corresponding events can be identified through a delayed time-coincidence
with the original muon event.
Isotopes with a longer half-life can be more problematic.
Finally, muon spallation in the nearby rock can generate a
penetrating, energetic neutron background that must be mitigated (see Sec.~\ref{sec:exp:background:neutrons}).

\subsubsection{Elements in the actinide decay chains}

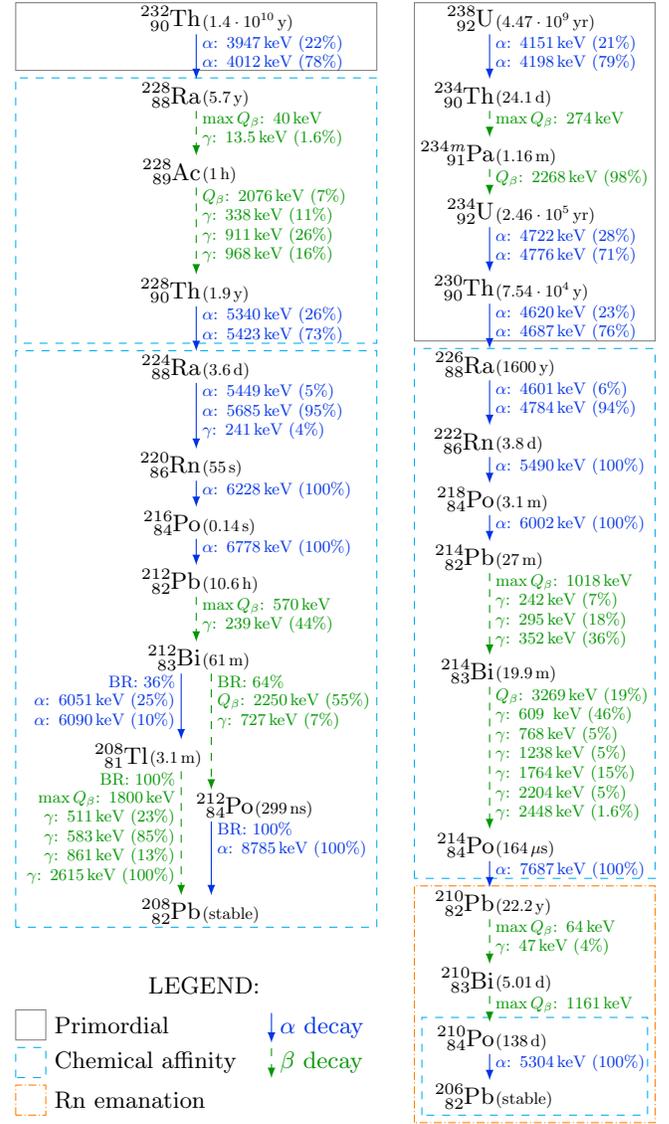
\begin{figure}[htbp]
  \centering
  \begin{tikzpicture}
    \dimendef\prevdepth=0
    \definecolor{greenish}{RGB}{00,150,00}
    \definecolor{blueish} {RGB}{00,50,230}
    \def\scale{0.7}
    \newcommand{\iso}[5]{
      \node[left, align=right]            at ([xshift=+5pt]#1) {\ce{^{#3}_{#4}#2}};
      \node[right, align=left, scale=\scale] at ([yshift=-1.35pt]#1)             {(#5)};
    }
    \newcommand{\alphaarrow}[3]{
      \draw[blueish, -latex, shorten >= +6pt, shorten <= +6pt](#1)--(#2);
      \node at ([yshift=-4pt]#1) [color=blueish, align=left, anchor=north west, scale=\scale, text width=82pt]{#3};
    }
    \newcommand{\alphaarrowleft}[4]{
      \draw[blueish, -latex, shorten >= +6pt, shorten <= +6pt](#1)--(#2);
      \node at ([yshift=-4pt]#1)  [color=blueish, align=right, anchor=north east, scale=\scale]{BR: #4\%};
      \node at ([yshift=-11pt]#1) [color=blueish, align=right, anchor=north east, scale=\scale, text width=82pt]{#3};
    }
    \newcommand{\alphaarrowright}[4]{
      \draw[blueish, -latex, shorten >= +6pt, shorten <= +6pt](#1)--(#2);
      \node at ([yshift=-4pt]#1)  [color=blueish, align=left, anchor=north west, scale=\scale]{BR: #4\%};
      \node at ([yshift=-11pt]#1) [color=blueish, align=left, anchor=north west, scale=\scale, text width=82pt]{#3};
    }
    \newcommand{\betaarrow}[3]{
      \draw[dashed, greenish, -latex, shorten >= +6pt, shorten <= +6pt](#1)--(#2);
      \node at ([yshift=-4pt]#1) [color=greenish, align=left, anchor=north west, scale=\scale, text width=83pt]{#3};
    }
    \newcommand{\betaarrowleft}[4]{
      \draw[dashed, greenish, -latex, shorten >= +6pt, shorten <= +6pt](#1)--(#2);
      \node at ([yshift=-4pt]#1)  [color=greenish, align=right, anchor=north east, scale=\scale]{BR: #4\%};
      \node at ([yshift=-11pt]#1) [color=greenish, align=right, anchor=north east, scale=\scale, text width=82pt]{#3};
    }
    \newcommand{\betaarrowright}[4]{
      \draw[dashed, greenish, -latex, shorten >= +6pt, shorten <= +6pt](#1)--(#2);
      \node at ([yshift=-4pt]#1)  [color=greenish, align=left, anchor=north west, scale=\scale]{BR: #4\%};
      \node at ([yshift=-11pt]#1) [color=greenish, align=left, anchor=north west, scale=\scale, text width=83pt]{#3};
    }
    \draw [cyan,dashed]   (0,2.6)    rectangle (4.8,10.27);
    \draw [cyan,dashed]   (0,10.37)  rectangle (4.8,13.9);
    \draw [gray]          (0,14.0)   rectangle (4.8,14.9);
    \draw [orange,densely dashdotted] (5.3,0)    rectangle (8.5,3.15);
    \draw [cyan,dashed]   (5.4,0.1)  rectangle (8.4,1.4);
    \draw [cyan,dashed]   (5.3,3.25) rectangle (8.5,10.3);
    \draw [gray]          (5.3,10.4) rectangle (8.5,14.9);
    \def\x{2.4};
    \def\y{80pt}
    \def\dx{0.7}
    \def\dy{30pt}
    \coordinate(Th232Pos)  at ( \x,     \y+338pt );
    \coordinate(Ra228Pos)  at ( \x,     \y+309pt );
    \coordinate(Ac228Pos)  at ( \x,     \y+280pt );
    \coordinate(Th228Pos)  at ( \x,     \y+235pt );
    \coordinate(Ra224Pos)  at ( \x,     \y+206pt );
    \coordinate(Rn220Pos)  at ( \x,     \y+169pt );
    \coordinate(Po216Pos)  at ( \x,     \y+147pt );
    \coordinate(Pb212Pos)  at ( \x,     \y+125pt );
    \coordinate(Bi212Pos)  at ( \x,     \y+96pt  );
    \coordinate(BiTlStart) at ( \x-0.2, \y+96pt  );
    \coordinate(BiPoStart) at ( \x+0.2, \y+96pt  );
    \coordinate(BiTlStop)  at ( \x-0.2, \y+59pt  );
    \coordinate(Tl208Pos)  at ( \x-\dx, \y+59pt  );
    \coordinate(Po212Pos)  at ( \x+\dx, \y+40pt  );
    \coordinate(BiPoStop)  at ( \x+0.2, \y+40pt  );
    \coordinate(Pb208Pos)  at ( \x,     \y       );
    \coordinate(TlPbStop)  at ( \x-0.2, \y       );
    \coordinate(PoPbStop)  at ( \x+0.2, \y       );
    \iso{Th232Pos}{Th}{232}{90}{$1.4\cdot10^{10}$\,y}
    \alphaarrow{Th232Pos}{Ra228Pos}{\A: 3947\,keV (22\%) \A: 4012\,keV (78\%)}
    \iso{Ra228Pos}{Ra}{228}{88}{5.7\,y}
    \betaarrow{Ra228Pos}{Ac228Pos}{$\max Q_{\beta}$: 40\,keV \G: 13.5\,keV (1.6\%)}
    \iso{Ac228Pos}{Ac}{228}{89}{1\,h}
    \betaarrow{Ac228Pos}{Th228Pos}{$Q_{\beta}$: 2076\,keV (7\%) \G: 338\,keV (11\%) \G: 911\,keV (26\%) \G: 968\,keV (16\%)}
    \iso{Th228Pos}{Th}{228}{90}{1.9\,y}
    \alphaarrow{Th228Pos}{Ra224Pos}{\A: 5340\,keV (26\%) \A: 5423\,keV (73\%)}
    \iso{Ra224Pos}{Ra}{224}{88}{3.6\,d}
    \alphaarrow{Ra224Pos}{Rn220Pos}{\A: 5449\,keV (5\%) \A: 5685\,keV (95\%) \G: 241\,keV (4\%)}
    \iso{Rn220Pos}{Rn}{220}{86}{55\,s}
    \alphaarrow{Rn220Pos}{Po216Pos}{\A: 6228\,keV (100\%)}
    \iso{Po216Pos}{Po}{216}{84}{0.14\,s}
    \alphaarrow{Po216Pos}{Pb212Pos}{\A: 6778\,keV (100\%)}
    \iso{Pb212Pos}{Pb}{212}{82}{10.6\,h}
    \betaarrow{Pb212Pos}{Bi212Pos}{$\max Q_{\beta}$: 570\,keV \G: 239\,keV (44\%)}
    \iso{Bi212Pos}{Bi}{212}{83}{61\,m}
    \iso{Tl208Pos}{Tl}{208}{81}{3.1\,m}
    \iso{Po212Pos}{Po}{212}{84}{299\,ns}
    \iso{Pb208Pos}{Pb}{208}{82}{stable}
    \alphaarrowleft{BiTlStart}{BiTlStop}{\A: 6051\,keV (25\%) \A: 6090\,keV (10\%)}{36}
    \betaarrowleft{BiTlStop}{TlPbStop}{$\max Q_{\beta}$: 1800\,keV \G: 511\,keV (23\%) \G: 583\,keV (85\%) \G: 861\,keV (13\%) \G: 2615\,keV (100\%)}{100}
    \betaarrowright{BiPoStart}{BiPoStop}{$Q_{\beta}$: 2250\,keV (55\%) \G: 727\,keV (7\%)}{64}
    \alphaarrowright{BiPoStop}{PoPbStop}{\A: 8785\,keV (100\%)}{100}
    \def\x{6.3};
    \def\y{10pt}
    \coordinate(U238Pos)   at ( \x,     \y+408pt );
    \coordinate(Th234Pos)  at ( \x,     \y+379pt );
    \coordinate(Pa234mPos) at ( \x,     \y+357pt );
    \coordinate(U234Pos)   at ( \x,     \y+335pt );
    \coordinate(Th230Pos)  at ( \x,     \y+306pt );
    \coordinate(Ra226Pos)  at ( \x,     \y+277pt );
    \coordinate(Rn222Pos)  at ( \x,     \y+248pt );
    \coordinate(Po218Pos)  at ( \x,     \y+226pt );
    \coordinate(Pb214Pos)  at ( \x,     \y+204pt );
    \coordinate(Bi214Pos)  at ( \x,     \y+161pt );
    \coordinate(Po214Pos)  at ( \x,     \y+95pt );
    \coordinate(Pb210Pos)  at ( \x,     \y+73pt  );
    \coordinate(Bi210Pos)  at ( \x,     \y+44pt  );
    \coordinate(Po210Pos)  at ( \x,     \y+22pt  );
    \coordinate(Pb206Pos)  at ( \x,     \y       );
    \iso{U238Pos}  {U} {238} {92}{$4.47\cdot10^9$\,yr}
    \alphaarrow{U238Pos}{Th234Pos}{\A: 4151\,keV (21\%) \A: 4198\,keV (79\%)}
    \iso{Th234Pos} {Th}{234} {90}{24.1\,d}
    \betaarrow{Th234Pos}{Pa234mPos}{$\max Q_{\beta}$: 274\,keV}
    \iso{Pa234mPos}{Pa}{234m}{91}{1.16\,m}
    \betaarrow{Pa234mPos}{U234Pos}{$Q_{\beta}$: 2268\,keV (98\%)}
    \iso{U234Pos}  {U} {234} {92}{$2.46\cdot10^5$\,yr}
    \alphaarrow{U234Pos}{Th230Pos}{\A: 4722\,keV  (28\%) \A: 4776\,keV (71\%)}
    \iso{Th230Pos} {Th}{230} {90}{$7.54\cdot10^4$\,y}
    \alphaarrow{Th230Pos}{Ra226Pos}{\A: 4620\,keV (23\%) \A: 4687\,keV (76\%)}
    \iso{Ra226Pos} {Ra}{226} {88}{1600\,y}
    \alphaarrow{Ra226Pos}{Rn222Pos}{\A: 4601\,keV (6\%) \A: 4784\,keV (94\%)}
    \iso{Rn222Pos} {Rn}{222} {86}{3.8\,d}
    \alphaarrow{Rn222Pos}{Po218Pos}{\A: 5490\,keV (100\%)}
    \iso{Po218Pos} {Po}{218} {84}{3.1\,m}
    \alphaarrow{Po218Pos}{Pb214Pos}{\A: 6002\,keV (100\%)}
    \iso{Pb214Pos} {Pb}{214} {82}{27\,m}
    \betaarrow{Pb214Pos}{Bi214Pos}{$\max Q_{\beta}$: 1018\,keV \G: 242\,keV (7\%) \G: 295\,keV (18\%) \G: 352\,keV (36\%)}
    \iso{Bi214Pos} {Bi}{214} {83}{19.9\,m}
    \betaarrow{Bi214Pos}{Po214Pos}{$Q_{\beta}$: 3269\,keV (19\%) \G: 609\, keV (46\%) \G: 768\,keV (5\%) \G: 1238\,keV (5\%) \G: 1764\,keV (15\%) \G: 2204\,keV (5\%) \G: 2448\,keV (1.6\%)}
    \iso{Po214Pos} {Po}{214} {84}{164\,$\mu$s}
    \alphaarrow{Po214Pos}{Pb210Pos}{\A: 7687\,keV (100\%)}
    \iso{Pb210Pos} {Pb}{210} {82}{22.2\,y}
    \betaarrow{Pb210Pos}{Bi210Pos}{$\max Q_{\beta}$: 64\,keV \G: 47\,keV (4\%)}
    \iso{Bi210Pos} {Bi}{210} {83}{5.01\,d}
    \betaarrow{Bi210Pos}{Po210Pos}{$\max Q_{\beta}$: 1161\,keV}
    \iso{Po210Pos} {Po}{210} {84}{138\,d}
    \alphaarrow{Po210Pos}{Pb206Pos}{\A: 5304\,keV (100\%)}
    \iso{Pb206Pos} {Pb}{206} {82}{stable}
    \def\scale{1.0}
    \node at (1.65,1.6) [align=left, anchor=south west]{LEGEND:};
    \draw [gray]          (0,1.1) rectangle (0.4,1.5);
    \draw [cyan,dashed]   (0,0.6) rectangle (0.4,1.0);
    \draw [orange,densely dashdotted] (0,0.1) rectangle (0.4,0.5);
    \node at (0.4,1.3) [align=left, anchor=west]{Primordial};
    \node at (0.4,0.8) [align=left, anchor=west]{Chemical affinity};
    \node at (0.4,0.3) [align=left, anchor=west]{Rn emanation};
    \coordinate(pos1) at (3.4,1.67);
    \coordinate(pos2) at (3.4,0.85);
    \coordinate(pos3) at (3.4,1.20);
    \coordinate(pos4) at (3.4,0.38);
    \alphaarrow{pos1}{pos2}{\A\ decay}
    \betaarrow{pos3}{pos4}{\B\ decay}
  \end{tikzpicture}
  \caption{\U\ and \Th\ decay chains. For each isotope, we report \A\ particles with intensity $I>1\%$,
    \G\ with $I>5\%$ or energy close to the Q-value of some \bb\ isotope,
    and all $\beta$ with a Q-value above 2\,MeV. The boxes highlight the chain parts that are
    typically found in equilibrium: in black the isotopes due to the primordial material radioactivity;
    in dashed blue the isotopes in equilibrium with its predecessor Ra isotopes, and $^{210}$Po;
    in dash-dotted orange the isotopes in equilibrium with $^{210}$Pb,
    caused by $^{222}$Rn emanation and the subsequent $^{210}$Pb accumulation.
}\label{fig:decaychains}
\end{figure}

\nubb-decay mimicking events can be induced 
by naturally occurring radiation from the decays of primordial elements in the
actinide decay chains. Such elements are found ubiquitously in all materials.
In particular, \U\ and \Th\ are the progenitors of long decay chains
made of 10 and 14 isotopes, respectively.
The actinides produce \A, \B, and \G\ radiation across a wide energy range:
\A\ particles between 4 and 9\,MeV;
\B\ radiation mostly concentrated below 2\,MeV,
with the exception of $^{214}$Bi that \B\ decays with an end-point of 3.3\,MeV;
and \G\ rays of various energies up to the $^{208}$Tl line at
2.6\,MeV\footnote{Rare branches yield some higher energy \G\ rays.}.
An experiment is essentially vulnerable to mimicking events coming from any
\A, \B, and \G\ particles or their coincidences with energies above the $Q$ value
of the \bb\ isotope used (Tab.~\ref{tab:isotopes}).
The \A\ particles can also undergo (\A,$n$) reactions and thus produce a neutron
background, discussed in Sec.~\ref{sec:exp:background:neutrons}.
Figure \ref{fig:decaychains} summarizes the \U\ and \Th\ decay chains, listing
all \A\ particles with intensity $>1\%$ and all \G\ with intensity $>5\%$
or energy close to the Q-value of some \bb\ isotope.
We also report all \B\ particles with an end-point $>2$\,MeV,
otherwise we just report the highest possible end-point.

Most experiments have the capability of identifying and suppressing
background from actinides via the study of event topology or particle identification techniques,
which are covered in detail in Sec.~\ref{sec:sec:activebkgsuppression}.
However the base levels of actinide backgrounds are set by the purity
of the employed materials, especially those closest to the detector.
The purity in turn depends on the material origin and fabrication history.
The \U\ and \Th\ chains feature isotopes with very different decay times and chemical properties.
In particular, Ra has a very different chemical behavior than U and Th,
hence it is common to find different concentrations of Ra and U/Th.
As a result, decay chains are often not in secular equilibrium, but are split
in correspondence to the Ra isotopes, as highlighted by the dashed blue boxes in Fig.~\ref{fig:decaychains}.
Additionally, both chains include isotopes of Rn,
an inert gas with high mobility and permeability that is emanated by natural
radioactivity in the surrounding environment.
When Rn decays near a component during handling and fabrication, its decay
progeny can become embedded in and contaminate the component surfaces.
Rn can also diffuse in from the
experiment infrastructure and contaminate the detector {\it in situ},
as is the case, e.g., for Rn emanated from the surface of large vessels
containing liquid scintillators or cryogenic liquids \cite{Wojcik:2017vux}.
\Rn\ from the \U\ chain is particularly relevant because it leads
to the accumulation of the long-lived $^{210}$Pb (T$_{1/2}=22.3$\,y).
Thus the last part of the \U\ chain is often out-of equilibrium
(orange dash-dotted box in Fig.~\ref{fig:decaychains}). Moreover, while the Pb
and Bi species can be cleaned off the surfaces relatively easily, $^{210}$Po
(T$_{1/2}=138$\,d) is difficult to remove without aggressive, surface-specific
techniques \cite{Guiseppe:2017yah}, and is thus often found to generate
background on its own, unsupported by $^{210}$Pb.

In order to reduce the backgrounds from natural radioactivity,
special care must be put into the selection, fabrication or purification of pure materials for
use in or near the detector. Material selection and actinide material purity
demonstration is performed using several primary assay methods:
mass spectrometry, \G\ spectroscopy, neutron activation analysis, and \A\ spectroscopy.

Mass spectrometry (MS) involves atomizing and ionizing the sample material followed
by electromagnetic separation of chemical species by their mass-to-charge ratio.
Inductively Coupled Plasma Mass Spectrometry (ICP-MS) is a common technique 
that can reach sensitivities of $\leq10^{-14}$\,g/g for \U\ and \Th\
using less than a gram of material \cite{LaFerriere:2015owa,Nisi:2017mgi}.
Atomization is performed by nebulizing liquid or dissolved samples, or by laser
ablation from surfaces of solid samples, followed by ionization by the ICP.
ICP-MS is advantageous for its short measurement time (hours) and the small
amount of material required, but is limited by isomeric interference and is
usually only sensitive to long-lived decay chain progenitors, which are present
in the sample in much larger quantities than their short-lived descendants.
Thus ICP-MS cannot detect whether a chain is out of equilibrium. Other MS
techniques include glow discharge mass spectrometry, thermal ionization mass
spectrometry, and accelerator mass spectrometry.

\G\ spectroscopy is performed with low-background high-purity germanium (HPGe) detectors
operated underground \cite{Laubenstein-2004,Baudis:2011am}. 
It is a non-destructive technique applicable to a variety of isotopes
and can reach sensitivities down to the $\mu$Bq/kg level \cite{Laubenstein-2017}.
The sensitivity, though, depends on the sample mass and measurement time:
typical measurements last for a few weeks with samples of 0.1--1\,kg.
The advantages of \G\ spectroscopy include the possibility to identify any
\G-emitting isotope, independently of its mass,
and the capability --- to some extent --- to independently measure the activity of separate parts
of a decay chain out of the secular equilibrium.

Neutron activation analysis (NAA) is a technique that combines
the activation of an isotope via the exposure to a high intensity neutron flux
and the subsequent measurement of the activated nuclei with \G\ spectroscopy \cite{greenberg-2011}.
Of particular relevance for actinide analysis are the production of $^{239}$Np
(T$_{1/2}=2.4$\,d) from $^{238}$U, and $^{233}$Pa (T$_{1/2}=27$\,d) from $^{232}$Th.
Knowing the neutron flux and cross section for the neutron activation cross section ---
or measuring their product with a reference sample ---
it is possible to reconstruct the original concentration of the target from the
decay rates of the activated nuclei.
NAA sensitivity can be superior to that of \G\ spectroscopy \cite{Clemenza-2018}
but like MS is only sensitive to long-lived decay chain progenitors.
Moreover, it can require a non-trivial sample preparation
and is subject to possible backgrounds from the neutron activation of stable nuclei
present in the sample itself. The latter consideration makes NAA inappropriate
for assay of most metals.

Finally, \A\ spectroscopy is useful exclusively for measuring superficial
concentrations at depths
shallower than the $\alpha$ range in the measured material, i.e., $O(10)$\,$\mu$m.
It can be performed with surface barrier detectors or large ionization chambers,
whose main limitations are the sensitive surface size and energy resolution, respectively.
The best sensitivity achieved by an \A\ counter is at the level of $30$\,nBq/cm$^2$ \cite{Warburton-2004},
which is an order of magnitude worse than the values required
by, e.g., calorimetric \nubb-decay experiments \cite{cupid-cdr}.

Strict procedures are also followed to avoid contaminating the materials through exposure to Rn.
Sensitive parts are typically stored or even assembled in radon-free environments.
Small volumes such as storage vessels or glove boxes are flushed with boil-off nitrogen from large
liquid nitrogen dewars \cite{Wojcik:2017vux}, while larger environments such as clean rooms
can be flushed with Rn-free air obtained from dedicated radon abatement systems \cite{Benato:2017kdf}.
Rn emanation from material within the detector is especially problematic
for Xe-based experiments, due to a \G\ line from $^{214}$Bi at 2448\,keV, very close to the \Xe\ \Qbb.
Continuous Xe purification has been demonstrated via adsorption on activated charcoals \cite{K-2012}
or cryogenic distillation \cite{Aprile:2017kop}.
Similarly, Rn suppression by $\geq3$ orders of magnitude was also demonstrated
via distillation on $n$-dodecane, a common admixture in liquid scintillator \cite{Keefer:2013eaa}.

In addition to the maximization of the detector radio-purity,
the actinide purity of the surrounding components and laboratory environment
must be kept under control as well.
In this case, high-energy \G\ rays are the most worrisome component due to their
long attenuation lengths.
\bb-decay detectors must therefore be completely enveloped by a material
capable of efficiently absorbing \G\ radiation without inducing further background.
This can be achieved in several ways, including: a set of concentric passive
(non-instrumented) layers of shielding material with increasing
radio-purity, typically Pb and Cu \cite{Alduino:2017qet,Abgrall:2013rze};
a high-purity cryogenic liquid such as Ar possibly instrumented to detect the scintillation light
produced by incoming radiation \cite{Agostini:2017hit};
for liquid scintillator experiments, the division of the detector medium
into an inner region loaded with the \bb\ isotope and an outer region with no
isotope working as an active shield \cite{Xe_enr,Andringa:2015tza}. These
shielding layers are designed to be thick enough to eliminate external radiation
as a concern.

\subsubsection{Anthropogenic radioactivity}

Background can be induced by anthropogenic radioactivity, in particular as a result of nuclear accidents or nuclear weapon tests.
The great majority of these isotopes are $\beta$ emitters,
and in some case are the progenitor of a decay chain.
In order to represent a potential background source for \nubb-decay experiments,
the decay chains must include an isotope with a $Q$-value greater than \Qbb,
and a dominant half-life on the same order as an experiment's lifetime.
Table~\ref{tab:manmadeisotopes} lists some examples of potentially worrisome isotopes with $Q_{\beta}>2$\,MeV
reported by \textcite{fukushimareport}. Of these,
only $^{108\text{m}}$Ag has been detected so far \cite{Xe_enr}.
The actual relevance of these isotopes as potential backgrounds must be assessed for each experiment separately.
While a pure $\beta$ emitter such as $^{144}$Pr could be worrisome for most experiments,
an isotope that also emits $\gamma$ rays (e.g., $^{108\text{m}}$Ag) could be tagged 
with event topology reconstruction capabilities.

\begin{table}[bp]
  \caption{Anthropogenic isotopes. Isotopes belonging to the same chain are grouped between horizontal lines.}
  \label{tab:manmadeisotopes}
  \begin{tabular}{rcccl}
    \toprule
    \smallspace
    Isotope & Half-life & $Q_{\beta}$ [keV] & Detected & Notes \\
    \colrule
    \smallspace
    $^{88}$Y           & 107\,d  & 3008 & No  & Several $\gamma$ lines \\
    \colrule
    \smallspace
    $^{90}$Sr          & 28.8\,y & 546  & No  &  \\
    $^{90}$Y           & 64\,h   & 2279 & No  & Pure $\beta$ emitter \\
    \colrule
    \smallspace
    $^{110\text{m}}$Ag & 250\,d  & 3008 & Yes & Several $\gamma$ lines \\
    \colrule
    \smallspace
    $^{134}$Cs         & 2\,y    & 2059 & No  & Several $\gamma$ lines \\
    \colrule
    \smallspace
    $^{144}$Ce         & 285\,d  & 319  & No  &  \\
    $^{144}$Pr         & 17.3\,m & 2997 & No  & Pure $\beta$ emitter \\
    \botrule
  \end{tabular}
\end{table}

\subsubsection{Neutrons}
\label{sec:exp:background:neutrons}

In the previous sections we have mentioned several processes producing neutrons.
The actual background induced in a \nubb-decay experiment depends on the neutron flux
and energy spectrum, on the location of neutron emission,
and on the employed materials.
In the present context, neutrons can be divided in two groups:
above surface neutrons originating from cosmic rays in the atmosphere,
and underground neutrons produced by muon spallation,
\an\ reactions on light nuclei, and spontaneous fission reactions, mainly from $^{238}$U.
Above ground neutrons represent the dominant cause of cosmogenic activation
in detector materials prior to their installation underground,
previously discussed in Sec.~\ref{sec:exp:background:cosmic}.
Underground neutrons can be further divided between external neutrons generated in the rock
or in the concrete walls, and internal neutrons generated
inside or next to the detector.
Neutrons from \an\ and fission reactions have energies $\lesssim10$\,MeV \cite{Wulandari:2003cr},
while those from spallation can reach several GeV \cite{Mei:2005gm}.
The activity of underground neutrons from \an\ and fission reactions
is only of order $\sim$1\,n/yr/g, but the high mass of rock within a scattering length of
the neutrons yields a flux 2--3 orders of magnitude
higher than that of neutrons from spallation \cite{Wulandari:2003cr}.

The flux of neutrons from \an\ and fission reactions is efficiently suppressed
by neutron shielding of moderate thickness, located outside the $\gamma$ shielding.
One possible strategy is to enclose
the experiment in a thick layer (a few tens of cm) of neutron moderator such as 
polyethylene, with the innermost layer (a few cm) comprised of a material with
high neutron absorption cross section (e.g., boric acid or borated
polyethylene \cite{Abgrall:2013rze,cuore-prl2017}).
The outer layer slows down the neutrons to thermal energies, while the inner one captures them.
Alternatively, a $\geq$1\,m layer of water can be used both for moderation and absorption:
in this way, a water tank can simultaneously act as a neutron shield and a muon
veto detector \cite{Ackermann:2012xja}. In the case of liquid scintillator
detectors, the outermost scintillator region serves as a very effective neutron
moderator, providing active tagging in addition to high neutron capture
capability. 

While external neutrons with energies $\lesssim10$\,MeV are efficiently suppressed with a neutron shield,
high energy neutrons can still reach the detector.
Additional neutrons can also be produced within the neutron shield
by muons or, again, \an\ and fission reactions taking place in the $\gamma$ shield,
the active material, or in calibration sources \cite{Baudis:2015sba}.
These neutrons can undergo elastic and inelastic scattering,
or be captured to produce stable or unstable nuclei
and possibly yield prompt $\gamma$ de-excitations.
The interactions induce a variety of signatures that strongly depend
on the detector technology and employed materials.
The most worrisome for \nubb-decay experiments are the in-situ activation of long-lived isotopes
in or next to the detector, and inelastic scatterings yielding penetrating $\gamma$ rays
with energies comparable to \Qbb.

Finally, one possible technique to minimize the neutron induced background
consists in embedding in the $\gamma$ shielding or in the detector medium
a material with high-cross section for neutron capture,
and possibly that produces events with a signature that does not mimic a \nubb-decay event.
An example could be $^{6}$Li, which undergoes the $^6\text{Li}+n\rightarrow \alpha+^3\text{H}$
reaction with a Q-value of 4.8\,MeV\cite{Poda:2017jnl}.

\subsubsection{Neutrinos}

\begin{figure}[tbp]
  \includegraphics[width=\columnwidth]{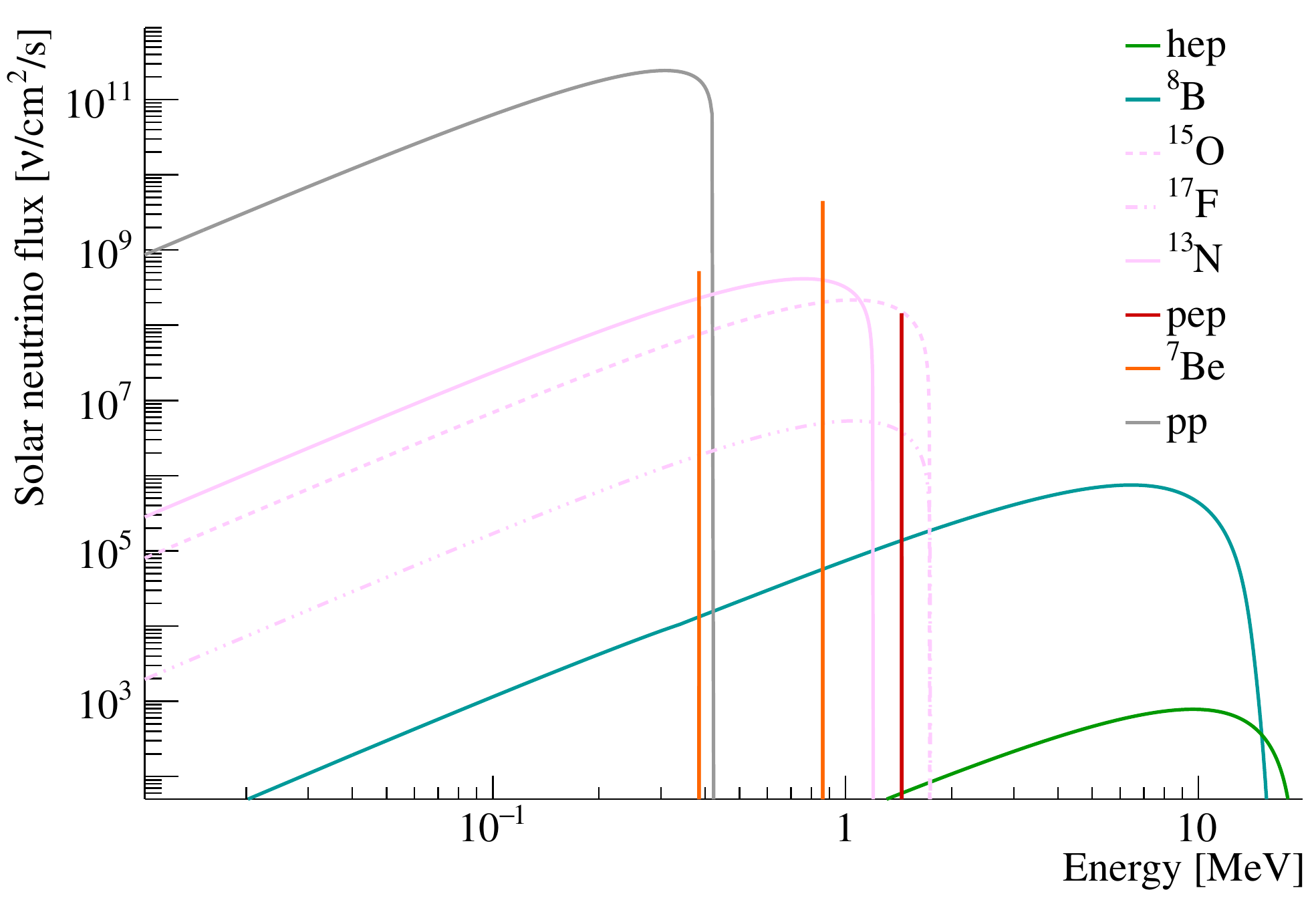}
  \caption{Solar neutrino spectra.
    Data from \cite{Agostini:2018uly,Bergstrom:2016cbh,Bahcall:1994cf,Bahcall:1996qv,Bahcall:1987jc}.  
    The curves labeled hep, pep, and pp correspond to neutrinos emitted in helium-proton, proton–electron–proton, and proton-proton fusion, respectively.
    } 
    \label{fig:solarneutrinos}
\end{figure}

Neutrinos represent a potential source of irreducible background for \nubb-decay experiments.
Sources with appreciable neutrino fluxes include solar neutrinos, atmospheric neutrinos,
geoneutrinos, reactor neutrinos and the diffuse supernova neutrino background (DSNB).
Due to their higher flux at energies below 20\,MeV, solar neutrinos are the most worrisome background source.
Their spectrum is composed of several contributions corresponding to the primary
nuclear reactions in the sun (Fig.~\ref{fig:solarneutrinos}).
Solar neutrinos can undergo two types of interactions in a \nubb-decay experiment,
elastic scattering (ES) and charged current (CC) interactions \cite{Elliott:2004hr}:
\begin{alignat}{3}
  &\text{ES:\quad} &&\nu + \text{e}^-          &&\rightarrow\ \nu + \text{e}^- \\
  &\text{CC:\quad} &&^{\text{Z}}\text{A} + \nu &&\rightarrow\ ^{\text{Z}+1}\text{A} + e^- ~[+ \gamma(\text{s})] + Q_{\nu} \label{eq:invbetadecay}\\
  &                &&^{\text{Z}+1}\text{A}     &&\rightarrow\ ^{\text{Z}+2}\text{A}+\beta^- + \nu ~[+ \gamma(\text{s})] + Q_{\beta} \label{eq:delayedbeta}
\end{alignat}

In ES, a neutrino scatters off an electron in the detector.
The electron is scattered non-isotropically, and its energy spectrum is a continuum
up to the end-point of the solar neutrino spectrum, $\sim19$\,MeV.
Only neutrinos with energy $E_{\nu}>$\Qbb\ contribute,
and given the low flux of hep neutrinos, in practice only $^8$B neutrinos are relevant (Fig.~\ref{fig:solarneutrinos}).
The expected background per unit sensitive mass is $\sim2\cdot10^{-7}$\,counts/keV/kg/yr \cite{Elliott:2004hr, deBarros:2011qq}.
For current and next-generation experiments this is negligible for experiments
in which the active material is mostly made of the \bb\ isotope,
but becomes significant for liquid scintillator experiments with dissolved
sources \cite{Elliott:2004hr,Andringa:2015tza,deBarros:2011qq}.
A partial suppression of the ES background might be achievable exploiting the signal directionality \cite{Askins:2019oqj},
at a non-negligible cost in terms of signal efficiency.

In CC interactions, the \bb\ isotope $(A,Z)$ undergoes inverse \B\ decay
to the ground state or an excited state of the $(A,Z\!\!+\!\!1)$ isotope (Eq.~\ref{eq:invbetadecay}),
the intermediate isotope of the \bb\ decay to $(A,Z\!\!+\!\!2)$.
Since $(A,Z\!\!+\!\!1)$ is heavier than $(A,Z)$, the reaction has a threshold of $E_t = m_{A,Z+1}-m_{A,Z}$,
so neutrinos with energy $E_\nu \simeq E_t +$\Qbb\ pose a possible background in this prompt event.
In some cases the intermediate nucleus can then capture an electron and decay back to the original \bb\ parent isotope,
but it more often undergoes \B$^-$ decay to the same final state of the \bb\ decay, $(A,Z+2)$ (Eq.~\ref{eq:delayedbeta}),
releasing an energy $Q_{\beta} > Q_{\beta\beta}$ that can pose a delayed background. 
The actual relevance of CC interactions as a background depends on the \bb\ isotope
and the corresponding value of $E_t$, the emission of de-excitation \G\ rays
in the two involved reactions that could modify the event topology, 
and the half-life of the intermediate state, which could allow a time correlation analysis.
Without applying any of these event identifications, the intermediate nucleus \B\ decay
yields a background of $10^{-3}$--$10^{-1}$\,events/keV/ton$_{\text{iso}}$/yr \cite{Ejiri:2013jda,Ejiri:2017vyl}.

Other neutrino sources do not represent a significant background source
primarily due to their low flux. 
However the presence of antineutrinos in these sources requires consideration of
additional inverse $\beta$ interactions that could give a background, particularly
at sites with appreciable reactor neutrino fluxes.
In the case of atmospheric and DSNB neutrinos,
the CC interaction energies are also so high that the nucleus is often left in a
highly excited state, leading to background signatures similar to in-situ
cosmogenic activation but with much lower production rate.

\subsubsection{\nnbb\ decay}

The only intrinsic and irreducible background for \nubb\ decay is the concurrent \nnbb-decay channel.
The two processes induce a similar event topology,
with the exception of the energy signature: a peak at \Qbb\ for \nubb\ decay,
and a continuum from zero to \Qbb\ for \nnbb\ decay (Fig.~\ref{fig:signature}).
The detector resolution results in some of the highest energy \nnbb-decay events
reconstructing with energies at \Qbb. 
Although the event topologies differ in detail --- in the energy
distributions and angular correlations between the emitted electrons
\cite{Kotila:2012zza} --- those differences can only be leveraged at high
statistics with tracking detectors capable of making those measurements.
Thus the relevance of the \nnbb-decay background depends primarily on the
energy resolution and the \nnbb-decay half-life.
In practice, \nnbb-decay contributes significantly to the background
only for experiments with a resolution of some percent.
Additionally, if the \nnbb-decay rate is high compared to the desired \nubb-decay sensitivity,
\nnbb-decay events can pile-up and contribute to the background at \Qbb.
In practice, this is relevant only for cryogenic calorimeters
using \Mo\ as candidate isotope \cite{CUPIDInterestGroup:2020rna}.

\subsection{Signal and background discrimination techniques}
\label{sec:sec:activebkgsuppression}

In the previous sections we highlighted some of the features that distinguish
each possible background component from the \nubb-decay signal. In this
section we collect and summarize the experimental techniques available to
discriminate the two. Although many of these techniqes have been mentioned
in the previous sections, our aim is to achieve a concise listing. We organize
the discussion according to the key features of the signal. 
The primary signature is a peak at \Qbb\ in the sum energy spectrum.
\nubb-decay events must also be homogeneously distributed in time and space,
with a rate proportional to the fraction of target isotope in the active material.
The electrons are subject to a localized energy deposition, whose dimension
depends on the electron attenuation length: $O(1-10)$\,mm for solids and liquids,
and $O(10)$\,cm for high-pressure gases. The events are uncorrelated with any
other physical processes, and the final state includes the presence a particular
daughter nucleus.

As described above, energy is the only necessary and sufficient observable for a discovery,
hence energy resolution is crucial to minimize the background level in the vicinity of \Qbb.
Of particular concern are the irreducible \nnbb-decay contribution
that extends up to \Qbb, and emissions of the \U\ and \Th\ decay chains 
(\A, \B, or \G\ particles)
with energies greater than \Qbb.
Often the background at \Qbb\ can be approximated as flat.
If not, a spectral fit over a larger energy region is required
to properly constrain the background at \Qbb\ using the information obtained
from the rest of the spectrum.
Since the lifetime of an experiment spans several years,
calibrating the energy scale and monitoring its stability over time
is fundamental for avoiding any degradation of energy resolution in the physics spectrum,
and for a precise characterization of the detector response.

The expected \nubb-decay signal is uniform in the volume containing the isotope.
The same is true for some background processes --- e.g., \nnbb\ decay, neutrino, and often neutron reactions --- 
but does not hold for others, especially those induced by natural or
anthropogenic radioactive contaminants
located outside the detector.
Thus the detector medium can act as a self-shield, with the inner part subject
to a lower background than the outer one.
This is a natural feature of monolithic experiments, while for granular experiments
it can be approximated by dividing the detectors into concentric layers.

The electrons emitted by \nubb\ decay carry a directional correlation due to the
angular momentum exchanged by the mediating mechanism. However the direction of
any one electron emitted in sequential decays are uncorrelated.
On the other hand, some backgrounds, for example the elastic scattering of solar
neutrinos with electrons, are not isotropic.
Directional reconstruction, e.g., with the detection of Cherenkov light,
is therefore useful for suppressing these backgrounds.

The event topology of a \nubb\ decay is clearly defined for each detector technology:
an energy deposition contained in $O(1-1000)$\,mm$^3$ in a solid or liquid detector;
a track of $O(10-30)$\,cm length with two blobs at its extremes in a high-pressure gas TPC;
a pair of distinguished electron tracks with a common origin in a low-pressure tracking detector.
Depending on the detector spatial resolution, several particles might be distinguishable:
muons induce long tracks that cross the detector medium, or hit multiple detectors of a granular experiment,
and generate a signal in muon veto detectors;
\G\ rays have a longer range and can undergo Compton scattering, inducing multiple energy depositions
at different locations independently of the detector technology;
\A\ particles have a shorter range easily identifiable in gas detectors;
\B\ particles produce a track with a single blob in a high-pressure TPC,
or a single track in a tracking detector.
These signatures can also be combined and thus facilitate their identification,
e.g., in the case of a radioactive isotope decaying via \A\ or \B\ decay
with the subsequent emission of de-excitation \G\ rays.

For some readout channels such as scintillation and Cherenkov light,
different particles induce a different detector response. Therefore an additional means
of background suppression is particle identification via signal shape analysis.
A common strategy is having multiple read-out channels: one optimized for energy reconstruction,
the other for particle discrimination.
Examples are scintillation experiments with Cherenkov readout for \A\ and single \B\ identification,
or cryogenic calorimeters with scintillation light readout for \A\ vs \B/\G\ discrimination.

\nubb\ decays are homogeneous in time and uncorrelated with anything else.
Conversely, some backgrounds are identifiable due to their specific time correlations.
This is the case for delayed coincidences between the decays of several isotopes
in the \U\ and \Th\ chain (e.g., the Bi-Po sequences, see Fig.~\ref{fig:decaychains}),
between the decay of a cosmogenically activated isotope (e.g., $^{68}$Ga in Ge)
and the detection of its parent muon in the muon veto,
or between inverse \B\ decay and the subsequent \B\ decay in solar neutrino CC
reactions.

Finally, daughter nucleus tagging is an additional tool for background
suppression, which distinguishes \bb\ decays (but not exclusively \nubb\ decays)
from anything else except solar neutrino CC reactions. Another background
characterization method is abundance scaling,
where different measurements with enriched vs non-enriched materials
or loaded vs non-loaded active material allow
an experiment to isolate backgrounds that are not related to the presence of
the \bb\ decay isotope.

\subsection{Statistical analysis and sensitivity}
\label{sec:exp:stat}

\subsubsection{Signal extraction}
\label{sec:exp:stat:signal}

As discussed in the previous sections, all \nubb-decay experiments measure multiple observables for each event.
Some observables are related to the amount and spatial distribution of the energy deposited within the detector.
Others are related to the timing and type of particles involved in the event.
The value of several of these observables is well-defined for \nubb-decay events.  
For instance, each event should have energy equal to the decay Q-value,
and be contained within a small region of the detector.
Background events will also have specific features, resulting in characteristic observable values.

All experiments in the field use a multivariate analysis to extract the sought-after \nubb-decay signal.
The observables define the basis of a multi-dimensional parameter space,
in which signal and background events are distributed according to multivariate probability distributions.
Since \nubb-decay events have well-defined features, the bulk of their
probability distribution will be restricted to a small volume of the parameter space.
On the other hand, most of the background events will be outside of this small volume, populating other regions.
Thus the signature of a possible \nubb-decay signal is an excess of events compared to the background expectation in a narrow region of the multi-dimensional parameter space.
We will refer to this volume with a maximum signal-to-background ratio as the
{\it sensitive volume}. The rest of the parameter space is effectively used to
constrain the background contribution to the sensitive volume.

The signal and background probability distributions are often well separated for one or more observables.
In such cases, it is advantageous to apply a cut on such observables,
discarding background data without a significant reduction of the signal-detection efficiency,
while decreasing the dimensionality of the parameter space, and also reducing
systematic uncertainty due to imperfect knowledge of the distributions in the observables.
These considerations make applying cuts often advantageous even when there is
some overlap, between signal and background, although in such cases the reduced
dimensionality and systematic uncertainty must be weighed against any loss of
statistical precision. For observables where signal and background strongly overlap, a
full multivariate fit is unavoidable. However in many experiments, the
signal-background separation is good for most of the observables,
and the multivariate analysis can be simplified into a 1-dimensional fit of the
energy spectrum with negligible degradation of sensitivity.

The analysis techniques of the field at present are rather established and uniform.
Most experiments report frequentist maximum likelihood fits based on likelihood ratio tests \cite{Zyla:2020zbs}.
Monte-Carlo parametric-boostrapping methods are often used to compute the test statistic probability distributions
or to confirm their behavior when asymptotic formulae are assumed \cite{Cowan:2010js}.
Given the low counting rate and the fact that the parameter of interest is constrained to non-negative values,
the test statistic distribution can significantly differ from a chi-square function \cite{Algeri:2019arh}.
While frequentist methods have historically been dominant \cite{Cousins:1994yw},
recently most experiments also report results 
based on Bayesian methods, with inference 
deriving from the marginalized posterior distribution.
Given the lack of a strong signal to date, the choice of prior distribution
continues to significantly affect posterior probabilities.

\subsubsection{Discovery and exclusion sensitivity}
\label{sec:exp:stat:sens}

The reach of \nubb-decay experiments is traditionally quoted in terms of discovery and exclusion sensitivity, two statistical concepts belonging to frequentist inference.
The discovery sensitivity corresponds to the expected number of signal events
for which an experiment has 50\% chance to observe an excess of events over the
background at 99.73\% confidence level (CL).
The exclusion sensitivity corresponds to the expected number of signal events
that an experiment has 50\% chance of excluding at 90\% CL.

The discovery and exclusion sensitivity confidence levels are less stringent
compared to other fields --- e.g., the particle accelerator community --- due to the lack of a ``look elsewhere'' effect (the \nubb\ peak must occur at \Qbb) and simpler-to-control systematic uncertainties \cite{NsacDbdSC2014,NsacDbdSC2015}. In particular, the CL required for a discovery is equivalent to
excluding 3$\sigma$ two-sided fluctuation of a Gaussian random variable, and not one-sided 5$\sigma$ fluctuation as for accelerator experiments.
Other sensitivity definitions have been used \cite{Caldwell:2006yj,CUORE:2017tpp}, 
including Bayesian concepts based on Bayes factors and posterior distributions,
but these are not common in the field at present.

\begin{figure}[tb]
  \includegraphics[width=\columnwidth]{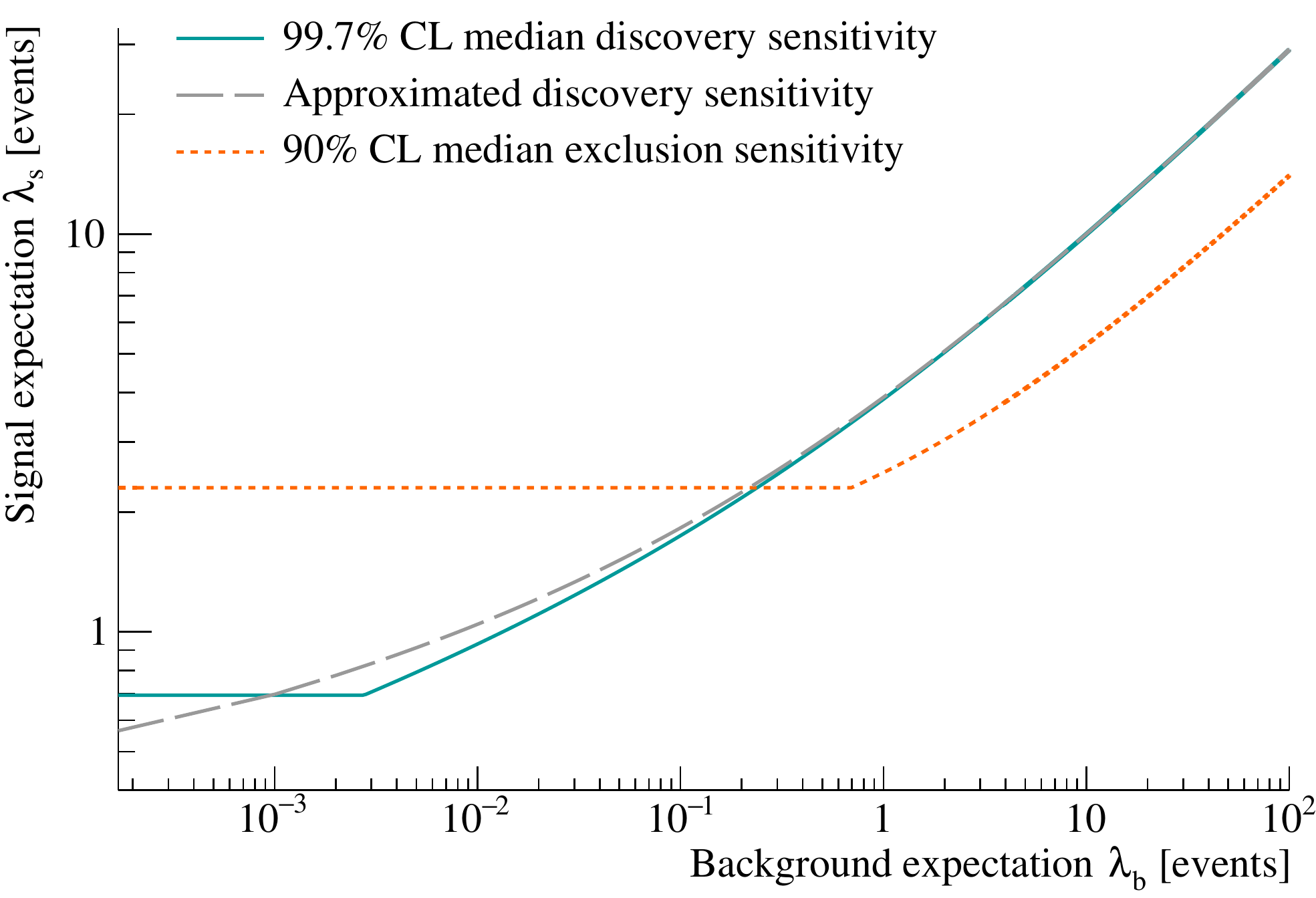}
  \caption{Median 99.7\% CL discovery sensitivity and median 90\% exclusion sensitivity as a function of the expected number of background events.
    The discovery sensitivity shows the signal event expectation at which an experiment has 50\% chance to observe a 99.7\% CL excess of events over the background.
    The exclusion sensitivity is instead the signal event expectation that can be excluded at 90\% CL with 50\% probability, assuming that there is no signal.
    The plot shows also the approximated discovery sensitivity extracted using
    the asymptotic formulae from \textcite{Cowan:2010js}.}
  \label{fig:cts_sens} 
\end{figure}

A precise evaluation of the expected number of signal events ($\lambda_s$) fulfilling the sensitivity definitions above requires calculations considering probability distributions in the multivariate space and experiment-specific information.
However, it can be approximated by considering a counting analysis in the
sensitive volume, with known background expectation $\lambda_b$. Uncertainties
on $\lambda_b$ can usually be neglected, as experiments are able to constrain its value using large background-dominated regions of the multivariate parameter space.
As both the signal and background events are generated by Poisson random processes,
the discovery sensitivity can be calculated by solving this system of equations: 
\begin{equation}
  \begin{cases}
  P(X\leq x | \lambda_b) \geq 99.73\%\\
  P(X\geq x | \lambda_b+\lambda_s) \geq 50\%\,,
  \end{cases}
  \label{eq:disc}
\end{equation}
where $P(X\leq x|\lambda)$ is the Poisson probability of observing a number of events $X$ smaller or equal to $x$ given an  expectation $\lambda$.
For a given $\lambda_b$ value, the system has a unique solution that can be
found by calculating the minimum $x$ satisfying the first equation,
substituting it in the second equation, and then finding the minimum $\lambda_s$
that satisfies the resulting inequality.
The exclusion sensitivity can be similarly computed by solving:
\begin{equation}
  \begin{cases}
  P(X\leq x | \lambda_b) \geq 50\%\\
  P(X\geq x | \lambda_b+\lambda_s) \geq 90\%\,.
  \label{eq:excl}
  \end{cases}
\end{equation}
Although the Poisson mass function is discrete, an actual multivariate fit
operates with non-integer number of events in the form of the probability
distribution weights for each event.
We can reproduce this behavior by interpolating the Poisson mass function with a
 normalized upper incomplete gamma function, and re-define the probability in the systems above as:
\begin{equation}
P(X>x|\lambda) \doteq \frac{\Gamma(x+1, \lambda)}{\Gamma(x+1)}\,.
\label{eq:smoothCDF}
\end{equation}
This definition leads to the discovery and exclusion sensitivity shown in Fig.~\ref{fig:cts_sens}.
Also shown in Fig.~\ref{fig:cts_sens} is an approximation of
Eqs.~\ref{eq:disc} from \textcite{Cowan:2010js} using elementary functions.

The discovery sensitivity degrades rapidly as the expected number of background events increases:
the greater $\lambda_b$, the greater $\lambda_s$ must be to create an excess incompatible with background fluctuations.
For large enough values, the background fluctuations become Gaussian and the
sensitivity scales only as $\sqrt{\lambda_b}$. 
For instance, when $\lambda_b=100$, a 3$\sigma$ discovery 
sensitivity requires $\lambda_s\geq3\sqrt{100} = 30$.
Conversely, the lower the background, the lower the signal expectation needs to
be for a discovery.
For any $\lambda_b \leq -\ln(99.73\%) \sim 0.0027$, the experiment has more than 99.73\% probability of observing no background events, and the observation of a single event becomes a discovery
\footnote{For a truly zero background experiment, one event is enough to claim a discovery.  In a similar fashion, encountering a unicorn is enough to claim its existence, provided that we have a template of a unicorn to which to compare the observed candidate.}.
In this ``background-free'' regime, the discovery sensitivity saturates: 
the first of Eqs.~\eqref{eq:disc} is always solved for $x=0$, and so the second
one is solved for $\lambda_s \lesssim -\ln(50\%) \sim 0.69$. 
In between these two regimes, for $\lambda_b < 1$, the sensitivity is not
independent of $\lambda_b$ but only degrades weakly with increasing $\lambda_b$. Experiments in
this ``quasi-background-free'' regime reap most of the benefits of a
background-free experiment, but still require a few signal counts to claim an observation.

The exclusion sensitivity behaves similarly to the discovery sensitivity, but it
saturates for larger background expectations, at $\lambda_b= -\ln(50\%)$.
Below this threshold, the experiment always has $\geq$50\% probability to observe no background events, and a further reduction of the background expectation cannot improve the median upper limit on the signal strength.
The first of Eqs.~\ref{eq:excl} is thus always solved for $x=0$, and
the second one yields $\lambda_s=\ln(1-90\%)\sim2.3$\,events.

The saturation of both sensitivities is connected to the properties of the 
Poisson probability, and is thus an intrinsic feature of the frequentist median sensitivity. 
This behavior can be problematic when the sensitivity is used as a figure of merit to optimize or compare experiments, and alternative sensitivity definitions have been recently proposed \cite{Bhattiprolu:2020mwi}.

The expected number of signal and background events in an experiment can be computed starting from
two effective parameters, the
{\it sensitive background} (\senbkg) and {\it sensitive exposure} (\senexp).
As they are connected to the sensitivity, these parameters capture well the performance of an experiment.
The sensitive exposure is the product of: the number of moles of isotope in the
active fiducial detector volume, the live time, and the signal detection
efficiency, i.e., the probability for a \nubb-decay event to occur in the
sensitive volume.
The sensitive background is the number of background events in the sensitive volume after all analysis
cuts, divided by \senexp.
Using these definitions, the expected number of signal and background counts in
the sensitive volume is given by:
\begin{equation}
  \lambda_s(\Thl) = \dfrac{\ln 2 \cdot N_{A} \cdot \senexp}{\Thl} \qquad
   \text{and} \qquad \lambda_b = \senbkg\cdot \senexp
\label{eq:senscounts}
\end{equation}
where $N_A$ is Avogadro's number and \Thl\ is the half-life of the decay.
Given that $\lambda_s$ depends on \Thl, the discovery and exclusion sensitivities on the expected number of events
can be directly translated into sensitivities on the \nubb-decay half-life.
\Thl\ sensitivities are the most common parameter reported by the experiments.

Next-generation searches aim to monitor tons of material for a decade,
reaching sensitive exposures at the level of $\senexp\sim 10^{5}-10^{6}$\,\molyr.
Such an exposure gives the possibility to observe a handful of signal events
even  for \nubb-decay half-life values of $10^{27}-10^{28}$ years.
As illustrated by Fig.~\ref{fig:cts_sens}, a requirement for discovery is
that the number of background events is similar to the number of expected signal events.
Thus  experiments aim at reaching backgrounds at the level of
$\senbkg<10^{-4}-10^{-5}$\,events per mole of material per year.
The concepts proposed to achieve this incredibly challenging performance are described in the next section.
   \section{Recent and future experiments}
  \label{sec:prj}
  A broad experimental program has been mounted in the last two decades to search for \nubb\ decay.
  Very diverse technologies have been developed and tested, leading to experiments
  with half-life sensitivities up to $10^{26}$ years. Thanks to these
  achievements, a number of new projects are being prepared with the goal of increasing the sensitivity by up to two orders of magnitude, opening the window to new energy frontiers and conclusively testing the scenario in which \nubb\ decay is mediated by inverted-ordered neutrinos \cite{Agostini:2021kba}.

  In this section, we discuss recent and future phases of existing experiments.
  In Sec.~\ref{sec:prj:land}, we review the experimental landscape and use the experiments' key performance parameters to evaluate their strength, strategy, and sensitivity.
  We then focus on the detection concept and technical aspects of each project.
  Experiments based on high-purity germanium detectors are reviewed in Sec.~\ref{sec:prj:ge}, time-projection chambers in Sec.~\ref{sec:prj:xe}, large liquid scintillator detectors in Sec.~\ref{sec:prj:ls}, cryogenic calorimeters in Sec.~\ref{sec:prj:bo}, and 
  tracking calorimeters in Sec.~\ref{sec:prj:tr}.
  New technologies that are currently being tested and newly proposed experimental designs are summarized in Sec.~\ref{sec:prj:rd}. 
  Sec.~\ref{sec:prj:land} should be accessible to all readers, while the other sections listed above are intended for more expert readers.

  \subsection{Experimental landscape}
  \label{sec:prj:land}

  Each experiment is characterized by a set of key performance parameters that
  are captured by the concepts of sensitive exposure (\senexp) and sensitive
  background (\senbkg) defined in Sec.~\ref{sec:exp:stat:sens}. The sensitive
  exposure and background are directly connected to the half-life sensitivity, and
  carry valuable information on the strategy pursued by each project. 
  Indeed, different combinations of \senexp\ and \senbkg\ can give the same
  sensitivity, and exposure increase can be traded for background reduction or vice versa.

\begin{figure*}[]
    \includegraphics[width=\textwidth]{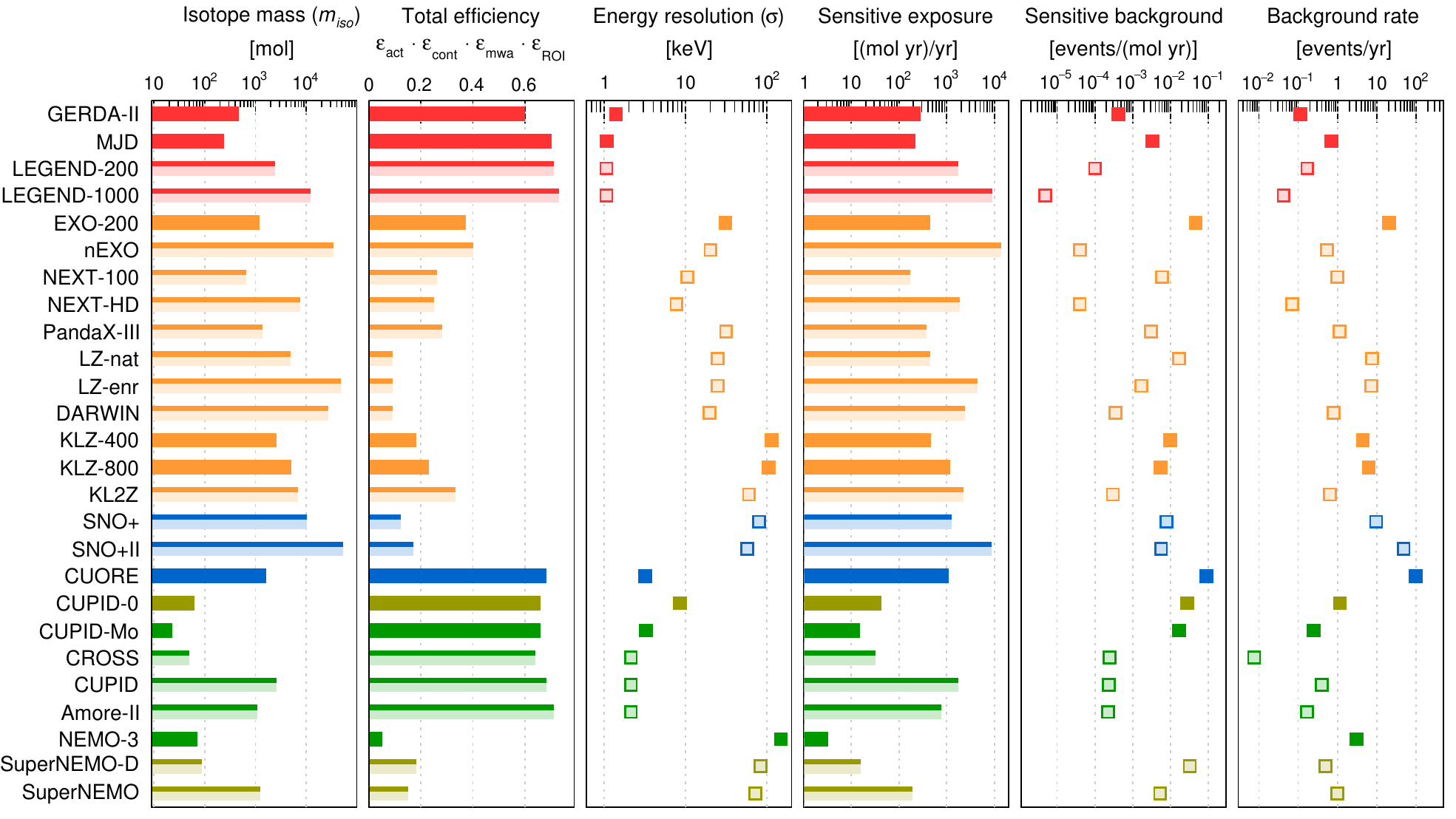}
    \caption{Fundamental parameters driving the sensitive background and exposure,
      and consequently the sensitivity, of recent and future phases of existing
      experiments (see Eq.~\ref{eq:senscounts}).
      Red bars are used for \Ge\ experiments, orange for \Xe, blue for \Te,
      green for \Mo, and sepia for \Se.
      Similar exposures are achieved with high mass but poorer energy resolution and efficiency by gas and liquid detectors,
      or with small mass but high resolution and efficiency by solid state detectors.
      The sensitive exposure is computed for one year of livetime.
      Ligher shades indicate experiments which are under construction or proposed.
      }
    \label{fig:pars} 
  \end{figure*}

  The sensitive exposure and background are effective parameters whose values
  are often not intuitive: they refer to the detector performance in the sensitive volume, where the signal-to-background ratio is maximal and drives the experimental sensitivity.
  We calculate the value of \senexp\ starting from the product of isotope mass $m_{iso}$ and data taking time,
  and correct it for a number of efficiencies:
  the active (or fiducial) fraction of the target mass $\varepsilon_{\text{act}}$,
  the probability that the energy deposited by the decay is fully contained within the detector $\varepsilon_{\text{cont}}$,
  and the multivariate analysis efficiency to tag events in the sensitive volume $\varepsilon_{\text{mva}}$. 
  Although $\varepsilon_{\text{mva}}$ would conceptually include the efficiency for a signal to fall in the energy region of interest (ROI) dominating the sensitivity,
  we separate this contribution and also quote the 
  energy resolution ($\sigma$) as well as the width of the effective ROI in units of $\sigma$,
  assuming a Gaussian approximation.
  To calculate \senbkg, we extract the rate of background events in the sensitive volume from the experiments' specifications.
  The values of these parameters and efficiencies are listed in Tab.~\ref{tab:summary} and shown graphically in Fig.~\ref{fig:pars}.
  When the value of a parameter cannot be computed from published specifications, we report effective values reproducing the nominal sensitivity of the experiment.
  Details of these approximations are discussed in the following
  subsections.
\begin{turnpage}
  \begin{table*}[tbp]
    \caption{Fundamental parameters driving the sensitive background and exposure of recent and future phases of existing experiments.
      The last two columns report the discovery sensitivity on the \nubb-decay
      half-life for 10 years of livetime,
      and the corresponding sensitivity on \mbb\ for the range of NMEs specified in Tab.~\ref{tab:NMEs}.
      For completed experiments, sensitivities are computed using the reported final exposure.
      MJD, KLZ, and SuperNEMO-D refer to the {\sc Majorana Demonstrator}, KamLAND-Zen, and the SuperNEMO Demonstrator, respectively.}
    \label{tab:summary}
     \tabcolsep=5pt
     \def\arraystretch{1.1}
\begin{tabular}{lll cc c C{4.5mm}C{4.5mm}C{4.5mm} D{.}{.}{3.1} cC{5mm} r c ccc}
      \toprule
      \multicolumn{2}{l}{\multirow{3}{*}{Experiment}}
      & \multirow{3}{*}{Isotope}
      & \multirow{3}{*}{Status}
      & \multirow{3}{*}{Lab}
      & \multicolumn{1}{c}{$m_\text{iso}$}
      & $\varepsilon_{\text{act}}$
      & $\varepsilon_{\text{cont}}$
      & $\varepsilon_{\text{mva}}$
      & \multicolumn{1}{c}{$\sigma$}
      & ROI
      & $\varepsilon_{\text{ROI}}$
      & \multicolumn{1}{c}{$\senexp$}
      & $\senbkg$
      & \multicolumn{1}{c}{$\lambda_b$}
      & \Thl
      & \mbb\\

      & 
      & 
      & 
      & 
      & \multicolumn{1}{c}{\multirow{2}{*}{[mol]}}
      & \multirow{2}{*}{[\%]}
      & \multirow{2}{*}{[\%]}
      & \multirow{2}{*}{[\%]}
      & \multicolumn{1}{c}{\multirow{2}{*}{[keV]}}
      & \multirow{2}{*}{[$\sigma$]}
      & \multirow{2}{*}{[\%]}
      & \multicolumn{1}{c}{\multirow{2}{*}{$\left[\frac{\molyr}{yr}\right]$}}
      & \multicolumn{1}{c}{\multirow{2}{*}{$\left[\frac{\text{events}}{\text{mol}\cdot\text{yr}}\right]$}}
      & \multicolumn{1}{c}{\multirow{2}{*}{$\left[\frac{\text{events}}{\text{yr}}\right]$}}
      & \multirow{2}{*}{[yr]}
      & \multirow{2}{*}{[meV]}\\

      & 
      & 
      & 
      & 
      & 
      & 
      & 
      & 
      &
      &
      & 
      & 
      & 
      &
      & 
      & \\
      \colrule
      \multicolumn{8}{l}{\it High-purity Ge detectors (Sec.~\ref{sec:prj:ge})}\\
      &GERDA-II        & \Ge & completed    & LNGS    & $4.5\cdot10^{2}$& 88  & 91  & 79 & 1.4 & -2,2               & 95 & 273     & $4.2\cdot10^{-4}$ & $1.1\cdot10^{-1}$ & $1.2\cdot10^{26} $ & \phantom{0}93-222 \\
      &{\sc MJD}       & \Ge & completed    & SURF    & $3.1\cdot10^{2}$& 91  & 91  & 86 & 1.1 & -2,2               & 95 & 212     & $3.3\cdot10^{-3}$ & $7.1\cdot10^{-1}$ & $4.7\cdot10^{25} $ & 149-355  \\
      &LEGEND-200      & \Ge & construction & LNGS    & $2.4\cdot10^{3}$& 91  & 91  & 90 & 1.1 & -2,2               & 95 & 1\,684  & $1.0\cdot10^{-4}$ & $1.7\cdot10^{-1}$ & $1.5\cdot10^{27} $ & 27-63             \\
      &LEGEND-1000     & \Ge & proposed     &         & $1.2\cdot10^{4}$& 92  & 92  & 90 & 1.1 & -2,2               & 95 & 8\,736  & $4.9\cdot10^{-6}$ & $4.3\cdot10^{-2}$ & $1.3\cdot10^{28} $ & \phantom{0}9-21   \\[10pt]
      
      \multicolumn{4}{l}{\it Xenon time projection chambers (Sec.~\ref{sec:prj:xe})}\\
      &EXO-200         & \Xe & completed    & WIPP    & $1.2\cdot10^{3}$& 46  & 100 & 84 & 31  & -2,2               & 95 & 438     & $4.7\cdot10^{-2}$ & $2.1\cdot10^{+1}$ & $2.4\cdot10^{25} $ & 111-477           \\
      &nEXO            & \Xe & proposed     & SNOLAB  & $3.4\cdot10^{4}$& 64  & 100 & 66 & 20  & -2,2               & 95 & 13\,700 & $4.0\cdot10^{-5}$ & $5.5\cdot10^{-1}$ & $7.4\cdot10^{27} $ & \phantom{0}6-27   \\
      &NEXT-100        & \Xe & construction & LSC     & $6.4\cdot10^{2}$& 88  & 76  & 49 & 10  & -1.0,1.8           & 80 & 167     & $5.9\cdot10^{-3}$ & $9.9\cdot10^{-1}$ & $7.0\cdot10^{25} $ & \phantom{0}66-281 \\
      &NEXT-HD         & \Xe & proposed     &         & $7.4\cdot10^{3}$& 95  & 89  & 44 & 7.7 & -0.5,1.7           & 65 & 1\,809  & $4.0\cdot10^{-5}$ & $7.2\cdot10^{-2}$ & $2.2\cdot10^{27} $ & 12-50             \\
      &PandaX-III-200  & \Xe & construction & CJPL    & $1.3\cdot10^{3}$& 77  & 74  & 65 & 31  & -1.2,1.2           & 76 & 374     & $3.0\cdot10^{-3}$ & $1.1\cdot10^{+0}$ & $1.5\cdot10^{26} $ & \phantom{0}45-194 \\
&LZ-nat          & \Xe & construction & SURF    & $4.7\cdot10^{3}$& 14  & 100 & 80 & 25  & -1.4,1.4           & 84 & 440     & $1.7\cdot10^{-2}$ & $7.5\cdot10^{+0}$ & $7.2\cdot10^{25} $ & \phantom{0}64-277 \\
      &LZ-enr          & \Xe & proposed     & SURF    & $4.6\cdot10^{4}$& 14  & 100 & 80 & 25  & -1.4,1.4           & 84 & 4302    & $1.7\cdot10^{-3}$ & $7.3\cdot10^{+0}$ & $7.1\cdot10^{26} $ & \phantom{0}20-87 \\
      &Darwin          & \Xe & proposed     &         & $2.7\cdot10^{4}$& 13  & 100 & 90 & 20  & -1.2,1.2           & 76 & 2\,312  & $3.5\cdot10^{-4}$ & $8.0\cdot10^{-1}$ & $1.1\cdot10^{27} $ & 17-72             \\[10pt]
    
      \multicolumn{4}{l}{\it Large liquid scintillators (Sec.~\ref{sec:prj:ls})}\\
      &KLZ-400         & \Xe & completed    & Kamioka & $2.5\cdot10^{3}$& 44  & 100 & 97 & 114 & \phantom{-0.}0,1.4 & 42 & 450     & $9.8\cdot10^{-3}$ & $4.4\cdot10^{+0}$ & $3.3\cdot10^{25} $ & \phantom{0}95-408 \\
      &KLZ-800         & \Xe & taking data  & Kamioka & $5.0\cdot10^{3}$& 55  & 100 & 100& 105 & \phantom{-0.}0,1.4 & 42 & 1\,143  & $5.5\cdot10^{-3}$ & $6.2\cdot10^{+0}$ & $2.0\cdot10^{26} $ & \phantom{0}38-164 \\
      &KL2Z            & \Xe & proposed     & Kamioka & $6.7\cdot10^{3}$& 80  & 100 & 97 & 60  & \phantom{-0.}0,1.4 & 42 & 2\,176  & $3.0\cdot10^{-4}$ & $6.5\cdot10^{-1}$ & $1.1\cdot10^{27} $ & 17-71             \\
      &SNO+I           & \Te & construction & SNOLAB  & $1.0\cdot10^{4}$& 20  & 100 & 97 & 80  & -0.5,1.5           & 62 & 1\,232  & $7.8\cdot10^{-3}$ & $9.7\cdot10^{+0}$ & $1.8\cdot10^{26} $ & \phantom{0}31-144 \\
      &SNO+II          & \Te & proposed     & SNOLAB  & $5.1\cdot10^{4}$& 27  & 100 & 97 & 57  & -0.5,1.5           & 62 & 8\,521  & $5.7\cdot10^{-3}$ & $4.8\cdot10^{+1}$ & $5.7\cdot10^{26} $ & 17-81             \\[10pt]
   
      \multicolumn{4}{l}{\it Cryogenic calorimeters (Sec.~\ref{sec:prj:bo})}\\
      &CUORE           & \Te & taking data  & LNGS    & $1.6\cdot10^{3}$& 100 & 88  & 92 & 3.2 & -1.4,1.4           & 84 & 1\,088  & $9.1\cdot10^{-2}$ & $9.9\cdot10^{+1}$  & $5.1\cdot10^{25} $ & \phantom{0}58-270 \\
      &CUPID-0         & \Se & completed    & LNGS    & $6.2\cdot10^{1}$& 100 & 81  & 86 & 8.5 & -2,2               & 95 & 41      & $2.8\cdot10^{-2}$ & $1.2\cdot10^{+0}$  & $4.4\cdot10^{24} $ & 283-551           \\
      &CUPID-Mo        & \Mo & completed    & LSM     & $2.3\cdot10^{1}$& 100 & 76  & 91 & 3.2 & -2,2               & 95 & 15      & $1.7\cdot10^{-2}$ & $2.5\cdot10^{-1}$  & $1.7\cdot10^{24} $ & 293-858           \\
      &CROSS           & \Mo & construction & LSC     & $4.8\cdot10^{1}$& 100 & 75  & 90 & 2.1 & -2,2               & 95 & 31      & $2.5\cdot10^{-4}$ & $7.6\cdot10^{-3}$  & $4.9\cdot10^{25} $ & 54-160            \\
      &CUPID           & \Mo & proposed     & LNGS    & $2.5\cdot10^{3}$& 100 & 79  & 90 & 2.1 & -2,2               & 95 & 1\,717  & $2.3\cdot10^{-4}$ & $4.0\cdot10^{-1}$  & $1.1\cdot10^{27} $ & 12-34             \\
      &AMoRE-II        & \Mo & proposed     & Yemilab & $1.1\cdot10^{3}$& 100 & 82  & 91 & 2.1 & -2,2               & 95 & 760     & $2.2\cdot10^{-4}$ & $1.7\cdot10^{-1}$  & $6.7\cdot10^{26} $ & 15-43             \\[10pt]

      \multicolumn{4}{l}{\it Tracking calorimeters (Sec.~\ref{sec:prj:tr})}\\
      &NEMO-3          & \Mo & completed    & LSM     & $6.9\cdot10^{1}$& 100 & 100 & 11 & 148 & -1.6,1.1           & 42 &  3       & $9.4\cdot10^{-1}$ &  $3.0\cdot10^{+0}$ & $5.6\cdot10^{23} $ & 505-1485    \\ 
      &SuperNEMO-D     & \Se & construction & LSM     & $8.5\cdot10^{1}$& 100 & 100 & 28 & 83  & -4.2,2.4           & 64 &  15      & $3.3\cdot10^{-2}$ &  $5.0\cdot10^{-1}$ & $8.6\cdot10^{24} $ & 201-391 \\ 
      &SuperNEMO       & \Se & proposed     & LSM     & $1.2\cdot10^{3}$& 100 & 100 & 28 & 72  & -4.1,2.8           & 54 &  185     & $5.3\cdot10^{-3}$ &  $9.8\cdot10^{-1}$ & $7.8\cdot10^{25} $ & 67-131 \\ 
      \botrule

\end{tabular}
  \end{table*}
  \end{turnpage}

  \begin{figure*}[]
    \includegraphics[width=\textwidth]{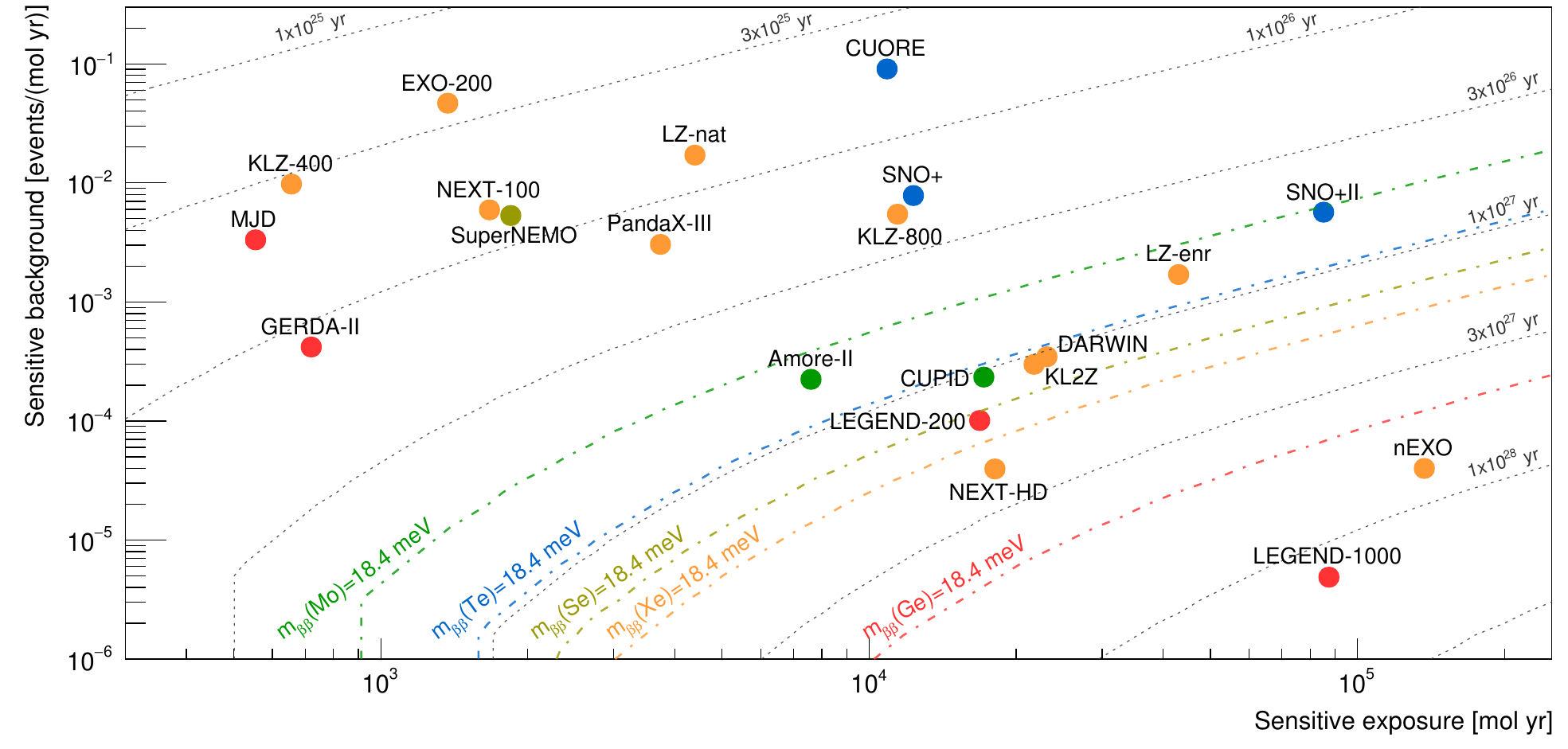}
    \caption{Sensitive background and exposure for recent and future experiments.
      The grey dashed lines indicate specific  discovery sensitivity  values on the \nubb-decay half-life.
      The colored dashed line indicate the half-life sensitivities required to
      test the bottom of the inverted ordering scenario for \Ge, \Xe, \Te\, \Mo, and \Se,
      assuming for each isotope the largest NME value among the QRPA calculations listed in Tab.~\ref{tab:NMEs}.
      A livetime of 10\,yr is assumed except for completed experiments, for which
      the final reported exposure is used.
      \jd{Pocar recommended some additional edits.}
      }
    \label{fig:t12} 
  \end{figure*}

  Figure~\ref{fig:pars} illustrates how each detection concept is characterized by specific parameter combinations. Liquid and gas detectors have
  large isotope masses and a relatively low signal detection efficiency due to the non-uniform background rate,
  with a small detector region driving the sensitivity.
  Solid state detectors operate a smaller isotope mass, but with higher efficiency and energy resolution.
  As a result, an isotope mass lower by a factor of 10 can be balanced by higher
  resolution and efficiency, yielding a similar sensitive exposure and sensitivity.

  Figure~\ref{fig:t12} shows a scatter plot of the sensitive exposure and background
  for the listed projects.
  Recent experiments populate the top left part of the plot, corresponding to
  exposures of thousands of mole-years --- i.e., tens or hundreds of kilograms of target mass operated for a few years --- and background rates between $10^{-3}$ and $10^{-1}$ events per mole-year. 
  To improve the sensitivity, future experiments need to either increase \senexp\ or reduce \senbkg. 
  Often a sequence of experimental upgrades with progressive incremental
  improvements is planned, ultimately leading to a combined factor of $\sim$100 improvement.

  These figures highlight the strengths and limitations of each detection concept,
  indicating the natural strategies to maximize the sensitivity, which are
  most evident in the \senexp/\senbkg\ ratios.
  For example, \Te\ experiments have large \senexp/\senbkg\ values
  (the blue markers are systematically above the other points in Fig.~\ref{fig:t12}).
  Given the high natural abundance of \Te, for them it is more efficient to increase the exposure rather than reducing the background.
  Conversely, \Ge-based experiments have small \senexp/\senbkg\ values.
  For them, reducing the background is easier than increasing the target mass,
  as their strengths are good energy resolution and advanced event reconstruction capabilities.
  Experiments based on other isotopes have intermediate \senexp/\senbkg\
  values, suggesting some flexibility in finding a trade off between the two quantities.
   
  Although the sensitive exposure and background are valuable parameters to characterize an experiment, the reach of an experiment is not fully captured by the \Thl\ sensitivity.
  The phase space factor also plays a strong role, and the nuclear structure of the
  isotopes deeply affects the connection between
  \Thl\ and the underlying decay mechanism. For instance, assuming the decay is
  mediated by the exchange of light Majorana neutrinos, the discovery power of an
  experiment is better quantified by the effective Majorana mass sensitivity.
  We hence include in the table and figures values for the \mbb\
  sensitivities, whose ratios provide a good figure of merit also assuming several
  other decay mechanisms.  We will discuss in detail the discovery power of the
  experiments in Sec.~\ref{sec:dis}.

  \subsection{High-purity Ge semiconductor detectors}
  \label{sec:prj:ge}

\begin{figure}[tb]
  \centering
  \includegraphics[width=.8\columnwidth]{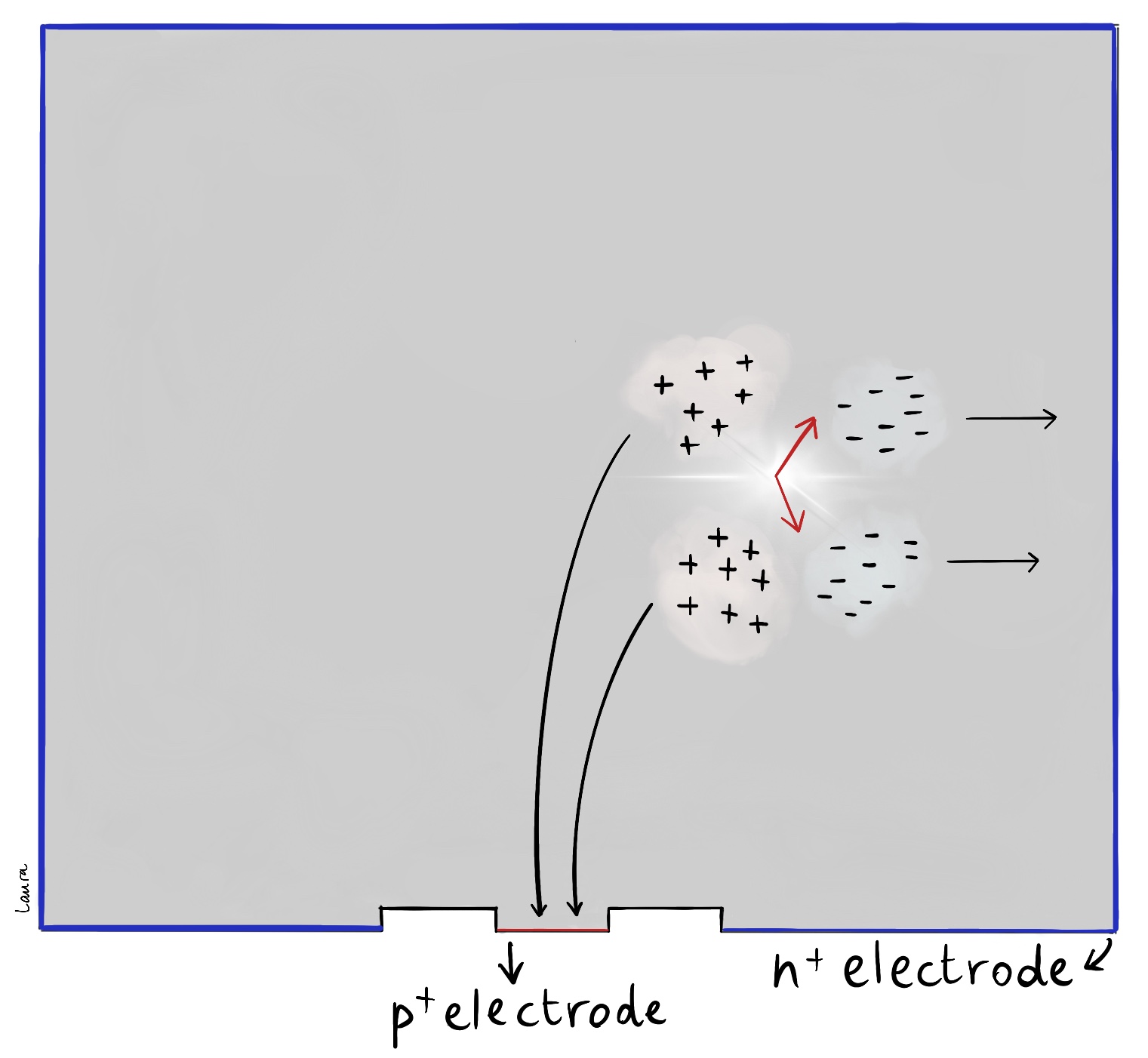}
  \caption{Artistic illustration of a high-purity Ge (HPGe) detector and its \nubb-decay detection concept. Electron and hole clusters created by ionization are collected to the electrodes by an electric field. Image courtesy of Laura Manenti.}
  \label{fig:ge-det}
\end{figure}

  High Purity Germanium (HPGe) detectors represent the longest-standing
  technology used for \nubb-decay searches \cite{AvignoneIII:2019spj}.
  The first \nubb-decay experiment based on Ge detectors was in 
  1967 \cite{Fiorini:1967in} and, since then, Ge-based experiments have stayed
  at the forefront of the field. Fig.~\ref{fig:ge-det} shows an example of the state-of-the-art model.

  HPGe detectors are semiconductor devices.  A detector consists of a single
  crystal grown by the Czochralski method \cite{DEPUYDT2006437}
  from Ge-material enriched up to 92\% in \isot{Ge}{76}.
  The detectors used by recent and future
  experiments are all p-type crystals, with two conductive electrodes obtained through B implantation (p+ electrode) and Li diffusion (n+ electrode).
  The semiconductor junction forms between the n+ electrode and the p-type crystal, and is extended throughout
  the whole detector volume by applying a reverse bias of a few thousands volts. 
  Electrons and holes produced within the crystal by ionization drift along the electric field, inducing a current. The current integral is proportional to the energy deposited within the detector, and its time-structure carries information on the event topology.
  The detector size is currently limited to 1-3\,kg, and multiple detectors need to be operated simultaneously to reach a competitive amount of isotope mass.
  These detectors are operated in ultra-low background environments, surrounded by
  shielding material and active veto systems.

  HPGe detectors feature high \nubb-decay detection efficiencies.
  The presence of conductive electrodes on the detector surface reduces the active volume fraction to $\varepsilon_{\text{act}}\sim90\%$
  and leads to energy loss for a fraction of the \nubb-decay events ($\varepsilon_{\text{cont}}\sim 90\%$).
  The \nubb-decay event tagging efficiency, $\varepsilon_{\text{mva}}\sim80-90\%$,
  is typically limited by signal-background discrimination  methods based on the analysis of the current time-structure.
  Given the low background level and high resolution, the optimal energy region of interest (ROI) for \nubb-decay searches is $\Qbb\pm2\sigma$, containing 95\% of the signal.
  Specific parameter values of \Ge\ experiments are listed in Tab.~\ref{tab:ge}.

  \begin{table}[htbp]
    \caption{Specific parameters of experiments using Ge detectors:
     total detector mass, fractional isotopic abundance, shielding strategy, and background
     index normalized over the entire detector mass. The background index is
     what is historically quoted by these experiments but, differently from our
     sensitive background, is not normalized over the signal detection
     efficiencies and detector resolution.
     The values are taken from
     \cite{GERDA:2020xhi,Majorana:2019nbd,LEGEND:2021bnm}, averaging over
     multiple datasets for GERDA-II and the {\sc Majorana Demonstrator}.}
    \label{tab:ge}
    \begin{tabular}{lccccccccc}
      \toprule
      Experiment & $m_{tot}$  & $f_{iso}$  &  Shield  & Background \\
                 & [kg]       & [\%]              &      & $\left[\dfrac{\text{events}}{\text{keV}\cdotp\text{kg}\cdotp\text{yr}}\right]$\\
      \colrule
      \smallspace
      GERDA-II         & 39   & 87 & liquid Ar &  $5.2\cdot10^{-4}$ \\
      {\sc MJD}        & 20   & 88 & Cu \& Pb  &  $6.0\cdot10^{-3}$ \\
      LEGEND-200       & 200  & 90 & liquid Ar &  $2\cdot10^{-4}$   \\
      LEGEND-1000      & 1000 & 91 & liquid Ar &  $1\cdot10^{-5}$   \\
      \botrule
    \end{tabular}
  \end{table}

  The GERDA experiment operated a compact array of about 40 detectors 
  in a cryostat containing 64\,m$^3$ of liquid argon (LAr) \cite{Agostini:2017hit}. 
  Several detector geometries were used during the experiment, giving an 
  average \isot{Ge}{76} mass of $\sim$34\,kg.
  The LAr acted as a passive shielding against natural radioactivity from any contamination outside the cryostat,
  and also attenuated background produced by radioactive isotopes in the
  materials near the detectors, such as the holders or cables. 
  The LAr was also used as an active veto system by detecting its scintillation
  light produced when a background event deposits energy in the argon volume. 
The average energy resolution achieved during the second
  phase of the experiment (GERDA-II) was $\sigma=1.4$\,keV, and the
  average background index was $5.2_{-1.3}^{+1.6}\cdot10^{-4}$\,\ctsper, corresponding to 
  $\senbkg = 4.2\cdot10^{-4}$\,\evmolyr \cite{GERDA:2020xhi}. 
  With these parameters, at present, GERDA has achieved the lowest
  sensitive background in the field.
  The remnant background composition was traced to U and Th in
  the material surrounding the detectors, and $\alpha$- and $\beta$-decaying isotopes on the
  detector surfaces \cite{GERDA:2019cav}. 
  The final result of GERDA is a constraint of $\Thl > 1.8 \cdot 10^{26}$\,yr at
  90\% C.L., consistent with the median upper limit expected for no signal, derived including also the data from Phase\,I of the experiment. 

  The {\sc Majorana Demonstrator} (MJD) \cite{Abgrall:2013rze} employed a compact array
  of up to 58 detectors comprised of both enriched and natural Ge.
  27\,kg of isotope were operated in enriched detectors~\cite{Arnquist:2022zrp}. 
  The HPGe crystals are deployed
  in two vacuum cryostats shielded from the environmental background by a
  layer of underground-electroformed copper, commercially obtained copper, and
  high-purity lead. The detectors incorporate a point-like p+ electrode providing very low-energy
  threshold and an excellent energy resolution of $\sigma=1.1$\,keV at
  \Qbb, which is currently the best in the field.
  With a sensitive background of $\senbkg=3.3 \cdot 10^{-3}$\,\evmolyr, 
  the experiment reported a limit of 
  $\Thl > 8.3 \cdot 10^{25}$\,yr  at 90\% C.L. with a limit setting sensitivity of 
  $\Thl = 8.1 \cdot 10^{25}$\,yr.
  The background is dominated by a distant source of thorium~\cite{Arnquist:2022zrp}.

  The next-generation Ge-based experiments will be realized in the framework of the
  LEGEND project \cite{LEGEND:2021bnm}, with two stages planned: LEGEND-200 and LEGEND-1000. 
  In the first, $\sim$200\,kg of Ge detectors will be operated in the 
  GERDA setup after upgrading part of the infrastructure.
  Compared to GERDA, a further reduction of the background is anticipated thanks
  to the use of larger-mass detectors (resulting
  in fewer cables and electronic components), improved light readout, and 
  materials with improved radiopurity, such as the electroformed copper developed for the {\sc Majorana Demonstrator}.
  An energy resolution equal to or better than the one achieved in the {\sc Majorana Demonstrator} is
  expected. 
  These improvements would bring the LEGEND-200 background to
  $2\cdot10^{-4}$\,\ctsper, less than a factor of 3 lower than what was achieved by GERDA. 
  With a sensitive background of $\senbkg = 1.0 \cdot10^{-4}$\,\evmolyr, 
  LEGEND-200 will achieve a discovery sensitivity 
  of $10^{27}$\,yr in 5~years of live time.
  For LEGEND-1000, a new infrastructure able to host 1\,ton of target mass will be
  realized. 
  A further twenty-fold background reduction is expected with the usage of underground
  argon, and lower radioactivity levels in cables and electronics.
  LEGEND-1000 expects a sensitive background of $\senbkg = 4.9 \cdot10^{-6}$\,\evmolyr, 
  leading to a discovery sensitivity of 
  $\Thl = 1.3 \cdot 10^{28}$\,yr after 10~years of operation.

  We note that during preparation of this manuscript plans were announced
  \cite{QYueTAUP2021} for a \nubb-decay-focused branch of the CDEX effort \cite{CDEX:2017pgl},
  culminating in a ton-scale \Ge\ experiment. At present public details
  for this project are insufficient for estimating its sensitivity.

  \subsection{Xenon time projection chambers}
  \label{sec:prj:xe}

\begin{figure}[tb]
  \centering
  \includegraphics[width=\columnwidth]{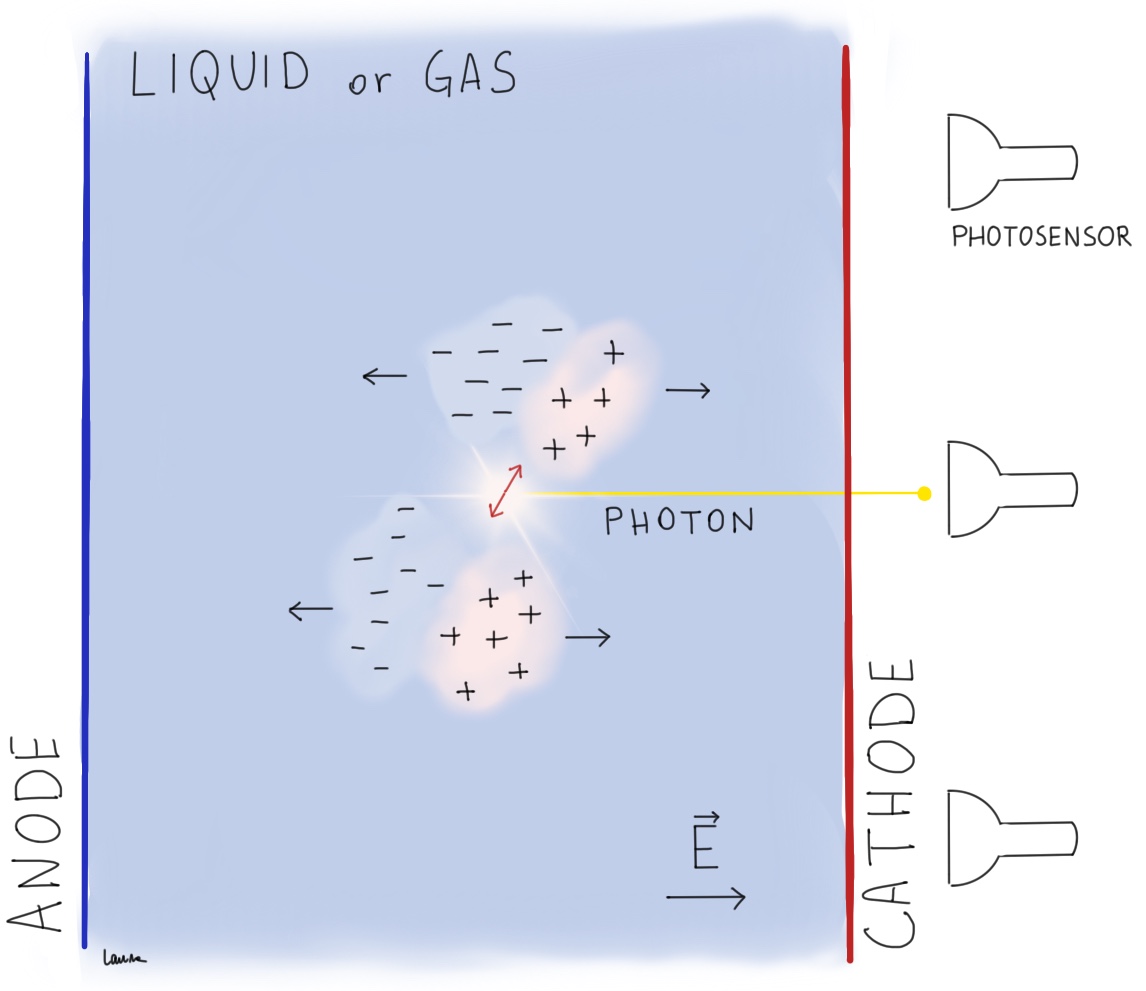}
  \caption{Artistic illustration of a Xe time projection chamber and its \nubb-decay detection concept. Electron and hole clusters created by ionization are collected to the electrodes by an electric field. In addition, scintillation light is detected by light sensors, providing the timing of the event. Image courtesy of Laura Manenti.}
  \label{fig:xe-tpc}
\end{figure}

  Time projection chambers (TPCs) were the first technology used to observe
  \nnbb\ decay in real time \cite{Elliott:1987kp} and have
  remained at the forefront of \nubb-decay searches ever since. In these
  detectors, a static electric field is applied in a region containing a liquid or
  gaseous medium. As shown in Fig.~\ref{fig:xe-tpc}, the electrons and ions liberated by ionizing radiation
  drift  to analyzing planes which reconstruct
  their number and position in the plane normal to the field.
The position along the field is derived from the drift durations.
The event reconstruction allows to discriminate spatially-localized \nubb-decay events from spatially extended ones, such as those produced by multiple Compton scattering.
  Depending on the spatial resolution, even the full 3D tracks of the two electrons emitted
  in \nubb\ decay can in principle be reconstructed, making it possible to
  discriminate them from from single $\beta$-decays, \G-ray scattering and
  absorption, or nuclear recoils from neutron scattering.

  TPCs are particularly well-suited to searches for the \nubb\
  decay of \Xe. The source itself is an inert noble element and can be used directly in
  TPCs in its purified form as a liquid, gas, or both. In either phase, Xe
  exhibits VUV scintillation emitted promptly with an energy deposition. Experiments able to detect the scintillation light signal can reconstruct the full 3D topology of the event using a single analyzing plane.
  The intensity of the scintillation light
  also provides a complementary measurement of energy, whose anti-correlation with
  the ionization signal can significantly enhance the energy resolution \cite{EXO-200:2003bso}.

  If the electric field is strong enough, the collision between drifting electrons
  and gas molecules results in the emission of secondary scintillation light, called electroluminescence (EL).
  Single-phase high-pressure gas TPCs (recently reviewed in \cite{Gomez-Cadenas:2019ges}) shape the field near the electrode to create a region where the incoming electrons produce EL.
  Dual-phase TPCs obtain the same result using a short gaseous EL layer at the top of the liquid volume.
  The EL signal gives a precise measure of the number of ionization electrons,
  improving the energy resolution. With a fine enough spatial resolution of the 
  light collection, the EL signal can also be used for track reconstruction.
  The energy resolution of experiments reading out the electroluminescence light
  is limited by fluctuations in the number of primary ionization electrons.
  These fluctuations are small and independent of fluid density up
  to about 0.6 g/cm$^3$ ($\sim$100 bar), but above that pressure 
  grow rapidly \cite{1997NIMPA.396..360B}. 

  Xe TPCs also potentially lend themselves to techniques for observation of the \bb\ daughter Ba
  ion, as first suggested by \textcite{Moe:1991ik}. Single-atom trapping and imaging was first
  achieved with Ba \cite{1980PhRvA..22.1137N}. Xe is a transparent fluid, offering
  multiple options for tagging based on fluorescence imaging techniques. The nEXO Collaboration is pursuing a strategy
  using a cryogenic probe \cite{Twelker:2014zsa} to electrostatically attract the
  Ba ion in the vicinity of a signal event and freeze it in a volume of Xe, then
  transport it to a fluorescence imaging stage capable of single-atom
  detection \cite{nEXO:2018nxx}.  The NEXT collaboration aims to transport
  Ba$^{++}$ ions to single-molecule fluorescence imaging (SMFI) sensors for
  example using RF carpets \cite{Brunner:2014sfa, NEXT:2021idl}. 
  Single-Ba-atom sensitivity with
  SMFI \cite{McDonald:2017izm} and an implementation applicable to high-pressure
  gas Xe TPCs \cite{Rivilla:2020cvm} have been demonstrated.  
  Both collaborations are still working to demonstrate that their tagging schemes can
  be achieved with sizeable efficiency and in an actual \nubb-decay experiment. 
  We do not discuss these techniques further in this review.

  Liquid Xe volumes operated in TPCs provide 
  self-shielding from external radiation, whose contribution to the background drops
  exponentially with the distance from the outer  Xe surface.  
  Of particular worry is the $^{214}$Bi gamma line at 2447.7 keV, just below
  the \Xe\ \Qbb\ at 2457.8 keV, and often not resolved.
  Xe TPC experiments use a multi-variate
  analysis to handle the varying background rate throughout the
  detector volume.  However, the sensitivity of the experiment is
  essentially driven by the innermost region of the detector, while the
  outer region is used primarily to constrain the background
  extrapolation to the detector center. A fiducial volume may be defined or 
  tuned to demark these regions, leading to 
  $\varepsilon_{act} = $~10--60\% depending on the enrichment level and the
  radioactivity of the structural materials.
  The background in that fiducial
  volume is then treated as an effective parameter, tuned 
  to reproduce the half-life sensitivities reported by the experiments.
  The most sensitive energy
  region of interest varies, depending on the background
  level and whether the $^{214}$Bi gamma line is resolved. Containment
  efficiencies are $\varepsilon_{act} \sim 100\%$ for
  liquid Xe TPCs after the effective fiducial volume cut, while they are
  typically 70--80\% for gaseous detectors.

  \begin{table*}[tbp]
    \caption{Specific parameters of Xe-TPC experiments: total mass, fractional isotopic abundance,
      phase, signal readout, effective background index in units of events per kg of mass in the fiducial volume,
      and the ratio $R$ between the effective background index used for our sensitivity calculation
      and the mean background quoted by the experiments, when available.
Values are taken from references \cite{EXO-200:2019rkq, nEXO:2018ylp,
      nEXO:2021ujk, NEXT:2012zwy, NEXT:2015wlq, NEXT:2020amj, Chen:2016qcd,
      LZ:2019qdm, DARWIN:2020jme}.
    \label{tab:XeTPCs}
    }
    \begin{tabular}{lcccccc}
      \toprule
      Experiment & $m_{tot}$  & $f_{iso}$ & Phase & Readout &
        Effective Background & $R$ \\ & [kg]       & [\%]              & & &$\left[\dfrac{\text{events}}{\text{keV}\cdotp\text{kg}\cdotp\text{yr}}\right]$ & \\
      \colrule
      EXO-200        & 161          & 81 & liquid     & LAPPDs + wires           & $1.8\cdot10^{-3}$ & 1 \\
      nEXO           & 5109         & 90 & liquid     & electrode tiles + SiPM s & $2.1\cdot10^{-6}$ & N/A \\
      NEXT-100       & 97           & 90 & gas        & SiPMs + PMTs             & $4.0\cdot10^{-4}$ & 1 \\
      NEXT-HD        & 1100         & 90 & gas        & SiPMs + PMTs             & $4.0\cdot10^{-6}$ & 1 \\
      PandaX-III-200 & 200          & 90 & gas        & Micromegas               & $1.0\cdot10^{-4}$ & 1 \\
LZ-nat         & 7\,000       & 9  & dual-phase & PMTs                     & $1.1\cdot10^{-4}$ & 0.4 \\
      LZ-enr         & 7\,000       & 90 & dual-phase & PMTs                     & $1.1\cdot10^{-4}$ & 0.4 \\
      DARWIN         & 39\,300      & 9  & dual-phase & PMTs                     & $3.4\cdot10^{-6}$ & 0.85 \\
      \botrule
    \end{tabular}
  \end{table*}

  The most sensitive Xe TPC experiment to date was EXO-200 \cite{Auger:2012gs}, which 
used liquid-phase enriched Xe, with 161\,kg of $^{136}$Xe. 
  The TPC employed two drift regions with a common cathode at the
  detector center. The ionization was read out via crossed wire planes at the
  anodes. The prompt scintillation light was read out by large-area avalanche
  photodiodes. The combined signal achieved an energy resolution of
  $\sigma = 28$\,keV at \Qbb\ \cite{EXO-200:2019rkq}, or 31\,keV when averaged
  over the full dataset. 
  Backgrounds and field non-uniformity near the
  detector edges required fiducialization, restricting the analysis to the innermost
  75\,kg of Xe. 
  An extensive screening campaign \cite{Leonard:2007uv, Leonard:2017okt} and a sophisticated analysis incorporating topological
  background discrimination \cite{EXO:2018bpx} led to 
  an averaged background level within the fiducial volume of $1.8\cdot10^{-3}$\,\ctsper,
  corresponding to $\senbkg=$4.7$\cdot 10^{-2}$\,\evmolyr,
  dominated by the $^{214}$Bi gamma line
  originating from $^{238}$U chain decays outside of the Xe volume.
The experiment reported an ultimate limit for \nubb\ decay of 
  $\Thl > 3.5 \cdot 10^{25}$\,yr  at 90\% C.L. with a sensitivity for limit setting of 
  $\Thl = 5.0 \cdot 10^{25}$\,yr \cite{EXO-200:2019rkq}. Our counting
  analysis described in Sec.~\ref{sec:exp:stat:sens} reproduces the EXO-200 limit sensitivity with no tuning required.

  nEXO \cite{nEXO:2018ylp,nEXO:2021ujk} build on EXO-200's technology and aims at 
  using $5$\,tons of Xe enriched to $90\%$ in $^{136}$Xe.
  The TPC design features one monolithic drift volume with
  ionization read out by silica tiles patterned with metallic electrode strips,
  and scintillation detection by an array of VUV-sensitive
  silicon photomultipliers on the TPC walls behind the field-shaping grid,
  yielding an energy resolution of $\sigma = 20$\,keV.
The effective background index that reproduces nEXO's published discovery sensitivity is
  $2.1\cdot10^{-6}$\,\ctsper, corresponding to $\senbkg = 4.0\cdot 10^{-5}$\,\evmolyr,
  a factor of $\sim$1000 improvement over EXO-200. 
nEXO is expected to achieve a discovery sensitivity of $\Thl = 7.4 \cdot 10^{27}$\,yr after 10~years of live time.

  NEXT \cite{Nygren:2009zz,NEXT:2009vsd} implements a high-pressure gaseous Xe TPC equipped
  with an EL region. Tracking information is obtained from a SiPM array
  at the anode, while PMTs at the cathode provide optimal energy resolution.
  NEXT-White \cite{NEXT:2018rgj}, a proof-of-principle detector 
with 5\,kg of Xe at 10\,bar,
  demonstrated an energy resolution of $\sigma = 10$\,keV at \Qbb\ \cite{NEXT:2019qbo}, and
  tracking performance capable of discriminating single and double electron track 
  events, retaining 57\% of signal events while rejecting background by a factor
  of 27 \cite{NEXT:2020try}. 
  A second stage, NEXT-100 \cite{NEXT:2012zwy,NEXT:2015wlq}, 
  with a pressure of 15\,bar and containing 87\,kg of \Xe,
  is currently under construction.
The NEXT-100 projected background is dominated by remnant events from U/Th in the PMTs
and other detector components, giving a background index of
  $4\cdot10^{-4}$\,\ctsper,
  or $\senbkg = 5.9\cdot 10^{-3}$\,\evmolyr.
NEXT-HD \cite{NEXT:2020amj}, a concept for a future ton-scale phase of NEXT, is a
  symmetric TPC with a common central cathode, large enough to accommodate 
  a full ton of \Xe\ in the form of enriched Xe gas at 15\,bar. The design of NEXT-HD substitutes PMTs with an
  all-SiPM light readout at both TPC ends, using wavelength-shifting fibers to enhance light
  collection further. Gas additives to reduce diffusion are expected to enhance
  both the energy resolution and tracking resolution relative to NEXT-100.
  The expected reduction in background index by a factor of 100 thus leads to an even
  large reduction in sensitive background, predicted to be $\senbkg = 4.0\cdot 10^{-5}$\,\evmolyr.
NEXT-BOLD aims to take this concept one step further by instrumenting the NEXT-HD TPC with Ba
  tagging capability \cite{Rivilla:2020cvm}, potentially achieving half-life
  sensitivity on the order of $10^{28}$ years.

  Another experiment based on the high-pressure gas Xe TPC technique is
  PandaX-III-200 \cite{Chen:2016qcd}.  The initial phase uses 180\,kg \Xe\ in a volume
  of enriched Xe gas at 10\,bar, deployed in a symmetric TPC with a common cathode. In contrast to the
  all-photon-based readout pursued by NEXT, PandaX-III-200 exclusively relies
  on an ionization-only readout of just the drift electrons using the Micromegas
  detector technology, where a high-field region near the anode provides avalanche
  amplification.  The expected energy resolution is $\sigma = 31$\,keV at \Qbb, while
  simulated topological discrimination based on track reconstruction predicts up to
  two orders-of-magnitude background reduction with 42\% signal
  efficiency \cite{Galan:2019ake}.  Backgrounds are dominated by U/Th contamination of the Micromegas readout plane.
The total background index goal is 10$^{-4}$\,\ctsper, giving $\senbkg=3\cdot 10^{-3}$\,\evmolyr.

  LZ \cite{LZ:2015kxe} and DARWIN \cite{DARWIN:2016hyl} both employ dual-phase natural-Xe
  TPCs with EL readout to perform direct searches for WIMP Dark Matter. These
  detectors also have sensitivity to \nubb\ decay even with natural Xe
  targets \cite{LZ:2019qdm,DARWIN:2020jme}. The instrumentation required for
  detection of the faint nuclear recoils from WIMPs naturally leads
  to higher external backgrounds than for a detector optimized for \nubb\ decay. With
  7\,tons of Xe (640\,kg \Xe) in the LZ inner vessel and and 40\,tons (3.6\,tons \Xe) 
  total in DARWIN, self-shielding reduces these
  background dramatically, but external $^{214}$Bi still dominates 
  in both experiments. Reproducing the LZ sensitivity requires an effective background
  index of 1.2$\cdot10^{-4}$\ctsper, giving $\senbkg$=1.7$\cdot10^{-2}$\,\evmolyr. A subsequent run
  with enriched Xe (90\% \Xe) would have enhanced sensitivity. DARWIN's larger
  mass affords it a lower effective background index of 3.4$\cdot10^{-6}$\,\ctsper, or
  $\senbkg$=3.5$\cdot10^{-4}$\,\evmolyr, with $^{137}$Xe $\beta$ decays representing an important background
  contribution.

  A summary of all TPCs significant experimental parameters is given in
  Tab.~\ref{tab:XeTPCs}.

  \subsection{Large liquid scintillators}
  \label{sec:prj:ls}

\begin{figure}[tb]
  \centering
  \includegraphics[width=\columnwidth]{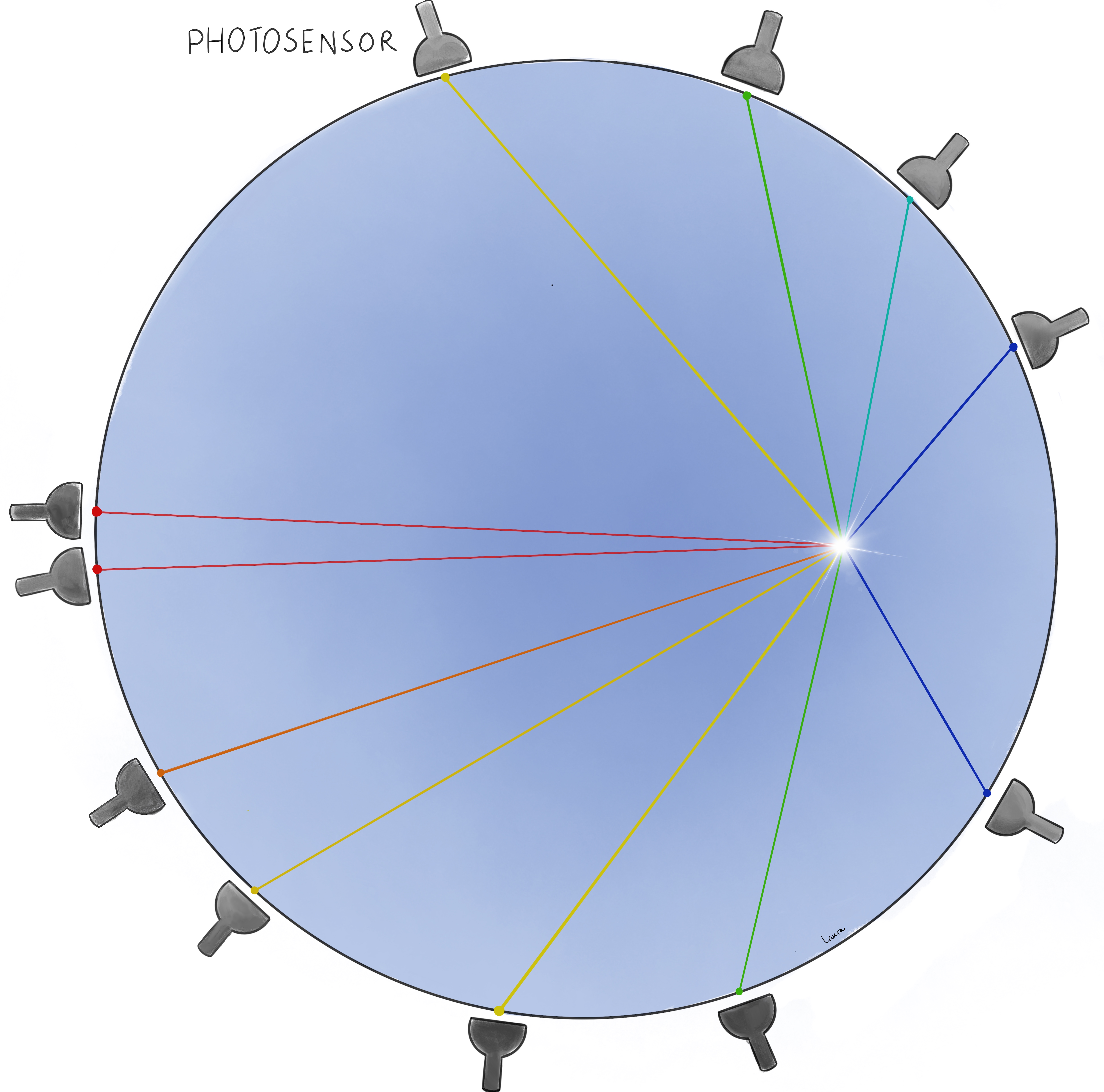}
  \caption{
  Artistic illustration of a large liquid scintillator detector and its \nubb-decay detection concept. The position of an event can be reconstructed through the time of flight of the scintillation photons. Image courtesy of Laura Manenti.}
  \label{fig:large-scintillators}
\end{figure}

  In what is perhaps the most successful departure from the ``source\,=\,detector''
  paradigm followed by most \nubb-decay experiments, large liquid scintillators
  offer the advantage of dissolving or loading vast amounts of isotope into the most
  sensitive regions of some of the lowest-radioactivity experiments ever
  constructed. With typical mass-loading fractions on the few percent level, a
  kton-scale scintillator detector can reach ton-year exposures with relative
  ease. Energy depositions within the detector generate scintillation photons,
  which are detected by PMTs viewing the target volume as shown in Fig.~\ref{fig:large-scintillators}. Event energy, position,
  and topology reconstruction is performed using the number, pattern, and
  timing of the detected photons. The position reconstruction is particularly
  important for these self-shielding detectors,
  whose inner volume has the lowest background and drives the sensitivity.
  The effective fiducial volume fractions range between $\varepsilon_{act} = 20$\%--80\%,
  due to a combination of self-sheilding and whether the target isotope is spread through the whole scintillator volume
  or confined to its central part.
  Containment efficiencies are maximal in the fiducial volume ($\varepsilon_{cont}=100\%$).

  The challenge for these detectors lies in their limited energy
  resolution due to the relatively low number of scintillation photons produced
  by energy depositions at \Qbb. Events due to \nnbb\ decay pose a problematic background in the energy region of interest, 
and the extraction of a \nubb-decay signal relies on an energy spectral analysis
  sensitive to distortion at the end-point of the \nnbb-decay energy distribution. Such an
  analysis requires a precise understanding of the detector response and energy
  reconstruction systematic effects. 
  The \nubb-decay background reduces the optimal energy region of interest to
  values above \Qbb, with an effective 40--60\% loss in detection efficiency.
  Like for the case of Xe TPCs, an effective background index for the fiducial
  volume was tuned to reproduce published experimental sensitivities.

  The presence of the large mass of liquid
  scintillator also increases the prevalence of solar neutrino backgrounds and
  cosmogenic activation products. The latter includes in
  particular $^{10}$C which is readily generated in organic liquid scintillators.  
  Vetoing schemes based on proximity to muon tracks and the detection of
  neutron capture gammas in delayed time coincidence can reduce these background
  contributions by roughly an order of magnitude at the cost of some exposure
  loss ($\varepsilon_{mva}=97\%$, see, e.g., \textcite{KamLAND-Zen:2016pfg}).

  KamLAND-Zen (KLZ) is an upgrade of the KamLAND apparatus \cite{Eguchi:2002dm} tailored
  to the search of \nubb\ decay: a nylon balloon is deployed in the
  active detector volume and filled with liquid scintillator in which
  enriched Xe has been dissolved.  A successful first phase
  deployment, KamLAND-Zen 400 (KLZ-400) with up to 340\,kg of \Xe, led to the strongest half-life
  limits for its time despite an unexpected background likely originating from
  fallout from the Fukushima nuclear
  disaster \cite{KamLAND-Zen:2012mmx,KamLAND-Zen:2016pfg}.  The second phase,
  KamLAND-Zen~800 (KLZ-800), is currently running with
  $\sim$680\,kg of \Xe\ redeployed in a larger, cleaner
  balloon. With just 1.6 years of data KLZ-800 produced a world-leading half-life limit
  $\Thl > 2.3 \cdot 10^{26}$\,yr  at 90\% C.L. with a limit setting sensitivity of 
  $\Thl = 1.3 \cdot 10^{26}$\,yr \cite{KamLAND-Zen:2022tow}.  
  The background measured in the KLZ-800 fiducial volume corresponds to 
  $\senbkg=5.5\cdot10^{-3}$\,\evmolyr.
The KLZ-800 sensitivity is well-reproduced by the background-dominated
  approximation.
The KamLAND-Zen collaboration is already preparing
  a follow-on phase, KamLAND2-Zen (KL2Z) \cite{Shirai:2017jyz}, in which
  $\sim$1\,ton of $^{136}$Xe
  will be deployed.  A major upgrade of the experiment is conceived
  for KL2Z to improve the energy resolution at \Qbb\ 
  from $\sigma=$114\,keV to 60\,keV.
The upgrade includes the
  installation of new light concentrators and PMTs with higher quantum
  efficiency as well as purer liquid scintillator.
A sensitive background reduction by a factor 20 over KLZ-800
  is expected for KL2Z, afforded primarily by the envisioned
  improvement in the detector resolution.
An effective background that is a factor of 0.45 times the predicted background
  reproduces the expected KL2Z sensitivity.

SNO+~\cite{Andringa:2015tza, SNO:2021xpa} is a follow up of the SNO experiment building on the SNO
  infrastructure~\cite{Boger:1999bb}.
  It is a multi-purpose neutrino experiment, with a $^{130}$Te-based \nubb-decay search
  as one of its main physics goals.
  SNO's acrylic sphere will be filled with $\sim$780\,tons of liquid scintillator,
  loaded with tellurium, with the surrounding SNO cavern instrumented as a water
  Cherenkov active veto.  As of the time of
  writing, SNO+ is filled with liquid scintillator and taking data, with Te
  loading scheduled to commence soon.
  A multi-staged approach is foreseen \cite{christopher_grant_2020_4142683}.
  Initially $\sim$1.3\,tons of $^{130}$Te (0.5\% $^{\rm nat}$Te loading)
  will be used and an energy resolution of $\sigma=$80\,keV is expected.
  The predicted background corresponds to $\senbkg=7.8\cdot10^{-3}$\,\evmolyr, 
  and is dominated by $^8$B solar neutrino elastic scatters.
  The goal of a subsequent phase is to increase the $^{130}$Te mass to 6.6\,tons
  (2.5\% $^{\rm nat}$Te loading) and improve the energy resolution to 57\,keV.
  This is achievable thanks to an improvement
  of the light yield to $800$\,pe/MeV \cite{KleinPrivate}.
  The predicted background corresponds to $\senbkg=5.7\cdot10^{-3}$\,\evmolyr.

  A summary of the relevant parameters for KamLAND-Zen and SNO+
  is given in Tab.~\ref{tab:scintillators}.

  \begin{table*}[htbp]
    \caption{Specific parameters for liquid scintillator experiments:
      isotope, total mass, fractional isotopic abundance, fractional mass of the loaded material, effective background
      per kg (of isotope) in the fiducial volume, and the ratio $R$ of that to the 
      mean background in the fiducial volume. 
      Values are taken from references ~\cite{KamLAND-Zen:2016pfg,
      Gando:2020cxo, Shirai:2017jyz, Andringa:2015tza, SNO:2021xpa}.}
    \label{tab:scintillators}
    \begin{tabular}{lccccccccccc}
      \toprule
      Experiment & Isotope & $m_{tot}$ & $f_{iso}$ & Loading & Effective Background & $R$ \\
      $\begin{array}{c} \text{Experi} \\ \text{ment} \end{array}$ 
        & & [kg] & & [\%wt.] & \begin{scriptsize} $\left[\dfrac{\text{events}}{\text{keV}\cdotp\text{kg}\cdotp\text{yr}}\right]$ \end{scriptsize} & & \\
      \colrule
      KLZ-400 & \Xe & 378     & 0.91 & 2.9 &  $1.8\cdot10^{-4}$ & 1 \\
      KLZ-800 & \Xe & 745     & 0.91 & 3.0 &  $1.1\cdot10^{-4}$ & 1 \\
      KL2Z    & \Xe & 1000    & 0.91 & 2.7 &  $1.1\cdot10^{-5}$ & 0.45 \\
      SNO+I   & \Te & 3825    & 0.91 & 0.5 &  $2.3\cdot10^{-4}$ & 1 \\
      SNO+II  & \Te & 19\,125 & 0.91 & 2.5 &  $2.3\cdot10^{-4}$ & 1 \\
      \botrule
    \end{tabular}
  \end{table*}

  \subsection{Cryogenic calorimeters}
  \label{sec:prj:bo}

\begin{figure}[tb]
  \centering
  \includegraphics[width=\columnwidth]{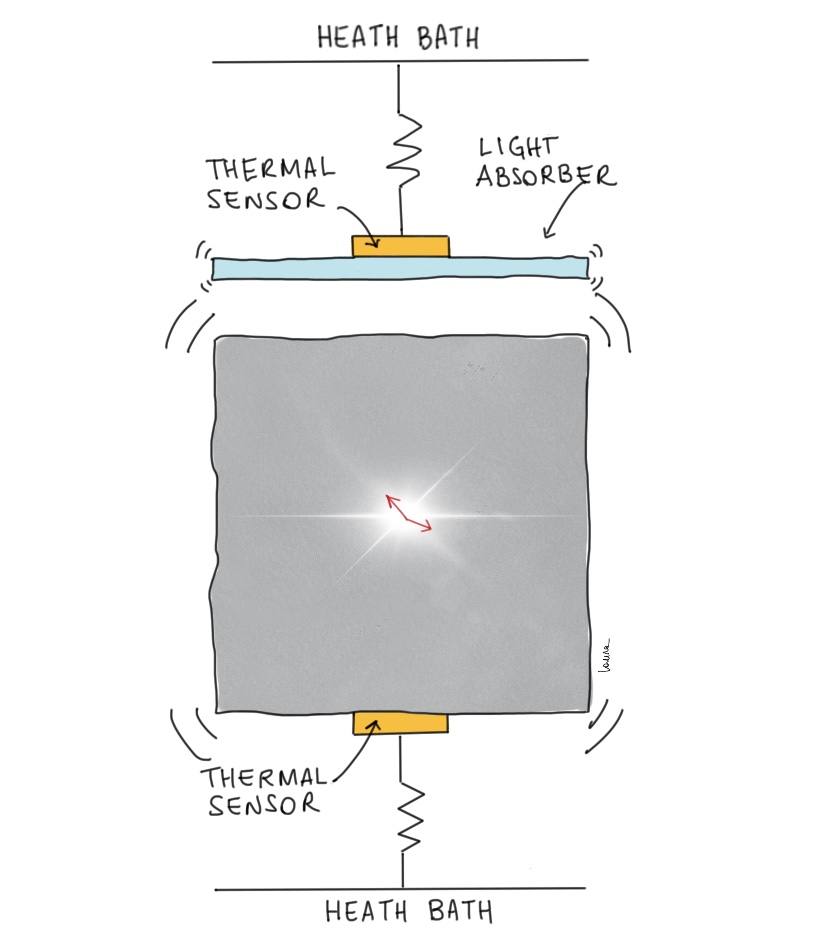}
  \caption{
  Artistic illustration of a cryogenic calorimeter and its \nubb-decay detection concept. Phonon and scintillation light signals are read-out through superconductive thermal sensors. Image courtesy of Laura Manenti.}
  \label{fig:cryo-calorimeters}
\end{figure}

  Cryogenic calorimeters, often referred to as bolometers,
  are one of the most versatile types of detectors for rare events searches.
  Their first development dates back to the 1980s and,
  since then, they have been successfully employed
  for \nubb-decay and dark-matter searches \cite{bolometer-saga}.
  Bolometers consist of crystals coupled to thermal sensors
  measuring the phonons induced by particles impinging on the crystal lattice,
  or the heat induced by phonon recombination (see Fig.~\ref{fig:cryo-calorimeters}).
Typically the crystals used in \nubb-decay experiments have masses between 0.2 and 0.8\,kg
  and are operated at 10--20\,mK. Their energy resolutions are typically in the range $\sigma =$~2--10\,keV,
  and the containment efficiency varies between 70--90\%,
  depending on the crystal type and size.

  Bolometers have an active volume fraction of 100\%,
  which makes them sensitive to background due to $\alpha$-decaying isotopes on
  their surfaces, or on the surfaces of nearby materials.
  In scintillating crystals, e.g., \ZS\ or \LMO, it is possible to discriminate
  $\alpha$ from $\beta/\gamma$ particles by their different light yields.
  The scintillation light is detected by a second bolometer
  placed in front of the crystal and consisting
  of a Ge or Si wafer instrumented with the NTD, TES or KID sensors discussed in Sec.~\ref{sec:exp:detection:eventreconstruction}.
Alternatively, surface events can be discriminated from bulk events via pulse-shape analysis
  using crystals with an Al-film coating, as is being pursued by CROSS \cite{cross}.
  In such devices, a ionizing particle interacting close enough to the coated surface
  will create quasi-particles that can be trapped in the superconductive Al layer for $O(1)$\,ms~\cite{cross}.
In all cases, the \nubb-decay tagging efficiency is $\varepsilon_{\text{mva}}\sim90\%$.

  Bolometric experiments feature high granularity,
  providing good suppression of the external \G\ backgrounds
  via the rejection of events releasing energy in multiple crystals.
The full absorption of phonons yields response times that can be as long as 0.1\,s with NTD sensors.
  Hence, the probability of having two \nnbb\ decays piling up is not negligible,
  especially when considering large crystals and isotopes with relatively short \nnbb-decay half-life.
  Techniques to mitigate this potential background are currently being developed.

  \begin{table}[tbt]
    \setlength{\tabcolsep}{4.5pt} \caption{Detection concept specific parameters for cryogenic bolometers:
      crystal molecule, total mass, fractional isotopic abundance, background per kg of total mass.
      All experiments except CUORE use a combined readout of heat and scintillation light.
      All experiments have an NTD readout, except for AMoRE-II, which uses MMCs.
      Values are taken from references \cite{CUORE:2021mvw, cupid0-final,
      Armengaud:2020luj, cross, cupid-cdr, Lee:2020rjh}.
      }
    \begin{tabular}{llcccccccc}
      \toprule
      Experiment  & Crystal  & $m_{tot}$ & $f_{iso}$  & Background     \\
                  &          & [kg]      & [\%]       & $\left[\dfrac{\text{events}}{\text{keV}\cdotp\text{kg}\cdotp\text{yr}}\right]$\\
                  
      \colrule
      \smallspace
      CUORE       & \natTeoo & 742  & 34\footnote{CUORE is using natural tellurium.} & $1.5\cdot10^{-2}$ \\
      CUPID-0     & \enrZS   & 9.65 & 96       & $3.5\cdot10^{-3}$ \\
      CUPID-Mo    & \enrLMO  & 4.16 & 97       & $4.7\cdot10^{-3}$ \\
      CROSS       & \enrLMO  & 8.96 & 98       & $1.0\cdot10^{-4}$ \\
      CUPID       & \enrLMO  & 472  & $\geq$95 & $1.0\cdot10^{-4}$ \\
      AMoRE-II    & \enrLMO  & 200  & 96       & $1.0\cdot10^{-4}$ \\
      \botrule
    \end{tabular}

\end{table}

  At present, the largest bolometric experiment is CUORE,
  operating $\sim$750\,kg of \Teoo\ crystals with natural isotopic composition
  (giving 206\,kg \Te) in a custom cryogen-free dilution refrigerator \cite{cuore-cryostat}.
  The \Teoo\ crystal detector technology is reviewed in \cite{Brofferio:2019yoc}.
  CUORE has demonstrated the feasibility of a ton-scale bolometric experiment,
  achieving an energy resolution of $\sigma=3.2$\,keV.
  With a background of $1.5\cdot 10^{-2}$\,\ctsper,
  corresponding to a sensitive background \senbkg$=9.1\cdot10^{-2}$\,\evmolyr,
  CUORE has set the most stringent constraints on \nubb\ decay of \Te: \Thl$>2.2\cdot10^{25}$\,yr at $90\%$~C.L., with an exclusion sensitivity of \Thl$=2.8\cdot10^{25}$\,yr \cite{CUORE:2021mvw}.

  In the coming years, a strong boost in sensitivity is expected with CUPID \cite{cupid-cdr},
  which will deploy enriched crystals with particle identification capabilities in the CUORE cryostat.
  Several projects have been realized to identify the optimal crystal and light detector technology.
  Among these, CUPID-0 operated abut 5\,kg of \Se\ in form of enriched \ZS\ scintillating crystals,
  demonstrating a background  of $3.5^{+1.0}_{-0.9}\cdot
  10^{-3}$\,\ctsper\ \cite{cupid0-final,cupid0-bkg}, a factor 3.3 times lower than that of CUORE.
  A limitation of \ZS\ crystals is their relatively poor energy resolution ($\sigma=8.5$\,keV)
  due to sub-optimal crystal purity.
  In parallel, CUPID-Mo\,\cite{Armengaud:2019loe} has collected data with 20 enriched \LMO\ crystals
  for a total isotope mass of 2.3\,kg. CUPID-Mo has demonstrated a resolution of $\sigma=3.2$\,keV,
  and $>$99.9\% \A\ rejection with $>$99.9\% acceptance of $\beta/\gamma$ events~\cite{Armengaud:2020luj,Armengaud:2019loe,cupidmo-dnp}.
  Finally, the CROSS collaboration is using Al-coated \LMO\ crystals
  to extensively investigate their surface/bulk discrimination capabilities,
  and is planning to deploy 32 enriched \LMO\ crystals for a total isotope mass of 4.7\,kg.
  The goal of the CROSS demonstrator is to reach a background $<10^{-4}$\,\ctsper~\cite{cross},
  which could boost the sensitiviy of bolometric experiments beyond the IO region.

  CUPID's baseline design is based on 250\,kg of \enrLMO\ instrumented with light readout
  in the CUORE cryostat.
Assuming achieved crystal quality
  and background levels, and the readout of scintillation light for particle discrimination,
CUPID projects a background of $10^{-4}$\,\ctsper, more than a factor 100 lower than CUORE.
  With a sensitive background \senbkg$=2.3\cdot10^{-4}$\,\evmolyr,CUPID will reach a $1.1\cdot10^{27}$\,yr discovery sensitivity with 10\,yr of live time.
  With additional purification of the crystal material,
  the use of light detectors instrumented with lower threshold and higher bandwidth sensors (e.g., TES),
  and the development of pulse shape discrimination techniques,
  CUPID can achieve a background of $2\cdot10^{-5}$\,\ctsper.
  Ultimately, a background level of $5\cdot10^{-6}$\,\ctsper\ is conceivable with the deployment
  of 1\,ton of isotope in a new cryostat featuring an active cryogenic shield \cite{bingo}.

  In parallel, the AMoRE collaboration has demonstrated the feasibility
  of using scintillating crystals with an MMC readout both on the phonon and photon channels for large experiments \cite{Lee:2020rjh}.
  The first phase of the experiment, AMoRE-I, is currently collecting data
  with a mix of \Mo-enriched \LMO\ and CaMoO$_4$ crystals for a total mass of $\sim$6\,kg,
  and is characterized by a background $<10^{-3}$\,\ctsper.
  The next phase, AMoRE-II, will make use of 200\,kg of \enrLMO\ crystals,
  for an isotope mass of 110\,kg.
  With a target background index of $<10^{-4}$\,\ctsper,
  corresponding to a sensitive background \senbkg$=2.2\cdot10^{-4}$\,\evmolyr,
  AMoRE-II will reach a discovery sensitivity of 6.7$\cdot10^{25}$\,yr
  with 10\,yr of data.

  \subsection{Tracking calorimeters}
  \label{sec:prj:tr}

\begin{figure*}[tb]
    \centering
    \includegraphics[width=\textwidth]{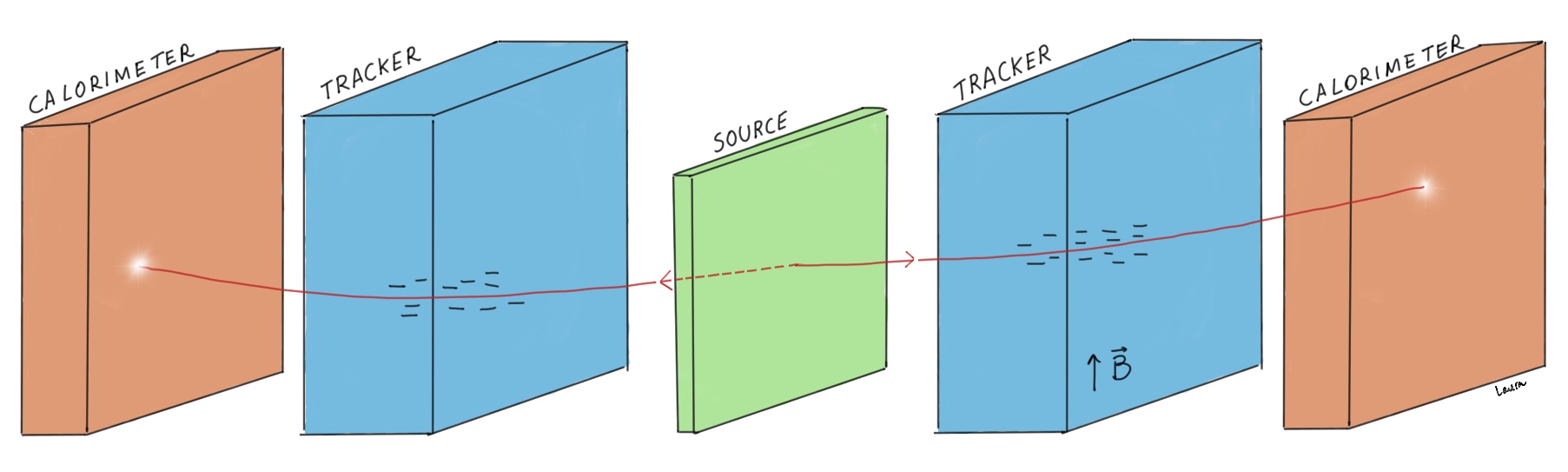}
    \caption{ Artistic illustration of a tracking-calorimeter detector and its \nubb-decay detection concept. The charge, momentum, and energy of the particles ejected by the source is measured through a combination of magnetic-field trackers and calorimeters. Image courtesy of Laura Manenti.} \label{fig:tracking-calorimeters}
  \end{figure*}

  Tracking calorimeters are the only actively-pursued detection concept in which 
  the \nubb-decay isotope material is decoupled from the detector. The
  source is in the form of a foil sandwiched by drift chambers with an applied
  magnetic field for discriminating electrons and positrons, beyond which lies
  calorimeters for measuring energy (see Fig.~\ref{fig:tracking-calorimeters}.
Due to the requirement that the foils be very thin to minimize energy losses
  prior to the electrons exiting the source,
  scaling up the isotope mass is particularly challenging for this technology.
  However, tracking calorimetry has the unique capability of precisely measuring properties of the decay kinematics
  such as single-$\beta$ energy spectra and opening angle distributions,
  which are connected to the Lorentz structure of the mechanism mediating the decay \cite{Ali07,SuperNemo10}.

  The most sensitive tracking calorimeter to date was NEMO-3, which set competitive constraints
  on a variety of target isotopes, particularly \Mo\ \cite{NEMO-3:2015jgm}.
  Its successor, SuperNEMO \cite{Piquemal:2006cd},
  builds on the same design principles and is currently in preparation. 
The SuperNEMO project is divided in two phases:
  a SuperNEMO Demonstrator (SuperNEMO-D) consisting of one module with 7\,kg of \Se,
  and a full-scale experiment consisting of multiple modules for a total \Se\ mass of 100\,kg.
  Future phases with different isotopes are still open.

  The energy resolution of a single calorimeter was $\sigma\sim$100\,keV for NEMO-3,
  and is expected to be $\sim$50\,keV for SuperNEMO thanks to 
  improved light collection and the use of PMTs
  with higher quantum efficiency.
  Some fraction of the energy emitted in a \bb\ decay event
  is inevitably released in the passive source foil:
  as a result, the \nubb-decay signature is peaked below \Qbb,
  and features a low-energy tail,
  significantly overlapping with the \nnbb-decay continuum spectrum.
  The optimal ROI strongly depends on the expected number of background events at the end of the data taking:
  it corresponds to [-1.6,1.1]\,$\sigma$ for NEMO-3,
  and to a $\sim$4\,$\sigma$ range around the degraded peak below \Qbb\ for SuperNEMO and SuperNEMO Demonstrator.

  In tracking calorimeters, the reconstructed event kinematics can be used to suppress most backgrounds,
  at the price of a lower signal efficiency.
  This was 11\% in NEMO-3, and is expected to reach 28\% in SuperNEMO
  thanks to the improved spacial resolution of the tracker.
  The most significant residual backgrounds come from $^{222}$Rn in the tracker,
  and the \B\ decays of $^{208}$Tl and $^{214}$Bi on the source foil.
  SuperNEMO aims to suppress the last two by a factor 50 and 30, respectively,
  and has partially achieved it so far \cite{Povinec:2017trz,calvez}.
  In our calculation, we use the design value for SuperNEMO of $9.8\cdot10^{-5}$\,\ctsper,
  and the experimentally measured contaminations for the
  Demonstrator \cite{calvez}, giving $\senbkg = 5.3\cdot10^{-3}$\,\evmolyr.
  Given the particular shape of the \nubb- and \nnbb-energy distributions, a spectral fit has higher sensitivity compared to a simple counting analysis. By reducing the background to 20\%, we match the sensitivity quoted by the collaboration, which corresponds to a 10-yr discovery sensitivity of $9\cdot10^{24}$ and $8\cdot10^{25}$\,yr for SuperNEMO-D and SuperNEMO, respectively.

  \subsection{Other detector concepts}
  \label{sec:prj:rd}

  Several additional projects exist that use technologies other than the ones discussed so far.
  Some technologies have already lead to proof-of-principle experiments,
  which however are not yet competitive in terms of sensitivity.
  In most cases, the projects are still at an early R\&D phase,
  and a significant effort is required to demonstrate that the underlying
  technology can be scaled to a \nubb-decay experiment
  capable of covering the inverted ordering region or beyond.
  In Tab.~\ref{tab:othertechnologies} we list a selection of such projects
  appearing in the literature,
  highlighting their isotope of choice (where defined) and key features.

  \begin{table*}[tb]
    \caption{Other detector concepts. Existing experiments are marked with a dagger.}
    \label{tab:othertechnologies}
    \begin{tabularx}{\textwidth}{l p{45pt} X}
      \toprule
      Project & Isosope(s) & Detector technology, main features, and references \\
      \colrule

      \multirow{3}{*}{CANDLES$^{\dagger}$} & 
      \multirow{3}{*}{$^{48}$Ca}           & 
      Array of scintillator crystals suspended in a volume of liquid scintillator. \\
      & & Possible operation as cryogenic calorimeters. \\
      & & \textcite{Yoshida:2009zzd,CANDLES:2020iya} \\
      \colrule
      \smallspace
      \multirow{3}{*}{COBRA$^{\dagger}$}                       &
      $^{70}$Zn,  & 
      CdZnTe semiconductor detector array.\\
      & $^{114,116}$Cd, & Room temperature; multi-isotope; high granularity. \\
      & $^{128,130}$Te & \textcite{COBRA:2015agm,Zuber:2001vm,Ebert:2015rda,Arling:2021usd}\\

      \colrule
      \smallspace
      \multirow{3}{*}{Selena}      &
      \multirow{3}{*}{$^{82}$Se}   & 
      Amorphous $^{\text{enr}}$Se high resolution, high-granularity CMOS detector array.  \\
      & & 3D track reconstruction ($O(10 \mu\text{m})$ resolution); room temperature; minimal shielding. \\
      & & \textcite{Chavarria:2016hxk}\\

      \colrule
      \smallspace
      \multirow{3}{*}{N$\nu$DEx}    & 
      \multirow{3}{*}{$^{82}$Se}    & 
      High-pressure gaseous $^{82}$SeF$_6$ ion-imaging TPC. \\
      & & $\lesssim1\%$ energy resolution; precise signal topology; possible multi-isotope. \\
      & & \textcite{Nygren:2018ewr,Mei:2020sgh}\\

      \colrule
      \smallspace
      \multirow{3}{*}{R2D2}       & 
      \multirow{3}{*}{$^{136}$Xe} &
      Spherical TPC. \\
      & & Single readout channel; inexpensive infrastructure. \\
      & & \textcite{Bouet:2020lbp}\\

      \colrule
      \smallspace
      \multirow{3}{*}{AXEL}        & 
      \multirow{3}{*}{$^{136}$Xe}  & 
      High-pressure TPC operated in proportional scintillation mode. \\
      & & High energy resolution; possible positive ion detection. \\
      & & \textcite{Obara:2019tnh}\\

      \colrule
      \smallspace
      \multirow{3}{*}{JUNO}      & 
      \multirow{3}{*}{---}       & 
      Isotope loaded liquid scintillator. \\
      & & 20 ktons of scintillator; multi-isotope; multi-purpose.\\
      & & \textcite{Zhao:2016brs,JUNO:2021vlw}\\

      \colrule
      \smallspace
      \multirow{3}{*}{NuDot}  &
      \multirow{3}{*}{---}    & 
      Liquid scintillator with quantum dots or perovskites as wavelength shifter for Cherenkov light. \\
      & & Discriminate directional backgrounds; multi-isotope. \\
      & & \textcite{Gooding:2018jok}; \textcite{Graham:2019zqb}; \textcite{Winslow:2012ey}; \textcite{Aberle:2013zza}\\

      \colrule
      \smallspace
      \multirow{3}{*}{ZICOS}     & 
      \multirow{3}{*}{$^{96}$Zr} & 
      Zr-loaded liquid scintillator. \\
      & & Topology and particle discrimination via Cherenkov light readout. \\
      & & \textcite{Fukuda:2020hah,Fukuda:2016yjg}\\

      \colrule
      \smallspace
      \multirow{3}{*}{THEIA}    & 
      \multirow{3}{*}{---}      & 
      Water-based loaded liquid scintillator with Cherenkov light readout. \\
      & & Topology and particle discrimination; multi-isotope; multi-purpose; 25 ktons of water.\\
      & & \textcite{Askins:2019oqj}\\

      \colrule
      \smallspace
      \multirow{3}{*}{LiquidO}  &
      \multirow{3}{*}{---}      & 
      Opaque isotope-loaded liquid scintillator with wavelength shifting fibers for event topology.\\
      & & Room temperature; multi-isotope; multi-purpose. \\
      & & \textcite{Buck:2019tsa,Cabrera:2019kxi}\\
      \botrule
    \end{tabularx}
  \end{table*}

 \section{Prospects and expectations}
\label{sec:dis}

In this section we bring together our expectations from the theory
and experimental landscape, and address some of the key questions
related to \nubb\ decay.  Section~\ref{sec:dis:heading} summarizes
near-term prospects and how ongoing efforts are going to shape the
field.  In Sec.~\ref{sec:dis:learn}, we discuss what we would learn
from a discovery under different assumptions on the underlying
theory framework.  Section~\ref{sec:dis:prob} addresses the question
of how likely a discovery is in the next round of experiments, also in terms of nuclear and particle theory inputs.
Sec.~\ref{sec:dis:other} reviews other discovery opportunities of
\nubb-decay experiments not related to the lepton-number violating
\nubb\ decay.  Finally, Sec.~\ref{sec:dis:shift} speculates on the
neutrino's role as a possible catalyst for the next paradigm shift
in fundamental physics, which may lead us to a new theory beyond
the Standard Model of particle physics.

This section aims at addressing in a comprehensive way the most important questions of experts and nonexperts alike. Its content is largely covered by the previous sections, but it is here presented stressing the connections between theory and experiment, as well as between particle and nuclear theory. We refer the reader to the previous sections for more information and detailed lists of references.

\subsection{Where are we heading?}
\label{sec:dis:heading}

\subsubsection{Experiment}
In the next decade, several experiments will be constructed to search
for \nubb\ decay at new uncharted half-life scales using multiple
nuclei and different technologies. Three scenarios can unfold,
depending on the half-life of the process, and 
whether the decay exists at all.

The signal half-life could be just beyond current constraints,
at a scale of $10^{25}$--$10^{26}$ years, depending on the isotope.
In this first scenario, hundreds of \nubb-decay events will be
observed in each next-generation experiment.  The half-life
will be measured with statistical uncertainty at the level
of $\sim$10\%. Systematic uncertainties
in \nubb-decay experiments are typically $\lesssim$10\%, and will not
limit the accuracy of the measurement in this 
strong signal scenario.  These first measurements will likely be followed by a second round of experiments,
not developed to maximize the discovery sensitivity, but capable
of measuring properties of the decay kinematics, such as single-$\beta$ energy
spectra and opening angle distributions connected to the Lorentz
structure of the mediating mechanism(s).

In the second scenario, the \nubb-decay half-life is above the current
limits, but still within the reach of upcoming searches 
(i.e., $\Thl \approx 10^{26}$--$10^{28}$\,years). For this case, only tens of
events or fewer would be expected.
Measurements of the half-life will hence be affected by large
statistical uncertainties at the level of 30\% for 10 events, or even 100\% for a couple of events. If the signal
is at the edge of the detection sensitivities, only some of the
experiments may observe a signal, while others would set a limit.
Such an inconclusive situation would require further discovery-style
experimental investigation to confirm the discovery claims.

It is also possible that the \nubb-decay half-life is too small to
be discovered by next-generation experiments (i.e., $\Thl >
10^{28}$\,years), or  the process does not exist at all. In this
case, the forthcoming searches will push the half-life constraints
by two order of magnitudes, excluding a significant part of the
parameter space of interest and ruling out specific models.
Further technological developments will then be needed to
realize affordable next-to-next generation experiments with scaled
sensitivity.

\subsubsection{Nuclear theory}

The extraction of beyond-Standard-Model physics information from
half-life measurements relies on NME
calculations, which currently differ from each other by about a
factor of three. NME calculations might also all be affected by 
systematic offsets.
Promising developments in ab-initio methods and chiral EFT will
reduce these uncertainties. Calculations may still disagree
due to the different approximations made, but systematic effects
(``$g_A$ quenching'', the short-range NME contribution, etc.) are expected
to be under better control.

As the decay rate depends on the NME squared, nuclear theory uncertainties will likely remain larger than
experimental statistical or systematic uncertainties on the half-life, representing the main
limitation in the interpretation of future results. 
These uncertainties will be smaller in nuclei with simpler nuclear structure
for which calculations are more robust: an ideal example is
$^{48}$Ca. Among the most interesting isotopes, NMEs would probably
be less reliable for nuclei with more complex structure,
such as $^{100}$Mo and $^{150}$Nd.

NMEs for other beyond-Standard-Model mechanisms will likely carry similar uncertainties
as for light neutrino exchange. Calculations of these matrix
elements do not pose different challenges, as even light neutrino
exchange has a short-range component. Nonetheless, a careful treatment
of short-range physics is probably more relevant in these scenarios, where \nubb\ decay is usually mediated by the exchange of heavy particles.

\subsubsection{Particle theory}

At present, we lack reliable theory predictions for the \nubb-decay rate, the origin of the matter-antimatter asymmetry, and the
neutrino mass values.  A large number of beyond-Standard-Model
theories have been proposed, but none can be tested with available
data, and might not be testable even considering next-decade
experiments.  We neither have a model for lepton number violation
nor a theory of lepton masses, and its establishment does not seem
close.  From this point of view,  \nubb-decay searches are among
the most promising sectors to guide future theory developments, and, vice versa,
the searches could benefit from theory breakthroughs.

Despite the parameter space broadness, we can identify clear milestones
for the experimental program. The Holy Grail for next-generation
experiments is to reach the bottom of the inverted ordering parameter
space, i.e., $\mbb=18.4\pm1.3$\,meV. This natural goalpost was
immediately identified after the discovery of neutrino oscillations,
boosting enormously the community's efforts.

We propose $\mbb\approx8-10$\,meV as the next target for the field.
As discussed in Sec.~\ref{sec:par:mbb:predictions},
there is an accumulation of theoretical motivation to explore \mbb\ values at
this mass scale, which corresponds to the mass scale measured in solar neutrino
oscillations ($\sqrt{\Delta m^2_{\text{\tiny sol}}} = 8.6\pm0.1$\,meV), and which
is pointed to by classes of models 
focusing on the coarse structure of the mass matrix ($m_{\beta\beta}\approx 
\sqrt{ \Delta m^2_{\mbox{\tiny atm}} } \times \theta_{\mbox{\tiny C}}\approx10$\,meV).
This scale is interesting also from the experimental point of view: it is almost
in the middle of the parameter space remaining after reaching the bottom of the
inverted ordering, and can constitute a challenging, and yet conceivable goal
for the next-to-next generation of \nubb-decay experiments. It is also the
vicinity of the minimum that would be imposed on \mbb\ by cosmological
observations if $\Sigma$ is measured to be just below its current upper
bounds.

An ultimate goal would be
to reach the floor of the normal ordering parameter space for vanishing $m_1$, 
$\mbb \sim |U_{e2}^2| \sqrt{\Delta m^2_{12}} - |U_{e3}^2| \sqrt{\Delta m^2_{32}} = 1.5$\,meV.
Barring flavor symmetries or strongly destructive interference with alternative
exchange mechanisms that would force $\mbb$ to be vanishingly small, experiments
with sensitivity to this normal-ordering floor would be virtually guaranteed
to detect \nubb\ decay if the Standard Model neutrino is 
a (dominantly) Majorana particle. Quasi-background-free kiloton experiments would be needed for this endeavor.

\subsection{What would we learn from a discovery?}
\label{sec:dis:learn}

\subsubsection{Model-independent consequences} 

Regardless of 
the mechanism mediating the decay, and of the uncertainties in the NMEs, a \nubb-decay observation would constitute
the discovery in a laboratory experiment of a process that creates matter without creating antimatter.
This ``Little Bang'' would prove that lepton number is not a conserved
quantum number, and that neutrinos can transform into antineutrinos.

The violation of lepton number is directly observable in
\nubb\ decay, as two new leptons are created without the creation of any
antiparticles.
The possibility for a neutrino to transform into an antineutrino
and vice versa would be proven indirectly. 
The \nubb\ decay operator, together with quantum fluctuations, provides a
non-zero neutrino-antineutrino transformation channel.
However, in the absence of precise theory, its size cannot be predicted.
From this point of view, an observation of \nubb\ decay guarantees only that the Majorana mass is not null. 
Although it is not favored by the best-motivated models, one cannot rule out the possibility that its value is so small that it does not have any practical consequence. 
In this case, the neutrino would phenomenologically behave as a Dirac particle, and theory inputs would still be needed to connect \nubb\ decay with the origin of neutrino masses.

\subsubsection{Model-dependent consequences}

Experiments measure the decay half-life, and  NMEs are needed to connect it with the
underlying beyond-Standard-Model mechanism.
Multiple mechanisms can  contribute to the \nubb-decay rate, which is proportional to the squared sum of amplitudes for 
all contributions. While both constructive and destructive interference are
possible, a complete cancellation between unrelated mechanisms would require
fine-tuned models.

Half-life measurements or bounds on different nuclei provide
information on the underlying mechanism. For instance,
measuring a half-life of $10^{27}$\,years for \Ge\ would
imply an expected \Mo\ half-life of \mbox{(1--3)}$\times10^{26}$\,years if the decay is dominated by the exchange of light neutrinos. Likewise, similar half-life ranges will be predicted for the decay of other isotopes, and also for \nubb\ decays to excited states.
Incompatible half-life measurements could hence prove the existence
of other mechanisms driving the decay, assuming accurate calculations on the NMEs~\cite{Deppisch06,Gehman07,Simkovic10b,Graf:2022lhj}. Indeed current NME uncertainties would severely
limit our ability to pin down the specific mechanism(s). Furthermore, these kinds of
analyses are sensitive to correlations between calculated NMEs as pointed out by \textcite{Lisi:2022nka}.

Measurements of the decay kinematics, which provide information on the Lorentz
structure of the mediating mechanism,  
could conclusively rule-out classes of models or corroborate others.
However these properties, as well as decays to excited states, are hard to measure. Experimental
efforts beyond the next decade might be needed to collect this
information if \nubb\ decay is not discovered in experiments currently underway
or starting soon.

\subsubsection{Assuming light-neutrino exchange}

If the decay is dominantly mediated by the exchange of light neutrinos,
a comparison of the measured \mbb\ with other data would provide
new insights on neutrino physics. An observation of \mbb\ below the
minimum value allowed for the inverted ordering would imply
that neutrino masses follow the normal ordering.  Vice versa, should
the inverted ordering be established by neutrino oscillation
experiments, the non-observation of \nubb\ decay in next-generation experiments
would rule out Majorana neutrino masses.

Galaxy surveys and measurements of the cosmic microwave
background will measure the value of the sum of the neutrino
masses $\Sigma$ in the next decade. Such a measurement would not
only set an indirect upper bound on \mbb, but could also provide a
lower bound. In particular, if $\Sigma$ is measured to be above 70--80\,meV,
then $\mbb$ must be larger than $\sqrt{\Delta m^2_{12}}$, suggesting
exciting discovery prospects for next-generation \nubb-decay
searches.

In the near term, direct measurements of the effective kinematic neutrino mass $m_{\beta}$ will
explore a parameter space that is already excluded by \nubb\ decay and
cosmology.  Thus a signal in those experiments, as well
as other inconsistencies among neutrino data sets, would strongly
point towards new physics beyond the 3-flavor neutrino oscillation  and
$\Lambda$CDM paradigm.

A measurement of \mbb\ is currently the only conceivable way
to obtain information on the values of the Majorana phases in the PMNS matrix, through a global analysis
with oscillation measurements. However, only one relative phase can be measured.
In addition, constraints on this relative phase
can be extracted only if the experimental and nuclear
theory uncertainties are strongly reduced below their
current levels.

\subsection{What are the odds of a discovery?} \label{sec:dis:prob}

\subsubsection{Model-independent considerations}

\begin{figure*}[t]
  \centering
  \includegraphics[width=\textwidth]{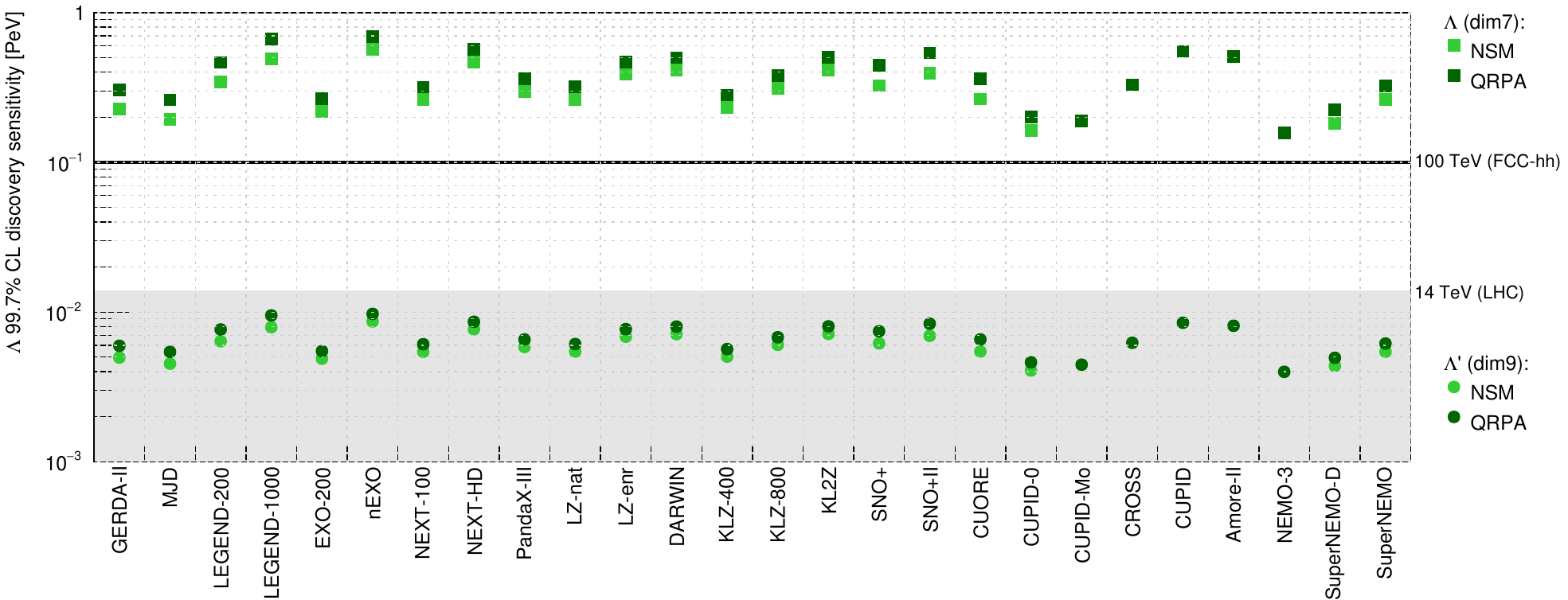}
  \caption{Discovery sensitivities of current- and next-generation \nubb-decay
  experiments for exchange mechanisms dominated by effective operators of
  dimension 7 and 9.  
  Values of $\Lambda$ smaller than the marked values are tested at higher CL. 
  The grey band corresponds roughly to the reach
  of modern accelerator experiments. The black line indicates an ambitious goal
  for future circular colliders such as the FCC-hh \cite{Golling:2016gvc}.}
  \label{fig:dis:constraints} 
\end{figure*}

A wide variety of particle theory
models predict \nubb\ decay. In most of them, unconstrained model
parameters prevent a precise prediction of the decay rate. At best, these models provide a lower limit on the half-life,
which sets a target for the experiments.  The master formula
in Eq.~\ref{eq:master} connects the
half-life to effective operators representing classes of models.
In general, the half-life is proportional to the energy scale of
the physics responsible for the decay, taken to some power which depends
on the dimension of the operator.  Operators above dimension 5
typically correspond to energy scales close to or beyond those explored by
accelerator experiments.  Figure~\ref{fig:dis:constraints}
shows that accelerator and \nubb-decay experiments are complementary,
and highlights how the reach of \nubb-decay searches can even exceed that of
accelerators for mechanisms other than light neutrino exchange. Note that, for reference, Fig.~\ref{fig:dis:constraints} just shows estimates for the most favored dimension-7 and dimension-9 operators. In general there can be order of magnitude differences for other operators suppressed e.g. by ratios of the nuclear-over-electroweak or chiral-over-nuclear scales.

We also indicate in Fig.~\ref{fig:dis:constraints} the energy scale of
100\,TeV. This is a round value, suggestive of a possible ambitious target for
next-generation colliders, but also a scale at which new flavor- and beyond-the-Standard-Model physics could
manifest.
Due to the large variety of possible decay mechanisms, one can
consider \nubb\ decay as a generic search for new physics,
similar to accelerator ones, where the decay half-life
plays the role of the collision energy. Increasing the half-life
sensitivity implies exploring uncharted parameter space, where
a discovery can happen at any time.

\subsubsection{Assuming light-neutrino exchange} 

The \nubb-decay mechanism requiring the least new physics
is light neutrino exchange, which only needs the Standard Model
neutrino to be a massive Majorana particle. From a general point of view this is
a particularly important mechanism, as it
is the only one driven by a dimension-5 operator, i.e., the Weinberg operator. 
Further, it is uniquely connected to neutrino masses
and is the dominant decay contribution in many models.  In
this scenario, the decay rate depends on the effective
Majorana mass \mbb, which is a function of the neutrino oscillation
parameters, Majorana phases, the lightest neutrino mass eigenstate,
and the neutrino mass ordering.

The oscillation parameters have been measured precisely,
nonetheless we have no information on the Majorana phases,
and the mass ordering has not been determined.
Although global fits show a preference for normal-ordered
masses,
we need to wait for the next
decade experiments --- i.e., JUNO, DUNE, and HyperKamiokande --- with the requisite
sensitivity to establish the neutrino mass ordering. These unknowns lead to
uncertainties on the \mbb\ value.

\begin{figure*}
  \includegraphics[width=\textwidth]{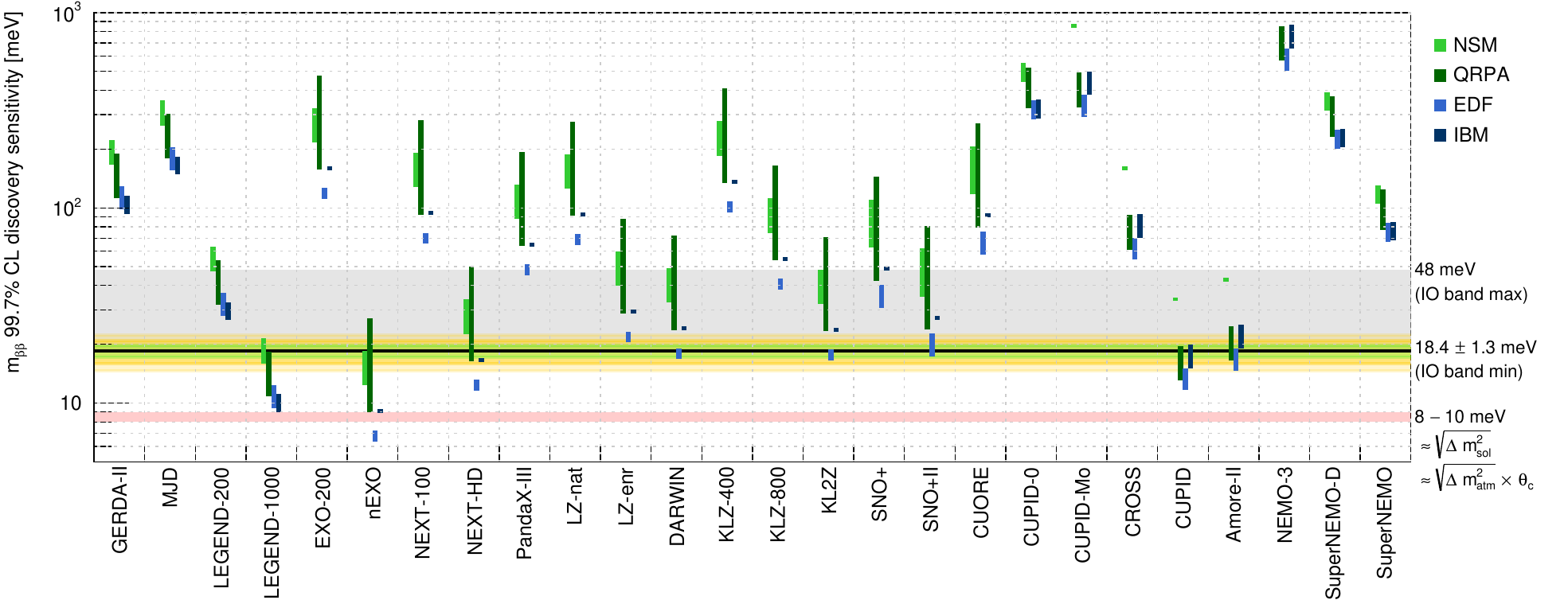}
  \caption{Discovery sensitivities of current- and next-generation \nubb-decay
  experiments for exchange dominated by effective operators of dimension 5, i.e.,
  the light neutrino exchange.  
  Values of \mbb\ larger than the marked values are tested at higher CL. 
  The grey band indicates the range of \mbb\
  values for inverted-ordered neutrino masses and vanishing values of the lightest neutrino mass.
  The minimum value of \mbb\ for
  the IO and its 1$\sigma$, 2$\sigma$, and 3$\sigma$ uncertainty bands are indicated by the
  black, green, orange, and yellow bands, respectively. The red band between
  8-10\,meV indicates a future goal for \nubb\-decay experiments motivated by
  theoretical and experimental considerations (see discussion in Sec.~\ref{sec:par:mbb:predictions}).}
  \label{fig:dis:constraints_mbb} 
\end{figure*}

Assuming that neutrino masses follow the normal-ordering, any
half-life value beyond the current upper limits is allowed. However,
for the inverted-ordering case, the half-life
has a lower bound corresponding to $\mbb=18.4\pm1.3$\,meV. Figure~\ref{fig:dis:constraints_mbb} shows the \mbb\
sensitivity of future \nubb-decay experiments.
The proposed experimental endeavor will fully test the inverted
ordering parameter space, guaranteeing a discovery if this is the
true scenario, and offering exciting discovery opportunities
also assuming the normal ordering. In fact, since the current best bounds on \mbb\ are
$\sim$160--180\,meV --- assuming the least favorable NMEs --- reaching 18.4\,meV means probing 80\%--90\%
of the currently allowed range for the normal ordering. 

It should be mentioned that the parameter space for \mbb\ might not
be equiprobable.  A theoretical prejudice for normal-ordered masses and
vanishing $m_1$ would prefer smaller values of \mbb, for example.
New symmetries predicting specific values for
the Majorana phases or the existence of new particles such as sterile
neutrinos could favor other corners of the parameter space, or even reduce
or open it. In addition,
Bayesian analyses assuming flat priors on the Majorana phases and a log-flat
prior on $\Sigma$ favor
\mbb\ values close to the current constraints, providing exciting
prospects for the field regardless of the mass ordering.

\subsubsection{Impact of nuclear physics}

How likely a discovery is in the next decade strongly
depends on systematic uncertainties on NME
calculations.  A broad effort to reduce uncertainties is ongoing
within the nuclear theory community.

Ab initio approaches offer a promising avenue: by incorporating wider nuclear
correlations and currents, measured $\beta$-decay rates can now be reproduced
without the ``quenching'' required by previous studies --- an ad hoc reduction of
calculated matrix elements.  The first ab initio matrix elements for \nubb-decay
nuclei, supported by studies in lighter systems, indicate a mild suppression by
tens of percent with respect to the lower values in Tab.~\ref{tab:NMEs}. This
suggests that current \nubb-decay rate predictions may have to be reduced, but
only moderately. Efforts are underway to improve the quality of the results, to
include missing momentum-dependent operators --- a key difference between $\beta$
and \nubb\ decay --- and to extend them to heavier nuclei.

The recently recognized short-range term can contribute significantly to the NME.
A first ab-initio study in $^{48}$Ca suggests that including this physics
increases the NME by about 40\% percent.
A similar enhancement has been found in heavier \nubb-decay nuclei with the NSM
and with QRPA.
Lattice QCD studies are underway to test whether this claimed enhancement is robust.
If so, the impact of the new term may balance the longer half-life values
anticipated due to the inclusion of the ``quenching'' physics.

Even if these systematic contributions to NMEs were fully resolved,
discrepancies remain between results obtained with different many-body methods.
Tests against nuclear structure data can gauge the quality of each calculation.
In addition, novel measurements of nuclear observables correlated with
\nubb-decay NMEs such as second-order Gamow-Teller or electromagnetic
transitions can provide insights on each method's strengths and weaknesses.

\subsection{What else can be discovered by \nubb-decay experiments?}
\label{sec:dis:other}

The unprecedented combination of ultra-low background, high-exposure, high
energy resolution, and multivariate analysis capabilities in modern
\nubb-decay experiments offers exciting discovery opportunities
beyond the primary target of observing \nubb\ decay. This includes searches not only for
other L-violating processes, such as neutrinoless electron
capture \cite{Blaum:2020ogl} or neutrinoless quadruple-beta
decay \cite{Guzowski:2018neg}, but also for completely decoupled physics.

The existence of new particles and fields, the violation of fundamental
principles, and non-standard interactions can each affect, in a
characteristic way, the distribution of the summed energy of the
electrons emitted in \bb\ decays.  Historically, searches for new
particles focused on massive and massless bosons called Majorons,
the Goldstone bosons that arise from the spontaneous breakdown of the
global $B-L$ symmetry. 
Searches for the violation of fundamental
principles have focused on Lorentz invariance, the Pauli Exclusion Principle,
and CPT symmetry.  We refer to \textcite{Bossio} for a
comprehensive review of this topic. Future searches will have high-sensitivity to additional physics, for instance exotic currents \cite{Deppisch:2020mxv}, and
light exotic fermions such as sterile neutrinos or $Z_2$-odd
fermions \cite{Bolton:2020ncv,Agostini:2020cpz}.

In addition to distortions on the energy distribution, next-generation \nubb-decay experiments will be highly sensitive to
numerous beyond-Standard-Model processes which could generate events with
well defined energy depositions and/or time
correlations. These searches include
B-violating tri-nucleon decay \cite{Majorana:2018pdo,EXO-200:2017hwz}
and charge-violating electron decay \cite{Majorana:2016hop}.
Dark matter candidates such as
WIMPS \cite{GERDA:2020emj,NEMO-3:2020mcq,Liu:2019kzq,Majorana:2016hop} and
axions \cite{Xu:2016tap,Majorana:2016hop} can also be identified
through an excess of events with a well-defined energy distribution
or time-modulation. New searches have been proposed
for inelastic boosted dark matter \cite{COSINE-100:2018ged} and
fermionic dark matter \cite{Dror:2019dib}, and constraints have already been placed fractional-charge lightly ionizing particles \cite{Majorana:2018gib}.

\subsection{What will be the next paradigm shift?}
\label{sec:dis:shift}

For half a century, the Standard Model of particle physics has
been the field's paradigm. The discovery of the Higgs boson, immediately recognized by the 2013 Nobel Prize in physics, was its crowning achievement. At the same time, we have known for almost two decades that this model is incomplete and needs to be extended, at least to incorporate massive neutrinos.

Extensions inspired by the very same symmetry principles that underlie the Standard Model have been explored in the framework of gauge theories, which
include the so-called ``Grand Unification'' models. These
theories have however not been confirmed in spite of the extensive experimental efforts to observe proton decay in the 1980s and 1990s.
Some intrinsic features of the Standard Model, such as
CP-symmetry in strong interactions, or the nature of radiative
corrections in the Higgs sector, have in turn suggested the possible
existence of new particles, e.g., axions or supersymmetric particles. Searches for these new particles have also been unsuccessful so far.

In the mean time, cosmological observations have led to the development of a
Standard Model of cosmology, $\Lambda$CDM. Its very name invokes the
existence of two forms of matter that cannot be found in the Standard Model of
particle physics: dark matter and dark energy.  Furthermore,
theoretical cosmology has proved unable to account for the cosmic baryon
excess.

Finally, several experimental anomalies have
emerged, the most recent of which is the measurement of the anomalous magnetic
moment of the muon \cite{Muong-2:2021ojo}. These anomalies could also point to some missing
piece of the Standard Model.

Nonetheless, the only unequivocal manifestation of
physics beyond the Standard Model supported by laboratory experiments is the evidence of neutrino oscillation, 
recognized by the 2015 Nobel Prize in
physics as a proof that neutrinos are massive. This suggests
that the importance of further studies on the neutrino mass should
not be underestimated. The most promising theoretical option is
that the mass type is exactly the one proposed by Majorana.
Its experimental demonstration is a concrete and well-defined goal
to strive for in the exploration of physics beyond the Standard Model.

The best way to probe the Majorana nature of neutrinos is to measure the rate of
neutrinoless double beta decay --- i.e., the rate at which electron
pairs are created in certain nuclear decays --- an observation which would
lead to a profound change in our understanding of matter. Although we
do not have an established theory that can guide us safely in the
next steps, we are just starting a pioneering exploration of the next two, uncharted orders of magnitude, an exciting journey that will bring us a step closer to unlock the secrets of the universe.

\begin{acknowledgements}

We are extremely grateful to 
Frank Deppisch, 
Steve Elliott, 
Jon Engel,
Juan Jose Gómez Cadenas,
Andrea Pocar, 
and the three anonymous referees of RMP
for invaluable comments on our manuscript, and to 
our artist, Laura Manenti, for preparing for us such beautiful and communicative figures

We would also like to thank 
Laura Baudis, 
Chamkaur Ghag, 
Giorgio Gratta,
Josh Klein,
Fabian Kuger,
and
Pia Loaiza,
for their help during the preparation of the manuscript.
Finally, we thank
Chiara Brofferio, 
Vincenzo Cirigliano,
Stefano Dell’Oro, 
Francesco Iachello, 
Aldo Ianni,
Eligio Lisi, 
Stefano Pirro,  
Matteo Viel,
and
Christoph Wiesinger
for valuable scientific discussions.

This work has been supported by the Science and Technology Facilities Council, part of U.K. Research and Innovation (Grant No. ST/T004169/1), by the UCL Cosmoparticle Initiative, by the EU Horizon2020 research and innovation program under the Marie Sk\l{}odowska-Curie Grant Agreement No. 754496, by the ``Ram\'on y Cajal'' program with grant RYC-2017-22781, and grants CEX2019-000918-M and PID2020-118758GB-I00 funded by MCIN/AEI/10.13039/501100011033 and, as appropriate, by ``ESF Investing in your future'',  the Italian Research Grant Number 2017W4HA7S ``NAT-NET: Neutrino and Astroparticle Theory Network'' under the program PRIN 2017 funded by MIUR, and by the U.S.~DOE Office of Nuclear Physics under Grant Number DE-FG02-97ER41020.

\end{acknowledgements}
\input{main_expanded.bbl}

\end{document}

%% file: main_expanded.bbl
%